%% file: chronos.tex
\documentclass[12pt, twoside, a4paper]{report}

\usepackage[margin=1in]{geometry}
\usepackage{comment}
\usepackage[utf8]{inputenc} 
\usepackage[T1]{fontenc} 
\usepackage{tabularx}
\usepackage{array}
\usepackage{graphicx}
\graphicspath{{Figures/}{./}} 
\usepackage{palatino} 
\usepackage[numbers,sort&compress]{natbib}
\usepackage{tabularx}
\usepackage{subfigure}
\usepackage{comment}
\usepackage{multirow}
\usepackage{verbatim}
\usepackage{amsmath} 

\usepackage{hyperref}
\usepackage{url}
\usepackage{bm}

\usepackage[autostyle=false,style=english]{csquotes}

\newcolumntype{Y}{>{\centering\arraybackslash}X}

\input{packages}
\usepackage{pdflscape}
\hypersetup{colorlinks=true,linkcolor=black,citecolor=blue!35!black,urlcolor=blue!35!black,pdftitle={CHRONOS Science Program},pdfauthor={CHRONOS Collaboration},pdfsubject={White Paper Ver.2.1, RUP-26-2}}

\newcites{personal}{List of Publications}

\begin{document}
\raggedbottom

\input{FrontMatter/cover}

\pagenumbering{roman}


\clearpage

\phantomsection
\addcontentsline{toc}{part}{Abstract}
\include{FrontMatter/abstract}

\phantomsection
\addcontentsline{toc}{part}{Acknowledgements}
\include{FrontMatter/acknowledgements}


\fancyhf{}
\pagestyle{fancy}
\setlength{\headheight}{15pt}

\renewcommand{\chaptermark}[1]{\markboth{#1}{}}
\fancyhead[L]{\small\nouppercase\leftmark}
\fancyfoot[C]{\thepage}



\tableofcontents
\clearpage



\pagenumbering{arabic}
\fancyhead[R]{}

\include{Chapters/Overview}
\include{Chapters/Collaboration}
\include{Chapters/Sensitivity}
\include{Chapters/Science_roadmap}
\include{Chapters/Quantum}
\include{Chapters/Dark_matter}
\include{Chapters/Earthquake}
\include{Chapters/Yukawa}
\include{Chapters/GW_Science}
\include{Chapters/Cosmology}
\include{Chapters/Interferometer}
\include{Chapters/Directional_Response}
\include{Chapters/Input_Optics}
\include{Chapters/Output_Optics}
\include{Chapters/IFO_control}
\include{Chapters/VIS}
\include{Chapters/Substrate}

\include{Chapters/Coating}
\include{Chapters/Cryogenic}
\include{Chapters/Photon_Pressure_Actuator}
\include{Chapters/GCAL}
\include{Chapters/Prototype}
\include{Chapters/Analysis}
\include{Chapters/Calibration}
\include{Chapters/Summary}


\clearpage

\bibliographystyle{unsrt}
\begingroup
\setlength{\emergencystretch}{1em}
\bibliography{chronos}
\endgroup




\end{document}

%% file: packages.tex
\usepackage{xcolor}
\usepackage{graphicx}
\graphicspath{{Figures/}{./}} 
\usepackage{array}
\usepackage{tabularx}
\usepackage{booktabs}
\usepackage{enumerate}
\usepackage{siunitx}

\usepackage[resetlabels]{multibib} 

\usepackage[T1]{fontenc}
\usepackage{lmodern}
\usepackage{palatino} 
\usepackage{microtype} 
\usepackage{hyperref} 

\usepackage{parskip}
\usepackage{fancyhdr}
\usepackage[explicit]{titlesec}

\titleformat{\chapter}[display]
  {\normalfont\centering}
  {\chaptername\ \thechapter}
  {1ex}
  {\Huge #1}

\usepackage{ntheorem}
\usepackage{theorem}
\usepackage[acronym]{glossaries}

\usepackage{tikz}
\usepackage{epigraph}
\usepackage{pifont}       
\usepackage{fontawesome}

\usepackage{lipsum} 
\usepackage{fancyhdr}

%% file: FrontMatter/cover.tex
\thispagestyle{empty}
\begin{tikzpicture}[remember picture,overlay]
\node[anchor=north east,inner sep=0,font=\normalfont\normalsize]
  at ([xshift=-25.4mm,yshift=-20mm]current page.north east) {RUP-26-2};
\end{tikzpicture}

\begin{center}
\vspace*{1.0in}
White Paper\\[0.01in]
\textit{Ver.2.1}\\
\textit{Last Update: 2026 Oct. 14}\\[0.3in]

\noindent\rule{0.85\textwidth}{1.0pt}\\[0.15in]

{\Huge \bfseries CHRONOS Science Program}\\[0.1in]

\noindent\rule{0.85\textwidth}{1.0pt}\\[0.35in]
{Yuki Inoue, Masashi Hazumi, Hsiang-Yu Huang, Vivek Kumar, Ippei Obata, Mario Juvenal S Onglao III, Daiki Tanabe, Ta-Chun Yu}\\[0.1in]

on behalf of CHRONOS collaboration\\[3.5in]

\includegraphics[scale=0.3]{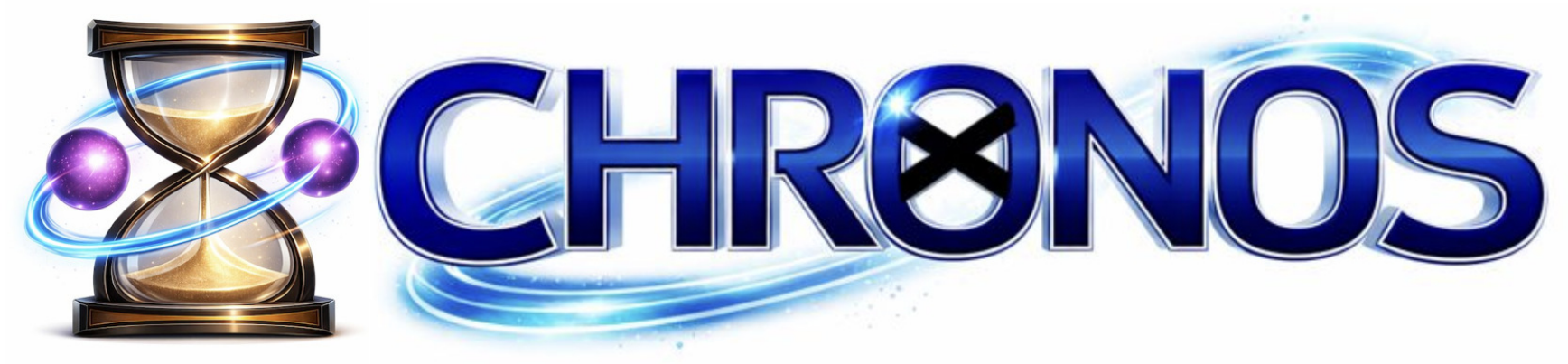}

\end{center}

%% file: FrontMatter/abstract.tex
\chapter*{Abstract}

CHRONOS (Cryogenic sub-Hz cROss torsion-bar detector with quantum
NOn-demolition Speed meter) is a proposed cryogenic torsion-bar interferometer
for precision measurements in the 0.1--10~Hz band. A Sagnac speed-meter
readout, power and signal recycling, and balanced homodyne detection provide
control of the optical response and quantum measurement correlations.
This White Paper describes the science program, detector design, calibration,
and staged experimental development.

The reference sensitivity uses a multi-input quantum model, including
rear-mirror vacuum fields and measurement--back-action correlations, combined
with classical noise. Two operating points are optimized for complementary
power-law-integrated sensitivity objectives. Gravitational-wave forecasts
use this common noise budget and state their source and observing assumptions.
The program includes intermediate-mass black-hole binaries, stochastic
backgrounds, quantum measurement, and terrestrial gravity perturbations.
Dark-matter applications are discussed as response and feasibility studies.
A differential gravitational calibrator provides a proposed Yukawa-gravity
test whose systematic floor is improved by subtracting auxiliary harmonics.
These are conditional projections: experimental reach requires validation
of the noise, transfer functions, environmental backgrounds, and systematic
uncertainties described here.

%% file: FrontMatter/acknowledgements.tex
\chapter*{Acknowledgments}
We thank S.Takano, M.Ando and T.Namikawa for providing the TOBA data and CMB data. We also thank K.W.Ng for academic advice during the preparation of this manuscript. We thank M.Hasegawa, T.Kanayama, S.Matsushita, H.Murakami, M.Inoue, and R.Shibuya for support to establish CHRONOS team. Y.I. acknowledges support from NSTC, CHiP and Academia Sinica in Taiwan under Grant No.114-2112-M-008-006-, and No.AS- TP-112-M01.

%% file: Chapters/Overview.tex
\chapter{Overview of CHRONOS}
\label{chap:overview}

\section{Scientific Background and Objectives}

The direct detection of gravitational waves from compact binary mergers by Advanced LIGO and Advanced Virgo
\cite{Abbott2016Discovery,LIGO2015}
opened a new observational window in astronomy and fundamental physics. KAGRA subsequently extended the global detector network into cryogenic interferometry
\cite{kagra_history}.
Current ground-based detectors primarily operate above $\sim 10~\mathrm{Hz}$, while the future space-based observatory LISA will explore the millihertz band
\cite{amaroseoane2017lisa}.
The intermediate sub-Hz regime, approximately $0.1$--$10~\mathrm{Hz}$, remains one of the least accessible frequency bands for precision measurement because seismic motion, Newtonian noise, suspension dynamics, and thermal noise become increasingly severe
\cite{Harms2015}.

This frequency range is nevertheless scientifically compelling. Macroscopic mechanical systems provide a platform for studying quantum measurement and measurement back action; controlled gravitational sources can probe weak gravitational interactions and possible deviations from Newtonian gravity; and sufficiently sensitive instruments could ultimately access gravitational-wave sources beyond the reach of present terrestrial observatories. In gravitational-wave astronomy, the sub-Hz band is particularly relevant to intermediate-mass black-hole binaries, the early inspiral of stellar-mass compact binaries, and stochastic gravitational-wave backgrounds, and may provide an important link between terrestrial and space-based observations
\cite{sesana2016multiband}.

CHRONOS, the Cryogenic sub-Hz cROss torsion-bar detector with quantum NOn-demolition Speed-meter,
is being developed to address this common experimental challenge
\cite{Inoue_CHRONOS_Main_2026}.
Rather than optimizing an instrument for a single science target, CHRONOS is conceived as a cryogenic torsion-bar interferometric platform for precision measurements in the sub-Hz band. Its architecture combines freely suspended torsion-bar test masses, Sagnac interferometric readout, vibration isolation, cryogenic operation, and calibrated force and torque actuation within a common experimental framework.

The same detector is intended to support three closely connected scientific directions:
\textbf{quantum measurement, precision gravity, and gravitational-wave science}.
These programs share a common requirement: the ability to measure extremely small forces, torques, and motions of macroscopic test masses over long timescales.

CHRONOS should therefore not be regarded simply as a lower-frequency version of an existing gravitational-wave detector. Its broader aim is to establish the experimental techniques required for precision interferometry with macroscopic mechanical systems in a frequency regime where conventional terrestrial measurements become increasingly difficult. Quantum measurement, precision gravity, and gravitational-wave detection are treated as successive and complementary applications of a common experimental platform.

\section{Measurement Principle}

The CHRONOS detector is based on the torsion-bar antenna concept
\cite{Ando2010,McManus_2016,Inoue_CHRONOS_Main_2026}.
Unlike conventional Michelson-type gravitational-wave interferometers, which primarily measure translational displacement of suspended test masses, a torsion-bar detector measures angular motion around a torsional degree of freedom.

Several terrestrial experiments have explored this concept in the sub-Hz regime. The torsion-bar antenna (TOBA)
\cite{Ando2010}
and the Torsion Pendulum Dual Oscillator (TorPeDO)
\cite{McManus_2016}
have demonstrated key elements of torsion-based low-frequency sensing. By exploiting a torsional degree of freedom, the mechanical resonance can be placed at very low frequency, providing a natural route toward sensitivity in the sub-Hz band.

The torsion-bar architecture is central to the CHRONOS strategy. Torsional motion provides a low-frequency mechanical degree of freedom that is sensitive to weak torques and can be monitored continuously using an optical interferometer. The test masses are supported by low-loss suspensions and are ultimately operated under cryogenic conditions to suppress thermal noise.

A defining feature of CHRONOS is the use of a Sagnac speed-meter interferometer
\cite{Chen2003,DanilishinKhalili2012,Inoue_CHRONOS_Main_2026}.
In a conventional position measurement, quantum radiation-pressure back action becomes increasingly important at low frequencies. A speed-meter interferometer instead measures a quantity related to the test-mass velocity by sampling its motion at different times. This modifies the low-frequency quantum back-action response and provides a route toward quantum nondemolition measurement
\cite{Chen2003,DanilishinKhalili2012}.

The same suspended torsion bars can also be used as probes of controlled gravitational fields. Known time-dependent gravitational torques can be applied using dedicated source masses, allowing the measured response to be compared directly with gravitational models. At later stages, the same detector architecture can be extended toward measurements of tidal motion induced by gravitational waves.

\section{Science-driven Development}

The development of CHRONOS follows a staged science roadmap. Rather than constructing the final detector in a single step, the project progressively establishes the measurement techniques required for increasingly demanding science goals. Each stage is intended both to produce scientific results and to validate technologies required for the subsequent stages.

The initial program focuses on precision measurements of torsional motion and on characterizing the detector response to controlled actuation. A central objective is the demonstration and characterization of the velocity-type response of the Sagnac interferometer. This provides the experimental basis for studying measurement imprecision, radiation-pressure back action, and their correlations, and for developing quantum nondemolition measurement techniques in a macroscopic mechanical system
\cite{Chen2003,DanilishinKhalili2012,Inoue_CHRONOS_Main_2026}.

The next stage extends CHRONOS toward precision gravity. Controlled gravitational sources can generate predictable periodic forces and torques on the suspended test masses, allowing direct comparison between the measured response and gravitational models. Rotating gravitational field calibrators provide both a calibration reference and a means to search for deviations from Newtonian gravity. In particular, CHRONOS targets Yukawa-type corrections to gravity at meter-scale interaction lengths.

In the longer term, these techniques will be integrated toward sub-Hz gravitational-wave observations. This requires low-frequency vibration isolation, cryogenic suspension, stable long-duration control, high-sensitivity interferometric readout, and reliable calibration to operate as a single system. CHRONOS approaches this goal incrementally, using quantum-measurement and precision-gravity experiments to establish the technologies required for gravitational-wave detection in the sub-Hz band.

Such an instrument would complement both terrestrial interferometers and future space-based detectors such as LISA
\cite{amaroseoane2017lisa}.
In particular, access to the intermediate frequency range may enable long-duration tracking of compact binaries and contribute to multiband gravitational-wave astronomy
\cite{sesana2016multiband}.

\section{Detector Concept}

The detector design is driven directly by these science requirements. The interferometer, mechanical system, control, calibration, and analysis are therefore treated as components of a single measurement architecture rather than as independent subsystems.

The optical system is centered on a Sagnac interferometer. Counter-propagating beams sample the test-mass motion at different times, producing the characteristic speed-meter response
\cite{Chen2003}.
Power recycling and signal recycling are incorporated to enhance the optical response, following concepts developed for advanced interferometric gravitational-wave detectors
\cite{meers1988,mizuno1993,Inoue_CHRONOS_Main_2026}.
The input optics provide stable laser light together with the modulation and spatial-mode conditioning required for interferometer sensing and control.

The optical configuration must simultaneously provide high sensitivity to the relevant torsional degrees of freedom and sufficient control authority for stable operation. Its design is therefore optimized together with the recycling cavities, readout configuration, and interferometer control system
\cite{Inoue_CHRONOS_Main_2026}.

Environmental isolation is equally critical. Ground translation and rotation can couple into torsional motion through suspension asymmetries and cross-couplings among mechanical degrees of freedom. CHRONOS therefore uses a hierarchical vibration-isolation architecture combining room-temperature seismic isolation, multistage mechanical filtering, and cryogenic suspension stages.

Isolation and control must be designed together. Soft mechanical degrees of freedom are desirable for low-frequency isolation, whereas lock acquisition, alignment, damping, and recovery from large disturbances require sufficient active-control authority. The suspension and interferometer control systems must therefore suppress unwanted motion while minimizing the coupling of control noise into the science observables.

Cryogenic operation is introduced primarily to reduce thermal noise and provide a stable environment for long-duration precision measurements. The test-mass substrates, suspension materials, thermal links, and cooling architecture affect mechanical loss, resonant behavior, optical performance, actuation, and control. These elements must therefore be optimized as part of the integrated detector design.

\section{Noise and Sensitivity}

Sensitivity in the sub-Hz regime is determined by several classes of noise, including quantum noise, suspension and substrate thermal noise, coating thermal noise, seismic noise, and Newtonian noise. Newtonian noise is particularly severe at low frequencies because fluctuations in the surrounding mass distribution produce gravitational forces that cannot be removed by mechanical isolation alone
\cite{Harms2015}.

The Sagnac speed-meter topology is intended to improve the low-frequency quantum-noise response by suppressing radiation-pressure back action relative to a conventional position-meter configuration
\cite{Chen2003,DanilishinKhalili2012}.
Cryogenic operation reduces thermal noise, while the vibration-isolation and suspension systems are designed to suppress direct seismic coupling.

The baseline CHRONOS design targets a strain sensitivity of approximately
\begin{equation}
	\sqrt{S_h^{\rm tot}}\simeq8.31\times10^{-19}~\mathrm{Hz}^{-1/2}
\end{equation}
at $2~\mathrm{Hz}$ for the minimum-PI reference point. The alternative
logarithmic-mean point gives $7.92\times10^{-19}~\mathrm{Hz}^{-1/2}$ there.
The common parameters, full noise curves, complex transfer functions, and
optimization assumptions are specified in Chap.~\ref{chap:sensitivity}
\cite{InoueQuantumTorque2026}.

\section{Actuation, Calibration, and Reconstruction}

Controlled force and torque injection is a core element of CHRONOS. In addition to conventional suspension actuators, photon-pressure actuators provide contact-free excitation of the test masses through radiation pressure. They can be used to measure mechanical and optical transfer functions and to characterize the detector response without direct mechanical contact.

The Gravitational Field Calibrator provides an independent calibration reference based directly on Newtonian gravity. A controlled moving mass distribution generates a predictable periodic gravitational torque on the test masses. Comparing this excitation with the measured interferometric response provides a calibration method that is independent of conventional actuator models. The same system also enables direct searches for non-Newtonian gravitational interactions.

The final science observables are not obtained from a single photodetector output. CHRONOS is a closed-loop, multi-degree-of-freedom dynamical system comprising the interferometer, suspensions, sensors, actuators, and digital control system. The physical motion of interest must therefore be reconstructed from multiple sensing and control channels.

The analysis framework must determine the relevant mechanical and optical transfer functions, quantify cross-couplings, model the closed-loop detector response, and propagate calibration uncertainties to the reconstructed physical observables. Analysis is therefore an integral part of the detector design rather than a separate post-processing stage.

\section{Scope of this White Paper}

This White Paper presents the baseline framework for the CHRONOS science program and detector design.

It first introduces the CHRONOS Collaboration and the Science Roadmap that guides the development of the project. The principal science goals are then described, with particular emphasis on quantum measurement and precision-gravity experiments, including searches for Yukawa-type deviations from Newtonian gravity.

The subsequent chapters describe the detector systems required to realize these goals: the Sagnac interferometer, input optics, interferometer sensing and control, vibration isolation and suspension, test-mass substrates, cryogenic system, photon-pressure actuation, and gravitational calibration. The analysis framework then describes how signals from these subsystems are combined to reconstruct calibrated physical observables.

The design presented here should not be viewed as a collection of independent subsystem developments. In CHRONOS, the science goals define the observables; the detector architecture determines how they are measured; actuation and calibration establish their physical scale; and analysis reconstructs them from the response of the controlled detector system as a whole.

Within this integrated framework, CHRONOS is intended to evolve from a prototype-scale precision torsion experiment into a platform for quantum measurement, precision gravity, and gravitational-wave science. Its broader objective is to establish the experimental techniques required for precision measurement in the sub-Hz regime and to open this largely unexplored frequency range as a new observational window for fundamental physics.

%% file: Chapters/Collaboration.tex
\chapter{CHRONOS Collaboration}
\label{chap:chronos_collaboration}

\section{Overview of the CHRONOS Collaboration}
\label{sec:chronos_collaboration_overview}

CHRONOS is an international scientific collaboration involving multiple
universities and research institutes and is organized to support a long-term
research program in sub-Hz gravitational physics, quantum measurement, and
fundamental physics.

The collaboration brings together researchers from different countries,
institutions, scientific disciplines, cultural backgrounds, and career stages.
In order to support coordinated experimental development, detector operation,
data analysis, theoretical studies, and scientific dissemination, CHRONOS
maintains a common framework for governance, research integrity, management of
research outputs, and professional conduct.

The governance and operation of CHRONOS are defined primarily through three
complementary documents:

\begin{itemize}
	
	\item \textit{CHRONOS Organization and Governance}
	\cite{CHRONOS_Organization}, which defines the organizational structure,
	positions, responsibilities, authorities, and meeting structure of the
	collaboration;
	
	\item \textit{CHRONOS Collaboration Policy}
	\cite{CHRONOS_CollaborationPolicy}, which defines the rules governing
	scientific integrity, research records, intellectual contributions,
	authorship, data and software, experimental apparatus, internal
	information, publication, internal review, and external communication; and
	
	\item \textit{CHRONOS Code of Conduct}
	\cite{CHRONOS_CodeOfConduct}, which defines standards for interpersonal
	and professional behavior, fair treatment, respectful communication,
	prevention of discrimination and harassment, consultation and reporting,
	and protection against retaliation.
	
\end{itemize}

These documents have complementary functions and together establish the
organizational and professional framework of the CHRONOS Collaboration
\cite{CHRONOS_Organization,CHRONOS_CollaborationPolicy,
	CHRONOS_CodeOfConduct}.

The Collaboration Policy defines CHRONOS research outputs broadly. They include
data, analysis results, figures, manuscripts, presentation materials, software,
source code, hardware designs, technical documents, research procedures,
models, simulations, research records, research ideas, and other outputs
created, developed, acquired, analyzed, or shared within the collaboration
\cite{CHRONOS_CollaborationPolicy}.

\section{Organization and Governance}
\label{sec:chronos_governance}

The CHRONOS Collaboration consists of the Executive Committee, the Principal
Investigators and members of participating universities and research
institutes, and Emeritus Members. The Executive Committee coordinates the
collaboration as a whole, while each institutional PI is responsible for
organizing CHRONOS activities within the corresponding institution
\cite{CHRONOS_Organization}.

The Executive Committee is organized around six principal positions:

\begin{enumerate}
	\item Principal Investigator,
	\item Experiment Coordinator,
	\item Instruments Coordinator,
	\item Science Coordinator,
	\item Publication and Authorship Coordinator, and
	\item Membership Coordinator.
\end{enumerate}

The Principal Investigator chairs the Executive Committee and oversees
CHRONOS as a whole. The PI is responsible for maintaining coherence among
scientific, experimental, technical, and organizational activities, defining
the medium- and long-term direction of the collaboration, coordinating the
work of the coordinators, maintaining communication with participating
institutions, and representing CHRONOS in communications with external
institutions, funding agencies, and collaborators
\cite{CHRONOS_Organization}.

Matters that substantially affect the collaboration as a whole, including
major budgets, inter-institutional responsibilities, long-term strategy, and
serious organizational issues, are discussed by the Executive Committee.
Consensus is sought whenever practical
\cite{CHRONOS_Organization}.

While the collaboration remains relatively small, one person may hold more
than one organizational position. The detailed responsibilities and
authorities associated with the Executive Committee and the individual
coordinator positions are defined in the
\textit{CHRONOS Organization and Governance}
\cite{CHRONOS_Organization}.

\section{Coordination of Science, Experiment, and Instrumentation}
\label{sec:chronos_coordination}

CHRONOS distinguishes the coordination of scientific objectives,
experimental operation, and instrument development while maintaining close
interaction among these activities
\cite{CHRONOS_Organization,CHRONOS_CollaborationPolicy}.

\subsection{Scientific Coordination}
\label{subsec:chronos_science_coordination}

The Science Coordinator organizes and promotes the scientific program of
CHRONOS. The coordinator maintains the scientific goals and overall science
program of the collaboration, receives proposals for new research, data
analysis, simulations, and scientific studies, and evaluates their scientific
significance, originality, feasibility, and resource requirements
\cite{CHRONOS_Organization}.

The Science Coordinator also identifies duplication or overlap among research
topics, develops priorities for scientific activities, and defines the
measurements, sensitivity targets, and instrument-performance requirements
needed to achieve the scientific objectives of CHRONOS.

These scientific requirements are communicated to the Experiment Coordinator
and Instruments Coordinator so that experimental and technical development
remain aligned with the science program
\cite{CHRONOS_Organization}.

The Science Coordinator may establish scientific working groups when needed
and coordinates data analysis, signal modeling, noise studies, simulations,
sensitivity evaluations, and scientific collaboration with external
researchers
\cite{CHRONOS_Organization}.

\subsection{Experimental Coordination}
\label{subsec:chronos_experiment_coordination}

The Experiment Coordinator oversees CHRONOS experimental projects and on-site
activities and is responsible for planning, coordinating, and reporting
experimental activities
\cite{CHRONOS_Organization}.

The responsibilities include preparation of annual, medium-term, and
short-term experimental plans and coordination of installation, assembly,
testing, commissioning, and operation of the experiment.

The Experiment Coordinator defines schedules, tasks, required personnel,
milestones, and deliverables for experimental activities and coordinates work
among participating members and institutions. The coordinator also tracks
experimental progress, measurement results, technical problems, and follow-up
actions and reports the experimental status to the Executive Committee and
the collaboration
\cite{CHRONOS_Organization}.

Experimental operation, measurement planning, and approval of experimental
configurations are overseen by the Experiment Coordinator
\cite{CHRONOS_CollaborationPolicy}.

\subsection{Instrumentation Coordination}
\label{subsec:chronos_instrument_coordination}

The Instruments Coordinator oversees the development of the instruments,
subsystems, and facilities required for CHRONOS and leads the prioritization
of instrument-development activities
\cite{CHRONOS_Organization}.

This responsibility includes coordination of development plans for optics,
mechanics, vacuum systems, cryogenics, control systems, data acquisition, and
other technical systems.

Development activities are evaluated according to scientific importance,
technical necessity, urgency, feasibility, cost, staffing, and schedule.
The Instruments Coordinator maintains the overall list of instruments and
subsystems, tracks specifications, responsible persons, budgets, delivery
dates, risks, and progress, and ensures technical compatibility among
subsystems developed by different institutions
\cite{CHRONOS_Organization}.

The Instruments Coordinator also coordinates procurement, fabrication,
modification, outsourcing, transport, maintenance, and upgrades and clarifies
responsibility for the ownership, custody, storage, operation, and maintenance
of equipment
\cite{CHRONOS_Organization}.

Major decisions involving substantial budgets, significant institutional
commitments, or major changes to the overall research plan require discussion
by the Executive Committee and approval by the PI
\cite{CHRONOS_Organization}.

\section{Scientific Integrity and Research Records}
\label{sec:chronos_integrity}

Scientific integrity, transparency, reproducibility, and appropriate
preservation of research records are fundamental requirements of CHRONOS
research activities
\cite{CHRONOS_CollaborationPolicy}.

Members must conduct research honestly, accurately, transparently, and
ethically and must appropriately document the research process and relevant
decisions.

Research must be documented sufficiently to permit appropriate review,
verification, and, where practicable, reproduction. Relevant records include:

\begin{itemize}
	\item research methods and assumptions;
	\item original and processed data;
	\item code and software versions;
	\item analysis procedures and significant analysis decisions;
	\item uncertainty estimates;
	\item apparatus configurations and design changes; and
	\item revisions to research results and scientific conclusions.
\end{itemize}

Important research records must not be destroyed, concealed, altered, or
withheld for the purpose of obstructing verification, reproducibility,
internal review, or formal investigation
\cite{CHRONOS_CollaborationPolicy}.

Honest scientific errors are distinguished from research misconduct. When an
error is identified, members are expected to disclose it promptly and, where
appropriate, correct the relevant records, analyses, presentations, or
publications and notify affected collaborators
\cite{CHRONOS_CollaborationPolicy}.

\section{Intellectual Contributions and Attribution}
\label{sec:chronos_attribution}

Scientific, technical, organizational, and intellectual contributions within
CHRONOS must be recognized fairly and accurately
\cite{CHRONOS_CollaborationPolicy}.

Members must not improperly appropriate another person's research idea,
analysis strategy, experimental concept, design, software, text, figure,
calculation, research method, or research result.

Significant scientific and technical decisions, assigned responsibilities,
and major contributions should be documented whenever practicable
\cite{CHRONOS_CollaborationPolicy}.

Particular care is required to recognize the contributions of students,
postdoctoral researchers, and other early-career researchers. Unpublished
ideas, research outputs, or scientific strategies shared within CHRONOS may
not be used for external research without the appropriate CHRONOS approval
\cite{CHRONOS_CollaborationPolicy}.

\section{Authorship and Presentation Rights}
\label{sec:chronos_authorship}

Authorship, author order, presentation rights, research credit, and
acknowledgements are determined on the basis of actual contributions
\cite{CHRONOS_CollaborationPolicy}.

They must not be determined improperly solely on the basis of position,
seniority, institutional affiliation, control of research funding, employment
or supervisory authority, age, personal relationships, or access to
institutional resources.

A legitimate contributor must not be excluded from authorship or presentation
without a valid reason. Conversely, authorship must not be granted solely on
the basis of status, authority, or control of funding when no meaningful
contribution has been made
\cite{CHRONOS_CollaborationPolicy}.

Where practicable, authorship and author order should be discussed from an
early stage of a research project and reviewed as the work develops.

The selection of presenters must also be fair. In particular, appropriate
presentation opportunities should be provided to students and early-career
researchers who have actually performed or led the relevant work
\cite{CHRONOS_CollaborationPolicy}.

Disputes concerning authorship, author order, presentation rights, or
contributions are handled on the basis of documented actual contributions and
may be referred to the Publication and Authorship Coordinator and the
Executive Committee
\cite{CHRONOS_CollaborationPolicy}.

\section{Management of Data, Software, and Internal Information}
\label{sec:chronos_data_management}

CHRONOS data, software, designs, technical documents, analysis results,
unpublished outputs, meeting records, and internal discussions are managed as
collaborative research assets associated with CHRONOS research activities
\cite{CHRONOS_CollaborationPolicy}.

Members are required to comply with collaboration access-control,
data-management, and information-security requirements and to appropriately
document analysis procedures, code modifications, and significant technical
decisions.

Creators and contributors of software, code, designs, documents, and analyses
must be properly identified and recorded.

Data, software, and analysis results should be maintained in a reasonably
reproducible form, and records necessary for scientific verification must be
preserved
\cite{CHRONOS_CollaborationPolicy}.

When members change roles or leave CHRONOS, necessary research records, data,
code, and other research outputs must be appropriately transferred.

Unpublished CHRONOS information must not be used for unauthorized personal
benefit, to obtain an improper competitive advantage, to exclude another
contributor, or for an unauthorized external research project
\cite{CHRONOS_CollaborationPolicy}.

\section{Management of Experimental Apparatus}
\label{sec:chronos_apparatus}

CHRONOS experimental apparatus and technical systems are regarded as important
collaborative research assets
\cite{CHRONOS_CollaborationPolicy}.

These assets include optical systems, mechanical systems, electrical and
electronic systems, control systems, vacuum systems, cryogenic systems,
sensors, actuators, data-acquisition systems, and other experimental
facilities.

The Instruments Coordinator oversees instrumentation status, technical
development, and technical contributions assigned to participating
institutions. The actual operation of the experimental apparatus, measurement
planning, and experimental configurations are overseen by the Experiment
Coordinator
\cite{CHRONOS_CollaborationPolicy}.

Members must not make modifications that may materially affect the experiment
without the appropriate prior approval. This includes disassembly or
modification of apparatus, alteration of experimental configurations, removal
or replacement of components, modification of wiring or control systems, and
other operations that may affect ongoing measurements, calibration, data
acquisition, or analysis
\cite{CHRONOS_CollaborationPolicy}.

Where appropriate, modifications to the apparatus must be documented,
including the nature and reason for the change, the responsible person, the
implementation time, and the resulting configuration.

Physical access to equipment or the technical ability to operate it does not
by itself constitute authority to modify the equipment independently
\cite{CHRONOS_CollaborationPolicy}.

\section{Scientific Policy and Research Coordination}
\label{sec:chronos_scientific_policy}

The Science Coordinator oversees CHRONOS scientific policy, scientific
priorities, major scientific objectives, and the coordination of scientific
research
\cite{CHRONOS_CollaborationPolicy}.

Research topics, analysis activities, and scientific studies are coordinated
to avoid excessive duplication, unnecessary competition, or conflicts in
research planning.

Changes that may have a significant impact on CHRONOS scientific policy are
discussed by the Science Coordinator with the Executive Committee.

Individual members conduct research in a manner consistent with CHRONOS
scientific policy and approved research plans
\cite{CHRONOS_CollaborationPolicy}.

\section[Publication, Internal Review, and External Communication]{Publication, Internal Review,\protect\linebreak and External Communication}
\label{sec:chronos_publication}

Scientific and technical outputs produced within CHRONOS are subject to an
internal review and approval process before external release
\cite{CHRONOS_CollaborationPolicy}.

The scope of this process includes:

\begin{itemize}
	\item journal submissions;
	\item preprints;
	\item conference abstracts;
	\item oral presentations;
	\item posters;
	\item proceedings;
	\item theses;
	\item technical reports;
	\item press releases;
	\item communications with the media;
	\item public websites;
	\item software repositories;
	\item data repositories; and
	\item disclosure to external research groups or third parties.
\end{itemize}

Before external release, the relevant output must undergo the scientific,
technical, authorship, and publication review procedures established within
CHRONOS
\cite{CHRONOS_CollaborationPolicy}.

The Publication and Authorship Coordinator manages publications, conference
presentations, technical documents, and other public dissemination of CHRONOS
results. The coordinator records proposed outputs, coordinates writing teams
and internal reviewers, tracks review progress, and confirms that authorship
and author order reflect substantive contributions
\cite{CHRONOS_Organization}.

Where appropriate, the Science Coordinator participates in the review of
scientific content, while the Experiment Coordinator and Instruments
Coordinator participate in review of experimental and technical content
\cite{CHRONOS_CollaborationPolicy}.

The purpose of internal review is to ensure scientific validity, technical
consistency, appropriate attribution, and responsible communication of
CHRONOS results.

\section{Code of Conduct}
\label{sec:chronos_code_of_conduct}

CHRONOS considers a safe, respectful, fair, and professionally responsible
working environment to be an essential component of scientific collaboration
\cite{CHRONOS_CodeOfConduct}.

All members are expected to respect the dignity, expertise, experience,
opinions, and contributions of others and to treat participants fairly in
professional interactions and in access to research opportunities.

Members are expected to communicate constructively and courteously, including
when scientific or organizational disagreements arise, and to respect
differences in nationality, culture, language, institution, discipline,
career stage, and personal background
\cite{CHRONOS_CodeOfConduct}.

The Code of Conduct applies to all CHRONOS-related activities, including
in-person and online meetings, electronic communication, laboratories,
experimental facilities, conferences, workshops, schools, training
activities, outreach, travel, meals, social events, and activities undertaken
externally in the name of CHRONOS
\cite{CHRONOS_CodeOfConduct}.

\section{Fair Treatment and Professional Conduct}
\label{sec:chronos_fair_treatment}

All members must treat one another fairly and respectfully.

No participant may be unfairly disadvantaged in meeting participation,
access to information, assignment of responsibilities, professional
opportunities, or career development because of personal characteristics,
institutional affiliation, career stage, employment status, or differences in
available resources
\cite{CHRONOS_CodeOfConduct}.

CHRONOS does not tolerate harassment, sexual harassment, bullying,
intimidation, coercion, discrimination, or abuse of authority
\cite{CHRONOS_CodeOfConduct}.

Scientific and organizational disagreements are recognized as a normal part
of collaborative research. Criticism and disagreement should address
evidence, methods, assumptions, reasoning, technical issues, or
organizational decisions and should be expressed professionally and
constructively
\cite{CHRONOS_CodeOfConduct}.

Disagreement must not be used as a basis for personal attacks, humiliation,
intimidation, exclusion, harassment, discrimination, or retaliation.

\section{Responsibilities of Persons in Positions of Authority}
\label{sec:chronos_authority}

Persons exercising supervisory, managerial, evaluative, or organizational
authority within CHRONOS must recognize that their authority may affect the
research, education, employment, and career development of other members
\cite{CHRONOS_CodeOfConduct}.

They are expected to communicate roles, responsibilities, deadlines, and
expectations clearly, provide appropriate supervision and constructive
feedback, treat members fairly, and provide reasonable educational and
research opportunities to students and early-career researchers.

Authority must not be used to impose excessive or inappropriate workloads,
disadvantage a person for raising a concern or scientific disagreement, or
improperly influence employment, evaluation, recommendations, contract
renewal, or career progression
\cite{CHRONOS_CodeOfConduct}.

Relevant conflicts of interest must be disclosed, and appropriate action must
be taken when persons in positions of authority become aware of harassment,
bullying, discrimination, retaliation, or safety concerns.

\section{Consultation, Reporting, and Protection against Retaliation}
\label{sec:chronos_reporting}

General concerns under the Code of Conduct and alleged violations are overseen
by the Membership Coordinator
\cite{CHRONOS_CodeOfConduct}.

Members may seek advice or report concerns to the Membership Coordinator or,
where appropriate, to the PI of their participating institution.

Confidential consultation may be sought before a formal complaint is made.
Where necessary, the nature, seriousness, urgency, safety risks, conflicts of
interest, and applicable CHRONOS rules are assessed before determining the
appropriate procedure
\cite{CHRONOS_CodeOfConduct}.

CHRONOS prohibits retaliation against any person who, in good faith, seeks
advice, raises a concern, makes a report, provides information or evidence,
participates in a review, supports another member, or contacts an institution,
funding agency, ethics body, or other competent authority
\cite{CHRONOS_CodeOfConduct}.

Information concerning consultations, reports, complaints, reviews,
protective measures, and corrective actions is handled as confidentially as
reasonably possible, subject to the requirements of safety, fair review,
fact-finding, prevention of retaliation, legal obligations, and external
reporting
\cite{CHRONOS_CodeOfConduct}.

\section{Membership Coordination}
\label{sec:chronos_membership}

The Membership Coordinator manages membership records and serves as the
primary contact for membership matters, concerns under the Code of Conduct,
and proposals for organizational improvement
\cite{CHRONOS_Organization}.

The coordinator receives applications from new members, institutional
members, and Emeritus Members and maintains records of names, affiliations,
roles, contact information, and participation status.

New members are provided with the relevant organizational rules,
Collaboration Policy, publication and data requirements, and Code of Conduct.

The Membership Coordinator also receives consultations and reports concerning
the Code of Conduct and serves as an initial contact for concerns involving
harassment, research integrity, authorship, misuse of results, conflicts of
interest, or related matters
\cite{CHRONOS_Organization}.

The Membership Coordinator does not independently conduct formal
investigations, determine facts, or impose sanctions. Serious matters are
referred, where appropriate, to institutions, funding bodies, or other bodies
with suitable investigative authority
\cite{CHRONOS_Organization}.

\section{Participating Institutions and Members}
\label{sec:chronos_institutions}

Each participating university or research institute appoints an institutional
PI who serves as the principal point of contact between the Executive
Committee and members of that institution
\cite{CHRONOS_Organization}.

The institutional PI coordinates CHRONOS members within the institution,
supports and supervises students, postdoctoral researchers, technical staff,
and other members, and monitors the institution's contributions to
experiments, instrument development, and scientific research.

Institutional PIs also coordinate local budgets, facilities, personnel, and
safety procedures, communicate decisions of the Executive Committee to local
members, collect proposals, requests, and concerns, and coordinate with the
relevant CHRONOS coordinators
\cite{CHRONOS_Organization}.

Institutional procedures concerning research integrity, safety, intellectual
property, export control, and related obligations remain the responsibility of
the relevant participating institution.

At the same time, scientific participation is not restricted by institutional
boundaries. Members may participate in experiments, instrument development,
simulations, data analysis, scientific studies, publications, and working
groups across participating institutions
\cite{CHRONOS_Organization}.

\section{Emeritus Members}
\label{sec:chronos_emeritus}

Researchers who have made sustained contributions to CHRONOS and have retired
from an active academic or research position may be appointed as Emeritus
Members
\cite{CHRONOS_Organization}.

Emeritus Members may remain involved without a current institutional
affiliation and may participate in scientific and technical activities,
advise on research direction and technical development, mentor students and
early-career researchers, review papers and technical documents, and
contribute to the transfer of institutional knowledge and technical
experience.

Emeritus status alone does not automatically confer authorship, priority in
research activities, or unrestricted rights to use CHRONOS results. Such
rights remain determined by substantive contribution and the applicable
CHRONOS policies
\cite{CHRONOS_Organization}.

\section{Conflicts of Interest}
\label{sec:chronos_conflicts}

Members must disclose conflicts of interest that could reasonably affect
decisions concerning authorship, internal review, access to information, data
use, publication, project allocation, resource allocation, complaints, or
corrective action
\cite{CHRONOS_CollaborationPolicy}.

The PI, coordinators, and institutional PIs must not use their positions to
improperly control authorship, research outputs, access to data, research
opportunities, employment, recommendations, contract renewal, or career
progression.

Where an actual, potential, or perceived conflict of interest exists and
impartiality may reasonably be questioned, the person concerned must withdraw
from the relevant review, deliberation, or decision
\cite{CHRONOS_CollaborationPolicy}.

\section{Obligations after Leaving CHRONOS}
\label{sec:chronos_departure}

Certain obligations arising from participation in CHRONOS continue after a
member leaves the collaboration
\cite{CHRONOS_CollaborationPolicy}.

These include obligations concerning confidentiality, unpublished CHRONOS
information, data and software use, authorship and attribution, internal
review, and publication approval.

Leaving CHRONOS does not automatically create an independent right to use,
continue, present, or publish unpublished CHRONOS research outputs.

Published CHRONOS results may be cited and discussed according to normal
academic practice, but the contributions of CHRONOS or individual
collaborators must not be misrepresented as the independent or sole work of a
former participant
\cite{CHRONOS_CollaborationPolicy}.

\section{Relationship to Institutional and External Rules}
\label{sec:chronos_external_rules}

The internal rules of CHRONOS do not replace the applicable policies,
procedures, and legal requirements of participating institutions, host
institutions, research facilities, funding agencies, publishers, professional
societies, or public authorities
\cite{CHRONOS_CollaborationPolicy,CHRONOS_CodeOfConduct}.

Members remain subject to applicable requirements concerning research
integrity, research ethics, intellectual property, confidentiality,
collaboration agreements, employment, funding conditions, safety, and
applicable laws.

Where necessary, CHRONOS procedures and formal procedures conducted by
institutions, funding agencies, or other competent external authorities may
operate in parallel
\cite{CHRONOS_CollaborationPolicy}.

\section{Operating Principles}
\label{sec:chronos_operating_principles}

The governance of CHRONOS is based on the principles of transparency,
fairness, accountability, and coordination
\cite{CHRONOS_Organization}.

Important decisions and their reasons should be documented and communicated
to relevant members. Opportunities, responsibilities, budgets, authorship,
and access to research results should be handled according to clear and
consistent criteria.

Each organizational role is accountable for explaining decisions and
reporting progress within its assigned area. Matters spanning multiple roles
should be handled jointly and, where necessary, brought to the Executive
Committee
\cite{CHRONOS_Organization}.

Experimental safety, scientific integrity, respectful conduct, privacy, and
protection against retaliation take priority over schedule or operational
convenience
\cite{CHRONOS_Organization}.

Through these principles and the complementary governance documents described
above, the CHRONOS Collaboration aims to provide a sustainable framework in
which researchers and participating institutions can contribute fairly,
responsibly, and effectively to a long-term international experimental
program.

%% file: Chapters/Sensitivity.tex
\chapter{Sensitivity}
\label{chap:sensitivity}

The reference sensitivity in this White Paper uses the multi-input quantum
model of Ref.~\cite{InoueQuantumTorque2026}. It includes the vacuum fields
entering the main input and the rear ports of the ring-cavity end and middle
test masses, together with measurement--back-action correlations. Classical
noise is added to form the full detector noise budget. Two operating points
are retained: one minimizes the lowest value of a power-law-integrated (PI)
curve, and the other minimizes its logarithmic mean over 0.1--10~Hz.
Unless an alternative is explicitly identified, the minimum-PI point is the
reference configuration.

\begin{figure}[p]
\centering
\includegraphics[width=0.72\linewidth]{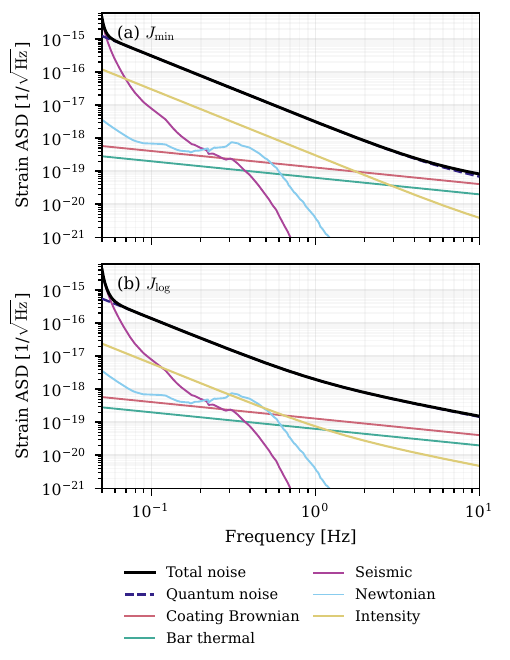}
\begingroup\linespread{1.1}\selectfont
\caption{Full strain noise budget reproduced from Fig.~7 of
Ref.~\cite{InoueQuantumTorque2026}. (a) Minimum-PI point; (b) minimum-logarithmic-mean
point. Black: total noise; dashed purple: quantum noise including correlations.
Coating Brownian, bar thermal, seismic, Newtonian, and intensity noise are
included in the PSD sum, also where a component lies below the plotting floor.
The displayed band is 0.05--10~Hz; the optimization band is 0.1--10~Hz.
The ordinate is an amplitude spectral density (ASD), in $\mathrm{Hz}^{-1/2}$.}
\label{fig:sensitivity_overview}
\endgroup
\end{figure}

Figure~\ref{fig:sensitivity_overview} replaces the earlier sensitivity estimate.
Quantum noise is the leading contribution over 0.1--10~Hz for these reference
points; coating and bar thermal noise contribute toward the upper end of the
band, particularly at the minimum-PI point. The relative importance of noise
sources must be read from this full budget rather than inferred from an ideal,
lossless speed-meter model. Table~\ref{tab:chronos_all_params} lists the shared
parameters, and Table~\ref{tab:sensitivity_operating_points} gives the two phase
settings and circulating powers.

\section{Reference parameters and strain calibration}
\label{sec:sensitivity_conventions}

Here $f$ is frequency in hertz, $\Omega=2\pi f$ is angular frequency in
radians per second, and $\hbar$ is the reduced Planck constant. Fourier
amplitudes use the time dependence $e^{+i\Omega t}$.
All spectra are one-sided power spectral densities (PSDs): $S_h(f)$ is the
equivalent strain-noise PSD and $S_\theta(f)$ is the equivalent input-angle
noise PSD. Their square roots are amplitude spectral densities (ASDs).
Let $\theta_X$ and $\theta_Y$ denote the externally driven signal rotations
of the two orthogonal bars before optical feedback. Their differential angle is
$\theta_{\rm sig}=\theta_X-\theta_Y=\eta h$, with $\eta=0.93656$ for the
adopted mass geometry. Here $h$ is the antenna-projected strain in the
free-rotation, long-wavelength limit. A specified source direction and
polarization require the corresponding antenna response. Thus
$S_h=S_\theta/|\eta|^2$; the angular and strain ASDs must not be interchanged.
The differential coordinate has inertia $I_d=I/2$, where $I$ is the inertia
of one bar. The standard quantum limit (SQL) is the uncorrelated
measurement-imprecision/back-action benchmark; its differential angular ASD is
$\theta_{\rm SQL,d}=\sqrt{4\hbar/(I\Omega^2)}$.

The mirror abbreviations are ITM (input test mass), ETM (end test mass),
MTM (middle test mass), PRM (power-recycling mirror), and SRM
(signal-recycling mirror). PR and SR denote power and signal recycling.

\begin{table}[tbp]
\centering\small
\begingroup\linespread{1.12}\selectfont
\caption{Shared parameters for both reference operating points, from
Ref.~\cite{InoueQuantumTorque2026}. Transmissions and reflectivities are power
quantities. The quoted circulating power in Table~\ref{tab:sensitivity_operating_points}
is for one propagation direction in a ring.}
\label{tab:chronos_all_params}
\begin{tabular}{lll}
\toprule
Quantity & Symbol & Value\\
\midrule
Laser wavelength & $\lambda_0$ & $1064~\mathrm{nm}$\\
Input power & $P_{\rm in}$ & $1~\mathrm W$\\
ITM power transmission & $T_I$ & $10^{-4}$\\
ETM and MTM power transmission & $T_E,T_M$ & $10^{-7}$ each\\
PRM / SRM power reflectivity & $R_p/R_s$ & $0.9/0.5$\\
Ring round-trip length & $L_{\rm rt}$ & $5.083~\mathrm m$\\
PR / SR path length & $L_{\rm PR}/L_{\rm SR}$ & $5.069/2.622~\mathrm m$\\
Single-bar mass & $M_{\rm bar}$ & $171.32~\mathrm{kg}$\\
Single-bar moment of inertia & $I$ & $19.86~\mathrm{kg\,m^2}$\\
Differential moment of inertia & $I_d$ & $9.930~\mathrm{kg\,m^2}$\\
Main-laser spot lever & $\ell$ & $0.54365~\mathrm m$\\
Main-laser spot separation & $2\ell$ & $1.0873~\mathrm m$\\
Test-mass temperature & $T$ & $8~\mathrm K$\\
Differential-angle strain coupling & $\eta$ & $0.93656$\\
\bottomrule
\end{tabular}
\endgroup
\end{table}

The reference calculation treats the nonreflected ETM and MTM light as
transmitted light and includes the associated independent rear-port vacuum
inputs. Absorption, suspension control, and departures from the specified
symmetric geometry require their own model. Residual laser-frequency and
actuator noise require separate engineering allocations; the six-component
reference budget is not an exhaustive list of technical noise. In particular, a large cancellation
of the driving torques does not alone establish a small calibrated readout
noise: mechanical feedback and the signal transfer function also enter.

\begin{table}[tbp]
\centering\small
\caption{Full strain ASDs at representative frequencies, from the archived
spectra underlying Ref.~\cite{InoueQuantumTorque2026}. Units are
$\mathrm{Hz}^{-1/2}$. These values provide a common numerical reference
for science and subsystem comparisons throughout this White Paper.}
\label{tab:reference_asd}
\begin{tabular}{rrr}\toprule
$f$ [Hz] & Minimum PI & Minimum logarithmic mean\\\midrule
0.1 & $3.078\times10^{-16}$ & $1.374\times10^{-16}$\\
0.5 & $1.237\times10^{-17}$ & $6.154\times10^{-18}$\\
1.0 & $3.136\times10^{-18}$ & $1.969\times10^{-18}$\\
2.0 & $8.306\times10^{-19}$ & $7.915\times10^{-19}$\\
3.0 & $4.037\times10^{-19}$ & $5.027\times10^{-19}$\\
10.0 & $8.235\times10^{-20}$ & $1.529\times10^{-19}$\\\bottomrule
\end{tabular}
\end{table}

\section{Sensing Function and final strain sensitivity}
\label{sec:sensitivity_final_expression}

The \emph{Sensing Function} describes how a small signal angle becomes a
measurable output. Let $\delta P_{\rm BHD}(f)$ be the signal-induced change
in the balanced-homodyne-detector (BHD) power difference. The complex angular
sensing function $H_\theta(f)$ maps $\theta_{\rm sig}$ to this power
difference; its magnitude is the gain in $\mathrm{W/rad}$ and its phase is
the output phase relative to the angular input. The corresponding
strain sensing function is $H_h=\eta H_\theta$, in watts per unit strain.
These optical functions precede the photodetector electronics and digitization.

Only the final readout expressions are retained below; their optical boundary
conditions and derivation are given in Ref.~\cite{InoueQuantumTorque2026}.
The dimensionless complex denominator $M(f)$ contains the optical return
paths and radiation-pressure feedback. The scalar $\mathcal D(f)$ is the
signal numerator after projection onto the detected power difference,
using $\theta_{\rm sig}/\theta_{\rm SQL,d}$ as signal normalization.
For independent vacuum input $q$, $\mathcal C_q(f)$ is a row with two
entries, the coefficients of its amplitude and phase quadratures in the
same output. These quadratures have unit one-sided symmetrized vacuum PSD;
$\dagger$ denotes complex-conjugate transpose. Both signal and
local-oscillator (LO) paths are included coherently in each projected
coefficient. These frequency-dependent quantities are calculated from the
reference model and the specified operating point, not independent fit constants.

The phase $\beta_{\rm sag}$ is the bare-Sagnac phase in that model.
Specifically, if $r_{\rm cav}$ is the complex ring amplitude reflectivity,
define its unwrapped phase $\beta_{\rm ring}$ by
$r_{\rm cav}=-|r_{\rm cav}|e^{-2i\beta_{\rm ring}}$, with
$\beta_{\rm ring}(0)=0$. For propagation time $\tau_{\rm link}$ along the
Sagnac paths outside the rings,
$\beta_{\rm sag}=2\beta_{\rm ring}+\Omega\tau_{\rm link}/2+\pi/2$.
With $S_h^{\rm quant}$ denoting the quantum strain-noise PSD,
the sensing function and equivalent quantum strain ASD are then
\begin{align}
 H_\theta(f)
 &\equiv\frac{\delta P_{\rm BHD}(f)}{\theta_{\rm sig}(f)}
 =\frac{e^{-i\beta_{\rm sag}(f)}\mathcal D(f)}
 {M(f)\theta_{\rm SQL,d}(f)},
 \qquad H_h(f)=\eta H_\theta(f),
 \label{eq:sensitivity_transfer}\\
 \sqrt{S_h^{\rm quant}(f)}
 &\simeq\frac{\theta_{\rm SQL,d}(f)}{|\eta|}
 \frac{\sqrt{\displaystyle\sum_q\mathcal C_q(f)\mathcal C_q^\dagger(f)}}
 {|\mathcal D(f)|}.
 \label{eq:theta_noise}
\end{align}
The ratio defining $H_\theta$ is evaluated for a small deterministic signal
with the additive noise set to zero. The retained inputs comprise one main
input, eight ETM-rear inputs, and eight MTM-rear inputs. For each input,
its signal- and LO-path amplitudes are combined coherently before the PSDs
of independent inputs are summed. The approximation denotes the omitted
input ports; their contribution was below 1\% in the configurations tested
in the reference, rather than identically zero.

The quantum components
$S_h^{\rm shot}$, $S_h^{\rm ba}$, and $S_h^{\rm cross}$ denote measurement
imprecision (shot noise), radiation-pressure back action, and their cross
term, respectively. Equivalently, the quantum PSD is
$S_h^{\rm quant}=S_h^{\rm shot}+S_h^{\rm ba}+S_h^{\rm cross}$.
The cross term is signed and can be negative. Shot and radiation-pressure
ASDs therefore cannot in general be added in quadrature without this term.
All three contributions use the same full signal response, including
optomechanical feedback. For classical noise, the superscripts
$\mathrm{coat}$, $\mathrm{bar}$, $\mathrm{seis}$, $\mathrm{NN}$, and
$\mathrm{int}$ denote coating Brownian, bar thermal, seismic, Newtonian,
and laser intensity noise. The superscript $\mathrm{tot}$ denotes their
total with quantum noise. Assuming independent classical contributions,
the final strain sensitivity is
\begin{equation}
 \sqrt{S_h^{\rm tot}}=
 \left[S_h^{\rm quant}+S_h^{\rm coat}+S_h^{\rm bar}
       +S_h^{\rm seis}+S_h^{\rm NN}+S_h^{\rm int}\right]^{1/2}.
 \label{eq:sensitivity_total}
\end{equation}
Every term is a
strain PSD in $\mathrm{Hz}^{-1}$, evaluated at the same operating point.

\begin{figure}[p]
\centering
\includegraphics[width=0.96\linewidth]{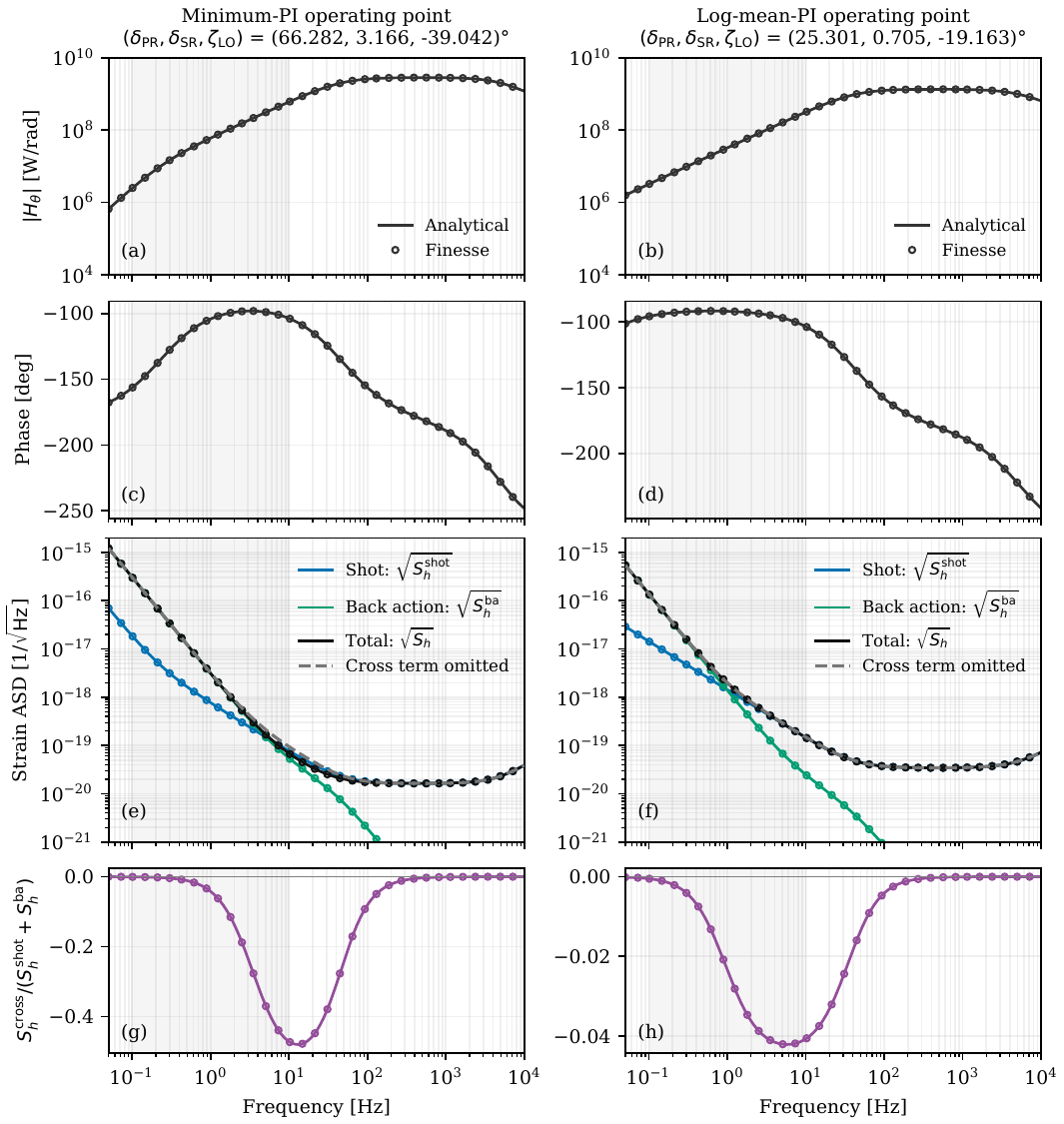}
\begingroup\linespread{1.1}\selectfont
\caption{\textbf{Sensing Function} $H_\theta$: the complex response from
external differential signal angle $\theta_{\rm sig}$ to balanced-detector
power difference $\delta P_{\rm BHD}$, together with quantum strain sensitivity,
reproduced from Fig.~5 of Ref.~\cite{InoueQuantumTorque2026}. Left: minimum-PI
point; right: minimum-logarithmic-mean point. The first two rows show the
sensing-function magnitude [W/rad] and phase; TF means transfer function.
The remaining rows show quantum-noise ASDs and the signed cross PSD normalized by the sum
of shot and back-action PSDs. The gray dashed ASD omits only the cross term.
Lines are the analytical model and circles are selected FINESSE~3 samples;
shading marks 0.1--10~Hz. The black ASD in this figure is quantum noise,
whereas the black curves in Fig.~\ref{fig:sensitivity_overview} include
classical noise as well.}
\label{fig:sensitivity_transfer}
\endgroup
\end{figure}

Figure~\ref{fig:sensitivity_transfer} presents the Sensing Function's magnitude and phase together
with the quantum-noise decomposition. At the minimum-PI point, the ETM and
MTM rear inputs each account for 44.84\% of the quantum PSD at 0.1~Hz in the
reference calculation. Omitting these inputs would underestimate the
quantum ASD there by a factor of 3.11. This residual back action explains
why the ideal speed-meter argument alone is insufficient for the noise budget.
The agreement with FINESSE~3 checks the calculated response; it does not
replace a stability assessment for a proposed controlled operating point.

\section{Optimization with the full noise budget}
\label{sec:sensitivity_optimization}

Let $\Omega_{\rm PI}(f)$ denote the dimensionless PI threshold on the
gravitational-wave energy density per logarithmic frequency, expressed as a
fraction of the present critical energy density. This $\Omega_{\rm PI}$ is
not the angular frequency $\Omega$. The PI construction is given in
Sec.~\ref{sec:gw_sgwb}. The phase settings are selected using PI curves formed from
Eq.~\eqref{eq:sensitivity_total}, following
Refs.~\cite{PhysRevD.88.124032,InoueQuantumTorque2026}. The reference assumes
one detector with known noise, one year of effective observation,
signal-to-noise ratio (SNR)~1, and a dimensionless signal-power response
factor $\mathcal R=1$. These assumptions define a comparison metric; they
are not the sensitivity of an independently validated stochastic-background
cross-correlation measurement.

For the fixed band $f_1=0.1~\mathrm{Hz}$ to $f_2=10~\mathrm{Hz}$, the two
objectives are
\begin{align}
 J_{\min}&=\min_{f_1\le f\le f_2}\Omega_{\rm PI}(f),\\
 J_{\log}&=\frac{1}{\ln(f_2/f_1)}\int_{f_1}^{f_2}
             \ln\Omega_{\rm PI}(f)\,d\ln f.
 \label{eq:sensitivity_PI_objectives}
\end{align}
The first favors the lowest PI threshold, while the second weights logarithmic
frequency uniformly; $\exp J_{\log}$ is the geometric mean of the PI curve.
The bandwidth itself is not optimized. The quoted operating points are
those calculated in Ref.~\cite{InoueQuantumTorque2026}. The tunable angles
$\delta_{\rm PR}$ and $\delta_{\rm SR}$ are the PRM and SRM microscopic
phase settings; $\zeta_{\rm LO}$ is the LO optical phase setting.
$P_{\rm ring}$ denotes circulating carrier power in one ring propagation
direction. The table reports angles in degrees; phase relations use radians.

\begin{table}[tbp]
\centering\small
\begingroup\linespread{1.12}\selectfont
\caption{Operating points selected from the same full noise budget and
0.1--10~Hz band \cite{InoueQuantumTorque2026}. $J_{\min}$ is the principal
reference; $J_{\log}$ is the alternative favoring the band-wide geometric mean.}
\label{tab:sensitivity_operating_points}
\begin{tabular}{lrr}
\toprule
Quantity & Minimum $J_{\min}$ & Minimum $J_{\log}$\\
\midrule
PR setting $\delta_{\rm PR}$ [deg] & 66.282 & 25.301\\
SR setting $\delta_{\rm SR}$ [deg] & 3.166 & 0.705\\
LO setting $\zeta_{\rm LO}$ [deg] & $-39.042$ & $-19.163$\\
$P_{\rm ring}$, one direction [W] & 1609.75 & 320.09\\
$\min\Omega_{\rm PI}$, one year & $4.0142\times10^{-4}$ & $5.9611\times10^{-4}$\\
Frequency of PI minimum [Hz] & 4.687 & 1.821\\
$\exp J_{\log}$, one year & $2.4034\times10^{-3}$ & $1.5067\times10^{-3}$\\
\bottomrule
\end{tabular}
\endgroup
\end{table}

The corresponding optical phases $\Phi_{\rm PR}$, $\Phi_{\rm SR}$, and
$\Phi_h$ for the PR, SR, and homodyne paths, respectively, obey
$\Phi_{\rm PR}=\pi/2-2\delta_{\rm PR}$,
$\Phi_{\rm SR}=\pi/2-2\delta_{\rm SR}$, and
$\Phi_h=\pi/2-\zeta_{\rm LO}$. The LO optical setting is independent of SR
setting and is distinct from the quadrature projection angle. In this
convention PR resonance occurs at $\delta_{\rm PR}=90^\circ$ and SR resonance
at $\delta_{\rm SR}=0^\circ$, including the Sagnac return phase.

A consistent optimization varies PR, SR, and LO settings together while
holding input power, optical geometry, mirror properties, bar inertia, and
spot lever fixed. At every trial, the carrier field, circulating power,
complex signal TF, quantum correlations, and intensity noise must be
recalculated before evaluating the full-noise PI objective. PR tuning
changes both internal power and the carrier phase relative to the LO; it
cannot be represented solely by a quadrature rotation at fixed power.
A practical search first explores bounded phase ranges from multiple starting
points, then refines the best candidates with the full frequency grid.
Convergence should be checked by increasing the frequency resolution and
repeating the search; the complex response and noise decomposition should be
cross-checked with an independent optical solver. The reported points are
numerically selected candidates, with no proof of global optimality.
A candidate setting must also satisfy independent power, thermal-load,
and control-stability requirements before experimental adoption.

\begin{figure}[tbp]
\centering
\includegraphics[width=0.70\linewidth]{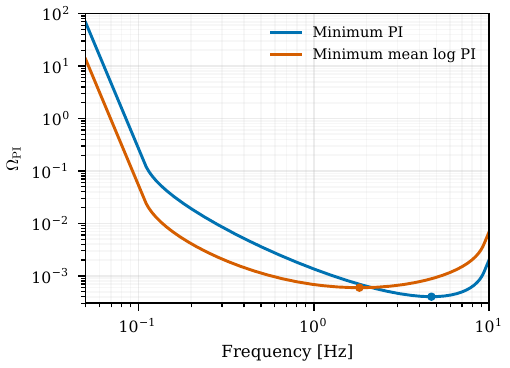}
\begingroup\linespread{1.1}\selectfont
\caption{PI comparison reproduced from Fig.~2 of
Ref.~\cite{InoueQuantumTorque2026}. Blue minimizes $J_{\min}$; orange minimizes
$J_{\log}$. Dots mark the minima. The curve is displayed from 0.05~Hz,
but integration and optimization use only 0.1--10~Hz. Both curves assume
one detector, known noise, one year, SNR~1, and $\mathcal R=1$.}
\label{fig:sensitivity_PI}
\endgroup
\end{figure}

The logarithmic-mean choice reduces the geometric mean by 37.3\% relative
to the minimum-PI choice, while raising the minimum by 48.5\%. It gives
lower full noise at low frequencies, whereas the minimum-PI choice performs
better toward the upper band edge. Selecting a setting for a particular
science target therefore requires its integrated signal statistic, not
only the lowest value of a strain ASD.

\section{Classical noise and design implications}
\label{sec:sensitivity_classical}

The classical spectra in Fig.~\ref{fig:sensitivity_overview} follow the
noise-budget and intensity-noise models used by the reference study
\cite{CHRONOSNoise2026,tanabe_chronos_intensity}. Their physical and
environmental assumptions are common to the two phase settings; intensity
noise is recalculated with the circulating power and signal TF at each point.
The relevant mitigation measures are summarized here without repeating the
derivations.

\subsection{Bar and coating thermal noise}
\label{app:bar_thermal}
\label{app:coat_brownian}
Thermal noise follows mechanical dissipation and temperature. Bar and
suspension losses couple through the angular and elastic response;
coating loss produces surface displacement sampled by the optical beam
\cite{Saulson1990,Levin1998,Harry2002}. Cryogenic operation, low-loss materials,
and suitable beam size are therefore required even where quantum noise is
larger. The test-mass and coating programs are described in
Chaps.~\ref{chap:test_mass_mirror} and~\ref{chap:optical-coatings}.
The reference curves do not establish that either thermal contribution
dominates the entire 1--10~Hz band.

\subsection{Seismic and Newtonian noise}
\label{app:seismic}
\label{app:rayleigh_nn}
Ground motion couples to yaw through suspension asymmetries and other
translation-to-rotation paths. Isolation and control must be assessed using
the corresponding measured transfer functions. Newtonian noise instead
arises from fluctuating gravitational fields and cannot be removed by
mechanical isolation; site characterization and environmental subtraction
are separate requirements \cite{Harms2015}. Their relative size is the one
shown in Fig.~\ref{fig:sensitivity_overview}, not a universal crossover
frequency valid for every site or operating point.

\section{Use of the reference sensitivity in science forecasts}
\label{sec:sensitivity_forecast_scope}

The reference ASD, an integrated SNR, a PI curve, and a systematic-error
floor describe different quantities. Forecasts must identify the operating
point, frequency band, exposure, response convention, and included noise.
The one-year PI values in Table~\ref{tab:sensitivity_operating_points} scale
as $T_{\rm obs}^{-1/2}$, where $T_{\rm obs}$ is effective observing time,
only with the same known-noise and response assumptions. This time is
distinct from the test-mass temperature $T$ in Table~\ref{tab:chronos_all_params}.
At ten years the corresponding minima are $1.2694\times10^{-4}$ and
$1.8851\times10^{-4}$, at the same frequencies.

Compact-binary reach, ideal Yukawa integration times, and the explicitly
defined earthquake template benchmarks in this White Paper are recalculated with the archived full-noise spectra of the reference study.
Other source-specific forecasts require their own calibrated signal model;
replacing a detector noise curve alone does not validate an earlier SNR. The Yukawa systematic-floor projections in
Chap.~\ref{chap:yukawa} use the harmonic-subtracted torque model and its stated
metrology assumptions; their finite-time noise must be propagated through
the same calibrated harmonic channels. A gravitational-wave strain ASD
alone also does not define a dark-matter coupling limit.

\section{CHRONOISE software}
\label{sec:chronoise}

CHRONOISE (CHRONOS Noise Estimation Software) is being developed as a common
environment for detector modeling, noise-budget evaluation, and science
forecasts. Its purpose is to keep optical settings, mechanical parameters,
calibration conventions, and derived sensitivity in a reproducible calculation,
so that an instrument change can be traced through to its scientific impact
\cite{CHRONOISESoftware2026}.

The current software combines FINESSE-based calculations with a staged
analytical quantum model, developed from mirror and ring-cavity responses
through the Sagnac interferometer, recycling cavities, coupled torsion bars,
and balanced readout. The analytical implementation retains all modeled vacuum
inputs; the compact expressions in this chapter use the reference paper's
17-input approximation. Comparison tools examine complex transfer functions,
shot and back-action noise, their signed correlation, and total quantum noise.
Agreement tests cover selected operating points and parameter variations;
they do not imply validation of every possible configuration.

The graphical interface provides parameter editing, frequency scans, component
noise budgets, transfer-function magnitude and phase, and mechanical-response
views. Science tools calculate known-noise PI curves and, with LALSuite,
compact-binary waveforms and matched-filter reach. Parameter scans and bounded
optimization support comparisons between operating points. Saved configurations,
numerical arrays, and calculation records allow those comparisons to be
reproduced. Optimization uses local exploration, population-based search, and
refinement; a completed evaluation budget is not proof of a global optimum.
Publication reference settings are explicit cases, rather than assumed GUI
defaults. The baseline results here do not include optional squeezing.

\begin{figure}[tbp]
\centering
\includegraphics[width=0.76\linewidth]{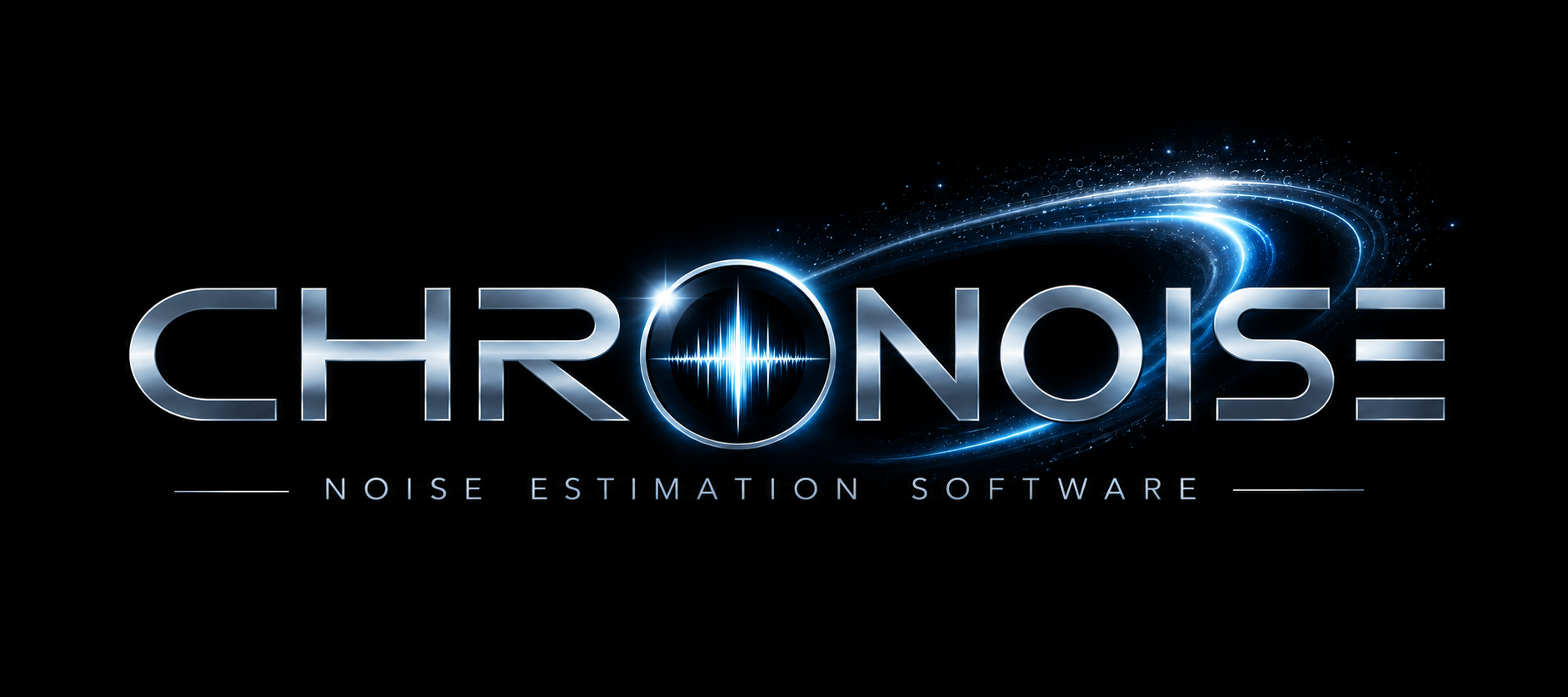}
\caption{Existing CHRONOISE software logo. The software links detector
parameters, noise and response calculations, and conditional science forecasts.}
\label{fig:chronoise_logo}
\end{figure}

%% file: Chapters/Science_roadmap.tex
\chapter{Science Roadmap}
\label{sec:science_roadmap}

The CHRONOS project is designed as a staged experimental program in which the scientific reach of the observatory expands progressively as the detector matures~\cite{Inoue_CHRONOS_Main_2026}. At each stage, scientifically impactful objectives are defined, with the aim of producing a continuous stream of scientific results throughout the development of the experiment~\cite{SPP-2026-3A-06}.

The experimental team, coordinated by the Experimental Coordinator, carries out the planned experimental program, tests the relevant hypotheses, and publishes the resulting experimental findings. In parallel, the theory team, under the coordination of the Science Coordinator, explores the scientific potential of CHRONOS and evaluates the feasibility of the proposed measurements and science cases. These activities are coordinated so as to maximize the scientific opportunities and contributions of all collaborators and to translate promising ideas into concrete experimental programs. The priorities for instrument development are defined by the Instruments Coordinator in accordance with the scientific and experimental requirements of each stage.

Rather than defining a single final science target, CHRONOS follows a three-phase roadmap:

\begin{enumerate}
    \item principle demonstration~\cite{Inoue_CHRONOS_Main_2026},
    \item gravity-gradient science~\cite{onglao2026prospectsobservinggravitygradientnoise}, and
    \item gravitational-wave science~\cite{SPP-2026-3A-06,Inoue_CHRONOS_Main_2026}.
\end{enumerate}

\begin{figure*}[t]
\centering
\includegraphics[width=\textwidth]{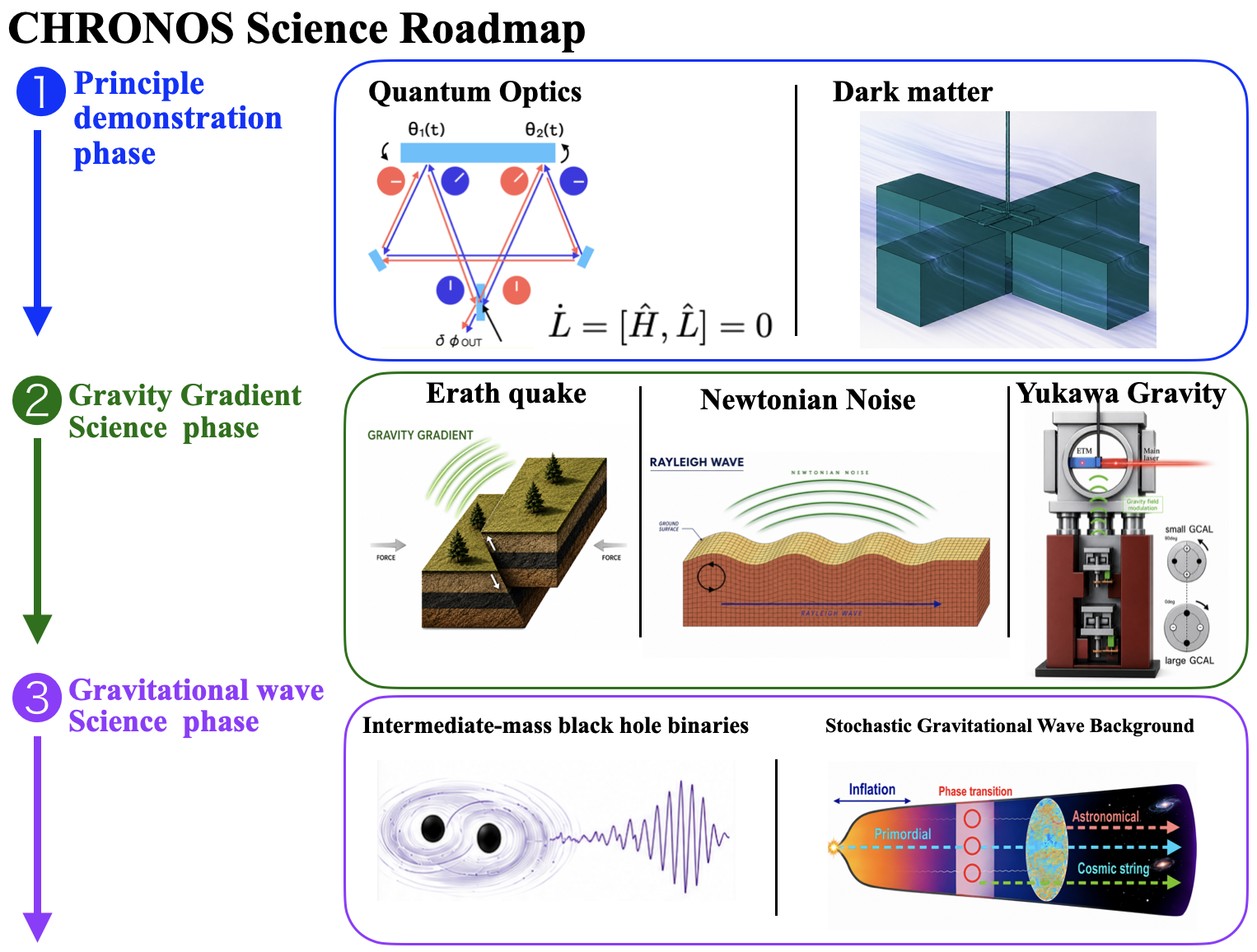}
\caption{
Science roadmap of the CHRONOS project.
The program is organized into three successive phases.
Phase~1 establishes the key measurement principles, including
speed-meter operation and quantum non-demolition measurements,
while also exploring new sensing concepts for dark-matter searches.
Phase~2 exploits the instrument for gravity-gradient science, including
earthquake-induced gravity perturbations, Newtonian-noise studies,
and laboratory searches for deviations from Newtonian gravity.
Phase~3 operates CHRONOS at its final sensitivity for sub-Hz
gravitational-wave observations, targeting stochastic gravitational-wave
backgrounds and intermediate-mass black-hole binaries.
Detailed science cases are discussed separately in the corresponding
chapters.
}
\label{fig:science_roadmap}
\end{figure*}

The overall concept is summarized in Fig.~\ref{fig:science_roadmap}. Each phase has a distinct experimental objective, while the achievements of each stage provide the technological and scientific foundation for the subsequent phase.

\section{Phase 1: Principle Demonstration}
\label{sec:roadmap_1}

The first phase focuses on establishing the key measurement principles and enabling technologies required for CHRONOS. In particular, this phase aims to demonstrate the operation of a torsion-based interferometric sensor and the optical configurations required for quantum measurements at low frequencies.

One of the primary objectives of this phase is the experimental demonstration of the speed-meter concept and quantum non-demolition (QND) measurement techniques in a macroscopic torsional system. Through these studies, we aim to establish the optical response, quantum-noise behavior, and control techniques ultimately required for CHRONOS to approach its fundamental sensitivity limits.

Within this phase, three successive experimental milestones are defined according to the technological maturity of the instrument and our understanding of its noise performance.

The first milestone is the realization of a speed-meter interferometer~\cite{Chen2003,5zyh-lg9b}. A Sagnac interferometer will be constructed, and its characteristic speed-meter response will be experimentally measured. In particular, the objective is to demonstrate the characteristic transfer function proportional to frequency, $f$, over the relevant frequency range. To our knowledge, such a speed-meter response has not yet been experimentally demonstrated in combination with a macroscopic torsion-bar system. Therefore, demonstrating the underlying measurement principle itself has scientific significance, even before the instrument reaches its ultimate displacement or strain sensitivity. An equally important objective at this stage is to establish the calibration procedure and reconstruct the calibrated strain-equivalent time series $h(t)$. These fundamental characterization measurements are intended to be performed at a relatively early stage of the experiment.

The second milestone is the demonstration of quantum non-demolition measurement~\cite{doi:10.1126/science.209.4456.547}. At this stage, the interferometer will be operated under active control and its noise spectrum will be measured quantitatively. Because the quantum-noise spectrum depends on both the optical power and the homodyne readout angle, these parameters will be systematically scanned to characterize the measurement imprecision, quantum back-action, and their correlations. From these measurements, we aim to experimentally demonstrate the QND property associated with the angular degree of freedom, corresponding to the condition
\[
[H,L]=0,
\]
where $H$ is the Hamiltonian of the measurement system and $L$ denotes the angular momentum of the torsional degree of freedom.

The third milestone is an experimental test of Ozawa's error--disturbance relation~\cite{Ozawa2003,Rozema2012,Erhart2012}. For this purpose, two dimensionless evaluation parameters are introduced. The Heisenberg indicator is defined as
\begin{equation}
\mathcal{H}(P,\xi;T)
=
\frac{
\epsilon_{\theta}(P,\xi;T)\,
\eta_{L}(P,\xi;T)
}{
C_{\theta L,T}
},
\label{eq:Heisenberg_indicator}
\end{equation}
while the Ozawa indicator is defined as
\begin{equation}
\mathcal{O}(P,\xi;T)
=
\frac{
\epsilon_{\theta}\eta_{L}
+\epsilon_{\theta}\sigma_{L}
+\sigma_{\theta}\eta_{L}
}{
C_{\theta L,T}
}.
\label{eq:Ozawa_indicator}
\end{equation}

The objective is to determine these quantities experimentally and demonstrate statistically that
\[
\mathcal{H}<1,
\qquad
\mathcal{O}>1.
\]
The condition $\mathcal{H}<1$ indicates a violation of the conventional Heisenberg error--disturbance product, whereas $\mathcal{O}>1$ confirms that the generalized Ozawa error--disturbance relation remains satisfied.

In parallel with these quantum-measurement studies, the instrument can also be used to investigate new sensing concepts for dark-matter searches. These measurements exploit the distinctive response of the CHRONOS configuration to weak forces or field-induced differential motions. Some candidate searches may be sensitive to the material composition of the mirrors or test masses, while others motivate tests of the dark-matter response of a speed-meter
configuration. A nonzero calibrated response must be established first,
as described in Chap.~\ref{chap:dark_matter}. Such science can therefore be pursued during the development of the detector, before CHRONOS reaches its ultimate gravitational sensitivity.

The scientific feasibility and appropriate timing of these measurements will be assessed and coordinated by the Science Coordinator. Based on these studies, the Instruments Coordinator will establish priorities for instrument development, taking into account both technological readiness and the availability of resources and funding. The Experimental Coordinator will then be responsible for organizing and executing the corresponding experimental program.

The primary role of Phase~1 is therefore to experimentally validate the measurement architecture on which the subsequent CHRONOS science program will be built.

\section{Phase 2: Gravity-Gradient Science}
\label{sec:roadmap2}

Once stable operation in the low-frequency band has been established, CHRONOS will move into a regime in which environmental and terrestrial gravitational fields can be measured directly. An important prerequisite for entering this second phase is that the interferometer reaches the engineering-run stage and is capable of acquiring stable observational data over extended periods. This requires the deployment of environmental monitoring systems, calibration of the detector response, and establishment of a reliable data-reconstruction pipeline capable of continuously producing trustworthy $h(t)$ data for at least approximately one hour.

Signals produced by gravity gradients are expected to be much larger than the targeted gravitational-wave signals and to appear prominently at low frequencies. By actively exploiting these signals, CHRONOS can generate unique scientific results well before reaching its ultimate gravitational-wave sensitivity. At the same time, reliable calibration in the low-frequency regime is essential for accurately reconstructing such signals. For this purpose, a gravity-field calibrator capable of generating a controlled gravitational-field modulation will become one of the key instruments in Phase~2.

In this phase, CHRONOS will observe gravitational signals associated with seismic motion and mass-density perturbations in the Earth and near the surface, while also carrying out precision gravity experiments using artificially generated gravitational fields. Phase~2 is organized around three major scientific objectives.

The first scientific objective is the observation of gravity perturbations induced by earthquakes~\cite{onglao2026prospectsobservinggravitygradientnoise}. Taiwan, where CHRONOS is planned to be installed, is a region of high seismic activity and therefore provides frequent opportunities to observe time-dependent gravitational fields associated with earthquakes. Signals that are normally treated as environmental disturbances in conventional gravitational-wave detectors can instead be exploited as scientific signals in CHRONOS. Directly targeting earthquake-induced gravitational-field variations in this way constitutes one of the distinctive science cases of CHRONOS.

The second scientific objective is the direct measurement and characterization of Newtonian noise~\cite{onglao2026prospectsobservinggravitygradientnoise}. Newtonian noise arises when environmental mass-density fluctuations, caused for example by seismic motion, atmospheric disturbances, groundwater motion, or other near-surface mass redistribution, exert direct gravitational forces on the test masses. Because this noise is gravitational in origin, it cannot be suppressed by conventional mechanical isolation. Its characterization and mitigation therefore remain active areas of research in current observatories such as LIGO and Virgo. Moreover, Newtonian noise is expected to become one of the dominant sensitivity limitations for future gravitational-wave detectors operating in the sub-Hz frequency band.

If CHRONOS can directly measure the spatial and temporal properties of Newtonian noise, these studies may contribute to the development of monitoring and mitigation techniques for future gravitational-wave observatories. For example, compact torsion-based gravity-gradient sensors could potentially be deployed as local sensors near the end stations of facilities such as LIGO or Virgo and used for environmental monitoring or feedforward Newtonian-noise subtraction. Related concepts have also been investigated in torsion-pendulum experiments such as TorPeDO. Evaluating, both theoretically and experimentally, whether CHRONOS-type sensors could serve as environmental gravity monitors for future gravitational-wave detectors therefore represents an important research direction within Phase~2.

The third scientific objective is the search for gravitational interactions beyond Newtonian gravity using dedicated gravity-field calibrators~\cite{doi:10.1126/science.209.4456.547,inoue2026probingyukawagravitymodulated}. By appropriately designing the multipole components of a rotating source mass, a gravity-field calibrator can realize a null configuration in which the Newtonian gravitational contribution to a selected observable is strongly suppressed. Such a configuration enables searches for small deviations from the Newtonian prediction while reducing the dominant conventional gravitational signal, thereby providing sensitivity to possible new interactions, including Yukawa-type forces. The updated forecast combines the Newtonian
null with signed $4f$ and $6f$ subtraction, applied to both the background
and the Yukawa response (Chap.~\ref{chap:yukawa}).

Furthermore, by varying the source-mass distribution, material composition, source-detector separation, rotation conditions, and geometrical configuration, a wide range of experimental conditions may be explored for different models of new physics. It is therefore important not to restrict the program to a single theoretical scenario, but rather to evaluate systematically which classes of new gravitational physics could be tested experimentally with CHRONOS. Under the coordination of the Science Coordinator, candidate science cases will be assessed in terms of sensitivity, feasibility, and required experimental configuration, with the aim of translating the most promising scenarios into concrete measurement programs.

In this way, Phase~2 combines geophysical studies based on naturally occurring gravity gradients with precision gravitational experiments based on artificially controlled gravitational fields. Through these activities, CHRONOS will deepen its understanding of the detector response at low frequencies, calibration, environmental gravitational fields, and gravity-gradient noise.

Phase~2 therefore serves not only as an independent program of gravitational and geophysical science, but also as a crucial bridge connecting detector characterization and precision gravitational physics to the eventual realization of sub-Hz gravitational-wave observations.
\section{Phase 3: Gravitational-Wave Science}
\label{sed:roadmap3}

Phase~3 represents the gravitational-wave observation phase of CHRONOS, to be carried out once the detector has reached its final design sensitivity. At this stage, CHRONOS will enter full scientific operation as a sub-Hz gravitational-wave observatory.

Through Phases~1 and~2, the suspension performance, optical sensitivity, low-frequency control, environmental-noise mitigation, calibration, and understanding of gravity-gradient noise will be progressively improved. Building on these developments, CHRONOS aims ultimately to achieve the sensitivity required to search for astrophysical and cosmological gravitational-wave signals. Unlike the earlier phases, which focus primarily on principle demonstrations and detector characterization, the principal objective of Phase~3 is long-duration scientific observation at the final detector sensitivity.

The first major scientific objective is the search for a stochastic gravitational-wave background (SGWB)~\cite{Inoue_CHRONOS_Main_2026,SPP-2026-3A-06}. The sub-Hz band targeted by CHRONOS lies between the frequency ranges explored by current ground-based interferometers and planned space-based gravitational-wave observatories, leaving a frequency region that remains comparatively unexplored. Phase~3 aims to achieve world-leading sensitivity in this observational gap and to open new parameter space for searches for stochastic gravitational-wave backgrounds. A longer-term upgrade objective is to test specified spectra below the
integrated Big Bang Nucleosynthesis (BBN) bound. The reference PI curve in
Chap.~\ref{chap:sensitivity} does not establish that capability; an integral
cosmological bound is not a pointwise detector threshold.

A robust search for a stochastic gravitational-wave background will also require careful consideration of joint observations with other detectors. With a single detector, it is difficult to completely distinguish a stochastic gravitational-wave signal from local instrumental noise and correlated environmental disturbances. Therefore, in addition to improving the sensitivity of CHRONOS itself, joint observations with other gravitational-wave detectors or related instruments, cross-correlation analyses, and characterization of correlated environmental noise will be essential. Prior to Phase~3, theoretical and experimental studies will therefore be required to identify the most effective detector combinations and to establish the necessary observing duration, frequency coverage, timing synchronization, data-sharing framework, and analysis methodology. Assessing the feasibility of a future joint-observation network is itself an important research objective for CHRONOS.

The second major scientific objective is the search for intermediate-mass black-hole (IMBH) binaries and the observational investigation of the intermediate-mass black-hole population~\cite{Inoue_CHRONOS_Main_2026,SPP-2026-3A-06}. The existence, abundance, and formation mechanisms of black holes between the stellar-mass and supermassive regimes remain incompletely understood. CHRONOS will approach this problem from a frequency band different from that of existing ground-based gravitational-wave observatories. By directly searching for gravitational waves from IMBH binaries, CHRONOS aims to provide observational evidence for this population and to place new constraints on black-hole formation histories and hierarchical growth scenarios.

Multiband observations with other gravitational-wave detectors may also provide important opportunities for IMBH science. If CHRONOS can observe the inspiral of an IMBH binary in the sub-Hz band before the signal enters higher-frequency detectors, combining these observations could improve source-parameter estimation and enable consistency tests of gravitational-wave propagation across widely separated frequency bands. The conditions required for such joint observations must therefore also be investigated in the context of future gravitational-wave detector networks.

These two major science targets provide key benchmarks for defining the sensitivity, long-term stability, calibration accuracy, observational duty cycle, and coordination requirements of the final CHRONOS configuration. The detailed scientific reach of each target is discussed separately in the corresponding science chapters.

Phase~3 therefore represents the culmination of the CHRONOS roadmap, integrating the technologies and understanding established through the principle demonstrations of Phase~1 and the low-frequency gravity science of Phase~2. At its final sensitivity, CHRONOS aims to open a new observational window for cosmological and astrophysical gravitational-wave science in the comparatively unexplored sub-Hz frequency band.

\section{Evolution of the CHRONOS Program}
\label{sec:roadmap_evolution}

The three phases should not be regarded as independent experiments. Instead,
they form a continuous progression in which each stage both delivers scientific
results and validates the technologies required for the next stage.

The principle-demonstration phase establishes the quantum measurement,
interferometric sensing, and control techniques. The gravity-gradient science
phase extends the experiment to precision measurements of real gravitational
fields and provides direct characterization of one of the dominant
environmental backgrounds in the sub-Hz band. These developments ultimately
converge in the gravitational-wave science phase, in which the full detector
concept is exploited for astrophysical and cosmological observations.

This staged strategy allows CHRONOS to produce scientifically meaningful results
throughout its development while progressively reducing the technical risk
associated with the final observatory. The roadmap therefore connects detector
development and scientific exploitation from the earliest prototype
measurements to the long-term goal of sub-Hz gravitational-wave astronomy.

%% file: Chapters/Quantum.tex
\chapter{Quantum measurement}
\label{sec:quantum_measurement}

The quantitative detector baseline is the full multi-input model in
Chap.~\ref{chap:sensitivity}. The ideal speed-meter response used below
organizes the staged measurement tests; it does not replace the calibrated
transfer function or imply complete cancellation of the rear-port vacuum
back action. All comparisons to detector sensitivity use the reference
operating points and angle-to-strain convention of that chapter.

CHRONOS provides a unique platform for investigating quantum
measurement with a macroscopic torsional mechanical system.
In addition to its primary scientific objectives as a low-frequency
precision interferometer, the combination of a torsion-bar suspension,
a Sagnac speed-meter topology, radiation-pressure actuation, and
homodyne readout enables a systematic study of the quantum measurement
process itself.

The quantum measurement program of CHRONOS is organized into three
successive scientific stages.
Each stage addresses a distinct question in quantum measurement and,
at the same time, establishes the experimental capability required for
the next stage.

The first stage is the experimental demonstration of the
\textit{speed-meter response}.
The central question is whether the CHRONOS interferometer measures
the velocity-like observable predicted for the Sagnac speed-meter
topology, rather than simply behaving as a conventional position
meter.
This can be tested through controlled excitation of the torsion bar
and measurement of the complex transfer function of the
interferometer.
Importantly, this first demonstration does not require operation at
the quantum-noise limit and can therefore be performed during the
early commissioning phase of CHRONOS.

The second stage is the demonstration of
\textit{quantum non-demolition (QND) measurement}.
Once the speed-meter response has been established, the experiment
moves from characterization of the classical optical response to
measurement of quantum noise.
By varying the optical power, the measurement strength can be changed,
allowing the competition between measurement imprecision and
radiation-pressure back-action to be investigated.
Balanced homodyne detection provides an additional degree of freedom:
the measured optical quadrature can be varied through the homodyne
angle, enabling the correlation between measurement imprecision and
quantum back-action to be controlled.
The objective of this stage is to demonstrate that the speed-meter
topology provides measurable quantum back-action suppression or
back-action evasion relative to a conventional position measurement.

The third stage extends these measurements to a fundamental test of
quantum measurement theory through
\textit{Ozawa's error--disturbance relation}.
In this stage, measurement imprecision and radiation-pressure
back-action are no longer treated only as contributions to the total
noise spectrum.
Instead, they are independently calibrated and interpreted as the
measurement error $\epsilon_\theta$ and the measurement disturbance
$\eta_L$ of the torsional degree of freedom.
Here, $\theta$ denotes the angular coordinate of the torsion bar and
$L$ denotes its conjugate angular momentum.
Together with the initial-state fluctuations $\sigma_\theta$ and
$\sigma_L$, these quantities allow a direct experimental evaluation
of the generalized error--disturbance relation.

For the canonical pair $(\theta,L)$, we define
\begin{equation}
    C_{\theta L}
    \equiv
    \frac{1}{2}
    \left|
        \left\langle
            [\theta,L]
        \right\rangle
    \right|.
    \label{eq:commutator_scale}
\end{equation}
For an ideal canonical angular coordinate and angular momentum,
$[\theta,L]=i\hbar$, and therefore $C_{\theta L}=\hbar/2$.
Using this scale, the conventional Heisenberg-type error--disturbance
relation can be written as
\begin{equation}
    \epsilon_\theta \eta_L
    \geq
    C_{\theta L}.
    \label{eq:heisenberg_edr_intro}
\end{equation}

We introduce the dimensionless Heisenberg indicator
\begin{equation}
    \mathcal{H}
    \equiv
    \frac{
        \epsilon_\theta \eta_L
    }{
        C_{\theta L}
    }.
    \label{eq:heisenberg_indicator_intro}
\end{equation}
The conventional Heisenberg-type error--disturbance relation therefore
corresponds to
\begin{equation}
    \mathcal{H}\geq1.
\end{equation}

Ozawa's generalized error--disturbance relation is given by
\begin{equation}
    \epsilon_\theta \eta_L
    +
    \epsilon_\theta \sigma_L
    +
    \sigma_\theta \eta_L
    \geq
    C_{\theta L}.
    \label{eq:ozawa_intro}
\end{equation}
Correspondingly, we define the dimensionless Ozawa indicator as
\begin{equation}
    \mathcal{O}
    \equiv
    \frac{
        \epsilon_\theta \eta_L
        +
        \epsilon_\theta \sigma_L
        +
        \sigma_\theta \eta_L
    }{
        C_{\theta L}
    }.
    \label{eq:ozawa_indicator_intro}
\end{equation}
Ozawa's generalized relation is therefore expressed simply as
\begin{equation}
    \mathcal{O}\geq1.
\end{equation}

These two dimensionless quantities provide a convenient experimental
representation of the error--disturbance measurement.
In particular, the regime
\begin{equation}
    \mathcal{H}<1,
    \qquad
    \mathcal{O}\geq1
    \label{eq:ozawa_target_region}
\end{equation}
is of particular interest.
Observation of this regime would demonstrate that the simple
Heisenberg-type error--disturbance product is not universally valid,
while the generalized Ozawa relation remains satisfied.

By varying the optical power and homodyne angle, CHRONOS can explore a
controlled family of quantum measurements.
The optical power controls the measurement strength and hence the
balance between measurement imprecision and radiation-pressure
back-action, whereas the homodyne angle controls the measured optical
quadrature and the correlation between these two noise contributions.
As a result, the quantities $\epsilon_\theta$, $\eta_L$,
$\sigma_\theta$, and $\sigma_L$, and consequently the indicators
$\mathcal{H}$ and $\mathcal{O}$, can be mapped over a controlled region
of the error--disturbance parameter space.

The three stages therefore form a progressive scientific roadmap,
\begin{equation}
    \text{Speed-meter response}
    \,\longrightarrow\,
    \text{QND measurement}
    \,\longrightarrow\,
    \text{Ozawa's error--disturbance relation}.
    \label{eq:quantum_science_roadmap}
\end{equation}

This staged approach is important experimentally.
The first stage establishes the relevant mechanical and optical
transfer functions.
The second stage establishes control and measurement of quantum
back-action and its correlations with measurement imprecision.
The third stage builds on both capabilities to reconstruct the
measurement error and disturbance in absolute physical units and to
evaluate the dimensionless indicators $\mathcal{H}$ and $\mathcal{O}$.
Thus, each stage provides not only an independent scientific result
but also the calibration and experimental foundation for the
subsequent stage.

The remainder of this section is organized as follows.
Section~\ref{sec:stage1_speedmeter} describes the demonstration of the
speed-meter response.
Section~\ref{sec:stage2_qnd} presents the strategy for demonstrating
QND measurement and quantum back-action suppression.
Section~\ref{sec:stage3_ozawa} describes the proposed test of Ozawa's
error--disturbance relation and the experimental reconstruction of
$\mathcal{H}$ and $\mathcal{O}$.
Finally, Sec.~\ref{sec:required_technologies} summarizes the key
technologies required to realize the complete quantum measurement
program.

\section{Stage 1: Demonstration of the Speed-Meter Response}
\label{sec:stage1_speedmeter}

The first stage of the CHRONOS quantum measurement program is the
experimental demonstration of the speed-meter response.
The purpose of this stage is to establish that the interferometer
responds to the velocity-like observable expected from the Sagnac
speed-meter topology, rather than simply behaving as a conventional
position meter.

Importantly, Stage~1 is a classical transfer-function measurement.
Operation at the quantum-noise limit is not required.
The experiment can therefore be performed during the early
commissioning phase using the torsion-bar suspension, the coil-magnet
actuator, and a mid-fringe interferometer lock.

For the torsional degree of freedom, let $\theta(\Omega)$ denote the
Fourier amplitude of the calibrated angular coordinate in radians,
and let $\Omega=2\pi f$ for frequency $f$ in hertz. For the main science
channel this coordinate is the differential bar angle; an individual-bar
measurement must be identified separately.
The interferometer readout is represented by the error signal
$d_{\rm err}(\Omega)$, which is the experimentally recorded sensing
channel in digital counts. This chapter's \emph{Sensing Function}
$H_{\theta\rightarrow d_{\rm err}}$ is the angle-to-error-signal response,
in counts/rad, evaluated for a deterministic small signal. If feedback
is active, its effect must be removed or included explicitly in the model.
We define this complex transfer function as
\begin{equation}
    H_{\theta\rightarrow d_{\rm err}}(\Omega)
    \equiv
    \frac{d_{\rm err}(\Omega)}{\theta(\Omega)}.
    \label{eq:theta_to_error}
\end{equation}

The superscripts PM and SM denote position-meter and speed-meter responses,
respectively. The PM optical/readout gain $G_{\rm opt}^{\rm PM}$ maps
angle to error counts. For a conventional position meter,
\begin{equation}
    H_{\theta\rightarrow d_{\rm err}}^{\rm PM}(\Omega)
    =
    G_{\rm opt}^{\rm PM}(\Omega),
\end{equation}
where $G_{\rm opt}^{\rm PM}(\Omega)$ represents the optical sensing
gain from angular displacement to the measured error signal.
In the frequency region well below the characteristic optical
frequencies, the position-meter response is approximately constant,
\begin{equation}
    H_{\theta\rightarrow d_{\rm err}}^{\rm PM}(\Omega)
    \simeq
    {\rm const.}
\end{equation}

In a speed meter, the same mechanical degree of freedom is sampled at
two different times separated by an effective optical delay $\tau$.
The measured signal is therefore proportional to the difference
between two successive measurements of the angular displacement,
\begin{equation}
    \theta(t+\tau)-\theta(t).
\end{equation}
Here, $\tau$ should be understood as the effective delay associated
with the Sagnac optical response, rather than simply as a single-pass
light-travel time.

We define the speed-meter response function as
\begin{equation}
    H^{\rm SM}(\Omega)
    \equiv
    e^{-i\Omega\tau}-1.
    \label{eq:H_SM}
\end{equation}

Let $G_{\rm opt}$ denote the remaining optical and readout-chain response
after factoring out the dimensionless delay response $H^{\rm SM}$.
The angle-to-error-signal transfer function can then be written as
\begin{equation}
    H_{\theta\rightarrow d_{\rm err}}(\Omega)
    =
    G_{\rm opt}(\Omega)
    H^{\rm SM}(\Omega),
    \label{eq:H_theta_derr_SM}
\end{equation}
where $G_{\rm opt}(\Omega)$ contains the remaining optical sensing
response of the interferometer, including any frequency dependence
associated with the optical cavities, recycling configuration, and
readout chain.

Therefore, the measured error signal is
\begin{equation}
    d_{\rm err}(\Omega)
    =
    G_{\rm opt}(\Omega)
    H^{\rm SM}(\Omega)
    \theta(\Omega).
\end{equation}

In the low-frequency limit,
\begin{equation}
    \Omega\tau\ll1,
\end{equation}
the speed-meter response becomes
\begin{equation}
    H^{\rm SM}(\Omega)
    =
    e^{-i\Omega\tau}-1
    \simeq
    -i\Omega\tau.
    \label{eq:H_SM_low_frequency}
\end{equation}

Consequently,
\begin{equation}
    H_{\theta\rightarrow d_{\rm err}}(\Omega)
    \simeq
    -i\Omega\tau
    G_{\rm opt}(\Omega).
    \label{eq:H_theta_derr_low_frequency}
\end{equation}

If the optical response $G_{\rm opt}(\Omega)$ varies slowly over the
measurement band, the characteristic low-frequency behavior is
therefore
\begin{equation}
    \left|
    H_{\theta\rightarrow d_{\rm err}}(\Omega)
    \right|
    \propto
    \Omega,
\end{equation}
with a phase shift corresponding to the time derivative of the angular
displacement.

This relation provides the direct connection between the theoretical
speed-meter response $H^{\rm SM}(\Omega)$ and the experimentally
measured transfer function
$H_{\theta\rightarrow d_{\rm err}}(\Omega)$.
In practice, the identification of the speed-meter response should be
based on a comparison of the full complex transfer function, including
both its amplitude and phase, with the complete optical model rather
than solely on the asymptotic
$|H_{\theta\rightarrow d_{\rm err}}|\propto\Omega$ behavior.

\subsection{Transfer-Function Measurement with the Coil Actuator}

The speed-meter response is measured by applying a controlled
sinusoidal excitation to the torsion bar using the coil-magnet
actuator.

We denote the digital control signal sent to the coil actuator by
$d_{\rm ctrl}(\Omega)$.
The actuator transfer function $H_{\rm coil}(\Omega)$ converts this
digital control signal into an applied torque,
\begin{equation}
    \tau_{\rm coil}(\Omega)
    =
    H_{\rm coil}(\Omega)
    d_{\rm ctrl}(\Omega).
\end{equation}

The torsion-bar motion is
\begin{equation}
    \theta(\Omega)
    =
    \chi_\theta(\Omega)
    \tau_{\rm coil}(\Omega),
\end{equation}
where $\chi_\theta(\Omega)$ is the mechanical angular susceptibility
of the torsional degree of freedom.

The measured interferometer error signal is therefore
\begin{align}
    d_{\rm err}(\Omega)
    &=
    H_{\theta\rightarrow d_{\rm err}}(\Omega)
    \theta(\Omega)
    \nonumber\\
    &=
    G_{\rm opt}(\Omega)
    H^{\rm SM}(\Omega)
    \chi_\theta(\Omega)
    H_{\rm coil}(\Omega)
    d_{\rm ctrl}(\Omega).
\end{align}

Thus, the directly measured control-to-error transfer function is
\begin{equation}
    H_{d_{\rm ctrl}\rightarrow d_{\rm err}}(\Omega)
    \equiv
    \frac{
        d_{\rm err}(\Omega)
    }{
        d_{\rm ctrl}(\Omega)
    }
    =
    G_{\rm opt}(\Omega)
    H^{\rm SM}(\Omega)
    \chi_\theta(\Omega)
    H_{\rm coil}(\Omega).
    \label{eq:dctrl_to_derr_SM}
\end{equation}

After removing the independently characterized actuator response and
the mechanical susceptibility,
\begin{equation}
    H_{\theta\rightarrow d_{\rm err}}(\Omega)
    =
    \frac{
        H_{d_{\rm ctrl}\rightarrow d_{\rm err}}(\Omega)
    }{
        \chi_\theta(\Omega)
        H_{\rm coil}(\Omega)
    },
    \label{eq:reconstruct_theta_to_derr}
\end{equation}
the experimentally reconstructed optical response can be directly
compared with the model
\begin{equation}
    H_{\theta\rightarrow d_{\rm err}}^{\rm model}(\Omega)
    =
    G_{\rm opt}(\Omega)
    H^{\rm SM}(\Omega).
\end{equation}

The essential test of Stage~1 is therefore whether the reconstructed
$H_{\theta\rightarrow d_{\rm err}}(\Omega)$ contains the
characteristic speed-meter factor
\begin{equation}
    H^{\rm SM}(\Omega)
    =
    e^{-i\Omega\tau}-1,
\end{equation}
in both amplitude and phase.

At low frequencies, the speed-meter response behaves as
\begin{equation}
    H_{\theta\rightarrow d_{\rm err}}^{\rm SM}(\Omega)
    \propto
    -i\Omega,
\end{equation}
whereas a conventional position meter approaches
\begin{equation}
    H_{\theta\rightarrow d_{\rm err}}^{\rm PM}(\Omega)
    \propto
    {\rm const.}
\end{equation}

The distinct frequency dependence and phase evolution of these two
responses provide the principal experimental signature of the
speed-meter behavior.
A successful Stage~1 demonstration will therefore require agreement
between the measured and modeled complex transfer functions over a
finite frequency band, rather than identification of a single
frequency point or amplitude scaling alone.
\section{Stage 2: Demonstration of QND Measurement}
\label{sec:stage2_qnd}
Stage~2 aims to demonstrate quantum non-demolition (QND) measurement using the CHRONOS Sagnac speed-meter topology. Following the classical characterization of the speed-meter response in Stage~1, Stage~2 focuses on the quantum measurement process, in particular the suppression of radiation-pressure back-action relative to a conventional position measurement.

\subsection{Quantum-Noise Measurement}

The interferometer is operated near the dark fringe, with balanced homodyne detection (BHD) providing the primary science readout. The calibrated torsional observable is written as
\begin{equation}
\theta_{\rm meas}
=
\theta_{\rm sig}
+
\theta_{\rm imp}
+
\chi_\theta \tau_{\rm BA},
\end{equation}
where $\theta_{\rm imp}$ is the measurement-imprecision noise, $\tau_{\rm BA}$ is the radiation-pressure back-action torque, and $\chi_\theta$ is the mechanical susceptibility of the torsional degree of freedom.

The corresponding quantum-noise spectrum is
\begin{equation}
S_{\theta}^{\rm quant}
=
S_{\theta}^{\rm imp}
+
|\chi_\theta|^2 S_{\tau}^{\rm BA}
+
2\,{\rm Re}
\left[
\chi_\theta S_{\theta\tau}
\right],
\label{eq:quantum_noise}
\end{equation}
where $S_{\theta\tau}$ is the imprecision--back-action cross-spectral density. In conventional interferometer terminology, these three terms correspond to shot noise, radiation-pressure noise, and their correlation, respectively.

For an uncorrelated position measurement, $S_{\theta\tau}=0$. In a homodyne measurement, however, the correlation can be controlled through the detected optical quadrature and therefore constitutes an essential element of the QND measurement.

\subsection{Power and Homodyne-Angle Scans}

Stage~2 uses the optical power $P$ and homodyne angle $\zeta$ as two independent experimental control parameters. The optical quadrature detected by the BHD is
\begin{equation}
\hat b_\zeta
=
\hat b_1\cos\zeta
+
\hat b_2\sin\zeta .
\end{equation}
Varying $\zeta$ changes the detected quadrature and hence the correlation between measurement imprecision and radiation-pressure back-action.

The optical power controls the measurement strength. In the quantum-noise-dominated regime,
\begin{equation}
\sqrt{S_{\theta}^{\rm imp}}
\propto P^{-1/2},
\qquad
\sqrt{S_{\tau}^{\rm BA}}
\propto P^{1/2},
\end{equation}
and therefore
\begin{equation}
\sqrt{S_{\theta}^{\rm RP}}
=
|\chi_\theta|
\sqrt{S_{\tau}^{\rm BA}}
\propto
|\chi_\theta|P^{1/2}.
\end{equation}

The measurement is performed over a set of optical powers and homodyne angles,
\begin{equation}
P=P_1,\ldots,P_N,
\qquad
\zeta=\zeta_1,\ldots,\zeta_M,
\end{equation}
yielding the two-dimensional data set
\begin{equation}
S_\theta(\Omega;P,\zeta).
\end{equation}

The power scan separates contributions according to their measurement-strength dependence: the imprecision contribution decreases with increasing power, whereas radiation-pressure back-action increases. The homodyne-angle scan provides an independent handle on the imprecision--back-action correlation through the measured optical quadrature. Technical noise that is approximately independent of $P$ and $\zeta$ can thereby be distinguished from these quantum-noise contributions.

The full data set is simultaneously compared with
\begin{align}
S_{\theta}^{\rm model}(\Omega;P,\zeta)
&=
S_{\theta}^{\rm imp}(\Omega;P,\zeta)
+
|\chi_\theta(\Omega)|^2
S_{\tau}^{\rm BA}(\Omega;P)
\nonumber\\
&\quad+
2\,{\rm Re}
\left[
\chi_\theta(\Omega)
S_{\theta\tau}(\Omega;P,\zeta)
\right].
\label{eq:stage2_model}
\end{align}
A global fit over the $(P,\zeta)$ parameter space therefore constrains the measurement imprecision, radiation-pressure back-action, and their correlation within a single quantum-measurement model.

\subsection{QND Signature and Success Criterion}

The principal QND signature in Stage~2 is the suppression of radiation-pressure back-action by the Sagnac speed-meter topology. For a conventional position meter,
\begin{equation}
S_{\theta}^{\rm shot}\propto P^{-1},
\qquad
S_{\theta}^{\rm RP}\propto P,
\end{equation}
which gives rise to the conventional quantum measurement trade-off. In a speed meter, sequential sampling of the mechanical degree of freedom suppresses the low-frequency coupling of radiation-pressure back-action to the measured observable.

The most direct test would be to operate the same apparatus in both speed-meter and position-meter configurations. However, implementing an equivalent position-meter configuration in CHRONOS would require substantial modifications to the optical layout and control system. The baseline Stage~2 test therefore compares the experimentally inferred speed-meter back-action,
\begin{equation}
S_{\theta}^{\rm RP,SM}(\Omega),
\end{equation}
with the prediction of a calibrated conventional position-meter model,
\begin{equation}
S_{\theta}^{\rm RP,PM,model}(\Omega),
\end{equation}
whose optical and mechanical parameters are constrained by independent measurements.

Stage~2 is considered successful when the measured spectra across the $(P,\zeta)$ parameter space are consistently described by the quantum-noise model and the inferred radiation-pressure back-action is suppressed or coherently evaded relative to the calibrated position-meter prediction.

Observation of
\begin{equation}
S_{\theta}^{\rm SM}(\Omega)
<
S_{\theta}^{\rm SQL}(\Omega)
\end{equation}
over a finite frequency band would provide a stronger demonstration of QND performance by surpassing the conventional position-meter standard quantum limit. However, sub-SQL operation is not adopted as a necessary condition for the Stage~2 milestone.

The measurement imprecision, back-action, and their correlation calibrated through the $(P,\zeta)$ scans provide the experimental basis for Stage~3, where they are quantitatively interpreted as the measurement error $\epsilon_\theta$ and disturbance $\eta_L$ and used to evaluate the Heisenberg and Ozawa indicators $\mathcal{H}$ and $\mathcal{O}$.
\section{Stage 3: Test of Ozawa's Error--Disturbance Relation}
\label{sec:stage3_ozawa}

The third stage of the CHRONOS quantum-measurement program aims at a
quantitative test of Ozawa's generalized error--disturbance relation
using the torsional degree of freedom as a macroscopic continuous
quantum-measurement system.

For the angular coordinate $\theta$ and its canonically conjugate
angular momentum $L$,
\begin{equation}
    [\theta,L]=i\hbar.
\end{equation}

For an experimentally defined measurement mode, we introduce
\begin{equation}
    C_{\theta L}
    \equiv
    \frac{1}{2}
    \left|
        \left\langle
            [\Theta_w,L_w]
        \right\rangle
    \right|,
\end{equation}
which reduces to
\begin{equation}
    C_{\theta L}=\frac{\hbar}{2}
\end{equation}
for a canonically normalized mode.

The conventional Heisenberg-type error--disturbance product is
characterized by
\begin{equation}
    \mathcal{H}
    \equiv
    \frac{
        \epsilon_\theta\eta_L
    }{
        C_{\theta L}
    },
\end{equation}
where $\epsilon_\theta$ is the measurement error of the angular
coordinate and $\eta_L$ is the measurement-induced disturbance of the
angular momentum.

Ozawa's generalized relation additionally depends on the intrinsic
fluctuations of the initial mechanical state,
$\sigma_\theta$ and $\sigma_L$,
and can be written as
\begin{equation}
    \mathcal{O}
    \equiv
    \frac{
        \epsilon_\theta\eta_L
        +
        \epsilon_\theta\sigma_L
        +
        \sigma_\theta\eta_L
    }{
        C_{\theta L}
    }.
\end{equation}

The conventional Heisenberg-type relation corresponds to
\begin{equation}
    \mathcal{H}\geq1,
\end{equation}
whereas Ozawa's relation requires
\begin{equation}
    \mathcal{O}\geq1.
\end{equation}

The scientifically distinctive target is therefore the regime
\begin{equation}
        \mathcal{H}<1,
        \qquad
        \mathcal{O}\geq1
    .
\end{equation}

Stage~2 provides the experimental basis for this program by measuring
and characterizing measurement imprecision, radiation-pressure
back-action, and their correlation.
Stage~3 extends this framework by additionally requiring a calibrated
determination of the intrinsic fluctuations of the mechanical state.

A complete experimental implementation therefore requires reliable
estimates of
\begin{equation}
    \epsilon_\theta,
    \qquad
    \eta_L,
    \qquad
    \sigma_\theta,
    \qquad
    \sigma_L,
\end{equation}
together with their statistical and calibration uncertainties.

The detailed methodology for reconstructing these quantities,
particularly the initial-state variances
$\sigma_\theta$ and $\sigma_L$ in the presence of active feedback and
quantum back-action, is currently under investigation.
Possible approaches include calibrated time-domain reconstruction,
spectral decomposition using controlled changes of the measurement
strength and readout quadrature, and independent reconstruction of the
mechanical state from multiple sensing and control channels.

At the present stage, CHRONOS therefore defines the Stage~3 goal at the
level of the experimentally testable error--disturbance relation,
while leaving the detailed state-reconstruction and estimation
procedure for future dedicated studies.

The development of this methodology will form part of the transition
from the Stage~2 QND experiment to a quantitative test of generalized
quantum measurement theory.
\section{Required Technologies and Minimum Implementation Path}
\label{sec:required_technologies}

The CHRONOS quantum-measurement program is designed as a staged
experimental development.
Each stage adds only the technologies required for the corresponding
scientific milestone.

The minimum progression is
\begin{equation}
\begin{aligned}
    \text{Stage~1: Speed-meter response}
    &\longrightarrow
    \text{Stage~2: QND measurement}
    \\
    &\longrightarrow
    \text{Stage~3: Error--disturbance test}.
\end{aligned}
\label{eq:minimum_implementation_path}
\end{equation}

The detailed measurement procedures are described in the preceding
sections.
Here we summarize only the additional experimental capabilities
required at each stage.

\subsection{Stage~1 Requirements}

Stage~1 requires the minimum system needed to measure the classical
speed-meter response.

The essential technologies are
\begin{enumerate}
    \item a torsion-bar suspension with a reproducible angular
    response;

    \item a Sagnac or equivalent two-time optical interaction with the
    same torsional degree of freedom;

    \item a coil-magnet actuator for controlled excitation;

    \item a linear interferometric readout, such as a mid-fringe
    sensing configuration;

    \item digital excitation and acquisition of
    $d_{\rm ctrl}$ and $d_{\rm err}$ for complex transfer-function
    measurements.
\end{enumerate}

Quantum-noise-limited operation, balanced homodyne detection, absolute
torque calibration, and state reconstruction are not required at this
stage.

\subsection{Stage~2 Requirements}

Stage~2 adds the technologies required to measure and separate quantum
noise and to demonstrate suppression of radiation-pressure
back-action.

The additional capabilities are
\begin{enumerate}
    \item stable dark-fringe operation;

    \item balanced homodyne detection of the science output;

    \item calibrated control of the optical power $P$;

    \item calibrated control of the homodyne angle $\zeta$;

    \item sufficient optomechanical coupling to resolve
    radiation-pressure back-action;

    \item calibrated angular reconstruction of the BHD science
    channel;

    \item acquisition and analysis of auto- and cross-spectral
    densities over multiple $(P,\zeta)$ configurations.
\end{enumerate}

Resonant optical enhancement and power recycling are introduced when
required to achieve sufficient circulating power and quantum
sensitivity.
Signal recycling remains optional and is introduced only if
frequency-dependent shaping of the optical response is needed.

A direct conversion of the CHRONOS interferometer to a conventional
position-meter configuration is not required.
The baseline comparison is made with a calibrated position-meter model
constrained by independently measured CHRONOS parameters.

\subsection{Stage~3 Requirements}

Stage~3 adds the capabilities required to convert the Stage~2
quantum-noise measurements into the absolute error--disturbance
quantities
$\epsilon_\theta$, $\eta_L$, $\sigma_\theta$, and $\sigma_L$.

The additional technologies are
\begin{enumerate}
    \item absolute calibration of the BHD angular response;

    \item PCAL calibration of the radiation-pressure torque;

    \item GCAL cross-calibration of the low-frequency torque scale;

    \item calibrated reconstruction of the torsion-bar angular
    displacement $\theta(t)$ and angular velocity $\dot{\theta}(t)$;

    \item two independent reconstruction pipelines for the mechanical
    state;

    \item cross-spectral and coherence validation of the two
    reconstructed signals;

    \item propagation of calibration, statistical, and model
    uncertainties into $\mathcal{H}$ and $\mathcal{O}$.
\end{enumerate}

The detailed reconstruction of
$\epsilon_\theta$, $\eta_L$, $\sigma_\theta$, and $\sigma_L$
is described in Sec.~\ref{sec:stage3_ozawa}.
The essential additional capability of Stage~3 is therefore absolute
calibration and independently validated reconstruction of the
torsional mechanical state.

\subsection{Technology Progression}

The staged implementation can be summarized as
\begin{equation}
\begin{aligned}
\text{Stage~1:}\quad
&
\text{torsion bar}
+
\text{two-time optical interaction}
+
\text{coil actuation}
+
d_{\rm ctrl}/d_{\rm err}
\\[3pt]
\longrightarrow\quad
\text{Stage~2:}\quad
&
\text{dark fringe}
+
\text{BHD}
+
P\text{ scan}
+
\zeta\text{ scan}
+
\text{quantum-noise decomposition}
\\[3pt]
\longrightarrow\quad
\text{Stage~3:}\quad
&
\text{PCAL/GCAL}
+
\text{two reconstruction pipelines}
+
\text{absolute calibration}
\\
&
\hspace{2cm}
\longrightarrow
\mathcal{H},\mathcal{O}.
\end{aligned}
\label{eq:technology_progression}
\end{equation}

\begin{table*}[t]
    \centering
    \caption{
    Summary of the principal technologies required for the three stages
    of the CHRONOS quantum-measurement program.
    }
    \label{tab:minimum_technology_matrix}
    \setlength{\tabcolsep}{5pt}
    \begin{tabular}{lccc}
        \hline
        Technology
        & Stage~1
        & Stage~2
        & Stage~3
        \\
        \hline

        Torsion-bar suspension
        & Required
        & Required
        & Required
        \\

        Two-time Sagnac interaction
        & Required
        & Required
        & Required
        \\

        Coil-magnet actuation
        & Required
        & Diagnostic
        & Diagnostic
        \\

        $d_{\rm ctrl}$ / $d_{\rm err}$ transfer-function measurement
        & Required
        & Useful
        & Useful
        \\

        Mid-fringe operation
        & Required
        & Not required
        & Not required
        \\

        Dark-fringe operation
        & Not required
        & Required
        & Required
        \\

        Balanced homodyne detection
        & Not required
        & Required
        & Required
        \\

        Input-power scan
        & Not required
        & Required
        & Required
        \\

        Homodyne-angle scan
        & Not required
        & Required
        & Required
        \\

        Quantum-noise decomposition
        & Not required
        & Required
        & Required
        \\

        Resonant optical enhancement
        & Not required
        & Conditional
        & Conditional
        \\

        Power recycling
        & Not required
        & Conditional
        & Conditional
        \\

        Signal recycling
        & Not required
        & Optional
        & Optional
        \\

        Absolute angular calibration
        & Not required
        & Required
        & Required
        \\

        PCAL torque calibration
        & Not required
        & Useful
        & Required
        \\

        GCAL torque cross-calibration
        & Not required
        & Not required
        & Validation
        \\

        Reconstruction of $\theta(t)$
        & Useful
        & Required
        & Required
        \\

        Reconstruction of $\dot{\theta}(t)$
        & Useful
        & Useful
        & Required
        \\

        Two independent reconstruction pipelines
        & Not required
        & Useful
        & Required
        \\

        Cross-spectrum/coherence validation
        & Not required
        & Useful
        & Required
        \\

        Evaluation of $\mathcal{H}$ and $\mathcal{O}$
        & Not required
        & Not required
        & Required
        \\

        \hline
    \end{tabular}
\end{table*}

In this staged strategy, Stage~1 establishes the speed-meter response,
Stage~2 establishes quantum back-action measurement and control, and
Stage~3 adds the absolute calibration and validated mechanical-state
reconstruction required for a quantitative test of quantum measurement
theory.

%% file: Chapters/Dark_matter.tex
\chapter{Dark-Matter Searches}
\label{chap:dark_matter}

\section{Scope and present objective}
\label{sec:dm_scope}

The technologies developed for gravitational-wave observatories provide a broad
platform for dark-matter searches.  Suspended test masses, long-baseline laser
interferometry, low-noise angular readout, stabilized lasers, optical cavities,
polarization optics, and continuously recorded auxiliary channels can respond
not only to gravitational waves, but also to weak forces and changes in the
properties of matter or light.  Existing interferometers have already been used
to search directly for scalar, vector, and tensor ultralight dark matter
\cite{Vermeulen2021Scalar,Abbott2022DarkPhoton,Abac2025Multimodel}, demonstrating
that the use of gravitational-wave instrumentation as a dark-matter detector is
experimentally well motivated.

This chapter does not present a projected sensitivity, an exclusion curve, or
a discovery reach for a particular dark-matter model.  It summarizes the
general detection principles used in conventional precision laser
interferometers: which physical quantity is perturbed, how that perturbation is
converted into an interferometric observable, what temporal signature is
expected, and which detector symmetries can suppress the response.  How these
principles might be implemented in CHRONOS is not yet fixed and is therefore
outside the present scope.  No particular CHRONOS degree of freedom, optical
port, readout scheme, or temporary optical configuration is assumed.

This distinction is important.  A gravitational-wave detector does not respond
to a particle label by itself.  It responds to displacement, velocity, torque,
optical phase, polarization, refractive index, laser frequency, or a transient
gravitational field.  The first task is therefore to map a dark-matter
interaction into generalized mechanical or optical coordinates of a
conventional interferometric detector.

\section{Generic signal description}
\label{sec:dm_generic_signal}

Ultralight bosonic dark matter with a large occupation number may be treated as
a classical field. Let $X_0$ be the amplitude of a local field component
$X$, $m_X$ its particle mass, $\omega_X$ its angular frequency, $\bm k_X$
its wave vector, and $\varphi_X$ its phase. Here $t$ is time, $\bm x$
position, $c$ the speed of light, and $\hbar$ the reduced Planck constant.
A local component can be represented schematically as
\begin{equation}
 X(t,\bm{x}) = X_0
 \cos\!\left(\omega_X t-\bm{k}_X\!\cdot\!\bm{x}+\varphi_X\right),
 \qquad
 \omega_X \simeq \frac{m_X c^2}{\hbar}.
 \label{eq:dm_classical_field}
\end{equation}
The Galactic velocity distribution gives the signal a finite linewidth and a
finite coherence time.  The laboratory motion through the Galactic field can
also produce daily and annual changes in the amplitude, phase, or apparent
direction.  The corresponding experimental target is therefore generally a
narrow spectral feature with a predictable coherence structure rather than an
exactly monochromatic line.
Writing $f=\omega_X/(2\pi)$ and $\Delta f$ for the signal linewidth,
for a virialized population with velocity dispersion $v\sim10^{-3}c$,
$\Delta f/f\sim10^{-6}$ and the coherence time is of order $1/\Delta f$
(up to convention-dependent factors of $2\pi$) \cite{Manley2021OptomechanicalDM}.
The conversion $m_Xc^2=4.1357\times10^{-15}~\mathrm{eV}\,(f/\mathrm{Hz})$
maps 0.1--10~Hz to $4.14\times10^{-16}$--$4.14\times10^{-14}$~eV.
This identifies an oscillation-frequency interval, not a coupling sensitivity.
The field amplitude also depends on the fraction of the local dark-matter
density carried by the candidate; a halo model and density assumption must
therefore accompany any eventual limit.

For a mechanical detector, the response may be written in the frequency domain
as
\begin{equation}
 \bm{q}(f)=\bm{H}_{\mathrm{mech}}(f)\,\bm{F}_{\mathrm{DM}}(f),
 \label{eq:dm_mechanical_response}
\end{equation}
where \(\bm{q}\) contains the translational and rotational coordinates,
\(\bm{H}_{\mathrm{mech}}\) is the calibrated mechanical response matrix, and
\(\bm{F}_{\mathrm{DM}}\) denotes generalized dark-matter forces and torques.  The
response in an arbitrary detector coordinate can then be expressed as
\begin{equation}
 d_I(f)=\bm{R}_{I}^{\mathrm{mech}}(f)\bm{q}(f)
       +d_I^{\mathrm{mat}}(f)+d_I^{\mathrm{pol}}(f)+n_I(f),
 \label{eq:dm_detector_response}
\end{equation}
where \(I\) labels any accessible detector coordinate,
\(\bm{R}_{I}^{\mathrm{mech}}\) maps motion into that coordinate,
\(d_I^{\mathrm{mat}}\) describes dimensional or refractive-index changes,
\(d_I^{\mathrm{pol}}\) describes polarization-dependent propagation, and
\(n_I\) is instrumental noise.  Equations
\eqref{eq:dm_mechanical_response} and \eqref{eq:dm_detector_response}, rather
than a sensitivity curve, define the present feasibility problem.

Macroscopic dark-matter structures instead produce transient signals.  A domain
wall may cause a temporary variation of fundamental constants, while a compact
clump passing near the instrument may accelerate the suspended masses through
gravity or an additional long-range interaction.  The event duration is set by
the size and velocity of the structure, and its trajectory determines the
relative arrival times and amplitudes in spatially separated sensors.

\section{Resonant and narrowband searches}
\label{sec:dm_resonant_searches}

A complementary strategy is to match the dark-matter oscillation frequency to
a mechanical, acoustic, optical, or polarization resonance.  Since the field
frequency $f_{\rm DM}$ is related to its particle mass $m_{\rm DM}$ by
using the Planck constant $h_{\rm P}=2\pi\hbar$ (distinct from GW strain),
\begin{equation}
 f_{\mathrm{DM}}=\frac{m_{\mathrm{DM}}c^2}{h_{\rm P}},
 \label{eq:dm_mass_frequency}
\end{equation}
each resonance defines a narrow dark-matter mass window.  Resonant searches are
therefore naturally suited to coherent ultralight fields, while broadband
transient searches remain more appropriate for domain walls and compact
clumps.

For a torsional mode with moment of inertia \(I\), resonant angular frequency
\(\omega_0\), and mechanical quality factor \(Q_{\mathrm{m}}\), the response to
a dark-matter torque $\tau_{\rm DM}$ is described by the angular
susceptibility $\chi_\tau$ (angle per unit torque). For Fourier angular
frequency $\omega=2\pi f$, the response is
\begin{equation}
 \theta(\omega)=\chi_{\tau}(\omega)\tau_{\mathrm{DM}}(\omega),
 \qquad
 \chi_{\tau}(\omega)=
 \frac{1}{I\left(\omega_0^2-\omega^2
 +i\omega\omega_0/Q_{\mathrm{m}}\right)}.
 \label{eq:dm_torsion_susceptibility}
\end{equation}
At resonance, the angular response to a coherent torque is enhanced by the
mechanical quality factor,
\begin{equation}
 \left|\chi_{\tau}(\omega_0)\right|
 =\frac{Q_{\mathrm{m}}}{I\omega_0^2}.
 \label{eq:dm_torsion_resonant_gain}
\end{equation}
The corresponding translational response is obtained by replacing \(I\) with
the effective mass of the mode.  Thus any translational, rotational, or
internal elastic mode with nonzero overlap with the dark-matter generalized
force is a potential narrowband sensor.  The relevant condition is mode
overlap, rather than the operational name assigned to an interferometer degree
of freedom.

The full steady-state gain requires coherent driving for a time long compared
with the amplitude ring-up time $2Q_{\rm m}/\omega_0$. If the field coherence
time or the dwell time at a tuning is shorter, the finite-duration response
must be calculated instead. A virialized field has
$Q_{\rm DM}\equiv f_{\rm DM}/\Delta f\sim(c/v)^2\sim10^6$; a substantially narrower mechanical
resonance samples only part of its spectrum \cite{Manley2021OptomechanicalDM}.
The resonance frequency and quality factor must remain stable over the search.

Resonant enhancement does not automatically multiply the final force
sensitivity by \(Q_{\mathrm{m}}\).  The signal displacement is enhanced near
resonance, which is valuable when optical readout noise is important, but the
thermal displacement noise is filtered by the same mechanical susceptibility.
In the thermal-force-limited regime, the advantage comes from low mechanical
dissipation, cryogenic operation, and long coherent integration rather than
from displacement amplification alone.  Resonant searches must therefore be
formulated in calibrated force or torque units as well as in displacement or
angle.

Several established response concepts illustrate this principle.
The torsional resonance of a composition-asymmetric bar can amplify the
oscillating torque from vector dark matter.  A differential optomechanical
resonator made from materials with different effective charges can similarly
search for vector forces \cite{Manley2021OptomechanicalDM}.  Longitudinal or
breathing modes of a test mass, resonant bar, suspension element, or reference
cavity can respond to scalar-induced strain; the AURIGA resonant-mass detector
has already been used for such a scalar search \cite{Branca2017AURIGA}.  These
internal modes extend the science case beyond the low-frequency rigid-body
modes normally used for gravitational-wave observation.

Optical resonance provides a separate form of enhancement.  A resonant optical
system can amplify a field-induced phase or polarization perturbation when its
eigenfrequency structure is matched to the signal.  For axion-like dark matter,
the frequency splitting between two polarization eigenmodes can be tuned so
that axion-induced conversion from the pumped polarization to the orthogonal
polarization is resonant \cite{Martynov2020AxionInterferometry}.  This principle
depends on the polarization eigenmode splitting and is distinct from ordinary
carrier resonance; its practical excitation and readout are left unspecified.

A resonant search need not be restricted to one fixed natural frequency.  In
principle the eigenfrequency may be varied through mechanical or dynamical
stiffness, or several mechanical, acoustic, optical, or polarization modes may
be combined.  Such tuning changes the response function but does not create
fundamental force sensitivity; additional dissipation and fluctuations must be
included in a concrete realization.

If a resonant coordinate is dynamically modified or actively controlled, its
physical susceptibility must be replaced by the corresponding effective
closed-loop susceptibility.  Information about an external generalized force
is then distributed between physical motion and applied control action.  The
invariant object is the external force or torque inferred from a complete
dynamical model, independently of a particular sensing or actuation scheme.

\section{Laser-interferometer detection principles}
\label{sec:dm_direct_searches}

\subsection{Scalar and dilaton dark matter}

A coherently oscillating scalar field may appear macroscopically as a temporal
variation of material dimensions, refractive index, elastic properties, or a
material-referenced optical frequency.  An optical system converts these
changes into phase only to the
extent that distinct propagation segments or materials have unequal response
coefficients.  Finite optical thickness and path asymmetry are examples of the
general mechanism by which common material expansion becomes differential
phase \cite{GroteStadnik2019,Vermeulen2021Scalar}; comparisons among dissimilar
material or geometric responses provide the same principle
\cite{Hall2022ReferenceCavity}.

The scalar response must be evaluated for the complete optical system.  An
identical fractional change of every relevant length and optical property may
be rejected as common mode, while unequal coefficients, finite propagation
times, and repeated sampling can create differential or frequency-dependent
responses.  The fundamental calculation is therefore to propagate the
scalar-induced perturbation of every transmissive and reflective component
through the optical transfer function.  This statement does not privilege a
particular output: any accessible coordinate with nonzero overlap may carry the
information.

\subsection{Vector and dark-photon dark matter}

Ultralight vector fields are candidates for ultralight dark matter. 
They can be regarded as gauge bosons of a $U(1)_D$ gauge symmetry, where $D$ denotes a charge such as $B$ (baryon number) or $B-L$ (baryon minus lepton number).
We consider vector dark matter coupled to the matter current through the interaction $\epsilon_D e J^{\mu}_D A_\mu$, where $\epsilon_D$ is the dimensionless coupling relative to the electromagnetic coupling $e$, $J_D^\mu$ is the matter four-current carrying charge $D$, and $A_\mu$ is the vector field. Greek indices label spacetime components and repeated indices are contracted. In this interaction expression we use natural units, $\hbar=c=1$. For vector-particle mass $m_A$, the field oscillates at $f=m_Ac^2/h_{\rm P}$ when SI units are restored, and induces a fifth force on test masses, resulting in a tiny oscillatory displacement of the mirrors in an optical cavity \cite{Michimura:2020vxn,Morisaki:2020gui}.
For electrically neutral atoms, the $B-L$ charge is the neutron number;
the relevant contrast is charge per unit mass. These gauged-matter-current
models are distinct from a dark photon coupled only through electromagnetic
kinetic mixing. In particular, $U(1)_{B}$ and $U(1)_{B-L}$ gauge bosons have been searched for using observational data from aLIGO, Virgo, and KAGRA \cite{LIGOScientific:2021ffg,KAGRA:2024ipf}.


For body element $a$, let $\bm r_a$ be its position relative to the bar
center of mass and $\bm F_a^{\rm DM}$ the force exerted by dark matter.
The corresponding generalized torque $\bm\tau_{\rm DM}$ is
\begin{equation}
 \bm{\tau}_{\mathrm{DM}}(t)
 =\sum_a \bm{r}_a\times\bm{F}_a^{\mathrm{DM}}(t).
 \label{eq:dm_torque}
\end{equation}
Equation~\eqref{eq:dm_torque} also exposes a central feasibility condition.  A
spatially uniform force proportional to mass has zero torque about the center
of mass, including for a geometrically asymmetric freely suspended bar.
A torque requires a dipole of effective charge relative to that center,
a field gradient, or a nontrivial suspension response. Composition contrast
can produce such a charge dipole. Mass or geometric asymmetry alone does not
evade the cancellation for a uniform mass-proportional force. Translational
and rotational coordinates must therefore be modeled separately.

\subsection{ALP dark matter}



\begin{figure}[thpb]
\begin{center}
    \begingroup
    \makeatletter\let\Gread@transgrouptrue\Gread@transgroupfalse\makeatother
    \includegraphics[clip, width=1.0\columnwidth]{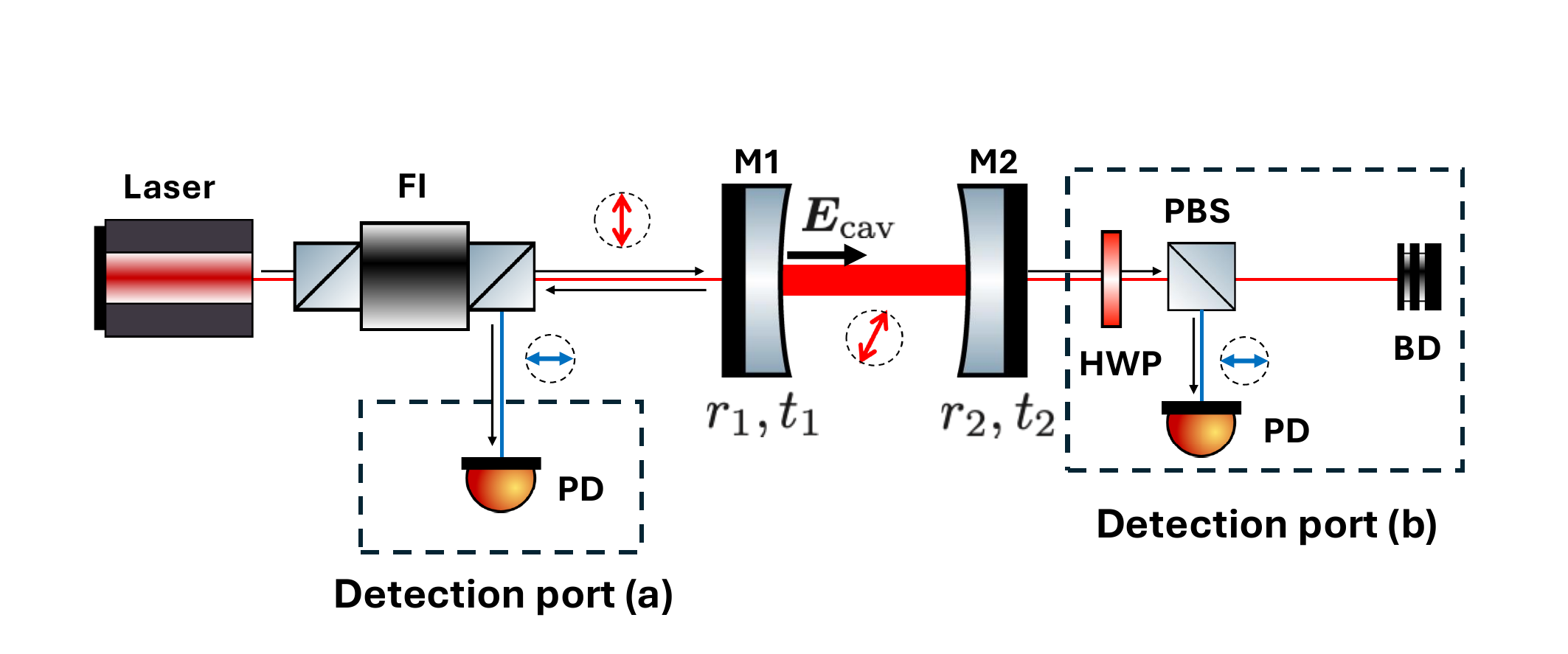}
    \endgroup
\caption{
Schematic diagram of a dark matter experiment. FI: Faraday isolator; M1(2): mirror with reflectivity $r_{1(2)}$ and transmittivity $t_{1(2)}$; $\bm{E}_{\rm cav}$: electric field inside the cavity at the reflection mirror M1; HWP: half-wave plate; PBS: polarizing beam splitter; BD: beam dump; PD: photodetector. The polarization plane of the incident laser rotates inside the resonant cavity due to the interaction between dark matter and photons. The laser polarization is separated by a polarizing beam splitter, and the new polarization state induced by dark matter is measured at the detection port near the cavity reflection mirror (a) or transmission mirror (b). This figure is adapted from \cite{Obata:2026qwx}.}
\label{fig: 143_CR_IObata}
\end{center}
\end{figure}

Axion-like particles (ALPs) arise in many extensions of the Standard Model,
including string-motivated scenarios \cite{Svrcek:2006yi,Arvanitaki:2009fg}.
Let $a$ be the ALP field, $m_a$ its particle mass, $g_{a\gamma}$ the
ALP--photon coupling, $F_{\mu\nu}$ the electromagnetic field-strength
tensor, and $\widetilde F^{\mu\nu}$ its dual. A possible interaction
Lagrangian density is
$\mathcal L_{a\gamma}=g_{a\gamma}aF_{\mu\nu}\widetilde F^{\mu\nu}/4$
in natural units. Its strength is model dependent. An oscillating ALP field
can produce opposite phase shifts for the two circular polarizations and
therefore an oscillating rotation of linear polarization, with SI frequency
$f=m_ac^2/h_{\rm P}$ \cite{DeRocco:2018jwe,Obata:2018vvr,Nagano:2019rbw}.
Figure~\ref{fig: 143_CR_IObata} illustrates a cavity polarimetry concept.
The signal depends on the complete polarization transfer function, including
mirror birefringence, propagation delays, eigenmode splitting, and losses.

Ring and linear cavities provide different polarization responses; neither
geometry is uniformly superior at every axion mass. The DANCE experiment
demonstrated a cavity search without an external magnetic field and reported
no significant signal in its first 24-hour data set \cite{Oshima:2023csb}.
LIDA and ADBC provide further experimental examples
\cite{Heinze:2023nfb,Pandey:2024dcd}. A 2025 KAGRA report describes installation
of polarization optics for an axion search \cite{MichimuraKAGRA2025}.
These results motivate a CHRONOS feasibility study; a projected CHRONOS
axion limit still requires a specific optical response and a measured
polarization-noise budget.

\subsection{Ultralight tensor fields}

A coherent massive tensor field can produce an effective oscillating strain or
a differential force with a polarization pattern distinct from ordinary scalar
or vector interactions.  Detector coordinates with different spatial
orientations provide complementary projections of the tensor structure.
Recent multi-model interferometer analyses demonstrate that scalar, vector,
and tensor hypotheses can be tested with separate response models
\cite{Abac2025Multimodel}.

The experimental challenge is model discrimination.  A tensor dark-matter line
can resemble a persistent gravitational-wave signal in a single channel.  Its
finite linewidth, spatial coherence, polarization response, daily modulation,
and consistency among differently oriented detectors must therefore be included
in the analysis.  A network can strengthen signal discrimination and test spatial coherence;
a single-detector search is also possible with an explicit response and
background model.

\subsection{Topological defects and transient variations of constants}

Scalar fields may form domain walls, strings, or other macroscopic structures.
Their passage through the apparatus can temporarily change material dimensions,
refractive indices, clock frequencies, or forces on test masses
\cite{GroteStadnik2019}.  Such a signal would appear as a transient with a
duration related to the defect thickness and Galactic transit speed.  The same
event could be observed sequentially by interferometers, optical cavities,
atomic clocks, or other precision sensors.

The response principle is coherent comparison of spatially separated detector
coordinates.  A distributed network can test whether relative arrival times
and amplitudes are compatible with passage of a common structure.  Because
local disturbances can also produce short transients, consistency with a
physically allowed propagation pattern and with independent environmental
information is essential to the dark-matter hypothesis test.

\subsection{Compact clumps and macroscopic dark matter}

A compact dark-matter clump or primordial compact object passing near a detector
can accelerate its test masses.  The leading observable is generally the
differential Newtonian acceleration; the gravitational time delay of the light
path may provide an additional contribution \cite{Baum2022DarkClumps,
Alvarez2026DarkClumpGlitches}.  The same trajectory may also produce a
body-dependent force in models with an additional long-range interaction.

Any coordinate sensitive to relative translation, rotation, or a local
gravitational reference may respond.  A spatially uniform acceleration is
rejected by a differential detector, whereas the field gradient
across the baseline or torsion bar remains observable.  The signal model must
therefore include the clump trajectory, the three-dimensional detector geometry,
and the suspension response.  Candidate events should be tested against
earthquakes, anthropogenic impulses, magnetic disturbances, and ordinary
interferometer glitches.

\subsection{Coherent scattering and dark-matter wind}

An exploratory possibility is the coherent transfer of momentum from a flux of
light dark-matter particles to macroscopic test bodies.  If two bodies have
different coherent-scattering form factors, their accelerations can differ and
produce a measurable torsional signal.  Proposed torsion-balance concepts use
equal-mass test bodies with deliberately different geometries to create this
difference \cite{Luo2025CoherentScattering}.  Similarly, different compositions
can probe interactions whose effective charge-to-mass ratios are not universal.

The response is nonzero only when the bodies differ in the relevant form factor
or effective charge-to-mass ratio.  Reversal of the asymmetry changes the sign
of the hypothesized differential response while leaving many backgrounds
unchanged, providing a general symmetry test independent of how the resulting
motion is observed.

\section{Candidate-to-observable map}
\label{sec:dm_platforms}

Table~\ref{tab:dm_measurement_methods} connects each dark-matter candidate to
the physical perturbation it may produce, the detector response coordinate,
and the minimum condition for a non-vanishing and distinguishable response.
It is a candidate-to-response map, not a prescription for a particular
measurement implementation.  Resonance is included only where it belongs to
the response principle; transient candidates instead require a response over
the inverse event duration.


\begin{table*}[p]
	\linespread{1.1}\selectfont
	\centering
	\small
	\caption{Dark-matter candidates and the corresponding physical response
		principles. The table is independent of a particular readout implementation
		and does not represent a sensitivity forecast.}
	\label{tab:dm_measurement_methods}
	
	\renewcommand{\arraystretch}{1.25}
	\setlength{\tabcolsep}{4pt}
	
	\begin{tabularx}{\textwidth}{
			@{}
			>{\raggedright\arraybackslash}p{0.16\textwidth}
			>{\raggedright\arraybackslash}X
			>{\raggedright\arraybackslash}X
			>{\raggedright\arraybackslash}X
			@{}
		}
		\hline
		\textbf{Dark-matter candidate} &
		\textbf{Physical effect in the apparatus} &
		\textbf{Response coordinate} &
		\textbf{Condition for nonzero response} \\
		\hline
		
		\textbf{Scalar or dilaton field} &
		Oscillation of material dimensions, refractive index, or a
		material-referenced optical frequency &
		Differential phase or excitation of a mechanical mode &
		Unequal material, geometric, or temporal response; absence of exact
		common-mode cancellation \\
		
		\textbf{Vector field or dark photon} &
		Oscillating, possibly composition-dependent force on suspended masses &
		Relative translation or rotation &
		Nonuniform force, charge dipole relative to the center of mass, or differential
		suspension response \\
		
		\textbf{Axion or axion-like field} &
		Opposite phase shifts of left- and right-circularly polarized light &
		Polarization rotation or differential polarization phase &
		Nonzero projection onto a polarization-sensitive eigenmode \\
		
		\textbf{Ultralight tensor field} &
		Coherent oscillating effective strain with a tensor antenna pattern &
		Oriented strain or mechanical coordinates &
		Nonzero contraction between the field tensor and detector response tensor \\
		
		\textbf{Topological defect} &
		Transient variation of constants, optical properties, or forces during
		passage &
		Time-dependent material, optical, or mechanical coordinates &
		Event duration within the response band and a consistent propagation
		pattern \\
		
		\textbf{Compact dark-matter clump} &
		Trajectory-dependent acceleration or field gradient &
		Differential displacement or torque &
		A resolvable gradient or body-dependent acceleration across the apparatus \\
		
		\textbf{Coherent particle scattering} &
		Geometry- or composition-dependent momentum transfer from a dark-matter
		wind &
		Differential force or torque &
		Different coherent form factors or effective charge-to-mass ratios \\
		
		\hline
	\end{tabularx}
\end{table*}

The table also clarifies why no single detector coordinate covers the full
program.  Scalar fields are converted by material and optical asymmetries,
vector fields by differential force or torque, axion-like fields by
polarization propagation, and macroscopic objects by transient gravitational
gradients.  A candidate interaction must produce a mutually consistent pattern
among phase, translation, rotation, polarization, frequency, and force
coordinates, regardless of the channels used to reconstruct them.

\section{Indirect dark-matter science with gravitational waves}
\label{sec:dm_indirect}

Gravitational-wave observations can also probe dark matter without a direct
interaction between the Galactic field and the detector.  These searches are
conceptually distinct from the instrumental measurements described above.  They
include searches for sub-solar-mass compact binaries as candidates for
primordial black holes, continuous gravitational waves from boson clouds around
rotating black holes, modifications of compact-binary evolution by a dense
dark-matter environment, lensing or propagation effects, and gravitational-wave
backgrounds associated with primordial-black-hole formation or dark-sector
cosmological processes
\cite{Abbott2022BosonCloud,Kacanja2026Subsolar,Eda2013DarkMatterSpike}.

These topics define a broader dark-matter science portfolio for gravitational-
wave astronomy.  They require separate studies of the source population,
waveform, detector response, and analysis band.  The direct instrumental
program and the astrophysical program should therefore remain distinct.

\section{Future work: applicability to CHRONOS}
\label{sec:dm_chronos_feasibility}

Any quantitative CHRONOS forecast should start from the full noise and
complex response in Chap.~\ref{chap:sensitivity}, then calculate the
candidate-specific force or optical coupling. A GW strain ASD cannot be
relabeled as a dark-matter coupling limit. Coherence time, scan dwell time,
field-density fraction, detector orientation, and the statistical confidence
construction must be specified. For a polarization search, the relevant
readout noise must be established independently of the GW strain budget.

The preceding sections summarize detection principles established for generic
precision interferometers.  Their applicability to CHRONOS has not yet been
demonstrated.  The next stage is therefore not to select a measurement channel,
but to determine, for each candidate class, whether the corresponding physical
response is observable at all in the CHRONOS instrument.  The required action
items are as follows.

\begin{enumerate}
 \item \textbf{Scalar and dilaton fields:} calculate the response to changes of
 material dimensions, refractive index, elastic properties, and reference
 frequency, including all common-mode and finite-propagation-time
 cancellations.

 \item \textbf{Vector fields and dark photons:} calculate the generalized
 forces and torques on the suspended bodies and determine whether the detector
 symmetries make every accessible differential response vanish.

 \item \textbf{Axion and axion-like fields:} propagate the predicted
 polarization-dependent phase through the optical system and determine whether
 any accessible detector coordinate retains a nonzero response.

 \item \textbf{Ultralight tensor fields:} derive the detector response tensor
 and determine whether the available spatial orientations provide a
 distinguishable projection of the predicted field.

 \item \textbf{Topological defects:} calculate the response to a propagating
 transient variation of material, optical, and mechanical quantities and
 determine the range of event durations that is not suppressed by the detector
 dynamics.

 \item \textbf{Compact clumps and macroscopic dark matter:} calculate the
 trajectory-dependent differential acceleration, torque, and optical
 propagation effects, and determine whether a resolvable spatial gradient
 remains after common acceleration is rejected.

 \item \textbf{Coherent particle scattering:} determine whether the relevant
 test bodies have sufficiently different form factors to produce a nonzero
 differential force or torque.

 \item \textbf{Cross-cutting assessment:} for every candidate above, identify
 the observable detector coordinates without assuming that the final science
 channel is used; derive the complete response function including mechanical,
 optical, and control dynamics; identify exact symmetry cancellations; and
 compare the surviving response with the relevant physical backgrounds.
\end{enumerate}

The result of this program should be an explicit classification of each
candidate as observable, conditionally observable, or unobservable with the
available CHRONOS configuration.  Only after that classification is complete
should a specific measurement implementation, calibration procedure, or
sensitivity projection be developed.

%% file: Chapters/Earthquake.tex
\chapter{Earthquake Gravity Signals}
\label{chap:earthquake}

\section{Introduction}

The sub-Hz frequency band targeted by CHRONOS provides access not only to
astrophysical gravitational-wave sources, but also to terrestrial
gravitational perturbations generated by seismic and atmospheric processes.
In this frequency range, fluctuations of the local gravitational field,
commonly referred to as gravity-gradient noise or Newtonian noise, are
expected to become an important environmental background for ground-based
gravitational-wave detectors~\cite{harms2015terrestrial,harms2013low}.

Unlike ordinary seismic disturbances, Newtonian noise couples directly to
the test masses through gravity. It therefore cannot be suppressed by
conventional mechanical isolation systems. This characteristic makes
gravity-gradient fluctuations one of the fundamental challenges for
extending ground-based gravitational-wave observations into the sub-Hz
regime.

For CHRONOS, however, terrestrial gravity perturbations are not only a
source of noise. The same sensitivity required for sub-Hz
gravitational-wave observation provides an opportunity to directly observe
time-dependent gravitational fields generated by geophysical processes.

A particularly interesting target is the prompt gravitational perturbation
generated by an earthquake. During an earthquake, rapid redistribution of
underground mass modifies the surrounding gravitational field. The
associated gravitational perturbation propagates essentially at the speed
of light and can therefore reach a detector before elastic seismic waves,
including the P wave~\cite{harms2015transient}.

This opens the possibility of observing an earthquake through gravity
before conventional seismic signals arrive. Such a measurement would
provide an observational channel fundamentally different from ordinary
seismology and could potentially complement existing earthquake
early-warning networks~\cite{https://doi.org/10.1029/2018JB016698,mcmanus2018earlyearthquakedetectiondual}.

More importantly, direct observation of the prompt gravity perturbation
would provide information about the earliest stages of earthquake rupture
and the evolution of the seismic moment itself. CHRONOS may therefore
operate not only as a sub-Hz gravitational-wave detector, but also as a
high-sensitivity terrestrial gravity-gradient observatory.

\section{Prompt Gravity Signal from an Earthquake}
\label{sec:earthquake_prompt_gravity}

An earthquake produces rapid deformation and displacement of material
inside the Earth. This redistribution of mass modifies the local Newtonian
gravitational field and produces a time-dependent gravitational
perturbation.

At the same time, the earthquake generates conventional elastic waves,
including P waves, S waves, and surface waves. These waves propagate
through the Earth with velocities of several kilometers per second~\cite{bormann2012seismic}.
The gravitational perturbation, in contrast, propagates effectively at the
speed of light.

As a result, a sufficiently sensitive gravity-gradient detector can respond
to the changing mass distribution before the seismic elastic wavefront itself
reaches the detector.

\begin{figure}[htbp]
    \centering
    \includegraphics[width=0.5\linewidth]{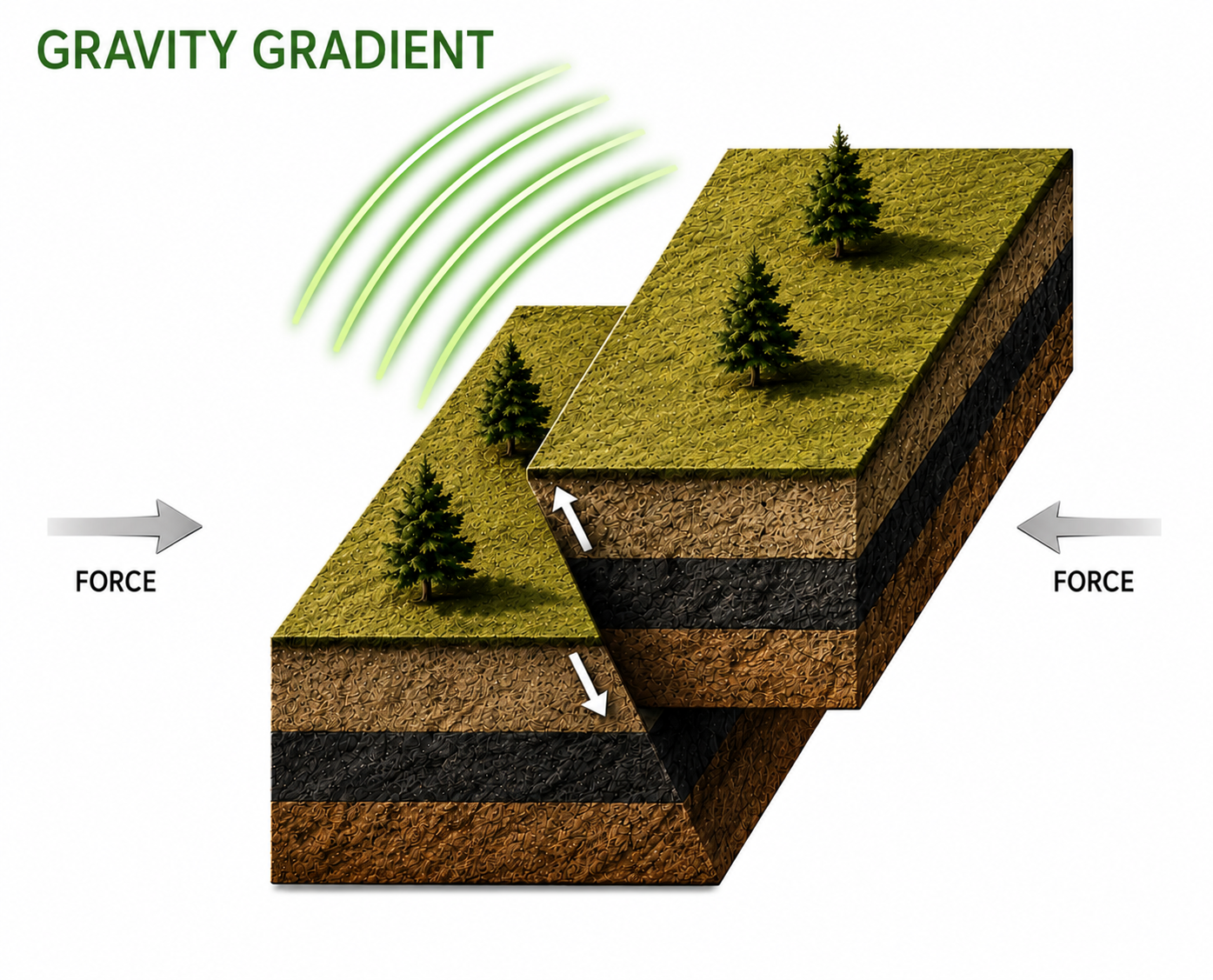}
    \caption{
    Concept of prompt gravity detection from an earthquake.
    Rapid redistribution of underground mass generates a time-dependent
    gravitational perturbation as well as conventional seismic waves.
    The gravity perturbation reaches CHRONOS before the seismic P wave.
    }
    \label{fig:earthquake_gravity_concept}
\end{figure}

Figure~\ref{fig:earthquake_gravity_concept} illustrates this basic concept.
The earthquake rupture generates both a prompt gravity perturbation and
conventional seismic waves. The prompt gravity signal reaches CHRONOS first,
followed by the P wave after a propagation delay determined primarily by the
source--detector distance and the seismic-wave velocity.

Within the seismic-source description adopted here, the density perturbation
associated with the earthquake can be expressed in terms of the seismic
potential $\phi_s$ as~\cite{harms2015transient}
\begin{equation}
    \delta \rho_{\rm EQ}
    =
    -\rho_0 \nabla^2 \phi_s ,
    \label{eq:earthquake_density}
\end{equation}
where $\rho_0$ is the unperturbed density of the surrounding medium.

Let $t$ be time since rupture onset, $r$ the source--detector distance,
$G$ Newton's gravitational constant, and $M_0(t)$ the seismic moment in
$\mathrm{N\,m}$. For a point double-couple source in a homogeneous,
unbounded medium, the root-mean-square (RMS) rotational gravity strain,
averaged over source and detector orientations, is~\cite{harms2015transient}
\begin{equation}
 h_{g,\mathrm{rms}}(r,t)
 =6\sqrt{\frac{14}{5}}\,\frac{G}{r^5}I_4[M_0](t),
 \qquad 0<t<t_P,\quad t_P=\frac{r}{v_P}.
 \label{eq:earthquake_prompt_signal}
\end{equation}
Here $v_P$ is the P-wave speed, $t_P$ its arrival time, and $I_4$ denotes
four integrations with zero initial conditions:
\begin{equation}
 I_4[M_0](t)=\frac{1}{6}\int_0^t(t-t')^3 M_0(t')\,dt'.
 \label{eq:seismic_moment_integral}
\end{equation}
The gravity-strain convention of Ref.~\cite{harms2015transient} uses half
the relative rotation of two ideal perpendicular bars. With the geometric
factor $\eta$ of Chap.~\ref{chap:sensitivity}, the free differential
rotation is $2\eta h_g$. The equivalent GW-strain input is therefore
$h_{\rm EQ}=2h_g$, so that $\theta_{\rm sig}=\eta h_{\rm EQ}$. This factor
of two is applied before comparison with the Chapter~3 noise. Gravity strain
here describes a Newtonian tidal response, not an emitted propagating GW.

For reproducible source waveforms, let $M_\infty$ denote the final moment,
$M_w$ the moment magnitude, and $T_s$ half the source duration. We adopt the
moment-rate shape and duration scaling of
Ref.~\cite{https://doi.org/10.1029/2018JB016698}:
\begin{align}
 M_\infty &=10^{1.5M_w+9.1}~\mathrm{N\,m}, &
 2T_s&=\left(\frac{M_\infty}{10^{16}~\mathrm{N\,m}}\right)^{1/3}\mathrm{s},
 \label{eq:earthquake_duration}\\
 \dot M_0(t)&=\frac{A_s M_\infty}{T_s}g(t/T_s), &
 A_s^{-1}&=\int_0^2g(u)\,du,\nonumber\\
 g(u)&=\begin{cases}u^2,&0\leq u<1,\\
 [1-(u-1)^2]^6,&1\leq u\leq2,\\0,&\text{otherwise}.
 \end{cases}
 \label{eq:earthquake_moment_rate}
\end{align}
The dimensionless constant $A_s$ normalizes the integrated moment to
$M_\infty$; $u=t/T_s$ is dimensionless time. The duration depends on
magnitude, independently of source distance.

\section{Coupling to the CHRONOS Torsion Bar}
\label{sec:earthquake_chronos_response}

Terrestrial gravity perturbations couple directly to the torsional degree
of freedom of CHRONOS~\cite{Ando2010,chronos_gcal}.

For a single-bar rotation $\theta(t)$, the rotational motion can be described by
\begin{equation}
    I\ddot{\theta}
    +
    \frac{I\omega_{\rm tb}}{Q}\dot{\theta}
    +
    I\omega_{\rm tb}^2\theta
    =
    \tau_{\rm ext},
    \label{eq:torsion_motion_earthquake}
\end{equation}
where $I$ is the moment of inertia of the torsion bar,
$\omega_{\rm tb}=2\pi f_{\rm tb}$ is the torsional resonant angular
frequency ($f_{\rm tb}$ is the resonance in hertz), $Q$ is the mechanical
quality factor, and $\tau_{\rm ext}$ is
the external gravitational torque.

Let $f$ denote Fourier frequency, $\Omega=2\pi f$ angular frequency,
and $\chi_\theta$ the torque-to-angle susceptibility in $\mathrm{rad/(N\,m)}$.
In the frequency domain,
\begin{equation}
    \theta(f)
    =
    \chi_\theta(f)\tau_{\rm ext}(f),
\end{equation}
where the torsional susceptibility is
\begin{equation}
    \chi_\theta(f)
    =
    \frac{1}
    {
    I
    \left(
    \omega_{\rm tb}^2
    -
    \Omega^2
    +
    i\Omega\omega_{\rm tb}/Q
    \right)
    },
    \qquad
    \Omega = 2\pi f.
    \label{eq:torsion_susceptibility_earthquake}
\end{equation}

Let $\Gamma_{ij}=\partial g_i/\partial x_j$ be the gravity-gradient tensor,
where $g_i$ is gravitational acceleration, and define the bar's mass second
moments $J_{ij}=\int\rho_{\rm bar}x_i x_j\,d^3x$. Here $x,y$ are horizontal
principal axes, $z$ is vertical, and $\rho_{\rm bar}$ is the bar density.
For one bar, the leading torque about $z$ is
\begin{equation}
    \tau_z
    =
    \left(
    J_{xx}-J_{yy}
    \right)
    \Gamma_{xy}.
    \label{eq:gravity_gradient_torque}
\end{equation}

The resulting angular response is therefore
\begin{equation}
    \theta(f)
    =
    \chi_\theta(f)
    \left(
    J_{xx}-J_{yy}
    \right)
    \Gamma_{xy}(f).
    \label{eq:gravity_gradient_response}
\end{equation}

The very low torsional resonance frequency of CHRONOS makes this rotational
degree of freedom particularly sensitive to slowly varying gravity-gradient
fields in the sub-Hz band.
This provides the fundamental detector response used to search for
earthquake-induced gravity perturbations.

\section{Detectability with CHRONOS}
\label{sec:earthquake_detectability}

The numerical forecasts use the same two full strain-noise spectra as
Chap.~\ref{chap:sensitivity}: the reference-paper quantum model combined
with CHRONOISE coating, bar-thermal, seismic, Newtonian, and intensity noise
\cite{InoueQuantumTorque2026,CHRONOISESoftware2026}. These archived operating
points are used consistently across the science chapters. Their ASDs at
0.5~Hz are $1.237\times10^{-17}$ and $6.154\times10^{-18}~\mathrm{Hz}^{-1/2}$
for minimum PI and minimum logarithmic mean, respectively.

\begin{figure}[tbp]
\centering
\includegraphics[width=\linewidth]{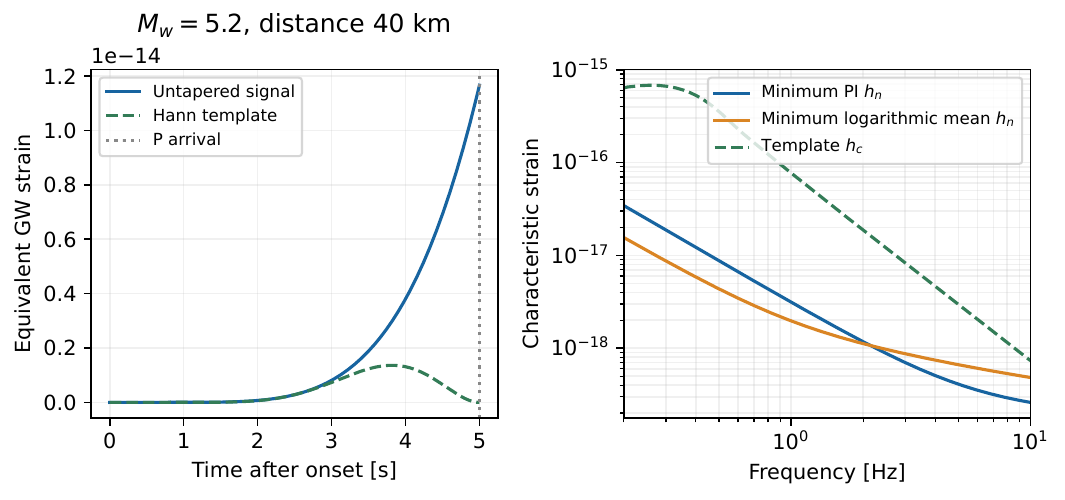}
\caption{Recalculated earthquake benchmark for $M_w=5.2$ at 40~km.
Left: RMS equivalent GW-strain input and its Hann-tapered pre-P template;
$t_P=5$~s. Right: characteristic template strain and the two Chapter~3
noise spectra, including the CHRONOISE classical noise budget. The integral
uses 0.2--10~Hz and gives $\rho_w=48.0$ (minimum PI) or 91.1
(minimum logarithmic mean), under the stated ideal source assumptions.}
\label{fig:earthquake_chronoise_waveform}
\end{figure}

We define a single-channel, pre-P template benchmark with
$v_P=8.0~\mathrm{km\,s^{-1}}$. Let $w(t)=\sin^2(\pi t/t_P)$ on
$0\leq t\leq t_P$ and zero elsewhere. Its Fourier amplitude is
\begin{equation}
 \widetilde h_w(f)=\int_0^{t_P}w(t)h_{{\rm EQ},\mathrm{rms}}(t)
                 e^{-2\pi i f t}\,dt.
 \label{eq:earthquake_template}
\end{equation}
The Hann taper suppresses endpoint discontinuities and is not amplitude
renormalized. Thus $\widetilde h_w$ has units $\mathrm{Hz}^{-1}$; it is
not a signal ASD. With one-sided strain-noise PSD $S_n(f)$, the template
norm, quoted below as the tapered pre-P SNR $\rho_w$, is
\begin{equation}
 \rho_w^2=4\int_{f_{\min}}^{f_{\max}}
       \frac{|\widetilde h_w(f)|^2}{S_n(f)}\,df,
 \qquad f_{\min}=\max(0.1~\mathrm{Hz},1/t_P),\quad
 f_{\max}=10~\mathrm{Hz}.
 \label{eq:earthquake_snr}
\end{equation}
The lower cutoff excludes periods longer than the available window and
stays inside the common science band. No multiple-channel gain is assumed.
Figure~\ref{fig:earthquake_chronoise_waveform} shows the waveform and the
characteristic strains $h_c=2f|\widetilde h_w|$ and $h_n=\sqrt{fS_n}$.
Table~\ref{tab:earthquake_snr_examples} gives representative results.

\input{Chapters/earthquake_snr_table}

These numbers are the noise-weighted norms of the specified tapered
signals. They do not establish real-time alert significance: a causal
search needs finite-window noise covariance, local transient backgrounds,
and a false-alarm calibration. The point-source calculation also omits
free-surface structure, finite faults, and the particular site's source
orientations. The approximation is especially uncertain when the rupture
extent is comparable to the source distance. The waveform, SNR grids,
source normalization, noise-file hashes, and integration checks are
included with the source data. Halving the time step and doubling Fourier
padding changes the four tabulated SNR pairs by less than $8\times10^{-5}$.

\section{Lead Time before Seismic-Wave Arrival}
\label{sec:earthquake_leadtime}

The most distinctive feature of prompt gravity detection is that the
gravitational perturbation is observed before the arrival of the seismic
P wave.

If the earthquake occurs at source distance $r$, the approximate lead time
relative to the P-wave arrival is
\begin{equation}
    \Delta t
    \simeq
    \frac{r}{v_P},
    \label{eq:leadtime_simple}
\end{equation}
where the propagation time of the gravitational perturbation itself is
negligible on this scale and $v_P$ is the effective P-wave velocity~\cite{bormann2012seismic}.

Assuming
\begin{equation}
    v_P
    =
    8.0~{\rm km\,s^{-1}},
\end{equation}
a source at $40~{\rm km}$ corresponds to a lead time of approximately
\begin{equation}
    \Delta t
    \simeq
    5~{\rm s}.
\end{equation}

Considering representative P-wave velocities between
$5.8$ and $13.7~{\rm km\,s^{-1}}$, the corresponding lead time for the
same source distance lies approximately between
\begin{equation}
    2.9~{\rm s}
    <
    \Delta t
    <
    6.9~{\rm s}.
\end{equation}

The interval $r/v_P$ is an upper bound set by propagation. An operational
warning time must subtract signal accumulation, processing, and communication
latencies. Detectability before P-wave arrival must be evaluated with the
truncated waveform, not with a full-event SNR.

This produces an important trade-off.
Nearby earthquakes generate stronger gravitational signals but provide
shorter warning times, whereas more distant earthquakes provide longer lead
times at the cost of significantly reduced signal amplitude.

The significance of the prompt gravity measurement should therefore not be
viewed solely in terms of the number of seconds of early warning.
The gravity signal provides information about the source through a channel
that is independent of seismic-wave propagation.
Its combination with conventional seismic data could therefore improve our
understanding of the earliest earthquake rupture process.

\section{Magnitude and Distance Dependence}
\label{sec:earthquake_magnitude_distance}

\begin{figure}[tbp]
\centering
\includegraphics[width=\linewidth]{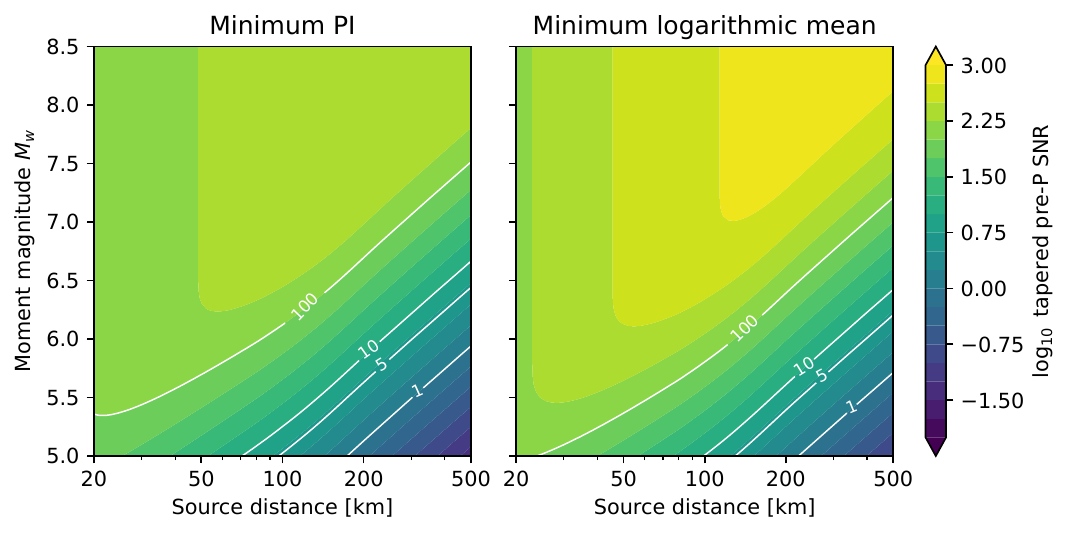}
\caption{Tapered pre-P SNR from the common CHRONOISE-based noise budget:
minimum PI (left) and minimum logarithmic mean (right). Colour gives
$\log_{10}\rho_w$; white curves label constant $\rho_w$. Each point uses
$t_P=r/(8~\mathrm{km\,s^{-1}})$, the RMS point-source model, and
Eq.~\eqref{eq:earthquake_snr}. These idealized maps do not include finite-fault
or site-specific orientation effects and are not alert-probability maps.}
\label{fig:earthquake_chronoise_snr}
\end{figure}

Figure~\ref{fig:earthquake_chronoise_snr} recalculates the magnitude--distance
map with the CHRONOISE-based noise used in Chap.~\ref{chap:sensitivity}.
Both panels use exactly the same source, strain conversion, pre-P window,
and frequency prescription; only the optical operating point changes.
The contours describe the tapered-template benchmark, rather than an
operational detection boundary.

Magnitude changes both the final moment and the rupture duration. In the
early part of the adopted self-similar source, $M_0(t)\propto t^3$ and
$M_\infty/T_s^3$ is constant; the pre-P signal can therefore approach a
magnitude-independent limit at fixed distance. Distance also changes the
available observation time, so the SNR cannot be obtained from an $r^{-5}$
amplitude rescaling alone. A physical detection rate requires finite-fault
waveforms and measured local backgrounds in addition to this comparison.

\section{Prospects for Earthquake Observation in Taiwan}
\label{sec:earthquake_taiwan}

The practical significance of the projected CHRONOS sensitivity can be
motivated by comparing candidate source parameters with historical
earthquake activity around Taiwan~\cite{CWA_EarthquakeCatalog}.

Using earthquakes with
\begin{equation}
    M_w \geq 5
\end{equation}
and source--detector distances below approximately $100~{\rm km}$ between
January 1995 and May 2026, the historical earthquake catalog~\cite{CWA_EarthquakeCatalog} gives an
average occurrence rate of approximately
\begin{equation}
    6.3~{\rm events~yr^{-1}}.
\end{equation}

Most of these earthquakes fall within the range
\begin{equation}
    5
   <
    M_w
    <
    6.
\end{equation}

The catalog motivates studies of moderate regional events as well as
rare large earthquakes; it does not by itself determine a detectable rate.

\begin{figure}[htbp]
    \centering
    \subfigure[]{
        \includegraphics[width=0.38\textwidth]{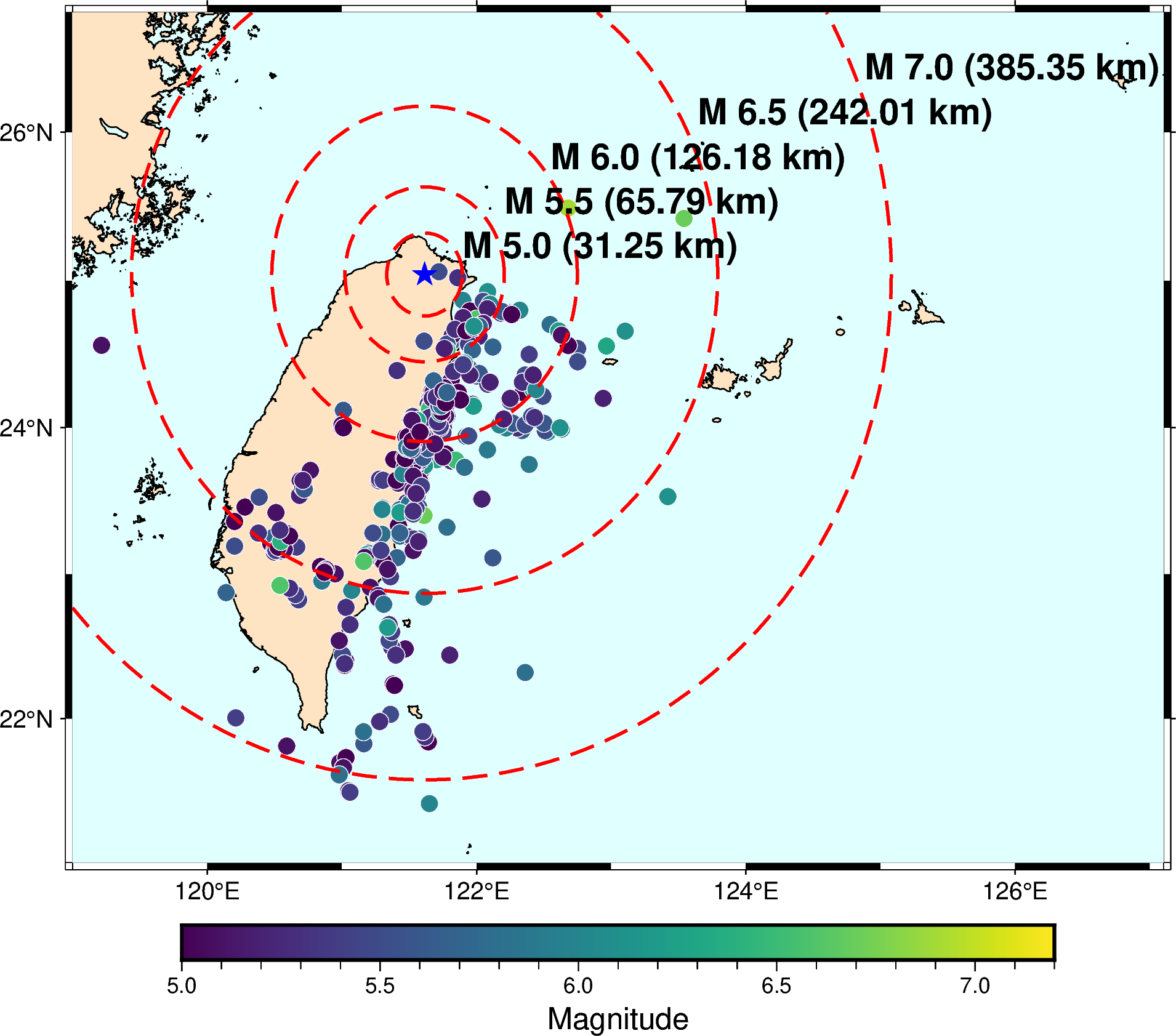}
        \label{fig:EQ_map}
    }
    \hfill
    \subfigure[]{
        \includegraphics[width=0.56\textwidth]{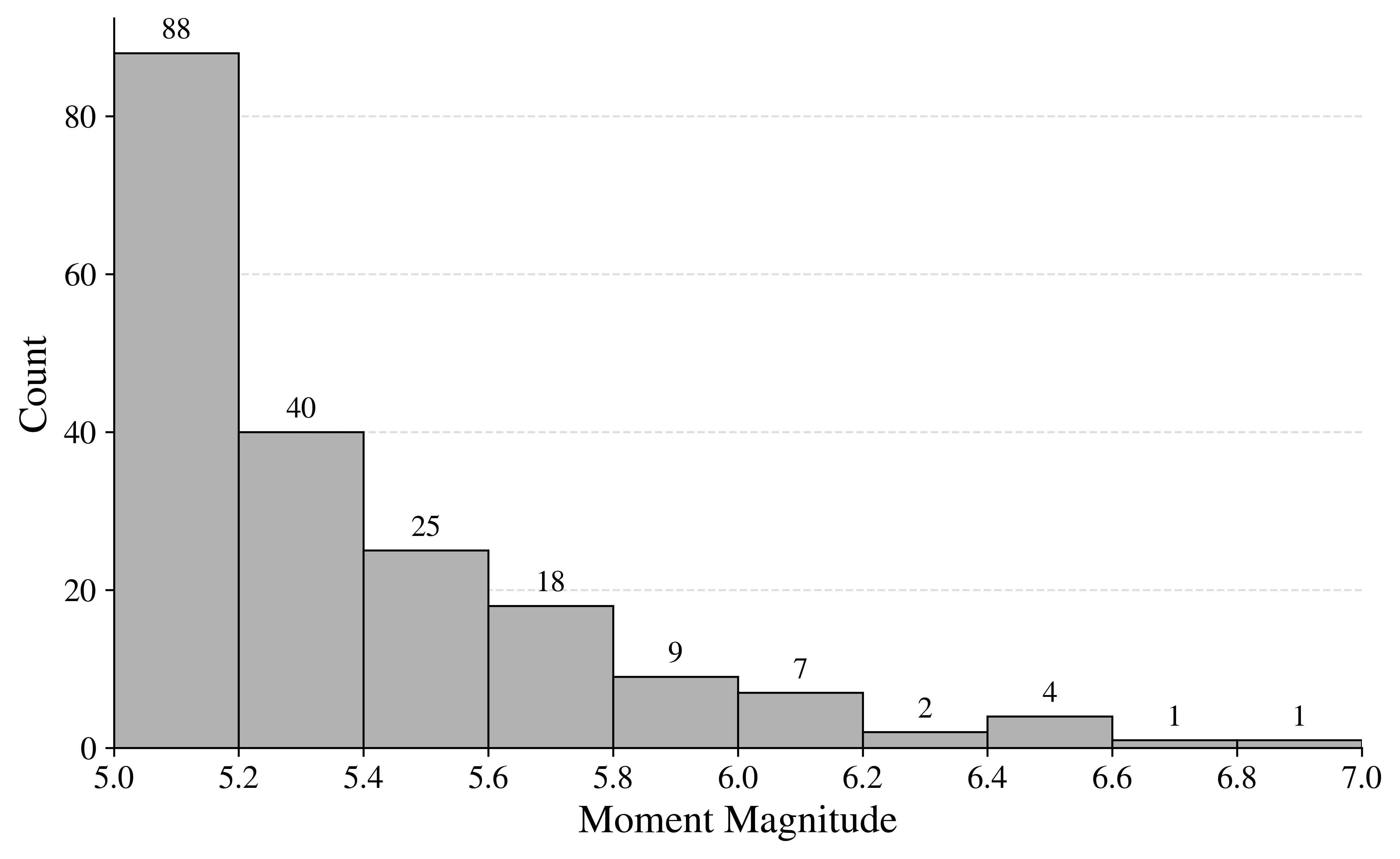}
        \label{fig:EQ_hist}
    }
    \caption{Historical earthquake population around Taiwan relevant to the
    candidate CHRONOS source region, together with the corresponding moment
    magnitude distribution.}
    \label{fig:earthquake_taiwan_catalog}
\end{figure}

Figure~\ref{fig:earthquake_taiwan_catalog} illustrates the distribution of
historical earthquakes around the proposed detector region together with
their magnitude distribution.

The catalog demonstrates that the region contains potential sources for
a prompt-gravity program. A rate forecast requires a verified
magnitude--distance selection function with the updated noise, local
backgrounds, detector uptime, and source orientations. Historical occurrence
alone does not establish repeated detectable events.

\section{Gravity-Gradient Background}
\label{sec:earthquake_newtonian_noise}

The observation of prompt earthquake gravity signals must be considered
together with the ambient terrestrial gravity-gradient environment.

Density fluctuations produced by seismic surface waves and atmospheric
pressure variations generate continuous fluctuations of the local
gravitational field.
These stochastic backgrounds constitute Newtonian noise for
gravitational-wave observations~\cite{harms2015terrestrial,fiorucci2018impact}.

\begin{figure}[tbp]
\centering
\includegraphics[width=0.88\linewidth]{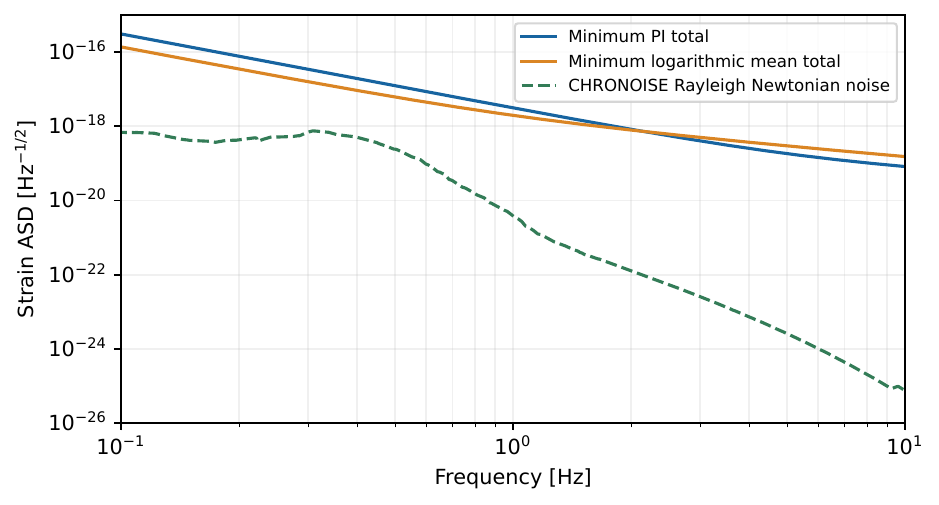}
\caption{CHRONOISE Rayleigh-wave Newtonian strain ASD (dashed green) and
the two full-noise spectra used throughout this White Paper. All curves
use the same differential-angle strain convention and the reference
0.1--10~Hz band. The Newtonian contribution is already included in both
totals; it must not be added a second time to the SNR noise.}
\label{fig:earthquake_nn_background}
\end{figure}

Figure~\ref{fig:earthquake_nn_background} uses the Newtonian component
from the same CHRONOISE noise data as the SNR calculation. It is the
Rayleigh-wave contribution for the reference environmental settings,
shared by both operating points. The ground-motion input is the archived
KAGRA spectrum multiplied by the reference amplitude factor $0.01$,
with an assumed detector depth of 200~m. This assumed spectrum is not a
measurement of the proposed Taiwan site. Atmospheric and local transient
gravity backgrounds require separate site-specific estimates.

The Rayleigh-wave gravity-gradient background depends on both the detector
depth and the seismic quietness of the site.
An underground location suppresses the gravitational coupling from surface
Rayleigh waves, particularly toward higher frequencies~\cite{harms2015terrestrial}.
However, the overall seismic activity of the site has a stronger effect on
the predicted background amplitude.

The gravity-gradient contribution depends on the measured seismic spectrum,
depth, density structure, and spatial correlations. Its comparison with
instrumental noise must use the updated budget in
Fig.~\ref{fig:sensitivity_overview}. A fixed crossover interval from an
earlier noise curve is not retained as a property of the revised baseline.

These results indicate that site characterization is an essential component
of the CHRONOS program.
Measurements of the local seismic field, subsurface structure, and spatial
correlations will be required both to predict the Newtonian-noise floor and
to distinguish transient earthquake signals from the stationary
gravity-gradient background.

The same measurements will also provide valuable information for developing
future Newtonian-noise subtraction techniques using environmental sensor
arrays~\cite{coughlin2016towards,harms2015newtonian}.

\section{Scientific Perspective}
\label{sec:earthquake_perspective}

The earthquake program illustrates a distinctive feature of sub-Hz
gravitational measurements with CHRONOS.

Terrestrial gravity fluctuations have traditionally been regarded primarily
as a limiting noise source for low-frequency gravitational-wave detectors.
For CHRONOS, however, these fluctuations can themselves become measurable
signals carrying information about geophysical processes.

The science opportunity is conditional on reconstructing a prompt-gravity
waveform above the full noise before seismic arrival. The updated sensitivity
and source-response calculation must be evaluated together; the geometric
lead-time estimates alone do not establish this capability.

The scientific importance of such a measurement extends beyond earthquake
early warning.
Because the gravitational perturbation directly traces changing mass
distribution rather than the subsequent propagation of elastic waves, it
provides an independent probe of the earthquake source.

Direct observation of prompt gravity signals could therefore establish a
new observational channel for studying the onset of earthquake rupture and
the evolution of the seismic moment.

At the same time, continuous measurements of the ambient gravity-gradient
field would provide direct information about Rayleigh-wave Newtonian noise,
the secondary microseism environment, and the site dependence of terrestrial
gravity fluctuations.

CHRONOS can therefore serve simultaneously as a sub-Hz
gravitational-wave detector and as a precision terrestrial
gravity-gradient observatory.
Earthquake-induced prompt gravity signals represent one of the clearest
examples of the interdisciplinary science enabled by extending precision
gravitational measurements into the sub-Hz regime.

%% file: Chapters/earthquake_snr_table.tex
\begin{table}[tbp]
\centering\small
\caption{Tapered pre-P template SNRs with the reference quantum plus CHRONOISE classical noise.
These use the one-channel RMS source benchmark and window specified in the text.}
\label{tab:earthquake_snr_examples}
\begin{tabular}{rrrrr}\toprule
$M_w$ & Distance [km] & $t_P$ [s] & Minimum PI & Logarithmic mean\\\midrule
5.2 & 40 & 5.0 & 48 & 91.1\\
6.0 & 100 & 12.5 & 76.5 & 161\\
7.0 & 100 & 12.5 & 257 & 527\\
8.0 & 300 & 37.5 & 309 & 652\\
\bottomrule\end{tabular}\end{table}

%% file: Chapters/Yukawa.tex
\chapter{Search for Yukawa-Type Gravity}
\label{chap:yukawa}

Let $r$ be the separation between a source mass $m$ and a test-body mass $M$,
and $G$ the Newtonian gravitational constant. The potential energy $V(r)$
includes a Yukawa correction of dimensionless relative strength $\alpha_Y$
and interaction range $\lambda$. CHRONOS can search for this correction
using controlled gravitational torques,
\begin{equation}
 V(r)=-\frac{GmM}{r}\left(1+\alpha_Y e^{-r/\lambda}\right),
 \label{eq:yukawa_potential_wp}
\end{equation}
This chapter adopts the revised differential-GCal analysis of
Ref.~\cite{InoueYukawaHarmonic2026}. Let $f_{\rm rot}$ denote the source-rotor rotation frequency. The key
improvement is to combine the $2f_{\rm rot}$ signal with calibrated $4f_{\rm rot}$ and $6f_{\rm rot}$
amplitudes, suppressing the leading geometrical uncertainties. The same
subtraction must be applied to the predicted Yukawa signal. The resulting
limits are conditional systematic-floor projections, not measured exclusions.

\section{Differential source and reference geometry}
\label{sec:yukawa_gcal_wp}
\label{sec:newtonian_cancellation_wp}

Two counter-rotating gravitational calibrators are placed at different
vertical distances below the detector. Their radii and separations are chosen
to cancel the nominal Newtonian $2f_{\rm rot}$ torque. The closer and farther
sources are denoted $S$ and $L$. For source index $i\in\{S,L\}$, $m_i$ is
the mass of each rotor endpoint, $b_i$ the rotor radius, $h_i$ the vertical
separation, and $\rho_i$ the transverse offset. The detector parameter $a$
is its quadrupole effective distance from the rotation axis. Define the
effective source--test-body separation $R_i$ and dimensionless expansion
parameter $\xi_i$ below. The function ${}_2F_1$ is the Gauss hypergeometric
function. The GCal torque and finite-size factor
$\eta_{\rm GCAL}$ are introduced in Chap.~\ref{chap:gcal}; the latter is
included once in each torque amplitude. With
$R_i^2=a^2+b_i^2+h_i^2+\rho_i^2$ and $\xi_i=2ab_i/R_i^2$, the exact null is
\begin{equation}
 \frac{m_L\xi_L^2}{R_L}\,{}_2F_1\!\left(\frac54,\frac74;3;\xi_L^2\right)
 =\frac{m_S\xi_S^2}{R_S}\,{}_2F_1\!\left(\frac54,\frac74;3;\xi_S^2\right).
 \label{eq:newton_null_wp}
\end{equation}
The distance ratio $\gamma\equiv h_L/h_S$ is fixed by this null at the nominal geometry. Uncertainty propagation
holds that operating position fixed; the null is not re-tuned separately
for each perturbed realization.

\begin{figure}[tbp]
\centering
\includegraphics[width=0.72\linewidth]{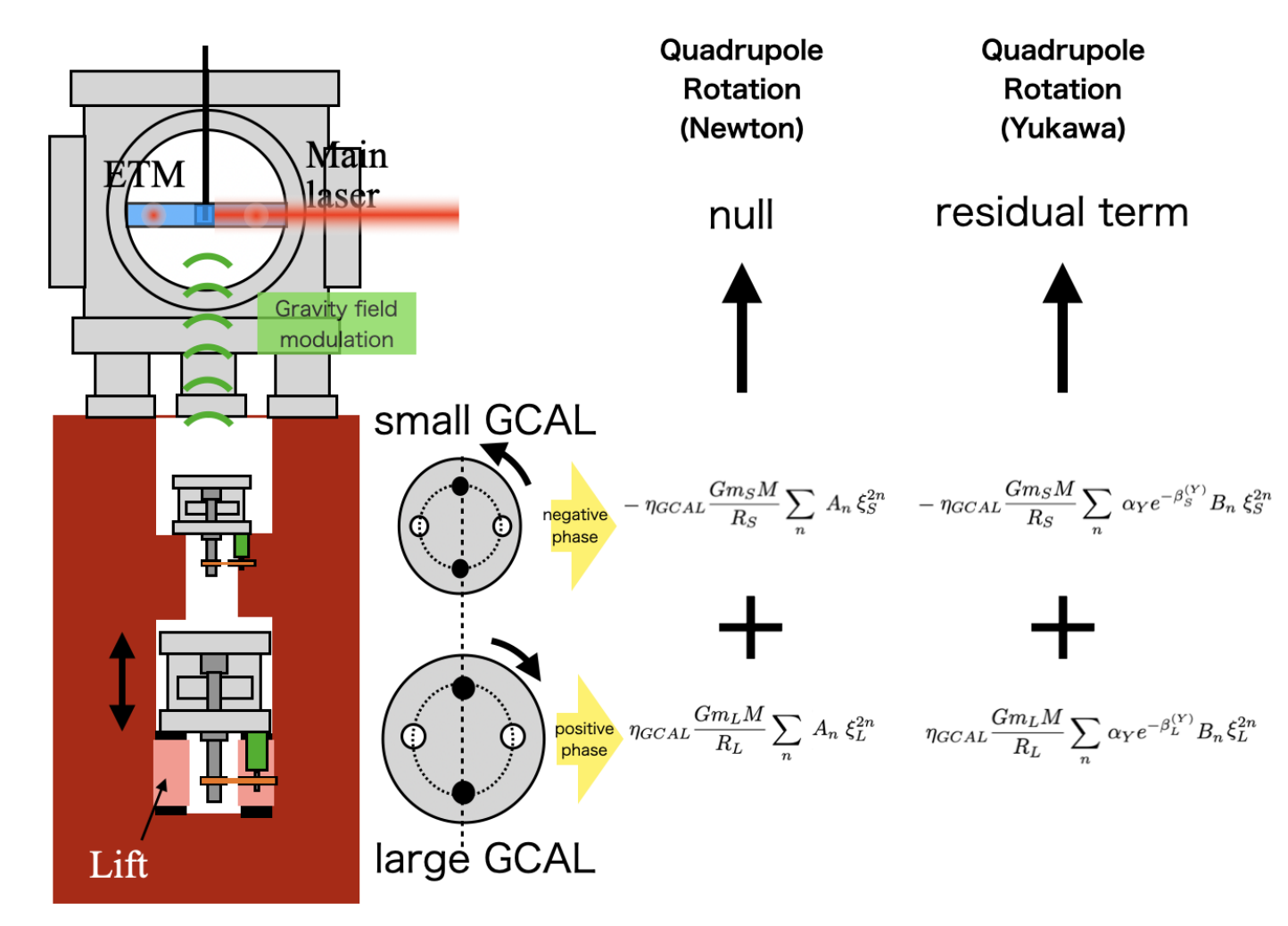}
\begingroup\linespread{1.1}\selectfont
\caption{Differential GCal concept, reproduced from Fig.~1 of
Ref.~\cite{InoueYukawaHarmonic2026}. The farther source position is adjusted
to null the Newtonian $2f_{\rm rot}$ torque. Auxiliary harmonics provide
additional information on departures from this nominal geometry.
The displayed series use positive integer index $n$, dimensionless Newtonian
coefficients $A_n$, and dimensionless Yukawa coefficients
$B_n(\beta_i^{(Y)})$, with $\beta_i^{(Y)}=R_i/\lambda$ for $i=S,L$.
The other geometric symbols are defined in the text.}
\label{fig:yukawa_concept_wp}
\endgroup
\end{figure}

The fiducial rotor frequency is $f_{\rm rot}=0.25~\mathrm{Hz}$, giving
measurement frequencies 0.5, 1.0, and 1.5~Hz. Table~\ref{tab:yukawa_geometry_wp}
records the geometry and uncertainty assumptions used for the quoted limits.
Here $M=85.66~\mathrm{kg}$ is the mass parameter of one effective end body,
not the full $171.32~\mathrm{kg}$ bar mass; $a=0.30213~\mathrm m$ is the
quadrupole effective distance, not the main-laser spot lever $\ell$.

\begin{table}[tbp]
\centering\small
\begingroup\linespread{1.12}\selectfont
\caption{Fiducial Yukawa-search parameters and absolute uncertainties from
Ref.~\cite{InoueYukawaHarmonic2026}. Fixing $h_S$ is an idealized prior,
not an experimentally established zero uncertainty.}
\label{tab:yukawa_geometry_wp}
\begin{tabular}{lrr}
\toprule
Parameter & Nominal value & Absolute uncertainty\\
\midrule
$m_S,m_L$ [kg], each & 1.8842 & $9.04416\times10^{-5}$\\
$h_S$ [m] & 8.000 & 0 (fixed prior)\\
$h_L$ [m] & 10.55258 & $5.00\times10^{-4}$\\
$b_S$ [m] & 0.250 & $2.50\times10^{-5}$\\
$b_L$ [m] & 0.500 & $2.50\times10^{-5}$\\
$\rho_S,\rho_L$ [m], each & 0.0010 & $1.00\times10^{-3}$\\
$M$ [kg] & 85.66 & $1.41339\times10^{-4}$\\
$a$ [m] & 0.30213 & $2.50\times10^{-5}$\\
$G$ [$\mathrm{m^3\,kg^{-1}\,s^{-2}}$] & $6.67430\times10^{-11}$ & $1.50172\times10^{-15}$\\
\bottomrule
\end{tabular}
\endgroup
\end{table}

\section{Harmonic subtraction and signal response}
\label{sec:yukawa_response_wp}
\label{sec:yukawa_harmonic_subtraction_wp}

Let $\bm p$ contain the source and detector parameters in
Table~\ref{tab:yukawa_geometry_wp}, with nominal values $\bm p_0$.
For harmonic index $q=1,2,3$, let $d_q$ be the measured signed, phase-referenced amplitude
at $2qf_{\rm rot}$, expressed in single-bar torque units. Write $N_q(\bm p)$
for the Newtonian prediction and $Y_q(\lambda)$ for the unit-strength
Yukawa torque, evaluated at $\alpha_Y=1$ and the nominal geometry.
Both predictions use the finite-size gravitational model of
Ref.~\cite{InoueYukawaHarmonic2026}. The nominal Newtonian amplitudes are
$N_{q,0}=N_q(\bm p_0)$; only $N_{1,0}$ is nulled.

Let $k_4$ and $k_6$ be dimensionless subtraction coefficients for the
$4f_{\rm rot}$ and $6f_{\rm rot}$ amplitudes. The residual torque $r$ and
its unit-strength Yukawa response $Y_{\rm eff}$ are
\begin{align}
 r&=(d_1-N_{1,0})-k_4(d_2-N_{2,0})-k_6(d_3-N_{3,0}),
 \label{eq:yukawa_residual_wp}\\
 Y_{\rm eff}(\lambda)&=Y_1(\lambda)-k_4Y_2(\lambda)-k_6Y_3(\lambda).
 \label{eq:yukawa_effective_wp}
\end{align}
Amplitudes are combined with their signs; subtracting the absolute values
would not implement this estimator. Let $J$ be the Newtonian torque
Jacobian, with entries $J_{qi}=\partial N_q/\partial p_i|_{\bm p_0}$ for
parameter $p_i$. A parameter name in the second subscript selects that
column of $J$. Then the $4f$ correction chooses
$k_4=J_{1,h_L}/J_{2,h_L}$ and $k_6=0$. When $6f$ is added, both coefficients
are recalculated from
\begin{equation}
 J_{1,p}-k_4J_{2,p}-k_6J_{3,p}=0,
 \qquad p\in\{h_L,b_S\}.
 \label{eq:yukawa_subtraction_coefficients_wp}
\end{equation}
These conditions remove the selected first-order parameter errors; they do
not remove all geometrical errors or nonlinear residuals.

Each harmonic must be calibrated using its own complex mechanical and
optical response before forming Eq.~\eqref{eq:yukawa_residual_wp}. In an
ideal orthogonal two-bar difference, the $4f$ torque cancels. Its auxiliary
measurement therefore requires an individual-bar or sum channel; it cannot
be recovered by simply reusing the differential $2f$ strain channel.

\section{Sensitivity after subtraction}
\label{sec:yukawa_sensitivity_wp}
\label{sec:yukawa_systematic_wp}

Let $\bm w=(1,-k_4,-k_6)^T$ be the subtraction-weight vector, where the
superscript $T$ denotes transpose, and $C_p$ the covariance matrix of
parameter errors. The Newtonian torque covariance is $C_N=JC_pJ^T$ at
first order. The corresponding variance of the residual systematic torque
is $\sigma_{\rm sys,lin}^2=\bm w^TC_N\bm w$. The quoted forecasts use the
nonlinear total checked with 50\,000 independent Gaussian Monte Carlo
realizations in Ref.~\cite{InoueYukawaHarmonic2026}; $\sigma_{\rm sys}$
denotes the resulting systematic torque uncertainty. Let $T$ be coherent
integration time in seconds and $S_\tau$ the one-sided, phase-referenced
harmonic torque-noise PSD matrix, in $(\mathrm{N\,m})^2/\mathrm{Hz}$.
The coupling thresholds $\alpha_{Y,\rm floor}$ and $\alpha_{Y,1\sigma}$
are the systematic floor and combined one-standard-deviation threshold.
Define $T_{\rm eq}$ as the time when statistical and systematic torque
variances are equal. The final expressions are
\begin{align}
 \alpha_{Y,\rm floor}(\lambda)
 &=\frac{\sigma_{\rm sys}}{|Y_{\rm eff}(\lambda)|},
 \label{eq:alpha_y_limit_wp}\\
 \alpha_{Y,1\sigma}(\lambda,T)
 &=\frac{\sqrt{\sigma_{\rm sys}^2+\bm w^TS_\tau\bm w/T}}
 {|Y_{\rm eff}(\lambda)|},
 \label{eq:yukawa_finite_time_wp}\\
 T_{\rm eq}&=\frac{\bm w^TS_\tau\bm w}{\sigma_{\rm sys}^2}.
 \label{eq:teq_wp}
\end{align}
Let $S_{\tau,2}$, $S_{\tau,4}$, and $S_{\tau,6}$ be the diagonal noise
PSDs of the $2f_{\rm rot}$, $4f_{\rm rot}$, and $6f_{\rm rot}$ channels.
For independent readout noise,
$\bm w^TS_\tau\bm w=S_{\tau,2}+k_4^2S_{\tau,4}+k_6^2S_{\tau,6}$.
Thus suppressing systematic error can amplify auxiliary measurement noise.
At $T=T_{\rm eq}$ the total threshold is $\sqrt2$ times the systematic
floor, and reaching within 10\% of that floor requires
$T_{10\%}=T_{\rm eq}/0.21$. The floor is not a finite-time confidence limit.

\begin{table}[tbp]
\centering\small
\begingroup\linespread{1.12}\selectfont
\caption{Harmonic-subtracted systematic-floor forecasts for the fiducial
geometry in Table~\ref{tab:yukawa_geometry_wp}, from
Ref.~\cite{InoueYukawaHarmonic2026}. The rows compare their respective floors,
not equal observation times. Auxiliary readout noise is neglected.}
\label{tab:yukawa_reach_wp}
\begin{tabular}{lrrr}
\toprule
Quantity & $2f$ & $2f+4f$ & $2f+4f+6f$\\
\midrule
Minimum $|\alpha_Y|$ & $1.99\times10^{-3}$ & $3.31\times10^{-4}$ & $1.88\times10^{-4}$\\
Range at minimum [m] & 3.38 & 5.08 & 5.98\\
Gain in minimum coupling & 1 & 6.0 & 10.6\\
$|\alpha_Y|$ at $\lambda=100$ m & 0.420 & $4.87\times10^{-4}$ & $2.56\times10^{-4}$\\
Large-$\lambda$ floor & $\propto\lambda^2$ & $4.89\times10^{-4}$ & $2.57\times10^{-4}$\\
\bottomrule
\end{tabular}
\endgroup
\end{table}

Without harmonic subtraction, the largest error contributions are $h_L$
and $b_S$. After canceling both at first order, source-mass uncertainties
and the effective distance $a$ become important. Better source metrology,
phase tracking, and auxiliary readout are therefore integral parts of the
sensitivity improvement.

\begin{figure}[tbp]
\centering
\includegraphics[width=0.98\linewidth]{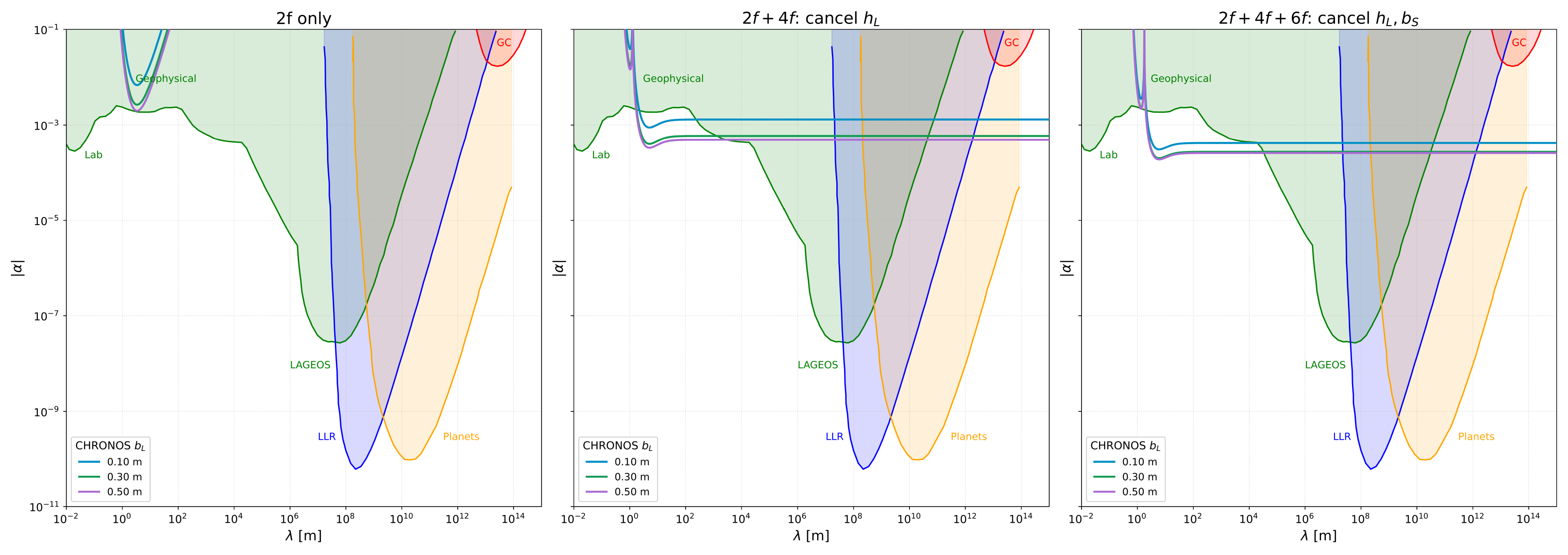}
\begingroup\linespread{1.1}\selectfont
\caption{Updated Yukawa projections reproduced from Fig.~3 of
Ref.~\cite{InoueYukawaHarmonic2026}. Left to right: $2f$, $2f+4f$, and
$2f+4f+6f$. Each panel compares $b_L=0.10,0.30,0.50$~m with $b_S=b_L/2$.
The corrections cancel $h_L$ and $(h_L,b_S)$, respectively. Curves use the
subtracted signal $Y_{\rm eff}$ as well as the subtracted systematic error,
and assume negligible auxiliary noise. Existing constraints retain the
confidence conventions of their original sources; the CHRONOS curves are
one-standard-deviation systematic-floor projections.}
\label{fig:yukawa_sensitivity_wp}
\endgroup
\end{figure}

\section{Noise normalization and integration time}
\label{sec:yukawa_baseline_wp}

The torque formulation above avoids mixing strain conventions between
studies. Here $\chi$ is the single-bar angular susceptibility in rad/(N m),
$\tau$ is the single-bar torque amplitude, $\eta_Y$ is the reference
geometry factor, and $F_{\rm eff}$ is its effective antenna factor.
In the free-rotor limit, $\chi=-1/[I(2\pi f)^2]$, with single-bar inertia $I$
and signal frequency $f$. The Yukawa reference defines the apparent strain of a single-bar
$2f$ torque $\tau$ as
$h_{\rm ref}=4|\chi\tau|/\eta_Y$, with $\eta_Y=0.936$ and
$|F_{\rm eff}|=1$. Chapter~\ref{chap:sensitivity} instead uses
$\theta_X-\theta_Y=\eta h$ with $\eta=0.93656$, giving
$h=2|\chi\tau|/\eta$ for opposite torques on the two bars.
Consequently $h=(\eta_Y/2\eta)h_{\rm ref}$; both noise and signal must be
converted by this same factor. It cancels from $\alpha_{Y,\rm floor}$ but
must be retained when computing integration times with a new noise curve.

The published ideal times assume
$\sqrt{S_{h,\rm ref}(0.5~\mathrm{Hz})}=10^{-17}~\mathrm{Hz}^{-1/2}$,
$I=19.86~\mathrm{kg\,m^2}$, and negligible $4f/6f$ readout noise. They are
listed as reference assumptions in Table~\ref{tab:yukawa_times_wp}; they
are not the noise level of the updated detector baseline. The updated
minimum-PI and logarithmic-mean noise values and the resulting times are
also provided there, with the same torque normalization and metrology
floor. These remain ideal projections until the auxiliary channels and
their correlated noise have been characterized.

\input{Chapters/yukawa_times_table}

\begin{figure}[tbp]
\centering
\includegraphics[width=0.98\linewidth]{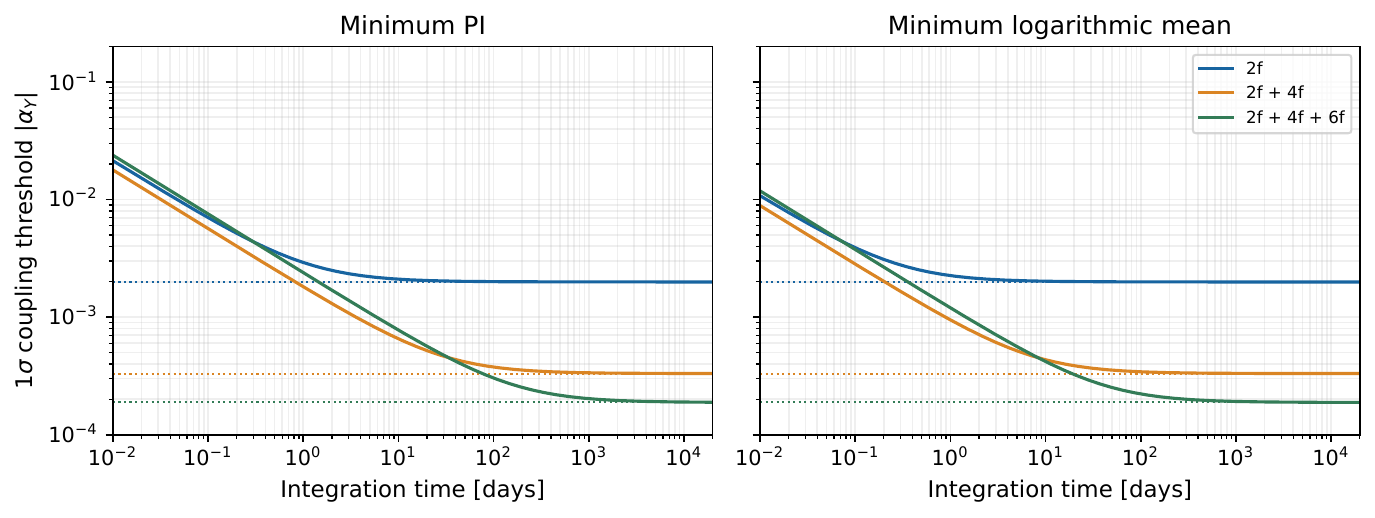}
\begingroup\linespread{1.1}\selectfont
\caption{Integration-time forecasts recalculated with the two full-noise
operating points in Chap.~\ref{chap:sensitivity}. Each estimator is evaluated
at the range of its own minimum in Table~\ref{tab:yukawa_reach_wp}.
Solid curves combine statistical and systematic errors; dotted lines show
the systematic floors from Ref.~\cite{InoueYukawaHarmonic2026}.
The strain conversion is applied to the floors before adding the detector
noise. Auxiliary $4f/6f$ measurement noise is neglected, so these are ideal
times rather than established experimental integration requirements.}
\label{fig:yukawa_integration_wp}
\endgroup
\end{figure}

\section{Long-range interpretation and priorities}
\label{sec:yukawa_science_wp}

At the exact Newtonian $2f$ null, $Y_1\propto\lambda^{-2}$ for large
$\lambda$, so the uncorrected coupling threshold grows as $\lambda^2$.
For the auxiliary channels $Y_{2,3}\to N_{2,3,0}$, and the subtracted
response tends to $-k_4N_{2,0}-k_6N_{3,0}$. This gives the finite plateaus
in Table~\ref{tab:yukawa_reach_wp}. Those plateaus depend on external
priors on $G$, source masses, and geometry: without those priors, an
infinite-range Yukawa term is degenerate with the gravitational normalization.
Zeros of $Y_{\rm eff}$ instead produce loss of sensitivity and poles in
the threshold curve.

The experimental program is therefore to validate the Newtonian null,
measure all participating complex harmonic responses, demonstrate the
predicted rejection of geometrical errors, and determine the statistical
and systematic covariance of the combined estimator. The projected gain
from subtraction is realized only if these measurements support the
assumptions used in Eqs.~\eqref{eq:alpha_y_limit_wp}--\eqref{eq:teq_wp}.

%% file: Chapters/yukawa_times_table.tex
\begin{table}[tbp]
\centering\small
\caption{Ideal integration times $T_{\rm eq}$ in days for the three estimators.
The reference-noise row is recomputed from the rounded Monte Carlo floors
with the original noise and strain convention.
The other rows use the full strain noise of Chap.~\ref{chap:sensitivity}:
$1.237\times10^{-17}$ and $6.154\times10^{-18}~\mathrm{Hz}^{-1/2}$ at 0.5~Hz.
The Monte Carlo floors are converted to the same convention. In all rows,
$T_{10\%}=T_{\rm eq}/0.21$ and auxiliary readout noise is neglected.}
\label{tab:yukawa_times_wp}
\begin{tabular}{lrrr}\toprule
Noise model & $2f$ & $2f+4f$ & $2f+4f+6f$\\\midrule
Reference noise & 0.187 & 4.74 & 26.1\\
Minimum PI & 1.14 & 29 & 160\\
Minimum logarithmic mean & 0.283 & 7.19 & 39.7\\
\bottomrule\end{tabular}
\end{table}

%% file: Chapters/GW_Science.tex
\chapter{Gravitational-Wave Science}
\label{chap:gw_science}

CHRONOS targets the 0.1--10~Hz interval between the principal bands of
space-based and conventional terrestrial laser interferometers. Its science
case includes binaries containing intermediate-mass black holes (IMBHs)
and stochastic gravitational-wave backgrounds (SGWBs). The forecasts here
use the full strain noise of Chap.~\ref{chap:sensitivity} for both optimized
operating points. They are conditional instrument projections, not measured
performance or guaranteed detection rates.

\section{Accessing the gravitational-wave frequency gap}
\label{sec:gw_frequency_gap}

Pulsar timing arrays probe nanohertz frequencies, LISA is designed for the
millihertz band, and conventional terrestrial interferometers principally
observe above roughly 10~Hz. These windows are not sharp boundaries.
Ground-based access to lower frequencies is limited by seismic coupling,
gravity-gradient noise, and mechanical dissipation. CHRONOS combines a
low-frequency torsional response with cryogenic operation and a Sagnac
speed-meter readout. Environmental gravity fluctuations still require site
characterization and mitigation.

The relevant sensitivity measure depends on the source. A compact-binary
search coherently accumulates a waveform; an SGWB search accumulates excess
or correlated power. Neither reach is determined by the lowest strain ASD
alone. The two PI-optimized settings need not be optimal for every binary
mass or every stochastic spectrum.

\section{Intermediate-mass black-hole binaries}
\label{sec:gw_imbh}

IMBHs are commonly associated with masses of roughly $10^2$--$10^5M_\odot$.
GW190521 provided evidence for an IMBH merger remnant
\cite{PhysRevLett.125.101102}; the abundance and formation history of this
population remain uncertain. Observations could test hierarchical mergers,
stellar-cluster formation channels, and the growth of massive black-hole
seeds, provided the detector selection function is established.

\subsection{Mass, frequency, and calibrated signal overlap}

Increasing binary mass shifts coalescence toward lower frequency. A useful
scale is the Schwarzschild test-particle innermost-stable-circular-orbit
(ISCO) frequency. Let $G$ be the gravitational constant, $c$ the speed of
light, $M_\odot$ the solar mass, $z$ the cosmological redshift,
$M_{\rm tot}$ the source-frame total mass, and $M_z=(1+z)M_{\rm tot}$
the detector-frame total mass. The observed frequency is
\begin{equation}
 f_{\rm ISCO,obs}\simeq\frac{c^3}{6^{3/2}\pi GM_z}
 \simeq4.4~\mathrm{Hz}\left(\frac{10^3M_\odot}{M_z}\right),
 \quad M_z=(1+z)M_{\rm tot}.
 \label{eq:fisco}
\end{equation}
This scale is not a precise merger or ringdown frequency for
comparable masses. Lower-mass binaries traverse the CHRONOS band in inspiral;
higher-mass systems can coalesce in the band or below it.

For the signal comparison, $\widetilde h(f)$ is the Fourier amplitude of
the projected strain and $S_n(f)$ is the full one-sided strain-noise PSD.
Define the dimensionless characteristic signal $h_c=2f|\widetilde h|$ and
noise $h_n=\sqrt{fS_n}$. The spin labels denote the mass-weighted projections
$\chi_{\rm eff}$ along the orbital angular momentum, defined precisely
below in Eq.~\eqref{eq:gw_effective_spin}.

\begin{figure}[tbp]
\centering
\includegraphics[width=0.97\linewidth]{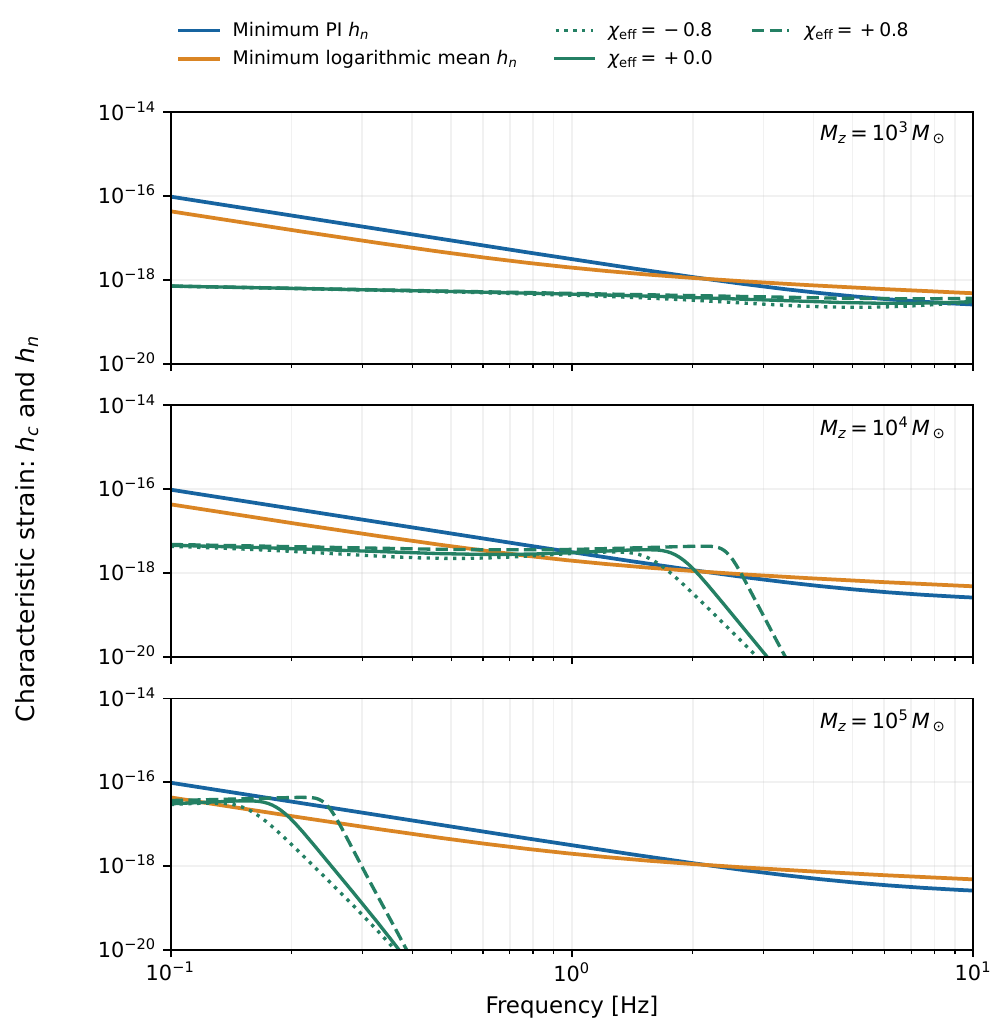}
\caption{Consistently normalized signal and noise comparison using the
updated detector spectra. Noise is $h_n=\sqrt{fS_n}$; binary signals are
$h_c=2f|\widetilde h|$. All signals are at luminosity distance 50~Mpc and
use IMRPhenomD, equal masses, face-on inclination, and antenna factor 1.
The three panels give detector-frame total masses. Green curves compare
equal component spins $\chi_{1z}=\chi_{2z}=\chi_{\rm eff}=-0.8,0,+0.8$;
blue and orange curves show the two full-noise spectra.
All quantities on the ordinate are dimensionless.}
\label{fig:gw_waveform_overlap}
\end{figure}

\subsection{Spin dependence and matched-filter SNR at 50 Mpc}

For a Fourier-domain projected strain $\widetilde h(f)$ and a one-sided
noise PSD $S_n(f)$, the matched-filter SNR $\rho$ is
\begin{equation}
 \rho^2=4\int_{0.1\,\mathrm{Hz}}^{10\,\mathrm{Hz}}
 \frac{|\widetilde h(f)|^2}{S_n(f)}\,df
 =\int\left(\frac{h_c}{h_n}\right)^2d\ln f.
 \label{eq:gw_snr}
\end{equation}
The noise is squared ASD, and the signal and noise use the same strain
calibration. Antenna or orientation factors must not be applied twice.

\begin{figure}[tbp]
\centering
\includegraphics[width=0.97\linewidth]{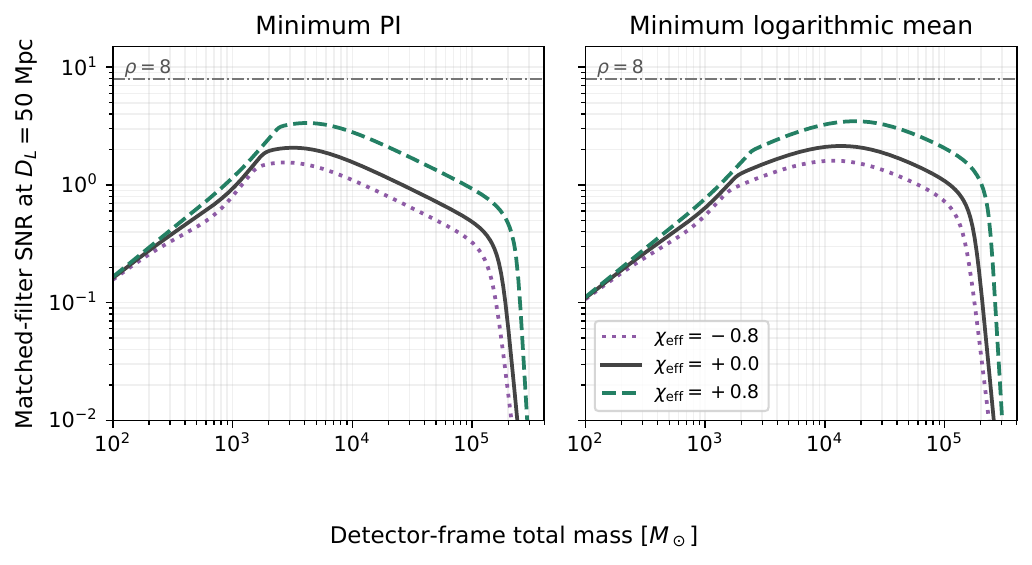}
\caption{CBC matched-filter SNR at 50~Mpc as a function of detector-frame
total mass, for the two operating points and the three equal-component
spin choices in Table~\ref{tab:gw_reach}. The horizontal line marks the
illustrative $\rho=8$ threshold. All curves use the full one-sided noise
PSD and integration over 0.1--10~Hz.}
\label{fig:IMBH_2p5}
\end{figure}

Figures~\ref{fig:gw_waveform_overlap} and~\ref{fig:IMBH_2p5} use LALSuite
IMRPhenomD \cite{Khan2016IMRPhenomD}, a dominant-quadrupole model for
quasi-circular binaries with spins parallel or antiparallel to the orbital
angular momentum. We compare equal-mass binaries at $D_L=50$~Mpc with
face-on inclination and effective antenna response 1. For binary component
$i=1,2$, let $m_i$ be its mass, $\boldsymbol S_i$ its spin angular momentum,
and $\widehat{\boldsymbol L}$ the unit vector along the orbital angular
momentum. The dimensionless aligned-spin component $\chi_{iz}$ and
mass-weighted effective spin $\chi_{\rm eff}$ are
\begin{equation}
 \chi_{iz}=\frac{c\,\boldsymbol S_i\cdot\widehat{\boldsymbol L}}{Gm_i^2},
 \qquad
 \chi_{\rm eff}=\frac{m_1\chi_{1z}+m_2\chi_{2z}}{m_1+m_2}.
 \label{eq:gw_effective_spin}
\end{equation}
We set $\chi_{1z}=\chi_{2z}=-0.8,0,+0.8$, so that these values also label
$\chi_{\rm eff}$. Positive spin denotes alignment with the orbit. These
examples lie within the equal-mass calibration range of IMRPhenomD
\cite{Khan2016IMRPhenomD}; they do not cover spin precession, unequal masses,
eccentricity, or higher multipoles. The full passage
through 0.1--10~Hz is included, without duty-cycle or finite observing-window
corrections. The mass is detector-frame mass and the distance is luminosity
distance; no source-frame conversion or population model is imposed.
The archived input spectra and derived values accompany this White Paper.

\input{Chapters/cbc_spin_table}

Aligned spins shift the merger and ringdown to higher frequencies at fixed
total mass, changing their overlap with the detector band. In the examples
computed here, the maximum SNR at 50~Mpc rises from 1.55 to 3.36 for the
minimum-PI point, and from 1.61 to 3.48 for the logarithmic-mean point, as
$\chi_{\rm eff}$ changes from $-0.8$ to $+0.8$. The mass of maximum SNR
also increases (Table~\ref{tab:gw_reach}). This is a result of the full
waveform integral, not a spin correction to the Schwarzschild ISCO estimate.

\subsection{Mass--distance SNR contours}
\label{sec:gw_mass_distance}

\begin{figure}[p]
\centering
\includegraphics[width=0.97\linewidth]{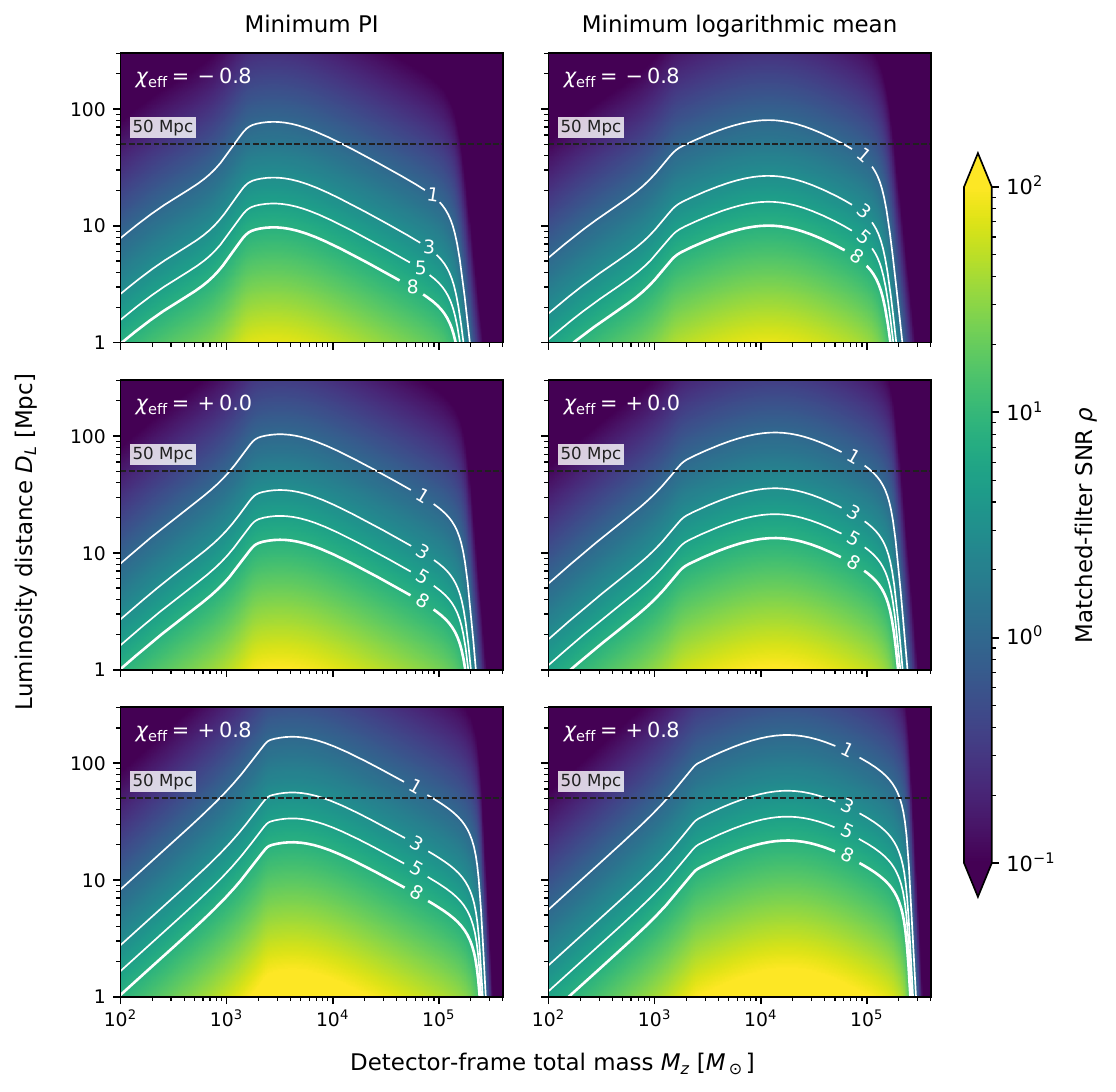}
\begingroup\linespread{1.1}\selectfont
\caption{Mass--distance contours of CBC matched-filter SNR. Columns show
the minimum-PI and minimum-logarithmic-mean operating points; rows show
$\chi_{1z}=\chi_{2z}=\chi_{\rm eff}=-0.8,0,+0.8$. The color scale gives
$\rho$ and white contours mark $\rho=1,3,5,8$, with the $\rho=8$ contour
emphasized. Dashed lines mark 50~Mpc. All panels use equal-mass, face-on
IMRPhenomD waveforms, antenna response 1, the full noise spectra of
Chap.~\ref{chap:sensitivity}, and integration over 0.1--10~Hz.
The horizontal axis is detector-frame total mass.}
\label{fig:gw_mass_distance}
\endgroup
\end{figure}

Figure~\ref{fig:gw_mass_distance} extends the calculation to luminosity
distance $D_L$. Define the reference distance $D_0=50$~Mpc and
$\rho_0(M_z,\chi_{\rm eff})$ as the SNR at that distance, using the
equal-mass, equal-spin waveforms and noise spectra specified above.
At fixed detector-frame total mass $M_z$ and spin, the waveform amplitude
scales as $D_L^{-1}$. For a target SNR $\rho_\star$, the distance contour
$D_L^{(\rho_\star)}$ therefore follows
\begin{equation}
 \rho(M_z,D_L,\chi_{\rm eff})=
 \rho_0(M_z,\chi_{\rm eff})\frac{D_0}{D_L},
 \qquad
 D_L^{(\rho_\star)}(M_z,\chi_{\rm eff})=
 D_0\frac{\rho_0(M_z,\chi_{\rm eff})}{\rho_\star}.
 \label{eq:gw_distance_scaling}
\end{equation}
The columns compare the two optimized operating points and the rows the
three spin choices. The contours show $\rho_\star=1,3,5,8$; the 50~Mpc
line is the slice plotted in Fig.~\ref{fig:IMBH_2p5}. Distance is varied
without changing detector-frame mass, so no source-frame mass or redshift
conversion is implicit in this map.

The
nonspinning $\rho=8$ horizons remain 12.9 and 13.4~Mpc; the aligned
$\chi_{\rm eff}=+0.8$ examples reach 21.0 and 21.7~Mpc, respectively.
Even these examples stay below $\rho=8$ at 50~Mpc. A real search
threshold depends on data quality, the template bank, and the measured
background. Neither this threshold nor $\rho=1$ establishes a detection
probability or an event rate. The projected distances concern nearby
systems; they do not imply direct observations of seed formation at high
redshift. A population interpretation requires inclination and sky averaging,
a source-frame mass distribution, and an observing-time selection function.

Multiband observations require adequate SNR in each participating instrument.
Low-mass binaries may evolve into the band of conventional terrestrial
detectors; more massive systems may connect a space-based inspiral to a
low-frequency coalescence. Residence times and early-warning intervals
depend on mass and entry frequency, and are not automatically months long
in the 0.1--10~Hz band.

\section{Stochastic gravitational-wave backgrounds}
\label{sec:gw_sgwb}

Unresolved compact binaries and early-Universe processes can produce SGWBs.
Their detectability requires a specified spectrum, amplitude, and response.
Let $\rho_{\rm GW}$ be the present GW energy density, $\rho_c$ the
critical energy density, and $H_0$ the present Hubble expansion rate. The
dimensionless energy-density spectrum $\Omega_{\rm GW}$ is defined by
\begin{equation}
 \Omega_{\rm GW}(f)=\frac{1}{\rho_c}\frac{d\rho_{\rm GW}}{d\ln f},
 \qquad\rho_c=\frac{3H_0^2c^2}{8\pi G},
 \label{eq:gw_omega_gw_def}
\end{equation}
where both densities are energy densities. The one-sided tensor-background
strain convention used here is
\begin{equation}
 S_h^{\rm GW}(f)=\frac{3H_0^2}{2\pi^2}\frac{\Omega_{\rm GW}(f)}{f^3}.
 \label{eq:omega_sh}
\end{equation}

\subsection{The common PI comparison metric}

Chapter~\ref{chap:sensitivity} uses a single-detector, weak-signal,
perfectly known-noise metric with response $\mathcal R=1$, SNR~1,
$H_0=67.4~\mathrm{km\,s^{-1}\,Mpc^{-1}}$, and integration over 0.1--10~Hz.
Let $T$ denote effective observing time in seconds, $\rho_\star$ the
target SNR, $f_{\rm ref}$ a reference frequency, and $\beta$ a power-law
slope. For a spectrum $\Omega_\beta(f/f_{\rm ref})^\beta$, $\Omega_\beta$
is its amplitude at $f_{\rm ref}$. Define $\Omega_n$ as the equivalent
noise energy-density spectrum, $\Omega_\beta^{\rm thr}$ as the threshold
amplitude, and $\Omega_{\rm PI}$ as the power-law-integrated envelope:
\begin{align}
 \Omega_n(f)&=\frac{2\pi^2}{3H_0^2}\frac{f^3S_n(f)}{\mathcal R},\\
 \Omega_\beta^{\rm thr}&=\frac{\rho_\star}{\sqrt{T}}
 \left[\int_{0.1\,\mathrm{Hz}}^{10\,\mathrm{Hz}}
 \frac{(f/f_{\rm ref})^{2\beta}}{\Omega_n^2(f)}\,df\right]^{-1/2},
 \label{eq:omega_beta_threshold}\\
 \Omega_{\rm PI}(f)&=\max_\beta\left[\Omega_\beta^{\rm thr}
 (f/f_{\rm ref})^\beta\right].
\end{align}
The envelope uses $-8\le\beta\le8$ and $f_{\rm ref}=1$~Hz
\cite{PhysRevD.88.124032,InoueQuantumTorque2026}. It is not a per-bin limit.
Peaked or broken spectra must be integrated with their actual shape.

The common one-year PI curves are shown once in
Fig.~\ref{fig:sensitivity_PI}. For ten years of effective observing time,
the same curves scale downward by $1/\sqrt{10}$ under the known-noise
assumptions above. The response $\mathcal R=1$ is a comparison convention,
not an independently calculated isotropic CHRONOS response.

The minimum-PI reference has $\min\Omega_{\rm PI}=4.0142\times10^{-4}$
for one year and $1.2694\times10^{-4}$ for ten years. Its reported pivot is
4.687~Hz. The logarithmic-mean setting gives $5.9611\times10^{-4}$ and
$1.8851\times10^{-4}$, respectively, with pivot 1.821~Hz. These values
replace the earlier ten-year estimate used in this chapter.

\subsection{From a comparison metric to an SGWB measurement}

A robust stochastic measurement requires control of noise bias. For two
channels with independent instrumental noise, let $C_{12}^{\rm GW}$ be
their GW cross-PSD and $\Gamma_{12}$ their dimensionless overlap response,
so that $C_{12}^{\rm GW}=\Gamma_{12}S_h^{\rm GW}$. Let $P_1,P_2$ be the
channel noise PSDs and $\rho_{12}$ the cross-correlation SNR. The weak-signal statistic is
\begin{equation}
 \rho_{12}^2=2T\int df\,
 \frac{|\Gamma_{12}(f)|^2[S_h^{\rm GW}(f)]^2}{P_1(f)P_2(f)}.
 \label{eq:sgwb_snr_omega}
\end{equation}
Here $P_1,P_2$ are one-sided noise PSDs and $\Gamma_{12}$ includes geometry,
angular averaging, and separation \cite{Romano2017}. Separate optical
readouts of shared mechanics need not have independent noise. The
single-detector PI benchmark must not be presented as a demonstrated
cross-correlation sensitivity without this calculation and a correlated-noise
assessment.

CMB constraints on the tensor-to-scalar ratio and direct interferometer
constraints apply at very different scales. For example, the BK18 bound
$r_{0.05}<0.036$ at 95\% confidence is specified at a pivot wavenumber
$0.05~\mathrm{Mpc}^{-1}$ \cite{BK18_Inoue}; extrapolation requires a primordial
spectrum and transfer history. NANOGrav's 15-year evidence does not establish
a unique cosmological origin or a hertz-band extrapolation \cite{NANOGrav2023}.
The baseline here does not establish detectability of nearly scale-invariant
inflationary vacuum fluctuations. Phase transitions, strings, or enhanced
primordial spectra require model-specific comparisons with the integrated
statistic and with existing bounds, as discussed in Chap.~\ref{chap:cosmology}.

%% file: Chapters/cbc_spin_table.tex
\begin{table}[tbp]
\centering\small
\caption{Spin dependence of the maximum CBC SNR at 50~Mpc. Each row is
maximized over detector-frame total mass for an equal-mass, face-on binary
with $\chi_{1z}=\chi_{2z}=\chi_{\rm eff}$. The final column rescales the same
waveform to the illustrative $\rho=8$ horizon.}
\label{tab:gw_reach}
\begin{tabular}{lrrrr}\toprule
Operating point & $\chi_{\rm eff}$ & $M_{z,\rm peak}$ & $\rho_{\max}(50~\mathrm{Mpc})$ & $D_{L,\max}^{\rho=8}$\\
 & & [$10^3M_\odot$] & & [Mpc]\\\midrule
Minimum PI & $-0.8$ & 2.77 & 1.55 & 9.7\\
 & $+0.0$ & 3.20 & 2.07 & 12.9\\
 & $+0.8$ & 4.16 & 3.36 & 21.0\\
Minimum logarithmic mean & $-0.8$ & 11.79 & 1.61 & 10.0\\
 & $+0.0$ & 13.62 & 2.14 & 13.4\\
 & $+0.8$ & 17.73 & 3.48 & 21.7\\
\bottomrule\end{tabular}
\end{table}

%% file: Chapters/Cosmology.tex
\chapter{Cosmology}\label{chap:cosmology}

Primordial gravitational waves are a unique observational
window into the physics of the early Universe. 
Unlike electromagnetic radiation, gravitational waves propagate
essentially freely once produced and therefore preserve direct
information about the physical processes occurring at extremely
high energy scales.
The detection of primordial gravitational waves would therefore
provide powerful insight into the origin of cosmic structure
and the fundamental physics governing the earliest moments
of the Universe.

One of the most compelling targets of modern cosmology is the
gravitational-wave background generated during the inflationary
epoch.
Quantum fluctuations of the spacetime metric during inflation
produce tensor perturbations that subsequently propagate as
relic gravitational waves.
The amplitude of these perturbations is commonly parameterized
by the tensor-to-scalar ratio $r$, which quantifies the
relative strength of tensor 
compared to 
scalar fluctuations.

The most sensitive probe of primordial tensor perturbations
is the observation 
of the cosmic microwave
background (CMB) polarization. 
Primordial gravitational waves generate a characteristic
B-mode polarization pattern in the CMB, providing a direct
observable signature of tensor perturbations.
Experiments such as \textit{Planck} and BICEP/Keck have
therefore placed increasingly stringent limits on the
tensor amplitude at cosmological scales.

While CMB observations probe gravitational waves at extremely
large cosmological scales, corresponding to frequencies near
$f \sim 10^{-17}\,\mathrm{Hz}$, 
all the other 
gravitational-wave
detectors probe dramatically higher frequencies.
Ground-based interferometers such as LIGO and Virgo are
sensitive to stochastic gravitational-wave backgrounds in
the frequency band around tens to hundreds of hertz, while
future space-based detectors will explore the millihertz
band.

One of the key quantities 
describing the primordial tensor spectrum
is the tensor spectral index $n_t$, which characterizes
the scale dependence of the gravitational-wave amplitude.
While many inflationary models predict a slightly
red-tilted spectrum, a variety of early-Universe scenarios
allow a positive tensor spectral index.
Such a blue-tilted spectrum corresponds to a gravitational-wave
energy density that increases toward higher frequencies.

Blue-tilted spectra are particularly 
interesting: the enhanced power of gravitational waves 
at high frequencies can bring primordial signals within the
sensitivity range of 
other detectors such as CHRONOS or interferometers 
even when the tensor amplitude at the CMB scales is small.
This makes gravitational-wave observations across multiple
frequency bands a powerful probe of early-Universe physics.

In this context, gravitational-wave detectors operating
in the sub-Hz frequency band can play a critical role.
This frequency range bridges the gap between space-based
and ground-based interferometers and provides a key window
for probing blue-tilted primordial spectra.
The CHRONOS experiment is designed to explore this
frequency band with high sensitivity, enabling
cosmological measurements that are complementary to
both CMB observations and higher-frequency
gravitational-wave detectors.

In this chapter, we describe the cosmological framework
used to connect primordial tensor perturbations with
present-day gravitational-wave observations and
investigate how multi-band measurements can constrain
blue-tilted spectra. 

\section{Basic principle of the analysis}
\label{sec:principle}

The stochastic gravitational-wave background 
provides a statistical probe of primordial gravitational
waves generated in the early Universe.
Unlike transient signals produced by individual
astrophysical events, the primordial background is
expected to consist of the superposition of a large
number of independent sources and is therefore
characterized statistically through its spectral
energy density.


Let $k$ be comoving wavenumber, $f$ observed frequency, $c$ the speed
of light, $a_0$ the present scale factor, and $H_0$ the present Hubble rate.
The dimensionless primordial tensor power spectrum $P_t(k)$ has amplitude
$A_t$ at pivot wavenumber $k_*$ and spectral index $n_t$:

\begin{equation}
P_t(k)
=
A_t
\left(
\frac{k}{k_*}
\right)^{n_t},
\end{equation}

where \(A_t\) is the tensor amplitude evaluated at the
pivot scale \(k_*\) and \(n_t\) denotes the tensor
spectral index.
The pivot scale is typically chosen as

\[
k_* \simeq 0.05\,{\rm Mpc}^{-1},
\]

which is commonly adopted in CMB analyses.

After their generation, primordial tensor perturbations
propagate through the expanding Universe.
Their evolution is described by the gravitational-wave
transfer function \(T_h(k)\), which accounts for the
horizon re-entry of each mode and the expansion history
of the Universe.

The present-day gravitational-wave energy density
spectrum is related to the primordial tensor spectrum
through

\begin{equation}
\Omega_{\rm GW}(f)
=
\frac{1}{12}
\left(
\frac{ck}{a_0 H_0}
\right)^2
P_t(k)
T_h^2(k),
\end{equation}

where $\Omega_{\rm GW}$ is the GW energy density per logarithmic
frequency divided by the critical energy density, as in
Sec.~\ref{sec:gw_sgwb}. The amplitude transfer $T_h$ is normalized to
unit primordial tensor amplitude.

The comoving wavenumber \(k\) is related to the observed
gravitational-wave frequency through

\begin{equation}
k = 2\pi f a_0/c .
\end{equation}

This relation provides the link between the primordial
tensor spectrum defined at cosmological scales and the
gravitational-wave energy density measured by
interferometric detectors.

Measurements of \(\Omega_{\rm GW}(f)\) at interferometer
frequencies therefore provide complementary information
to CMB observations and enable constraints to be placed
on the scale dependence of primordial tensor
perturbations.


\subsection{Transfer function in radiation- and matter-dominated eras}

The amplitude transfer function \(T_h(k)\) describes the evolution of
primordial tensor perturbations after horizon re-entry and
encodes the effect of the cosmological expansion history.
Modes entering the horizon during the radiation-dominated
epoch evolve differently from those entering during the
matter-dominated epoch.
This produces a scale-dependent modification of the
primordial tensor spectrum that must be included when
computing the present-day gravitational-wave energy
density.

For a compact energy-density approximation, define a different,
dimensionless transfer factor $\mathcal T_\Omega(k)$. It incorporates
both $T_h$ and the frequency prefactor above. Let $A_s$ denote the primordial
scalar curvature-power amplitude at $k_*$, and $\Omega_r$ the present
radiation density parameter. Then

\begin{equation}
\Omega_{\rm GW}(k)
=
\Omega_{\rm ref}
\left(
\frac{k}{k_*}
\right)^{n_t}
\mathcal T_\Omega^2(k),
\end{equation}

where the normalization set by CMB parameters is

\begin{equation}
\Omega_{\rm ref}
=
\frac{3}{128} \, r A_s \, \Omega_r .
\end{equation}

Here \(\Omega_r\) denotes the present radiation density
parameter and \(k_*\) is the pivot scale used in CMB
analyses.

The labels RD and MD denote radiation and matter domination. Define
$k_{\rm eq}$ as the comoving horizon wavenumber at matter--radiation equality.
The corresponding contributions to $\mathcal T_\Omega$ are approximated by

\begin{equation}
\mathcal T_\Omega^2(k)
=
\mathcal T_{\rm MD}^2(k)
+
\mathcal T_{\rm RD}^2 .
\end{equation}

For modes that re-enter the horizon during the
matter-dominated era, the transfer function scales as

\begin{equation}
\mathcal T_{\rm MD}^2(k)
=
\frac{1}{2}
\left(
\frac{k_{\rm eq}}{k}
\right)^2 ,
\end{equation}

where \(k_{\rm eq}\) is the comoving horizon scale at
the 
matter-radiation equality.

For modes entering during the radiation-dominated era,
the transfer function approaches a constant value,

\begin{equation}
\mathcal T_{\rm RD}^2
=
\frac{16}{9}.
\end{equation}

Combining these contributions yields the approximate
expression

\begin{equation}
\Omega_{\rm GW}(k)
=
\Omega_{\rm ref}
\left(
\frac{k}{k_*}
\right)^{n_t}
\left[
\frac{1}{2}
\left(
\frac{k_{\rm eq}}{k}
\right)^2
+
\frac{16}{9}
\right].
\end{equation}

This analytic expression provides a convenient model
for the primordial gravitational-wave spectrum over a
wide range of frequencies and is widely used in studies
of multi-band gravitational-wave cosmology.

\section{Cosmological model and assumptions}
\label{sec:model}

The cosmological analysis presented in this work is
performed within the framework of the standard
$\Lambda$CDM cosmology extended to include primordial
tensor perturbations.

The parameter set used in the analysis is

\begin{equation}
\Theta =
\{
\Omega_b,
\Omega_c,
h,
\tau,
n_s,
A_s,
r,
n_t
\},
\end{equation}

where $\Omega_b$ and $\Omega_c$ denote the present
baryon and cold dark-matter density parameters,
$h$ is the dimensionless Hubble parameter defined by

\[
H_0 = 100\,h\,{\rm km\,s^{-1}\,Mpc^{-1}},
\]

and $\tau$ is the optical depth to reionization.

The scalar perturbations are characterized by the
spectral index $n_s$ and the amplitude $A_s$ of the
primordial curvature power spectrum.

The tensor sector is described by the tensor-to-scalar
ratio $r$ and the tensor spectral index $n_t$.
The amplitude of the primordial tensor spectrum is
related to the scalar amplitude through

\begin{equation}
A_t = r A_s .
\end{equation}

In the simplest slow-roll inflation models, the tensor
spectral index satisfies the consistency relation
\(n_t=-r/8\), i.e., the spectrum is slightly red-tilted ($n_t<0$). 
In this work, however, we do not impose this relation
and instead treat $n_t$ as an independent free parameter.
This allows us to explore a wider class of early-Universe
scenarios and to test both red-tilted ($n_t<0$) and
blue-tilted ($n_t>0$) tensor spectra.

\subsection{Blue-tilted tensor spectrum}
\label{sec:blue_tilt}

While many slow-roll inflation models predict a
slightly red-tilted tensor spectrum, a variety of
early-Universe scenarios allow a positive tensor
spectral index ($n_t>0$).
Examples include inflationary models with
nonstandard dynamics, non-Bunch-Davies initial states,
particle production during inflation,
and alternative early-Universe scenarios such as
ekpyrotic cosmology
(see, e.g., Refs.~\cite{Smith2006,Contaldi2008,Cook2014}).
Generalized G-inflation provides the most general framework
of inflation models with second-order field equations, which
can accommodate the blue spectrum.\cite{Kobayashi_2011}.

A positive tensor spectral index implies that the
gravitational-wave energy density increases toward
higher frequencies.
Consequently, even a small tensor amplitude at
cosmological scales can 
correspond to 
an enhanced signal
in the frequency range probed by 
the other 
gravitational-wave detectors.

This property makes blue-tilted tensor spectra
particularly well suited to tests using
multi-band gravitational-wave observations.
By combining measurements from CMB experiments
and direct gravitational-wave detectors,
it becomes possible to constrain the spectral
tilt of primordial tensor perturbations across
many orders of magnitude in scale.

In this work we therefore treat $n_t$ as a free
parameter and investigate the resulting constraints
obtained from the combined analysis of CMB data
and gravitational-wave observations.
\section{Observables and expected sensitivities} 
\label{sec:observable}

Primordial tensor perturbations leave observable
signatures both in the cosmic microwave background (CMB)
and in the present-day stochastic gravitational-wave
background.

In CMB observations, tensor modes manifest themselves
primarily through the B-mode polarization generated by
gravitational waves at the time of 
recombination and reionization. 
These measurements constrain the amplitude of primordial
tensor perturbations at large cosmological scales and
are commonly expressed in terms of the tensor-to-scalar
ratio \(r\).

In contrast, 
gravitational-wave detectors 
probing 
the present-day gravitational-wave energy density
spectrum \(\Omega_{\rm GW}(f)\) 
are sensitive to 
much higher frequencies.
The strain sensitivity of interferometric detectors
therefore translates into constraints on the stochastic
gravitational-wave background.

The relation between the comoving wavenumber and the
observed gravitational-wave frequency is

\begin{equation}
k = 2\pi f a_0 ,
\end{equation}

which connects the primordial tensor spectrum defined
at cosmological scales to the gravitational-wave
energy density measured by interferometric detectors.

By combining CMB measurements of the tensor amplitude
with constraints on the stochastic gravitational-wave
background at interferometer frequencies, it becomes
possible to probe the scale dependence of primordial
tensor perturbations across a very wide range of
scales.

Figure~\ref{fig:sensitivity_PI} shows the common CHRONOS PI comparison metric;
its statistical definition is given in Sec.~\ref{sec:gw_sgwb}.
It uses the same two operating points as Chap.~\ref{chap:sensitivity}:
one-year minima $4.0142\times10^{-4}$ and $5.9611\times10^{-4}$,
or ten-year minima $1.2694\times10^{-4}$ and $1.8851\times10^{-4}$.
The assumptions are a single detector with known noise, SNR~1,
$\mathcal R=1$, and 0.1--10~Hz. A primordial spectrum should be integrated
against the noise, not tested solely by crossing a PI curve. These
conditional sensitivity curves are not cosmological posteriors;
an updated joint CMB--CHRONOS posterior requires a separate likelihood
analysis with an explicit signal and noise model.
\subsection{Bayesian framework and data combination}

The cosmological parameters are inferred within a
Bayesian statistical framework.
For a given observational data set \(D\) and a set of
cosmological parameters \(\Theta\), the likelihood
function is defined as

\begin{equation}
\mathcal{L}(D|\Theta).
\end{equation}

The posterior probability distribution of the parameters
is obtained using Bayes' theorem,

\begin{equation}
P(\Theta|D)
\propto
\mathcal{L}(D|\Theta)\,P(\Theta),
\end{equation}

where \(P(\Theta)\) denotes the prior distribution of the
model parameters.

A joint analysis can combine two classes of information: measurements of the cosmic microwave background
(CMB) and constraints on the stochastic gravitational-wave
background obtained from gravitational-wave
experiments.
Assuming that the two data sets are statistically
independent, the total likelihood can be written as

\begin{equation}
\mathcal{L}_{\rm total}
=
\mathcal{L}_{\rm CMB}
\,
\mathcal{L}_{\rm GW}.
\end{equation}

The CMB likelihood \(\mathcal{L}_{\rm CMB}\) is evaluated
using publicly-available likelihood packages for the
\textit{Planck} 2018 temperature and polarization data
together with the BICEP/Keck B-mode measurements.
These data primarily constrain the tensor-to-scalar ratio
\(r\), which probes primordial gravitational waves at
cosmological scales.

The second component of the likelihood,
\(\mathcal{L}_{\rm GW}\), incorporates observational
constraints on the stochastic gravitational-wave
background measured by interferometric detectors.
These measurements probe the present-day
gravitational-wave energy density spectrum
\(\Omega_{\rm GW}(f)\) at interferometer frequencies.

\subsection{Sensitivity and statistical uncertainty}
\label{sec:sensitivity}

For consistency with the single-detector reference metric, use a one-sided
strain PSD $S_n$ and a bin containing $T\Delta f_i$ independent positive
frequency samples. The negative-frequency samples of a real time series
are not independent additional measurements. In the weak-signal,
known-noise limit,
\begin{equation}
 \Omega_n(f)=\frac{2\pi^2}{3H_0^2}\frac{f^3S_n(f)}{\mathcal R},
 \qquad \sigma_i\simeq\frac{\Omega_n(f_i)}{\sqrt{T\Delta f_i}},
 \qquad \mathcal R=1.
\end{equation}
Here $\widehat\Omega_i$ is formed after subtraction of the known noise
bias. Averaging and Gaussian errors require sufficiently many independent
samples; spectral leakage and nonstationarity can invalidate this simple
approximation. Integrating these bin weights gives the single-detector PI
normalization of Chap.~\ref{chap:gw_science}. A two-detector
cross-correlation uses a different variance and an explicit overlap response,
as in Eq.~\eqref{eq:sgwb_snr_omega}.

\subsection{Likelihood for stochastic gravitational-wave constraints}
\label{sec:gw_likelihood}

The stochastic gravitational-wave spectrum is represented
by estimators

\begin{equation}
\hat{\Omega}_{\rm GW}(f_i),
\end{equation}

with statistical uncertainties \(\sigma_i\).

For a given cosmological parameter set \(\Theta\),
the theoretical prediction is

\begin{equation}
\Omega_{\rm GW}^{\rm th}(f_i;\Theta).
\end{equation}

Assuming independent Gaussian errors in each
frequency bin, the likelihood for the
gravitational-wave data is

\begin{equation}
\mathcal{L}_{\rm GW}
=
\prod_i
\frac{1}{\sqrt{2\pi}\sigma_i}
\exp
\left[
-
\frac{
\left(
\hat{\Omega}_{\rm GW}(f_i)
-
\Omega_{\rm GW}^{\rm th}(f_i;\Theta)
\right)^2
}{2\sigma_i^2}
\right].
\end{equation}

Equivalently,

\begin{equation}
\chi^2
=
\sum_i
\frac{
\left[
\hat{\Omega}_{\rm GW}(f_i)
-
\Omega_{\rm GW}^{\rm th}(f_i;\Theta)
\right]^2
}{\sigma_i^2},
\end{equation}

such that

\begin{equation}
\mathcal{L}_{\rm GW}
\propto
\exp\left(-\frac{\chi^2}{2}\right).
\end{equation}

In the absence of a significant detection,
\(\hat{\Omega}_{\rm GW}(f_i)\simeq 0\), leading to

\begin{equation}
\chi^2
=
\sum_i
\frac{\Omega_{\rm GW}^{\rm th}(f_i;\Theta)^2}
{\sigma_i^2}.
\end{equation}


\subsection{Parameter inference}
\label{sec:mcmc}

The cosmological parameters can be inferred using a Bayesian
parameter estimation framework.
Given the observational data \(D\) and the model parameters
\(\Theta\), the posterior probability distribution is
obtained from Bayes' theorem,

\begin{equation}
P(\Theta|D)
\propto
\mathcal{L}_{\rm total}(D|\Theta)\,P(\Theta),
\end{equation}

where \(\mathcal{L}_{\rm total}\) is the total likelihood
defined in Sec.~\ref{sec:gw_likelihood} and
\(P(\Theta)\) denotes the prior distribution of the
model parameters.

The posterior distribution can be explored using a
Markov Chain Monte Carlo (MCMC) sampling method.
A suitable implementation is the publicly available inference
framework \texttt{Cobaya}, which interfaces with the
Boltzmann solver \texttt{CLASS} to compute the
cosmological observables for each point in parameter
space.

For each proposed parameter set \(\Theta\), the
predicted gravitational-wave energy density spectrum
\(\Omega_{\rm GW}^{\rm th}(f)\) is evaluated and
compared with the observational constraints through
the likelihood function described in
Sec.~\ref{sec:gw_likelihood}.
The CMB likelihood is evaluated simultaneously,
allowing the primordial tensor parameters to be
constrained using both cosmological and
gravitational-wave observations.

Convergence of the MCMC chains is assessed using the
Gelman--Rubin statistic.
A proposed convergence criterion is

\begin{equation}
R - 1 < 0.01
\end{equation}

for all sampled parameters, together with effective sample-size and mixing
checks. A new joint posterior with the revised CHRONOS noise has not been
computed for this White Paper.

\section{Discussion}
\label{sec:discussion}

Future CMB polarization projects, such as
Simons Observatory~\cite{SimonsObservatory:2025avm}
and
LiteBIRD~\cite{LiteBIRD:2022cnt,LiteBIRD:2026gdf},
will provide significantly better sensitivity to the CMB B-mode
than BK18, aiming for $r \sim 0.001$.
We should run combined simulation studies using their expected sensitivities to evaluate CHRONOS's true value for primordial cosmology.

BBN observations provide the best constraints on the blue tilt. It would be significant if CHRONOS could provide better constraints than those.
Conversely, the sensitivity required to meet that constraint will serve as
important information when considering an upgrade to CHRONOS.

The potential synergy between CHRONOS and other projects is also of interest and warrants further examination.
For example, atomic interferometer systems, such as AION~\cite{Badurina_2020} and MAGIS~\cite{Abe_2021}, are sensitive to the 
mid-frequency band ranging from $\sim 0.01$ Hz to a few Hz,
the same frequency range as CHRONOS,
and cosmological studies utilizing these systems are underway. Examples include
first-order phase transitions in the early Universe, cosmic strings, and local effects of dark energy~\cite{Badurina_2020}.
We should conduct a similar analysis for CHRONOS to evaluate potential competition.
These studies also become important when considering a CHRONOS upgrade.

While the above focuses on competition, the
synergy resulting from simultaneous observations with atomic interferometers is also worth considering.
Both atom and optical interferometers have geometry-dependent responses.
CHRONOS can provide 
directional information (see Chapter~\ref{chap:Directional_Response}). 
Therefore, when there are signs of a signal, CHRONOS can provide valuable information for identifying its origin through complementary detector orientations and responses.


\section{Conclusion}
\label{sec:conclusion}

Primordial gravitational waves provide a unique probe
of the physical processes that occurred in the earliest
moments of the Universe.
Their spectrum encodes information about the dynamics
of inflation and other possible mechanisms of
gravitational-wave generation in the early Universe.

In this chapter, we have investigated cosmological
constraints on primordial gravitational waves using
a multi-band observational approach that combines
cosmic microwave background (CMB) measurements with
constraints on the stochastic gravitational-wave
background from direct gravitational-wave experiments.

While CMB observations probe tensor perturbations at
extremely large cosmological scales, gravitational-wave
detectors provide complementary sensitivity at much
higher frequencies.
Combining these observations therefore allows the
primordial tensor spectrum to be tested across an
exceptionally wide range of scales.

Particular attention has been given to blue-tilted
tensor spectra characterized by a positive tensor
spectral index \((n_t>0)\).
Such spectra arise in a variety of early-Universe
scenarios and can produce enhanced gravitational-wave
signals at interferometer frequencies.

Future gravitational-wave detectors operating in the
sub-Hz to Hz frequency range will play a particularly
important role in this program.
Experiments such as CHRONOS will explore a frequency
band that bridges the gap between space-based and
ground-based detectors, providing a key window for
probing blue-tilted primordial spectra.

The combination of cosmological observations and
gravitational-wave experiments therefore opens a new
observational window onto the physics of the early
Universe.
By probing the primordial tensor spectrum over an
unprecedented range of scales, future observations
have the potential to reveal new insights into the
origin of cosmic structure and the fundamental
physics governing the earliest stages of cosmic
history.

%% file: Chapters/Interferometer.tex
\chapter{CHRONOS Interferometer}

\section{Optical layout}
\label{sec:optical_layout}

\begin{figure}[t]
  \centering
  \includegraphics[width=0.9\textwidth]{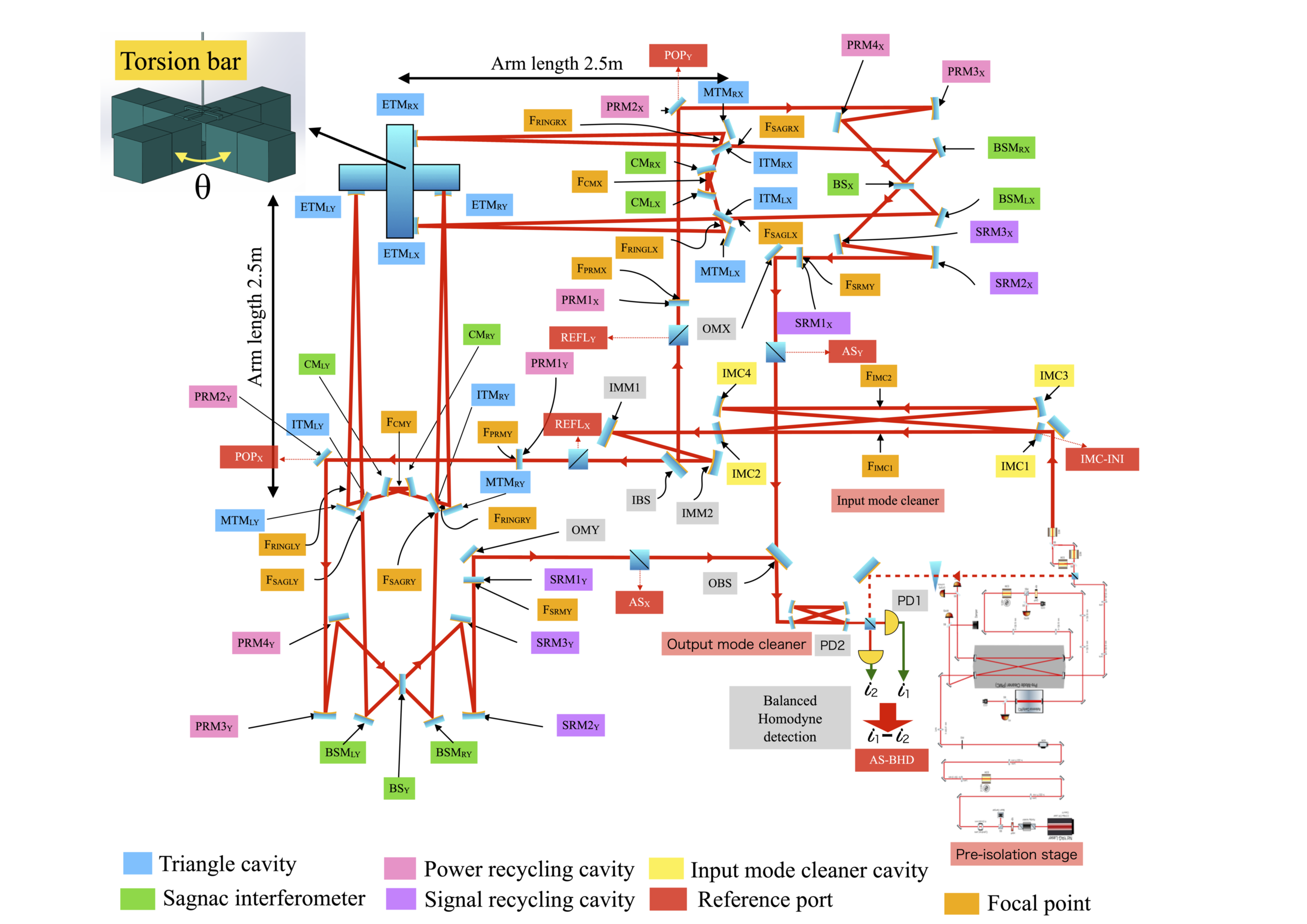}
  \caption{
  Optical layout of the CHRONOS interferometer.
  The laser stabilized by the PSL and IMC is split by the IBS and injected
  into two symmetric triangular Sagnac ring cavities.
  The returning beams are recombined at the OBS, and the signal is extracted
  through the SRM and OMC before being read out by balanced homodyne detection (BHD).
  }
  \label{fig:layout}
\end{figure}

Figure~\ref{fig:layout} shows the optical layout of the CHRONOS interferometer.
The instrument is based on a cross-shaped torsion-bar configuration,
with identical optical systems arranged symmetrically along the X and Y arms.
This configuration is optimized to extract the differential signal
associated with the rotational degree of freedom of the torsion bar,
while efficiently suppressing common-mode disturbances such as
seismic motion and laser intensity fluctuations.
Common-mode rejection is a central design principle in modern gravitational-wave
interferometers~\cite{LIGO2015,abbott2021ligo},
and is particularly crucial in the sub-Hz band,
where environmental noise sources are dominant.

The interferometer employs a single-frequency laser at a wavelength of 1064\,nm,
with an input power of 1\,W.
This wavelength is widely used in current ground-based gravitational-wave detectors,
owing to the maturity of laser and coating technologies~\cite{LIGO2015}.
The chosen power level represents a balance between achieving sufficient circulating power
for signal amplification and mitigating thermal effects such as optical absorption
and thermal lensing.

The input laser first passes through a Pre-Stabilized Laser (PSL) system,
where both frequency and intensity are actively stabilized.
In the sub-Hz regime, laser frequency noise couples directly to the interferometric signal,
making low-frequency stabilization at the PSL stage essential
\cite{abbott2021ligo}.
Furthermore, the relative intensity noise (RIN) must be sufficiently suppressed
to minimize residual coupling into the differential readout channel~\cite{tanabe_chronos_intensity}.
The detail of PSL system is summarized in Sec.~\ref{sec:IO}.

The beam is then spatially filtered by an Input Mode Cleaner (IMC).
The IMC operates as a high-finesse optical bow-tie cavity that transmits only the TEM$_{00}$ mode,
effectively removing higher-order spatial modes and beam jitter.
It also serves as a frequency filter,
suppressing high-frequency phase noise and unwanted modulation sidebands
\cite{LIGO2015,abbott2021ligo}.
As a result, the beam delivered to the interferometer achieves high mode purity
and optimal mode matching to the downstream cavities.

After the IMC, the beam is split equally into the X and Y arms
by the Input Beam Splitter (IBS).
Symmetric power splitting is essential for precise differential measurement,
as implemented in Michelson-type interferometers
\cite{LIGO2015}.

In each arm, the beam is injected into a triangular Sagnac ring cavity.
A defining feature of this configuration is that the interferometer measures
the \emph{velocity} of the test mass rather than its displacement.
This realizes a quantum nondemolition (QND) speed-meter configuration
\cite{Chen2003,Purdue2002,DanilishinKhalili2012}.
In such a topology, the circulating light accumulates a phase shift
proportional to the time derivative of the torsional motion.
This leads to suppression of radiation-pressure noise at low frequencies,
a key advantage for sub-Hz interferometry
\cite{Chen2003,DanilishinKhalili2012}.

After circulating in the ring cavity, the beams are recombined
at the Output Beam Splitter (OBS),
where the differential phase shift between the X and Y arms
is converted into intensity modulation.

The signal is further processed by a Signal Recycling Mirror (SRM),
which shapes the frequency response of the interferometer
by forming a coupled cavity with the main interferometer
\cite{meers1988,mizuno1993}.
Signal recycling allows tuning of the bandwidth and peak sensitivity,
a standard technique in advanced interferometric detectors
\cite{LIGO2015}.

An Output Mode Cleaner (OMC) removes residual higher-order modes
and unwanted radio-frequency sidebands prior to detection,
ensuring high readout purity.

The final readout is performed using balanced homodyne detection (BHD),
which enables selection of the optimal quadrature of the optical field
\cite{grote2013squeezing}.
BHD provides control over the trade-off between shot noise and radiation-pressure noise,
allowing quantum-noise-limited sensitivity in the measurement band
\cite{DanilishinKhalili2012}.

As a future upgrade, a silicon-based cryogenic configuration is under consideration.
In such a case, the laser wavelength would be shifted to 1550\,nm
to reduce optical absorption in silicon substrates and coatings
\cite{freise2010silicon,cole2013crystalline}.
Cryogenic silicon interferometers are promising candidates
for next-generation low-frequency gravitational-wave detectors.

For clarity, the following discussion focuses on the Y arm,
while the X arm has an identical optical configuration.

\subsection{Triangular Sagnac ring cavity}

The central optical element of each arm is a triangular Sagnac ring cavity
formed by the Input Test Mass (ITM), Middle Test Mass (MTM),
and End Test Mass (ETM).
Triangular ring cavities are widely used in interferometric systems
because they provide spatial separation between input and output beams
within a single plane and allow stable eigenmode control.
These ring cavities are arranged symmetrically on either side of the torsion bar,
and the light circulates unidirectionally along a closed triangular path.

In CHRONOS, while the left and right rings are geometrically symmetric,
the injection point is intentionally offset from the geometric center of the ring.
This asymmetry uniquely determines the circulation direction and optical path,
thereby ensuring stable Sagnac operation.
Such lifting of path degeneracy is a practical consideration
in Sagnac speed-meter implementations
\cite{Chen2003,Purdue2002}.
In addition, the asymmetric injection clarifies the separation of reflected
and transmitted beams, which is advantageous for routing control beams
and signal beams independently.

The ring cavity acts as a resonator that accumulates
the small optical path length variations induced by the torsional motion
of the bar as phase modulation.
Under resonant conditions, the phase shift acquired over one round trip
is effectively enhanced by the cavity buildup factor,
a standard property of optical resonators.
The triangular configuration enables a clear spatial separation of
the input, circulating, and returning beams within a single plane,
which significantly simplifies alignment control and mode matching.
Furthermore, the three-mirror geometry defined by ITM, MTM, and ETM
provides flexibility in accommodating geometrical constraints
such as installation footprint and beam height,
making it well suited for an instruments.

A key feature of the Sagnac topology is that the interferometer responds
to the \emph{velocity} of the test mass rather than its displacement.
This realizes a quantum nondemolition (QND) speed-meter configuration
\cite{Chen2003,Purdue2002,DanilishinKhalili2012}.
In such a topology, phase contributions from effectively counter-propagating
optical paths cancel for low-frequency common displacement components,
while the time-varying (velocity) component remains in the output signal.
This velocity response suppresses radiation-pressure noise at low frequencies,
which otherwise limits displacement-based interferometers
\cite{DanilishinKhalili2012}.
This property provides a fundamental advantage for CHRONOS
in the sub-Hz frequency band.

In this section, we focus on the conceptual role of the topology,
while a quantitative formulation of the speed-meter response
and the associated quantum noise will be discussed in later sections.

\subsection{Coupling between the torsion bar and the ring cavities}

Triangular Sagnac ring cavities are installed on both sides of the torsion bar.
When the bar undergoes a small angular rotation,
the effective optical path lengths in the left and right cavities
change with opposite signs.
This behavior follows directly from geometric phase considerations
in interferometric rotation sensing.

As a result, when the returning beams from the two cavities are interfered,
the torsional motion is extracted as a differential phase signal.
Differential readout is the standard strategy in precision interferometry
to suppress common-mode noise sources
\cite{LIGO2015,abbott2021ligo}.

Perturbations that act similarly on both sides,
such as laser intensity fluctuations and structural vibrations,
appear predominantly as common-mode signals
and are therefore strongly canceled in the differential channel.
Likewise, provided that the optical power and phase reference
supplied to each arm from the IBS are balanced,
a large fraction of laser-induced noise is intrinsically suppressed
\cite{LIGO2015}.
Consequently, the differential output is dominated by
the intrinsic angular response of the torsion bar.

In this way, the mechanical system (torsion bar) shown in Fig.~\ref{fig:bridge_cross}
and the optical system consisting of triangular Sagnac ring cavities
are clearly distinct subsystems.
The fundamental detection principle of CHRONOS is realized
through the coupling between these two systems.
In particular, the Sagnac topology naturally provides
a velocity-sensitive response proportional to the time derivative
of the torsional motion
\cite{Chen2003}.
This feature leads to suppression of radiation-pressure noise
at low frequencies, offering a key advantage for achieving
high sensitivity in the sub-Hz band.

\subsection{Dual-recycling configuration with PR-dominated operation}

The CHRONOS interferometer employs a dual-recycling configuration,
consisting of both power recycling (PR) and signal recycling (SR),
as originally proposed for advanced interferometric detectors
\cite{meers1988,mizuno1993}.
While both subsystems contribute to the interferometer performance,
their coupled response must be optimized together with the readout quadrature.

The PR system is implemented on the input side
using a set of mirrors (PRM1--PRM4) and the beam splitter (BS),
forming a coupled resonant system shared by the two Sagnac arms.
Power recycling increases the effective circulating power
without increasing the input laser power,
thereby improving the shot-noise-limited sensitivity
\cite{meers1988}.
Since shot noise scales inversely with the square root of the circulating power,
$\sqrt{S_h^{\rm shot}} \propto P_{\rm circ}^{-1/2}$ at fixed optical response,
the PR configuration directly determines the overall sensitivity scale
\cite{DanilishinKhalili2012}.

This role is particularly critical in CHRONOS,
where the input laser power is modest (1\,W)
and the target frequency band lies in the sub-Hz regime.
In this regime, achieving sufficient circulating power
without introducing excessive thermal or technical noise
requires careful optimization of the PR cavity parameters,
including reflectivity, cavity length, and mode matching
\cite{LIGO2015}.
The PR setting is therefore one of the jointly optimized parameters in
Sec.~\ref{sec:sensitivity_optimization}.

On the output side, a Signal Recycling Mirror (SRM) is employed
to form a signal-recycling cavity (SRC).
Signal recycling modifies the interferometer transfer function
by forming a coupled cavity with the main interferometer,
allowing tuning of bandwidth and peak sensitivity
\cite{mizuno1993,BuonannoChen2001}.
By adjusting the microscopic detuning phase of the SRM,
the frequency dependence of the response can be shaped,
enhancing sensitivity in selected frequency ranges.

The effect of SR tuning cannot be ranked independently of PR and LO settings.
Chapter~\ref{chap:sensitivity} evaluates their joint effect on the carrier,
complex signal response, and correlated quantum noise. Its two reference
operating points replace an optimization based only on a single frequency
or on an ideal lossless speed-meter response.

The output optical field is filtered by an Output Mode Cleaner (OMC),
which suppresses higher-order spatial modes and control sidebands
by tens of decibels
\cite{LIGO2015}.
The OMC acts as a spatial and spectral filter,
ensuring that only the fundamental mode carrying the signal sidebands
reaches the photodetector.
This reduces technical noise sources such as mode beating
and scattered-light-induced fluctuations,
and improves the fidelity of the detected signal.

The filtered output is read out using balanced homodyne detection (BHD),
in which the interferometer output field is mixed with a local oscillator (LO),
and the differential photocurrent is measured
\cite{grote2013squeezing,DanilishinKhalili2012}.
BHD enables measurement of an arbitrary optical quadrature,
allowing optimization of the quantum noise balance
between shot noise and radiation-pressure noise.
Compared to conventional DC readout schemes,
balanced homodyne detection offers improved suppression
of technical noise couplings at low frequencies.

The dual-recycling configuration is thus specified by a coupled PR, SR, and
LO operating point. The circulating powers and phase conventions used for
the science forecasts are given in Table~\ref{tab:sensitivity_operating_points}.
A quantitative evaluation of these subsystems
is presented in Chap.~\ref{chap:sensitivity}.

%% file: Chapters/Directional_Response.tex
\chapter{Directional Response of CHRONOS}
\label{chap:Directional_Response}

\section{Overview}

The response of CHRONOS to a gravitational wave (GW) depends not only on
the intrinsic strain amplitude of the incident wave, but also on its
direction of arrival and polarization.
This directional dependence is determined primarily by the geometry of
the two orthogonal torsion bars and can be described independently of the
mechanical, optical, control, and calibration transfer functions.

In this chapter, the directional response of CHRONOS is formulated using
a detector-tensor description based on the torsion-bar antenna framework
of Ref.~\cite{Eda2014TOBA}.
The baseline CHRONOS gravitational-wave channel is defined by the
differential horizontal torsional motion of two orthogonal bars.
We additionally consider an idealized multi-output configuration in which
two vertical rotational modes are measured together with the baseline
horizontal differential mode.

Throughout this chapter, the terminology ``$x$-bar'' and ``$y$-bar''
is used in a strict geometrical sense.
The $x$-bar is the torsion bar whose equilibrium longitudinal axis is
aligned with the local detector $x$ axis, while the $y$-bar is the
orthogonal torsion bar whose equilibrium longitudinal axis is aligned with
the local detector $y$ axis.
The corresponding rotational variables are labeled by the same axis
subscripts in order to make their physical meaning explicit.

Three complementary representations are used.
First, the detector geometry and angular coordinates are introduced.
Second, the intrinsic three-dimensional antenna pattern is shown in the
local detector frame.
Finally, the response is projected onto celestial coordinates for a
reference CHRONOS installation in Taipei, Taiwan.

\section{Detector Geometry and Coordinate System}

We define a right-handed local detector coordinate system
$(\hat{\bm x},\hat{\bm y},\hat{\bm z})$.
The $x$-bar is aligned with $\hat{\bm x}$ in its equilibrium configuration,
the $y$-bar is aligned with $\hat{\bm y}$, and
$\hat{\bm z}$ denotes the local vertical direction.

Thus,

\begin{equation}
	\text{$x$-bar} \parallel \hat{\bm x},
	\qquad
	\text{$y$-bar} \parallel \hat{\bm y}.
	\label{eq:bar_axis_definition}
\end{equation}

\begin{figure}[tbp]
	\centering
	\includegraphics[width=0.62\textwidth]{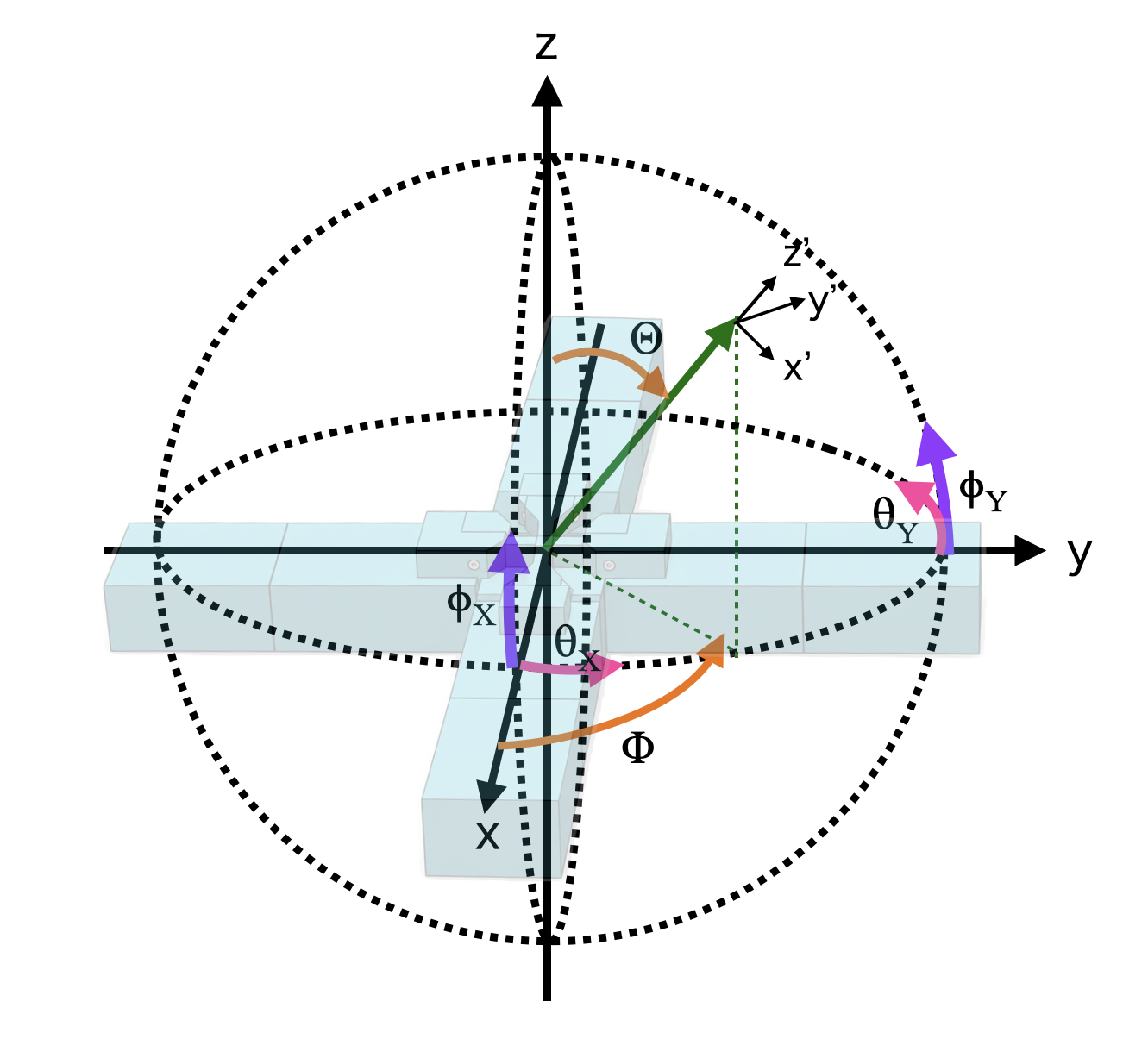}
	\caption{
		Coordinate system and rotational degrees of freedom used to describe
		the directional response of CHRONOS.
		The $x$-bar is the torsion bar aligned with the local detector $x$ axis,
		while the $y$-bar is the orthogonal torsion bar aligned with the local
		detector $y$ axis.
		The $z$ axis is vertical.
		The incident GW direction is specified by the polar angle $\Theta$
		and azimuthal angle $\Phi$.
		The horizontal torsional rotations of the $x$- and $y$-bars are denoted
		by $\theta_x$ and $\theta_y$, respectively.
		The corresponding vertical rotational degrees of freedom are denoted
		by $\phi_x$ and $\phi_y$.
		The primed coordinates $(x',y',z')$ define a wave-based frame associated
		with the incoming gravitational wave.
	}
	\label{fig:chronos_direction_geometry}
\end{figure}

The detector geometry and angular conventions are illustrated in
Fig.~\ref{fig:chronos_direction_geometry}.

The rotational coordinates are therefore defined as

\begin{align}
	\theta_x &: \text{horizontal torsional rotation of the $x$-bar}, \\
	\theta_y &: \text{horizontal torsional rotation of the $y$-bar}, \\
	\phi_x &: \text{vertical rotational mode of the $x$-bar}, \\
	\phi_y &: \text{vertical rotational mode of the $y$-bar}.
\end{align}

The GW tensor is written in transverse-traceless form as

\begin{equation}
	h_{ij}(t)
	=
	h_+(t)e^+_{ij}
	+
	h_\times(t)e^\times_{ij},
	\label{eq:dir_gw_tensor}
\end{equation}

where $h_+$ and $h_\times$ are the two tensor-polarization amplitudes and
$e^+_{ij}$ and $e^\times_{ij}$ are the corresponding polarization tensors.

Following Ref.~\cite{Eda2014TOBA}, the GW propagation direction in the
detector frame may be parameterized as

\begin{equation}
	\bm n
	=
	-
	\left(
	\sin\Theta\sin\Phi,\,
	\sin\Theta\cos\Phi,\,
	\cos\Theta
	\right).
	\label{eq:dir_propagation}
\end{equation}

The polarization angle $\Psi$ specifies a rotation of the polarization
basis about the propagation direction.

\section{Detector-Tensor Formulation}

For a detector output channel $N$, the purely geometrical GW response is

\begin{equation}
	h_N(t)
	=
	D_N^{ij}h_{ij}(t),
	\label{eq:dir_detector_contraction}
\end{equation}

where $D_N^{ij}$ is the detector tensor associated with that output.

Using Eq.~\eqref{eq:dir_gw_tensor},

\begin{equation}
	h_N(t)
	=
	F_{+,N}h_+(t)
	+
	F_{\times,N}h_\times(t),
	\label{eq:dir_strain_response}
\end{equation}

where

\begin{equation}
	F_{A,N}
	=
	D_N^{ij}e^A_{ij},
	\qquad
	A=+,\times.
	\label{eq:dir_antenna_tensor}
\end{equation}

Equation~\eqref{eq:dir_antenna_tensor} is adopted as the fundamental
definition of the CHRONOS directional response.

The complete measured signal can be written schematically as

\begin{equation}
	\tilde d_N(f,t)
	=
	H_N(f)
	\left[
	F_{+,N}(t)\tilde h_+(f)
	+
	F_{\times,N}(t)\tilde h_\times(f)
	\right]
	+
	\tilde n_N(f),
	\label{eq:dir_full_signal}
\end{equation}

where $H_N(f)$ contains the mechanical, optical, control, and calibration
responses.

\section{Baseline Horizontal Differential Output}

The baseline CHRONOS gravitational-wave readout is the differential
horizontal torsional motion of the two bars,

\begin{equation}
	s_{\rm I}
	=
	\theta_x-\theta_y.
	\label{eq:dir_output_I}
\end{equation}

Here, $\theta_x$ is the horizontal torsional rotation of the bar aligned
with the $x$ axis, while $\theta_y$ is that of the bar aligned with the
$y$ axis.

The detector tensor of this channel is

\begin{equation}
	D_{\rm I}^{ij}
	=
	\frac{1}{2}
	\left(
	\hat{x}^{i}\hat{y}^{j}
	+
	\hat{y}^{i}\hat{x}^{j}
	\right).
	\label{eq:dir_DI}
\end{equation}

Because the GW tensor is symmetric,

\begin{equation}
	h_{\rm I}
	=
	D_{\rm I}^{ij}h_{ij}
	=
	h_{xy}.
\end{equation}

The corresponding antenna-pattern functions are

\begin{align}
	F_{+,{\rm I}}
	&=
	-\frac{1+\cos^2\Theta}{2}
	\sin 2\Phi \cos 2\Psi
	+
	\cos\Theta
	\cos 2\Phi
	\sin 2\Psi,
	\label{eq:dir_Fplus_single}
	\\
	F_{\times,{\rm I}}
	&=
	\cos\Theta
	\cos 2\Phi
	\cos 2\Psi
	+
	\frac{1+\cos^2\Theta}{2}
	\sin 2\Phi
	\sin 2\Psi.
	\label{eq:dir_Fcross_single}
\end{align}

For visualization, we define the polarization-independent antenna power

\begin{equation}
	P_N
	=
	F_{+,N}^2
	+
	F_{\times,N}^2,
	\label{eq:dir_power}
\end{equation}

and the corresponding directional amplitude

\begin{equation}
	A_N
	=
	\sqrt{P_N}.
	\label{eq:dir_amplitude}
\end{equation}

For the baseline horizontal channel,

\begin{equation}
	P_{\rm I}
	=
	\cos^2\Theta\cos^2 2\Phi
	+
	\frac{1}{4}
	\left(1+\cos^2\Theta\right)^2
	\sin^2 2\Phi.
	\label{eq:dir_power_single}
\end{equation}

The polarization angle $\Psi$ cancels from
Eq.~\eqref{eq:dir_power_single}.

\section{Directional Nulls of the Single-Output Detector}

The baseline horizontal response is not isotropic and possesses directions
for which the response becomes weak or vanishes.

For a wave propagating in the detector horizontal plane,

\begin{equation}
	\Theta=\frac{\pi}{2},
\end{equation}

the antenna power reduces to

\begin{equation}
	P_{\rm I}
	=
	\frac{1}{4}\sin^2 2\Phi.
\end{equation}

Hence, the horizontal differential output has null directions for specific
azimuths.

By contrast, for a wave arriving along the local vertical direction,

\begin{equation}
	\Theta=0,
\end{equation}

the response reaches

\begin{equation}
	P_{\rm I}=1.
\end{equation}

Thus, the single-output CHRONOS response contains characteristic regions of
high and low sensitivity determined solely by the detector geometry.

\section{Multi-Output Configuration}

In addition to the baseline horizontal differential mode, the two bars can
in principle provide vertical rotational GW-sensitive outputs.

Using the axis-based notation introduced above, the three outputs are

\begin{align}
	s_{\rm I} &= \theta_x-\theta_y,
	\label{eq:dir_multi_sI}
	\\
	s_{\rm II} &= \phi_x,
	\label{eq:dir_multi_sII}
	\\
	s_{\rm III} &= \phi_y.
	\label{eq:dir_multi_sIII}
\end{align}

The notation explicitly identifies the physical bar associated with each
rotation:
$\phi_x$ is the vertical rotational mode of the $x$-bar, and
$\phi_y$ is the corresponding mode of the $y$-bar.

The associated detector tensors are

\begin{align}
	D_{\rm I}^{ij}
	&=
	\frac{1}{2}
	\left(
	\hat{x}^{i}\hat{y}^{j}
	+
	\hat{y}^{i}\hat{x}^{j}
	\right),
	\\
	D_{\rm II}^{ij}
	&=
	\frac{1}{4}
	\left(
	\hat{x}^{i}\hat{z}^{j}
	+
	\hat{z}^{i}\hat{x}^{j}
	\right),
	\label{eq:dir_DII}
	\\
	D_{\rm III}^{ij}
	&=
	\frac{1}{4}
	\left(
	\hat{y}^{i}\hat{z}^{j}
	+
	\hat{z}^{i}\hat{y}^{j}
	\right).
	\label{eq:dir_DIII}
\end{align}

The different tensor structures imply different antenna nulls.
Therefore, a direction that produces little response in
$s_{\rm I}$ may still produce a significant response in
$s_{\rm II}$ or $s_{\rm III}$.

For an idealized detector with equal strain sensitivity and statistically
independent noise in the three channels, the combined antenna power is

\begin{equation}
	P_{\rm multi}
	=
	P_{\rm I}
	+
	P_{\rm II}
	+
	P_{\rm III}.
	\label{eq:dir_multi_power}
\end{equation}

The corresponding directional amplitude is

\begin{equation}
	A_{\rm multi}
	=
	\sqrt{P_{\rm multi}}.
\end{equation}

This quantity is used here only as a geometrical measure of the potential
multi-output sky coverage.

The actual CHRONOS multi-output sensitivity would require a noise-weighted
combination,

\begin{equation}
	\rho^2
	\propto
	\sum_N
	\int
	\frac{
		\left|
		F_{+,N}\tilde h_+
		+
		F_{\times,N}\tilde h_\times
		\right|^2
	}{
		S_{h,N}(f)
	}
	\,df,
	\label{eq:dir_multi_snr}
\end{equation}

where $S_{h,N}(f)$ is the strain-equivalent noise spectrum of output $N$.

\section{Intrinsic Three-Dimensional Antenna Pattern}

\begin{figure}[tbp]
	\centering
	\includegraphics[width=\textwidth]
	{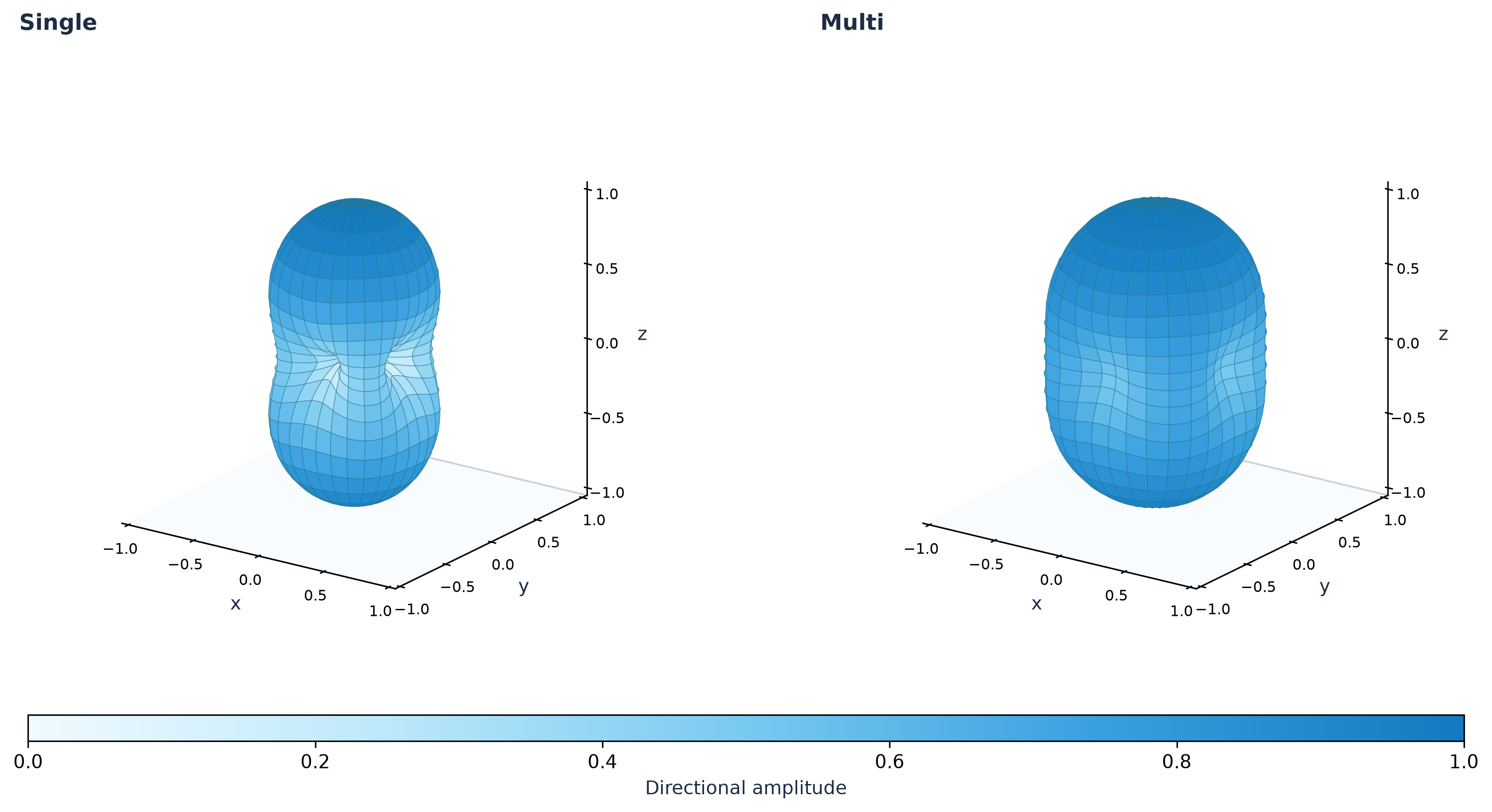}
	\caption{
		Three-dimensional directional amplitude of CHRONOS in the local
		detector frame.
		The left panel shows the baseline horizontal differential output
		$s_{\rm I}=\theta_x-\theta_y$ (Single), while the right panel shows the
		idealized three-output combination
		$(s_{\rm I},s_{\rm II},s_{\rm III})
		=(\theta_x-\theta_y,\phi_x,\phi_y)$ (Multi).
		The $x$-bar is aligned with the detector $x$ axis and the $y$-bar is
		aligned with the detector $y$ axis.
		The radial distance represents the polarization-independent directional
		amplitude $\sqrt{P_N}$.
		The positive $z$ direction corresponds to the local detector zenith.
	}
	\label{fig:chronos_direction_3d}
\end{figure}

Figure~\ref{fig:chronos_direction_3d} shows the intrinsic directional
amplitude of CHRONOS in the local detector frame.

The single-output antenna pattern exhibits pronounced weak-response regions.
These regions arise from the tensor structure of the horizontal
differential torsional mode.

The multi-output response is considerably more uniform because
$\phi_x$ and $\phi_y$ probe additional components of the GW tidal tensor.
Their null directions do not coincide with those of the baseline
$\theta_x-\theta_y$ channel.

For short-duration signals, this directional dependence directly affects
the instantaneous signal-to-noise ratio.
A source close to a directional null of the baseline output can therefore
appear significantly weaker than an intrinsically identical source located
near a maximum of the antenna pattern.

\section{Reference CHRONOS Site in Taipei}

To connect the local antenna geometry with celestial directions, we
consider a reference CHRONOS installation in Taipei, Taiwan, at

\begin{equation}
	\lambda_{\rm site}
	=
	25.0212^{\circ},
	\qquad
	\ell_{\rm site}
	=
	121.5379^{\circ},
	\label{eq:dir_taipei_coordinates}
\end{equation}

where $\lambda_{\rm site}$ and $\ell_{\rm site}$ denote geographical
latitude and longitude.

For the reference configuration used here, the horizontal orientation is

\begin{equation}
	\gamma=0^{\circ},
\end{equation}

where $\gamma$ denotes the angle between local East and the bisector of the
$x$- and $y$-bars.

The bar labels remain defined in the local detector frame:
the $x$-bar is always the bar parallel to the detector $x$ axis, and the
$y$-bar is always the orthogonal bar parallel to the detector $y$ axis,
independent of the detector's geographical orientation.

\section{Projection onto the Celestial Sphere}

The detector tensor can be transformed from the local detector frame to a
celestial frame by a rotation matrix $R(t)$,

\begin{equation}
	D_{N,\mathrm{cel}}^{ij}(t)
	=
	R^{i}{}_{k}(t)
	R^{j}{}_{l}(t)
	D_N^{kl}.
	\label{eq:dir_celestial_D}
\end{equation}

The celestial antenna functions are then

\begin{equation}
	F_{A,N}(t)
	=
	D_{N,\mathrm{cel}}^{ij}(t)e^A_{ij},
	\qquad
	A=+,\times.
	\label{eq:dir_celestial_F}
\end{equation}

\begin{figure}[tbp]
	\centering
	\includegraphics[width=\textwidth]
	{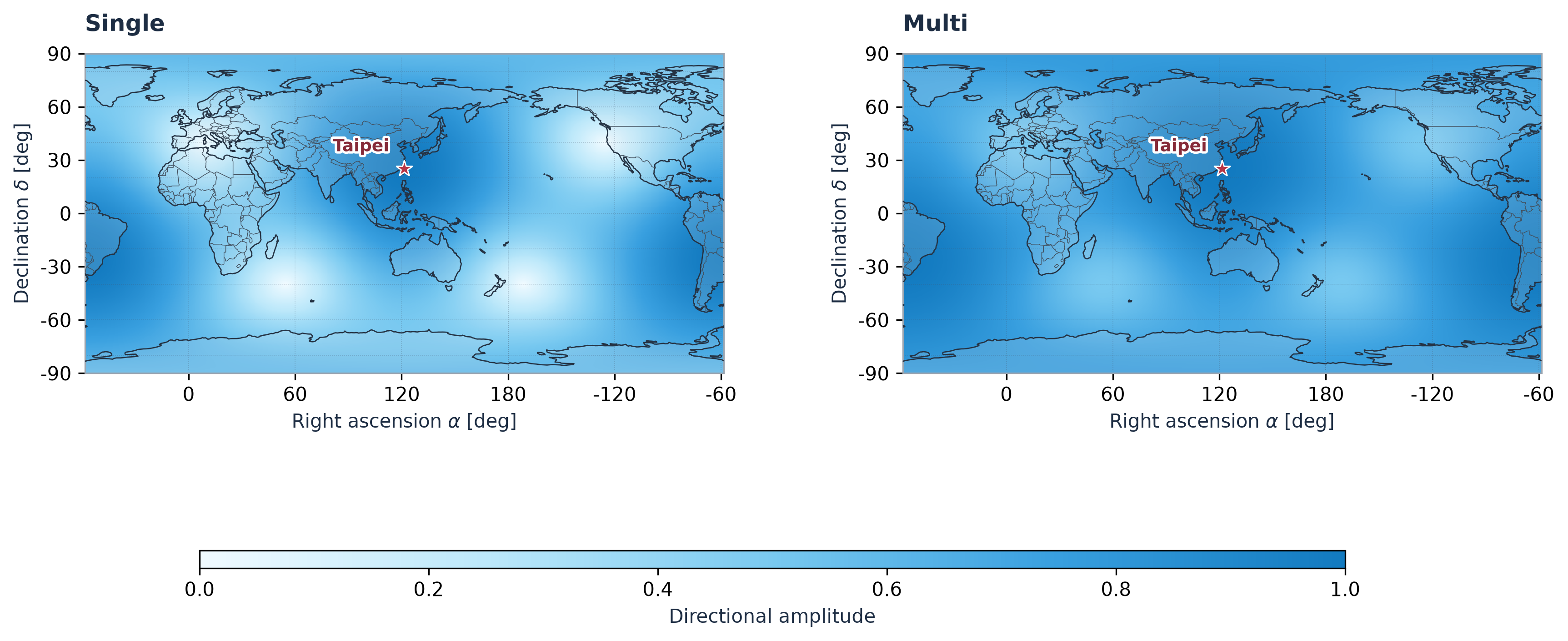}
	\caption{
		Instantaneous celestial directional response of CHRONOS for the
		Taipei reference configuration at
		$25.0212^{\circ}$ N and $121.5379^{\circ}$ E.
		The left panel shows the baseline horizontal differential output
		$s_{\rm I}=\theta_x-\theta_y$, while the right panel shows the
		idealized three-output combination including
		$\phi_x$ and $\phi_y$.
		The color scale represents the polarization-independent directional
		amplitude $\sqrt{P_N}$.
		The terrestrial map is overlaid only as a visual reference for the
		instantaneous detector orientation relative to the celestial sphere.
	}
	\label{fig:chronos_direction_sky}
\end{figure}

Figure~\ref{fig:chronos_direction_sky} shows the resulting instantaneous
directional amplitude for the Taipei reference configuration.

The terrestrial map should not be interpreted as placing the GW source on
the Earth's surface.
It only provides an intuitive reference for the instantaneous orientation
of the detector relative to celestial coordinates.

The single-output map displays extended regions of reduced sensitivity,
whereas the idealized multi-output configuration produces a more uniform
response over the sky.

The principal advantage of the multi-output configuration is therefore not
an increase of the maximum response, but a reduction of strong directional
nulls.

\section{Time Dependence due to Earth Rotation}

The sky pattern in Fig.~\ref{fig:chronos_direction_sky} represents an
instantaneous configuration.

For a fixed celestial source, the relative orientation between the source
and detector changes as the Earth rotates.
The antenna functions can therefore be expressed as

\begin{align}
	F_{+,N}(t)
	&=
	a_N(t)\cos 2\psi
	+
	b_N(t)\sin 2\psi,
	\\
	F_{\times,N}(t)
	&=
	b_N(t)\cos 2\psi
	-
	a_N(t)\sin 2\psi.
	\label{eq:dir_sidereal}
\end{align}

For short transient signals, this time dependence is usually negligible
during the event.

For signals lasting many hours or longer, however, Earth rotation produces
a characteristic sidereal amplitude modulation.
This is relevant to long-duration inspirals, continuous-wave searches, and
stochastic-background analyses.

For CHRONOS, the preferred implementation is to rotate the detector tensors
numerically according to Eq.~\eqref{eq:dir_celestial_D}, rather than using
fully expanded trigonometric expressions.

\section{Implications for CHRONOS Science}

For transient gravitational-wave signals, the observed strain is

\begin{equation}
	h_{\rm obs}
	=
	F_+h_+
	+
	F_\times h_\times.
\end{equation}

The detection distance therefore depends on source direction as well as on
the intrinsic source parameters.

For long-duration signals, Earth rotation sweeps the directional response
across the sky and can partially average instantaneous weak-response
directions.

For stochastic gravitational-wave-background searches, the detector tensor
enters naturally into the overlap-reduction function between detectors,

\begin{equation}
	\gamma(f)
	\propto
	\int d\hat{\bm n}\,
	\sum_A
	F^A_1(\hat{\bm n})
	F^A_2(\hat{\bm n})
	e^{2\pi i f\hat{\bm n}\cdot\Delta\bm x/c}.
	\label{eq:dir_orf}
\end{equation}

The same tensor formalism can also be extended to terrestrial gravity
signals.
For nearby sources, however, the plane-wave approximation must be replaced
by the appropriate near-field tidal model.

\section{Baseline and Future Multi-Output Operation}

The baseline CHRONOS gravitational-wave channel considered in this White
Paper is

\begin{equation}
	h_{\rm CHRONOS}
	\equiv
	h_{\rm I},
\end{equation}

corresponding to the differential horizontal torsional readout

\begin{equation}
	s_{\rm I}
	=
	\theta_x-\theta_y.
\end{equation}

The additional vertical modes

\begin{equation}
	s_{\rm II}=\phi_x,
	\qquad
	s_{\rm III}=\phi_y
\end{equation}

are treated as a possible future extension.

The Multi response shown in this chapter assumes, for illustration,

\begin{enumerate}
	\item independent readout of $\theta_x-\theta_y$, $\phi_x$, and $\phi_y$,
	\item calibration of all three outputs in strain-equivalent units,
	\item sufficiently low noise in the two vertical rotational channels,
	\item and characterized noise correlations between the outputs.
\end{enumerate}

Until these conditions are demonstrated experimentally, the CHRONOS
baseline sensitivity remains defined by the horizontal differential channel.

\section{Summary}

The directional response of CHRONOS is determined by the geometry of two
orthogonal torsion bars.

The $x$-bar is defined as the bar aligned with the detector $x$ axis, while
the $y$-bar is defined as the bar aligned with the detector $y$ axis.
Their horizontal torsional rotations are denoted by
$\theta_x$ and $\theta_y$, and their corresponding vertical rotational
degrees of freedom are denoted by $\phi_x$ and $\phi_y$.

The baseline gravitational-wave channel is

\begin{equation}
	s_{\rm I}
	=
	\theta_x-\theta_y,
\end{equation}

with detector tensor

\begin{equation}
	D_{\rm I}^{ij}
	=
	\frac{1}{2}
	\left(
	\hat{x}^{i}\hat{y}^{j}
	+
	\hat{y}^{i}\hat{x}^{j}
	\right).
\end{equation}

Its antenna pattern exhibits strong directional dependence and contains
several weak-response directions.

An idealized three-output extension,

\begin{equation}
	(s_{\rm I},s_{\rm II},s_{\rm III})
	=
	(\theta_x-\theta_y,\phi_x,\phi_y),
\end{equation}

provides complementary tensor responses and substantially improves
directional sky coverage.

The three-dimensional antenna pattern characterizes the intrinsic response
in the local detector frame, while the celestial projection describes the
instantaneous response for a specified detector site and orientation.

For the Taipei reference configuration, the baseline single-output response
shows substantial angular variation, while the idealized multi-output
configuration produces a considerably more uniform response.

The detector-tensor formulation established here provides a unified basis
for transient GW searches, long-duration signals, stochastic-background
analyses, and future multi-output extensions of CHRONOS.

%% file: Chapters/Input_Optics.tex
\chapter{Input Optics} \label{sec:IO}

The Input Optics system constitutes the pre-stabilization stage that delivers
spatially pure, frequency-stabilized, and intensity-stabilized laser light
to the main interferometer.
In CHRONOS, a torsion-bar speed-meter interferometer operating in the sub-Hz regime,
the requirements on low-frequency laser noise are particularly stringent.
Therefore, the Input Optics is not merely a beam-conditioning stage,
but a multi-layer stabilization architecture designed to independently
control three essential degrees of freedom:
spatial mode, frequency, and intensity.

\begin{figure}[t]
    \centering
    \includegraphics[width=0.95\linewidth]{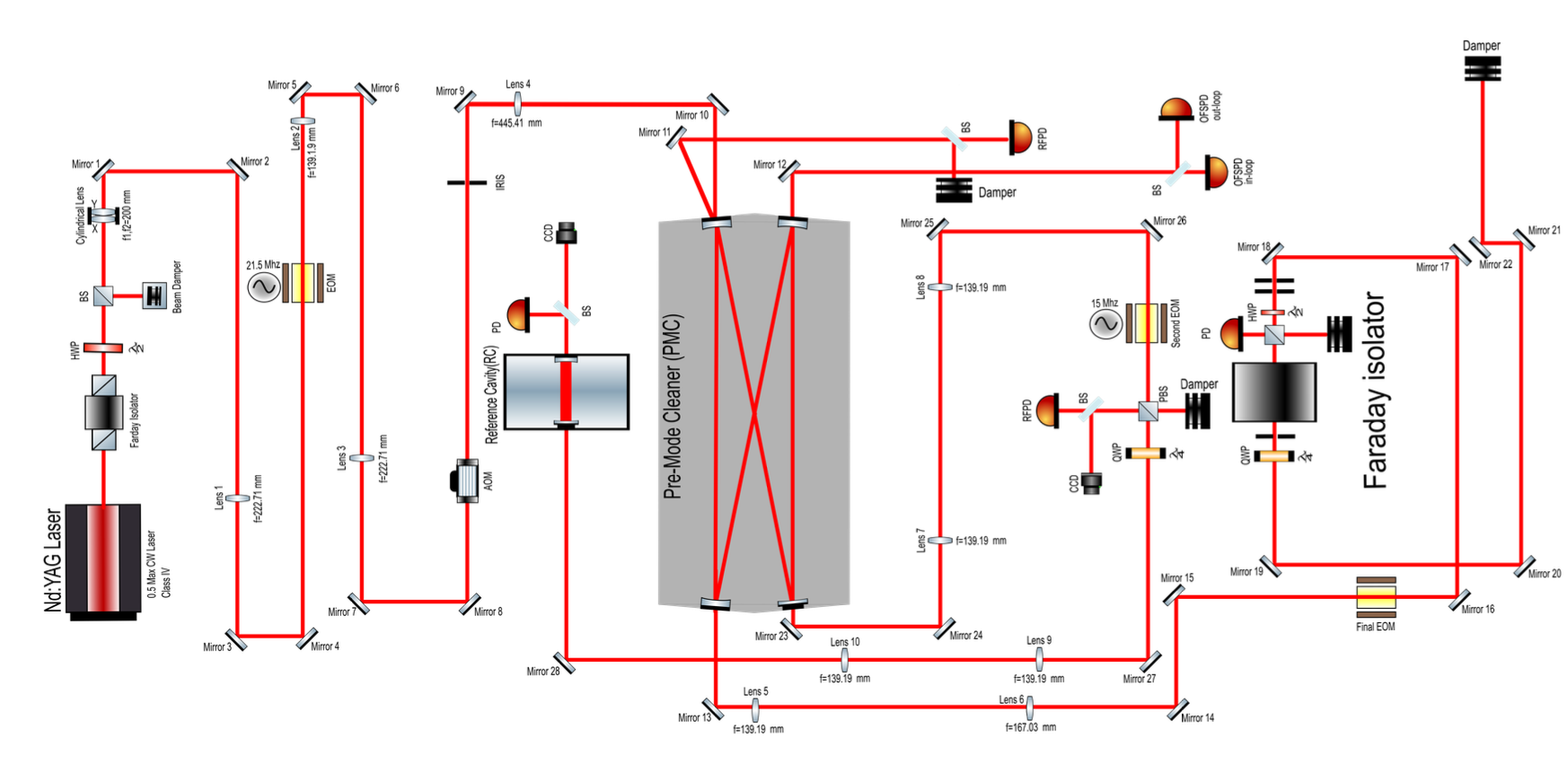}
    \caption{
    Optical layout of the CHRONOS Input Optics system.
    The seed laser output passes through the EOM, AOM,
    Input Mode Cleaner (IMC), and Reference Cavity (RC)
    before entering the main interferometer.
    }
    \label{fig:input_optics_layout}
\end{figure}

A schematic layout of the Input Optics system is shown in
Fig.~\ref{fig:input_optics_layout}.
Light emitted from the seed laser is conditioned through
electro-optic and acousto-optic modulation,
filtered by the Input Mode Cleaner (IMC),
stabilized in frequency by the Reference Cavity (RC),
and subsequently delivered to the main interferometer.

The system is composed of the following primary components:

\begin{itemize}
    \item Seed laser (Nd:YAG, 1064~nm)
    \item Steering mirrors and mode-matching optics
    \item Electro-Optic Modulator (EOM)
    \item Acousto-Optic Modulator (AOM)
    \item Input Mode Cleaner (IMC)
    \item Reference Cavity (RC)
\end{itemize}

The central element of the system is the Input Mode Cleaner (IMC),
configured as a four-mirror bow-tie ring cavity.
This traveling-wave configuration suppresses optical back-reflection
toward the laser source and mitigates parasitic interference
caused by scattered light.
Such suppression is especially critical in the sub-Hz frequency band,
where scattered light can couple ground motion into phase noise.

The IMC spatially filters the incoming beam by selecting the
TEM$_{00}$ eigenmode of the cavity,
thereby suppressing higher-order transverse modes
and improving spatial mode purity.
In a speed-meter interferometer,
spatial mode imperfections can couple into the velocity signal,
potentially degrading sensitivity.
Maintaining high spatial mode quality at the input stage
is therefore essential for preserving detector performance.

Both reflected and transmitted beams from the IMC are utilized
for downstream control systems.
The reflected beam provides the error signal required to maintain
the IMC resonance condition.
The transmitted beam is split into multiple paths:

\begin{itemize}
    \item A fraction is used for intensity stabilization (Intensity Stabilization servo),
    \item A fraction is directed to the Reference Cavity (RC) for frequency stabilization,
    \item The remaining portion serves as the input to the main interferometer.
\end{itemize}

The Reference Cavity (RC) functions as a high-stability optical frequency reference.
Laser frequency stabilization is achieved using the
Pound-Drever-Hall (PDH) technique,
which enables significant suppression of low-frequency frequency noise.
This stage is particularly important for CHRONOS,
where laser frequency noise can couple to the torsional degree of freedom
through residual asymmetries and cavity-length fluctuations.

Both the IMC and RC are equipped with active thermal stabilization systems
to minimize cavity-length fluctuations induced by environmental temperature variations.
Such stabilization further suppresses low-frequency drift
and enhances long-term operational stability.

In summary, the CHRONOS Input Optics system is designed as an integrated
stabilization architecture.
By simultaneously controlling spatial mode quality,
frequency noise, and intensity noise,
it provides a high-purity and ultra-stable laser source
required for sub-Hz gravitational-wave detection
with a torsion-bar speed-meter interferometer.

\section{Pre-Mode Cleaner (PMC)}
\begin{figure}[t]
\centering
\includegraphics[width=0.85\linewidth]{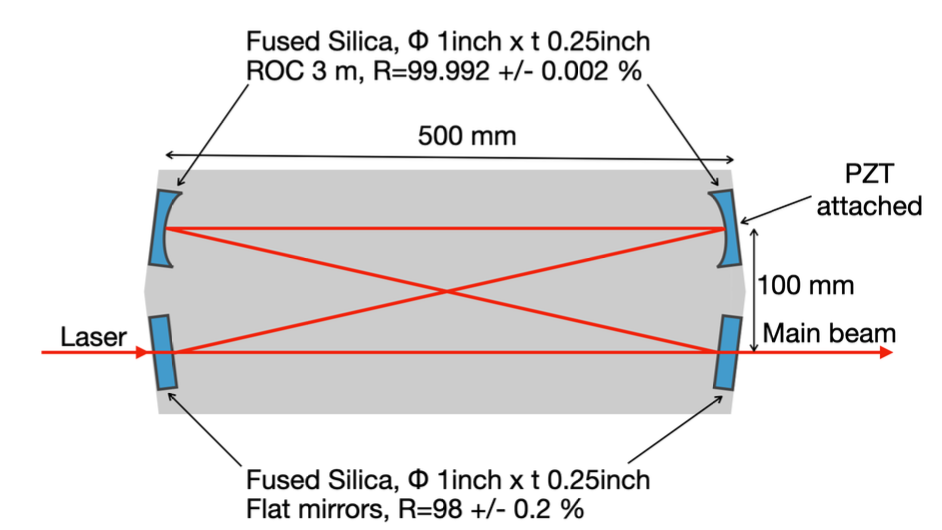}
\caption{
Schematic configuration of the pre-mode-cleaner (PMC) cavity.
The cavity consists of two concave high-reflectivity mirrors 
(ROC = 3~m, $R = 99.992\% \pm 0.002\%$) separated by 500~mm, 
and two flat mirrors ($R = 98\% \pm 0.2\%$), forming a folded linear cavity.
The mirror substrates are fused silica 
($\phi$1~inch $\times$ 0.25~inch).
One of the end mirrors is mounted on a piezoelectric transducer (PZT) 
for cavity length control.
The main transmitted beam is extracted from the output coupler.
}
\label{fig:PMC_structure}
\end{figure}

The Pre-Mode Cleaner (PMC) serves as the first optical filtering stage
in the CHRONOS Input Optics chain.
While often regarded as a simple pre-stabilization cavity,
its role is scientifically essential for enabling
ultra-low-frequency operation of the torsion-bar speed-meter interferometer.

Figure~\ref{fig:PMC_structure} shows the mechanical and optical structure
of the PMC.
It is implemented as a bow-tie traveling-wave ring cavity
with a total length of $L = 0.5~\mathrm{m}$.
The traveling-wave configuration suppresses back-reflection
into the laser source,
thereby preventing optical feedback-induced frequency noise,
which is particularly detrimental in sub-Hz interferometry.

\subsection{Scientific Role of the PMC}

The PMC fulfills several critical functions:

\paragraph{(i) Spatial Mode Filtering}

The cavity transmits only the TEM$_{00}$ eigenmode
and suppresses higher-order transverse spatial modes.
Residual spatial mode distortion couples to interferometer
alignment degrees of freedom and can convert beam jitter
into phase noise.
In a speed-meter configuration,
such coupling may contaminate the velocity signal,
making spatial purity at the earliest stage indispensable.

\paragraph{(ii) High-Frequency Intensity Noise Filtering}

Above the cavity pole frequency,
the PMC acts as a low-pass filter for laser intensity noise.
The cavity transfer function for amplitude fluctuations is

\begin{equation}
H(f) = \frac{1}{1 + i f / f_c},
\end{equation}

where $f_c$ is the cavity pole frequency.
This passive filtering reduces high-frequency relative intensity noise (RIN)
before the beam enters the active intensity stabilization servo,
thereby relaxing its control bandwidth requirements.

\paragraph{(iii) Preliminary Frequency Conditioning}

Although the Reference Cavity (RC) provides the primary frequency reference,
the PMC suppresses frequency fluctuations outside its linewidth.
The cavity resonance converts frequency noise into transmitted power fluctuations,
which are actively suppressed by the PDH locking loop.
This stage reduces the dynamic range required for the RC actuator
and improves lock acquisition robustness.

\paragraph{(iv) Isolation from Back-Scattered Light}

Back-scattered light from downstream optics,
including the IMC and main interferometer,
can re-enter the laser and produce excess phase noise.
The traveling-wave geometry of the PMC significantly reduces this feedback.
This is particularly important in the sub-Hz regime,
where scattered light beating with ground motion
can generate excess low-frequency noise.

\paragraph{(v) Control Noise Decoupling}

The PMC acts as an optical buffer between the laser source
and the IMC.
Without this intermediate stage,
frequency and amplitude noise would directly drive
the IMC control loop,
increasing control noise injection into the main interferometer.
By reducing upstream noise,
the PMC improves overall control-loop stability.
\subsection{Pound--Drever--Hall Frequency Discrimination}

Phase modulation sidebands are generated using an Electro-Optic Modulator (EOM).
The modulated beam is injected into the high-finesse Reference Cavity (RC),
and the reflected field is detected to obtain a frequency error signal.

The reflected beam is separated by a Polarizing Beam Splitter (PBS)
and detected with a radio-frequency photodetector (RFPD).
Using the Pound--Drever--Hall (PDH) method,
the detuning between the laser frequency and the cavity resonance
is extracted as

\begin{equation}
V_{\mathrm{err}}
\propto
\Im \left( E_{\mathrm{carrier}} E_{\mathrm{sideband}}^* \right).
\end{equation}

Near resonance, the PDH error signal is linear with frequency deviation,
allowing high-precision discrimination.
The slope of the discriminator is proportional to
the cavity finesse $\mathcal{F}$,
which enhances frequency sensitivity.

\subsection{Optical and Mechanical Parameters}

The cavity consists of two curved mirrors
(ROC = $3~\mathrm{m}$, $R = 99.992\%$)
and two flat mirrors ($R = 98\%$),
all made of fused silica.
One mirror is mounted on a piezoelectric actuator (PZT)
for cavity-length control using the Pound-Drever-Hall method.

The free spectral range is

\begin{equation}
\mathrm{FSR} = \frac{c}{L}
\approx 600~\mathrm{MHz},
\end{equation}

ensuring large longitudinal mode spacing and stable locking.

\subsection{Thermal Stability Considerations}

The housing is fabricated from 6061 aluminum alloy.
Although its high thermal conductivity promotes uniform temperature distribution,
its relatively large thermal expansion coefficient
($\alpha = 2.32 \times 10^{-5}~\mathrm{K^{-1}}$)
introduces cavity-length sensitivity to temperature fluctuations.

The induced cavity-length variation is

\begin{equation}
\delta L = \alpha L \delta T.
\end{equation}

A temperature fluctuation of $1~\mathrm{mK}$ yields

\begin{equation}
\delta L \approx 1.2 \times 10^{-8}~\mathrm{m}.
\end{equation}

Such fluctuations correspond to resonance-frequency drift,
which can couple to downstream frequency stabilization stages.
Therefore, active temperature control of the PMC housing
is implemented to suppress low-frequency thermal noise.

\subsection{Importance for Sub-Hz Operation}

In conventional high-frequency gravitational-wave detectors,
the PMC primarily serves as a spatial filter.
However, for CHRONOS operating in the sub-Hz regime,
the PMC plays a deeper role:
it suppresses laser noise that would otherwise
couple through residual asymmetries,
control loops,
or scattered light paths
into the torsional degree of freedom.

The PMC thus constitutes an essential noise-conditioning stage
that enables the realization of quantum-noise-limited
speed-meter sensitivity at ultra-low frequencies.

\section{Intensity Stabilization}

Laser Relative Intensity Noise (RIN) is one of the fundamental technical
noise sources in interferometric gravitational-wave detectors.
For CHRONOS, a torsion-bar speed-meter interferometer operating in the
sub-Hz regime, intensity fluctuations are particularly critical.
Laser power fluctuations couple directly to radiation-pressure torque,
which in turn drives angular displacement of the torsion bar.
Therefore, intensity stabilization at low frequencies is indispensable
for achieving the target sensitivity.

In the present system,
an Acousto-Optic Modulator (AOM) is used as the intensity actuator,
and an analog Optical-Follower Servo (OFS) provides feedback control.

\subsection{Physical Coupling of Intensity Noise}

A fluctuation in optical power $\delta P$ induces a fluctuation
in radiation-pressure force.
For a torsion-bar interferometer,
this force produces a torque about the rotational axis.

To leading order, the torque fluctuation is approximated as

\begin{equation}
\delta \tau
=
\frac{2 L}{c} \, \delta P ,
\end{equation}

where $L$ is the lever arm from the beam spot to the rotation axis,
and $c$ is the speed of light.

The angular response of the torsion bar is given by

\begin{equation}
\theta(\Omega)
=
\frac{\delta \tau(\Omega)}
     {I \left( \Omega_0^2 - \Omega^2 + i \Omega \Gamma \right)},
\end{equation}

where $I$ is the moment of inertia,
$\Omega_0$ the torsional resonance frequency,
and $\Gamma$ the damping rate.

At frequencies well above resonance ($\Omega \gg \Omega_0$),

\begin{equation}
\theta(\Omega)
\approx
\frac{2L}{c I \Omega^2} \, \delta P(\Omega).
\end{equation}

Thus, intensity fluctuations exhibit a $1/\Omega^2$ enhancement
in angular displacement at low frequencies,
making RIN suppression increasingly important
toward the sub-Hz band.

Expressed in terms of relative intensity noise,
$\mathrm{RIN} = \delta P / P$,
the induced angular noise becomes

\begin{equation}
\theta_{\mathrm{RIN}}(\Omega)
=
\frac{2 L P}{c I \Omega^2}
\, \mathrm{RIN}(\Omega).
\end{equation}

This relation explicitly shows that
radiation-pressure-induced angular noise scales linearly with optical power
and inversely with $\Omega^2$.
Consequently, even modest low-frequency RIN
can limit the CHRONOS sensitivity.

\subsection{Intensity Stabilization Architecture}

After passing through the AOM,
the laser beam enters the PMC.
The transmitted beam is split and detected by two photodiodes:

\begin{itemize}
    \item \textbf{OFSPD-in}: in-loop detector
    \item \textbf{OFSPD-out}: out-of-loop detector
\end{itemize}

The OFSPD-in signal provides the error signal
for the Optical-Follower Servo (OFS).
The OFS drives the AOM RF amplitude,
thereby modulating the diffraction efficiency
and stabilizing the output optical power.

The OFSPD-out detector is located outside the feedback loop
and independently monitors the achieved RIN suppression.
Since in-loop measurements are artificially reduced
by the loop gain,
out-of-loop verification is essential
to quantify the true stabilization performance.

\subsection{Optical-Follower Servo (OFS)}

The OFS consists of two primary subsystems:

\begin{itemize}
    \item Power Stabilization Servo (PSS)
    \item Interface chassis
\end{itemize}

The PSS processes the photodiode signal
and generates the control signal for the AOM.
The interface chassis routes diagnostic signals
to the data acquisition system,
allowing continuous monitoring of loop performance.

The closed-loop residual intensity noise is given by

\begin{equation}
S_{\mathrm{closed}}(f)
=
\frac{S_{\mathrm{free}}(f)}{1 + G(f)},
\end{equation}

where $G(f)$ is the open-loop gain.
In the frequency band where $|G(f)| \gg 1$,
the RIN is suppressed approximately by a factor $1/G(f)$.

\subsection{Control Loop Characteristics}

\begin{figure}[t]
    \centering
    \includegraphics[width=0.9\linewidth]{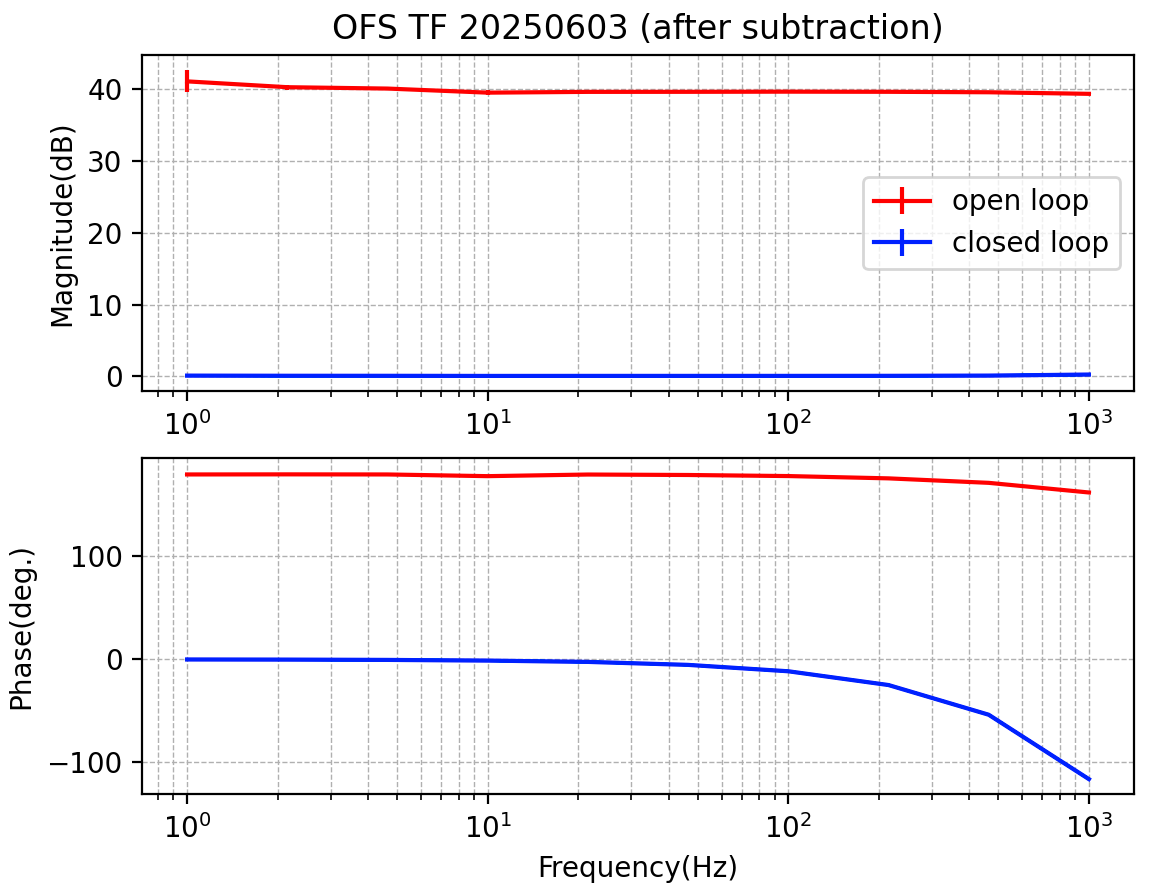}
    \caption{
    Measured transfer function of the intensity stabilization loop.
    Top: magnitude; Bottom: phase.
    Red: open-loop; Blue: closed-loop.
    }
    \label{fig:ofs_tf}
\end{figure}

The measured transfer function of the intensity stabilization loop
is shown in Fig.~\ref{fig:ofs_tf}.
An open-loop gain of approximately 30~dB
is achieved in the low-frequency band,
with sufficient phase margin to ensure stable operation.

The unity-gain frequency is chosen to balance
RIN suppression and control noise injection.
Excessive bandwidth would increase sensing noise coupling,
while insufficient bandwidth would leave low-frequency RIN unsuppressed.

\subsection{Noise Suppression Performance}

\begin{figure}[t]
    \centering
    \includegraphics[width=0.9\linewidth]{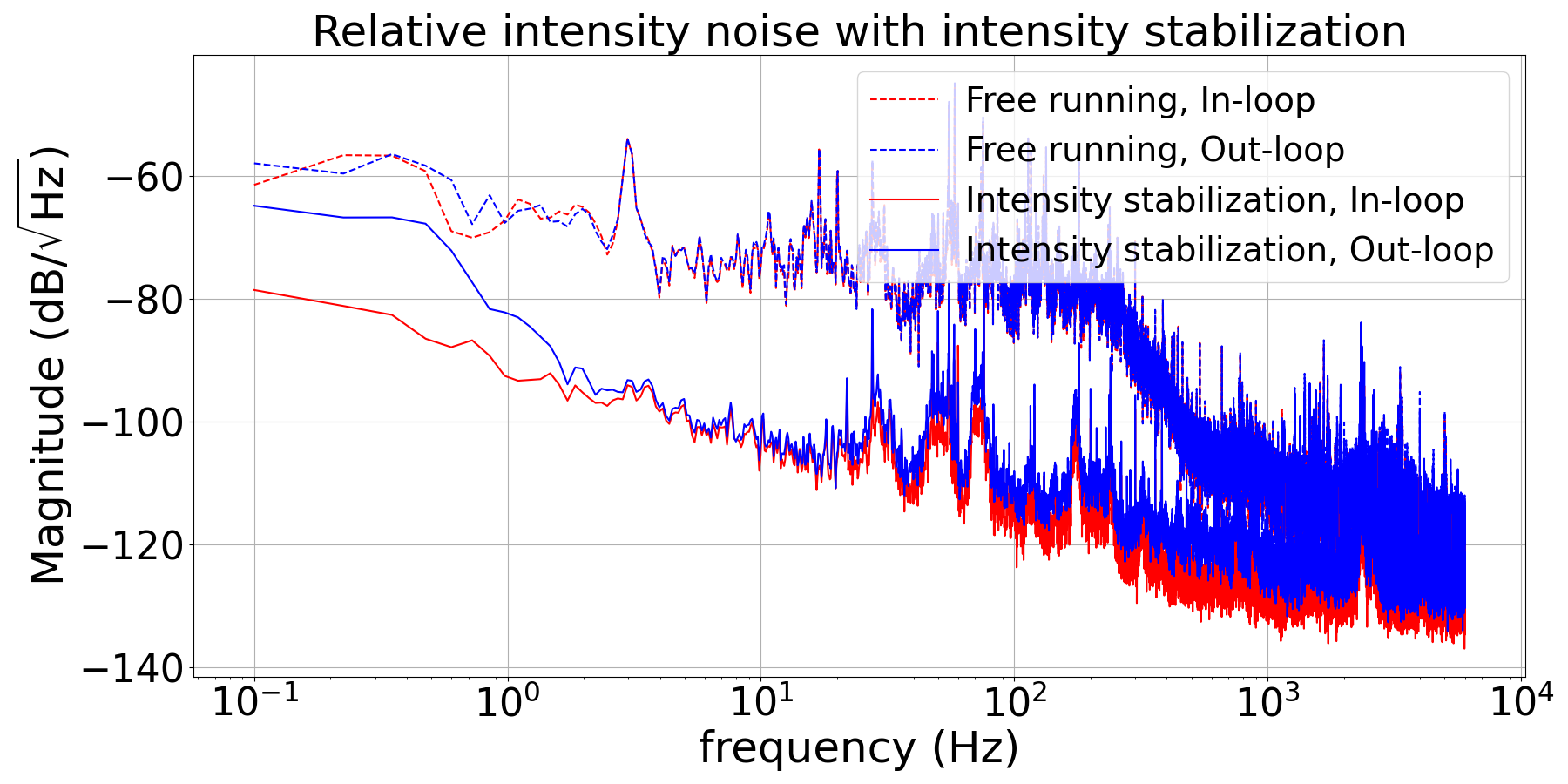}
    \caption{
    Relative intensity noise spectra.
    Dashed lines: open-loop.
    Solid lines: closed-loop.
    Red: OFSPD-in; Blue: OFSPD-out.
    }
    \label{fig:ofs_noise}
\end{figure}

The measured RIN spectra are shown in Fig.~\ref{fig:ofs_noise}.

In the open-loop configuration,
RIN at low frequencies is approximately
$-60~\mathrm{dB/Hz^{1/2}}$.
Closed-loop control achieves
20--30~dB suppression above 10~Hz.
The out-of-loop measurement confirms
that the achieved suppression is physical
and not an artifact of loop gain.

A 20~dB reduction in RIN corresponds directly
to a 20~dB reduction in radiation-pressure torque noise.
This improvement significantly relaxes
the low-frequency angular noise budget
and contributes directly to the sub-Hz sensitivity of CHRONOS.

\subsection{Role in Sub-Hz Speed-Meter Operation}

In conventional position-meter interferometers,
intensity noise primarily affects radiation-pressure force noise.
In CHRONOS, however,
the torsional degree of freedom enhances the low-frequency response,
making RIN coupling particularly severe.

Therefore, the intensity stabilization system
is not merely an auxiliary subsystem
but a fundamental component of the detector noise architecture.
Without sufficient RIN suppression,
radiation-pressure torque would dominate
the sub-Hz sensitivity band.

The achieved stabilization ensures that
laser intensity noise remains below
the torsion thermal noise and quantum radiation-pressure noise
across the operational frequency range.

\section{Frequency Stabilization}

Laser frequency noise is one of the dominant technical noise sources
in interferometric gravitational-wave detectors.
A fluctuation in laser frequency $\delta \nu$
is converted into phase noise through optical path-length asymmetry
$\Delta L$ as

\begin{equation}
\delta \phi
=
2\pi \frac{\delta \nu}{\nu} \Delta L ,
\end{equation}

where $\nu$ is the laser carrier frequency.
Even a small residual asymmetry $\Delta L$
can therefore transform frequency noise into measurable phase fluctuations.

In CHRONOS, a torsion-bar speed-meter interferometer operating in the sub-Hz regime,
low-frequency phase noise can couple to the rotational degree of freedom
through residual optical asymmetries,
imperfect common-mode rejection,
and control-loop cross couplings.
Because the torsional response enhances motion toward low frequencies
($\propto 1/\Omega^2$ above resonance),
stringent suppression of frequency noise is required in the sub-Hz band.

Although the Pound--Drever--Hall (PDH) technique
is also used to lock the Pre-Mode Cleaner (PMC),
its role there is limited to maintaining cavity resonance
for spatial filtering.
In contrast, the PDH locking to the Reference Cavity (RC)
defines the absolute frequency reference of the laser.
The frequency stabilization described in this section
is therefore fundamentally distinct from the PMC locking scheme.

\subsection{Reference Cavity as Absolute Frequency Standard}

\begin{figure}[t]
    \centering
    \includegraphics[width=0.75\linewidth]{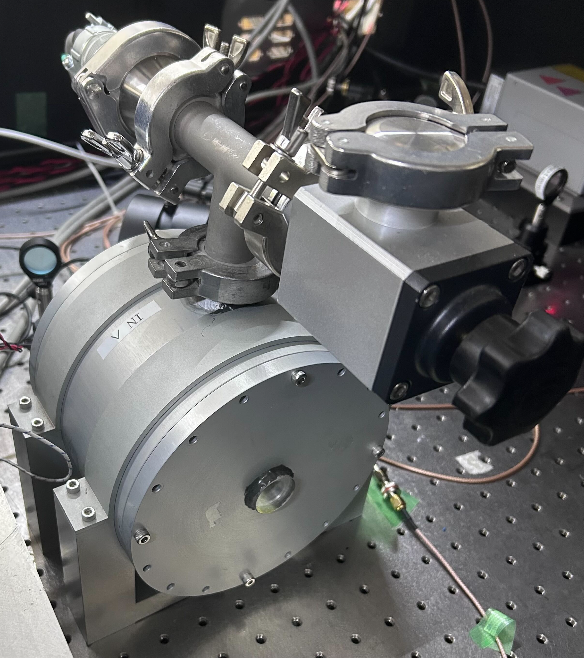}
    \caption{
    Reference Cavity (RC) assembly.
    A high-finesse optical resonator is enclosed
    within a cylindrical aluminum housing
    with thermal stabilization.
    }
    \label{fig:rc_photo}
\end{figure}

A photograph of the Reference Cavity assembly is shown in
Fig.~\ref{fig:rc_photo}.

The RC acts as a frequency reference.
Locking the laser to the cavity resonance transfers
the cavity-length stability to the laser frequency.

The fractional relationship between cavity length and frequency is

\begin{equation}
\frac{\delta \nu}{\nu}
=
-
\frac{\delta L}{L}.
\end{equation}

Therefore, mechanical vibration,
acoustic coupling,
and thermal expansion of the cavity spacer
directly determine the achievable frequency stability.

Temperature-induced length variations are given by

\begin{equation}
\delta L = \alpha L \delta T,
\end{equation}

where $\alpha$ is the thermal expansion coefficient.
Active thermal stabilization reduces long-term drift,
while mechanical isolation suppresses vibrational coupling.

The intrinsic linewidth of the cavity,
determined by its finesse,
sets the frequency discrimination bandwidth.
A higher finesse improves sensitivity but reduces capture range,
requiring careful optimization between stability and operational robustness.

\subsection{Multi-Stage Control Hierarchy}

Frequency stabilization is implemented using a hierarchical actuator scheme:

\begin{itemize}
    \item \textbf{High bandwidth}: fast phase correction via EOM
    \item \textbf{Intermediate bandwidth}: internal laser PZT actuator
    \item \textbf{Low bandwidth}: laser temperature control
\end{itemize}

This hierarchy enables broadband noise suppression
while maintaining large dynamic range.

The residual closed-loop frequency noise is

\begin{equation}
S_{\nu,\mathrm{closed}}(f)
=
\frac{S_{\nu,\mathrm{free}}(f)}{1 + G(f)},
\end{equation}

where $G(f)$ is the open-loop gain.
In the regime $|G(f)| \gg 1$,
frequency noise is suppressed approximately by $1/G(f)$.

Careful loop design is required to avoid
control-noise injection from sensing shot noise
or actuator electronics,
which may dominate at high frequencies.

\subsection{Frequency Noise Requirement for CHRONOS}

A preliminary engineering target for laser frequency noise in the
sub-Hz band is

\begin{equation}
\delta \nu_{\mathrm{req}}
\sim
10^{-2}
~\mathrm{Hz}/\sqrt{\mathrm{Hz}}.
\end{equation}

This allocation must be checked against the full transfer function and noise
in Chap.~\ref{chap:sensitivity}; it is not a universal strain-noise limit.
To estimate its impact, frequency noise can be converted into equivalent angular noise
through residual optical asymmetry:

\begin{equation}
\theta_{\nu}(\Omega)
=
\kappa_{\phi}
\frac{2\pi}{\nu}
\Delta L
\, \delta \nu(\Omega),
\end{equation}

where $\kappa_{\phi}$ represents the phase-to-angle coupling factor.
Even small asymmetries $\Delta L$
can produce measurable angular fluctuations
if frequency noise is insufficiently suppressed.

By locking the laser to the high-finesse RC,
the achieved frequency noise level satisfies the above requirement,
ensuring that frequency noise remains below
torsional thermal noise
and quantum radiation-pressure noise
across the operational band.

\subsection{Importance for Sub-Hz Speed-Meter Sensitivity}

In conventional position-meter interferometers,
frequency noise mainly couples through arm-length mismatch.
In CHRONOS, additional coupling pathways arise
through rotational sensing asymmetries
and control-loop cross couplings.

Because the torsional response enhances low-frequency motion,
frequency stabilization must be more stringent
than in high-frequency detectors.

Therefore, together with intensity stabilization,
frequency stabilization constitutes a core element
of the CHRONOS noise architecture.
It is not an auxiliary subsystem,
but a fundamental prerequisite
for achieving quantum-noise-limited sensitivity
in the sub-Hz regime.

\section{Summary of the Input Optics}

In this chapter, we have presented the design and implementation
of the Input Optics system for CHRONOS.
The Input Optics constitutes the pre-stabilization stage
that delivers laser light with controlled spatial mode,
intensity, and frequency
to the main torsion-bar speed-meter interferometer.

First, the Pre-Mode Cleaner (PMC)
selects the fundamental TEM$_{00}$ spatial mode
while suppressing higher-order transverse modes
and beam-pointing noise.
The bow-tie traveling-wave configuration
reduces optical back-reflection and scattered-light coupling,
providing a stable spatial reference
for downstream optical systems.
Although the PMC is locked using the PDH method,
its primary role is spatial filtering and preliminary conditioning,
rather than defining the absolute laser frequency.

Second, the Intensity Stabilization system
employs an Acousto-Optic Modulator (AOM)
and an Optical-Follower Servo (OFS)
to suppress Relative Intensity Noise (RIN).
In-loop and out-of-loop measurements confirm
a RIN reduction exceeding 20~dB
in the relevant frequency band.
Because radiation-pressure torque scales linearly with optical power
and the torsional response enhances low-frequency motion,
intensity noise directly contributes to angular displacement noise.
Therefore, RIN suppression is essential
for maintaining sub-Hz sensitivity.

Given that frequency noise couples to phase noise
through residual optical asymmetries,
and ultimately to angular motion,
this level of stabilization is necessary
to ensure that frequency noise remains below
torsional thermal noise
and quantum radiation-pressure noise.

Taken together,
the Input Optics system is not merely a beam-conditioning unit,
but a core stabilization architecture
that directly determines the achievable detector sensitivity.
The integrated control of spatial purity,
intensity stability,
and frequency stability
forms the foundation for sub-Hz gravitational-wave detection.

In a torsion-bar speed-meter interferometer,
low-frequency technical noise is inherently amplified
by the mechanical response of the test mass.
Therefore, the stringent laser stabilization described in this chapter
is a prerequisite for achieving
quantum-noise-limited performance in CHRONOS.

The design and implementation presented here
establish the technological basis
for stable and high-sensitivity operation
in the sub-Hz frequency band.

%% file: Chapters/Output_Optics.tex
\chapter{Output Optics}
\label{chap:outputoptics}

\section{Purpose}
\label{sec:oo_principle}

The Output Optics system forms the final optical stage between the main interferometer and the readout electronics, and is responsible for delivering a spatially clean, phase-controlled, and low-noise signal field to the detection chain. Consequently, the Output Optics is not simply a detection interface, but a carefully designed filtering and readout architecture that preserves the quantum-limited signal while suppressing unwanted optical and technical noise.

One key element of the subsystem is the Output Mode Cleaner (OMC), typically implemented as a compact optical cavity matched to the fundamental eigenmode of the interferometer output field. In contrast to the Input Mode Cleaner, which prepares the beam before it enters the interferometer, the OMC filters the field after it has accumulated imperfections inside the interferometer. Optical aberrations, mirror surface defects, residual misalignments, and scattering processes can all generate higher-order spatial modes at the output port. If left unfiltered, these modes produce excess noise in the detected signal. The OMC transmits the TEM\textsubscript{00} component that carries the science signal while strongly attenuating higher-order transverse modes and non-resonant optical sidebands.

Both the reflected and transmitted ports of the OMC are utilized for control and monitoring purposes. The reflected beam can be used to generate an error signal for maintaining cavity resonance and alignment, while the transmitted beam serves as the science channel input to the balanced homodyne detector. Stable resonance of the OMC is essential, as cavity length fluctuations can directly modulate the transmitted power and introduce excess noise at low frequencies.

Downstream of the OMC, the Balanced Homodyne Detection (BHD) system measures a selected quadrature of the filtered output field. The transmitted beam is interfered with a phase-stable local oscillator on a 50/50 beamsplitter, and the differential photocurrent from two photodiodes provides a low-noise readout of the interferometer signal. This approach enables flexible selection of the readout quadrature and improved rejection of common-mode intensity noise compared to direct detection.

The OMC and BHD assemblies are designed with high mechanical rigidity, low optical loss, and careful stray-light control to meet the stringent low-frequency noise requirements of CHRONOS. Provision is also made for thermal stabilization and potential future suspension upgrades, aimed at further reducing cavity-length and alignment fluctuations in the sub-Hz regime.

Therefore, the CHRONOS Output Optics system operates as a spatial filtering and precision readout architecture. By suppressing higher-order spatial modes, stabilizing the output field, and enabling low-noise quadrature measurement through balanced homodyne detection, it ensures that the detected signal faithfully represents the interferometer response required for sub-Hz gravitational-wave measurements.

\section{Output Mode Cleaner}
\label{sec:omc}

\subsection{Principle}
\label{sec:omc_principle}

The transverse profile of a laser beam can be expanded in terms of Gaussian spatial modes, specifically the Hermite-Gaussian modes. For CHRONOS, the desired spatial distribution is the fundamental transverse electromagnetic mode, TEM\textsubscript{00}. However, real optical components are never perfect, and surface roughness errors, coating irregularities, or slight angular misalignments of mirrors can redistribute optical power from the fundamental mode into higher-order transverse modes.

To mitigate this, an input mode cleaner is installed upstream of the main interferometer as described in previous chapters. Its role is to filter the incoming beam so that predominantly the TEM\textsubscript{00} mode is injected. Even with a well-prepared input beam, imperfections and misalignment within the main interferometer can still generate additional higher-order modes. For this reason, an OMC is positioned downstream of the main interferometer, at the detection port. The OMC selectively transmits the fundamental mode, which carries the gravitational-wave signal, while rejecting higher-order modes generated within the interferometer itself \cite{fritschel2004output,kumeta2015design}.

\subsection{Cavity Response}

For CHRONOS, the OMC is implemented as a four-mirror bow-tie ring cavity. Unlike a linear Fabry-P\'erot cavity, the light propagates around a closed optical path in a traveling-wave configuration. The resonance condition is determined by the round-trip phase,
\begin{equation}
    \phi(\omega)
    = \frac{\omega L_{\mathrm{rt}}}{c}
    = 2\pi q,
    \label{eq:omc_resonance}
\end{equation}
where $L_{\mathrm{rt}}$ is the total round-trip optical path length in vacuum, $c$ is the speed of light, and $q$ is an integer. The free spectral range (FSR), which gives the
frequency separation between adjacent longitudinal resonances, is therefore
\begin{equation}
    \mathrm{FSR}=\frac{c}{L_{\mathrm{rt}}},
    \label{eq:omc_fsr}
\end{equation}

The sharpness of the cavity resonance is characterized by its finesse,
\begin{equation}
    \mathcal{F}
    =\frac{\mathrm{FSR}}{\Delta\nu_{\mathrm{FWHM}}},
    \label{eq:omc_finesse}
\end{equation}
where $\Delta\nu_{\mathrm{FWHM}}$ is the full width at half maximum of the power resonance. For a high-finesse cavity with small round-trip power loss,
\begin{equation}
    \mathcal{F}
    \simeq
    \frac{2\pi}
    {T_{\mathrm{in}}+T_{\mathrm{out}}+\ell_{\mathrm{rt}}},
    \label{eq:omc_finesse_loss}
\end{equation}
where $T_{\mathrm{in}}$ and $T_{\mathrm{out}}$ are the power transmittances of the input and output coupling mirrors, and $\ell_{\mathrm{rt}}$ is the remaining round-trip power loss. The corresponding cavity pole frequency is
\begin{equation}
    f_{\mathrm{pole}}
    =\frac{\Delta\nu_{\mathrm{FWHM}}}{2}
    =\frac{\mathrm{FSR}}{2\mathcal{F}}.
    \label{eq:omc_pole}
\end{equation}
These relations show that increasing the finesse narrows the cavity linewidth and improves frequency discrimination, but also increases the importance of optical loss and resonance control \cite{Kogelnik:66,Black2001}.

The transmitted and reflected powers can be described using the cavity field response. Let $r_{\mathrm{in}}$ be the amplitude reflectivity of the input coupler, and let $\rho$ represent the amplitude retained by the remainder of the cavity during one round trip. The reflected field is
\begin{equation}
    \frac{E_{\mathrm{refl}}}{E_{\mathrm{in}}}
    =
    \frac{r_{\mathrm{in}}-\rho e^{i\phi}}
    {1-r_{\mathrm{in}}\rho e^{i\phi}},
    \label{eq:omc_reflected_field}
\end{equation}
where the phase convention is chosen such that the amplitude coefficients are real. The reflected power fraction is then
\begin{equation}
    \mathcal{R}(\phi)
    =
    \left|
    \frac{r_{\mathrm{in}}-\rho e^{i\phi}}
    {1-r_{\mathrm{in}}\rho e^{i\phi}}
    \right|^2.
    \label{eq:omc_reflected_power}
\end{equation}

For the transmitted field, define $\gamma$ as the round-trip amplitude retention and $g_{13}$ as the amplitude propagation factor from the input coupler to the output coupler, excluding their transmission coefficients. The cavity power transmittance is
\begin{equation}
    \mathcal{T}(\phi)
    =
    \frac{
    |t_{\mathrm{in}}t_{\mathrm{out}}g_{13}|^2
    }{
    (1-\gamma)^2
    +4\gamma\sin^2(\phi/2)
    },
    \label{eq:omc_airy}
\end{equation}
where $t_{\mathrm{in}}$ and $t_{\mathrm{out}}$ are amplitude transmission coefficients. In the high-finesse, low-loss limit, the on-resonance transmission and reflection reduce to
\begin{align}
    \mathcal{T}_0
    &\simeq
    \frac{4T_{\mathrm{in}}T_{\mathrm{out}}}
    {(T_{\mathrm{in}}+T_{\mathrm{out}}
    +\ell_{\mathrm{rt}})^2},
    \label{eq:omc_transmission_resonance}
    \\
    \mathcal{R}_0
    &\simeq
    \left(
    \frac{T_{\mathrm{in}}-T_{\mathrm{out}}
    -\ell_{\mathrm{rt}}}
    {T_{\mathrm{in}}+T_{\mathrm{out}}
    +\ell_{\mathrm{rt}}}
    \right)^2.
    \label{eq:omc_reflection_resonance}
\end{align}
These expressions illustrate the balance between input coupling, output coupling, and internal loss. In particular, the reflected carrier is minimized under the critical-coupling condition
$T_{\mathrm{in}}=T_{\mathrm{out}}+\ell_{\mathrm{rt}}$.
The transmitted power is therefore not determined by finesse alone, but also by the distribution of optical losses and the choice of coupling mirrors.

\subsection{Higher Order Mode Suppression}

The OMC is a resonant traveling-wave ring cavity matched to the fundamental spatial mode of the interferometer output. For a cavity resonant with TEM$_{00}$, the round-trip phase of a TEM$_{mn}$ mode may be written as
\begin{equation}
    \phi_{mn}
    =
    2\pi\frac{\delta\nu}{\mathrm{FSR}}
    -m\eta_x-n\eta_y,
    \label{eq:omc_hom_phase}
\end{equation}
where $\delta\nu$ is the frequency offset from the carrier resonance, and $\eta_x$ and $\eta_y$ are the round-trip Gouy phases in the two transverse directions. For a cavity without astigmatism, $\eta_x=\eta_y=\eta$, so the transverse-mode detuning depends on the total mode order $m+n$.
Higher-order transverse modes experience a different round trip Gouy phase and are therefore detuned from resonance with respect to the cavity. As a result, their transmission is strongly suppressed compared to the TEM\textsubscript{00} mode. In addition to spatial filtering, the cavity linewidth also provides a degree of frequency filtering, attenuating residual radio-frequency control sidebands that are not resonant with the cavity. Assuming approximately mode-independent optical loss, the
transmission of a higher-order mode relative to the resonant
fundamental mode is
\begin{equation}
    \frac{\mathcal{T}_{mn}}{\mathcal{T}_{00}}
    =
    \frac{1}{
    1+
    \dfrac{4\gamma}{(1-\gamma)^2}
    \sin^2\left(\dfrac{\phi_{mn}}{2}\right)
    }.
    \label{eq:omc_hom_suppression}
\end{equation}
The same cavity response describes the suppression of RF
sidebands by setting
$\delta\nu=\pm f_{\mathrm{RF}}$. Thus, the OMC length,
finesse, and Gouy phase must be selected together to avoid
accidental resonance of unwanted spatial modes or RF
sidebands \cite{kumeta2015design,fritschel2004output}. The OMC power transmittance of the $m^{th}$ spatial mode and the RF sideband is given, respectively, by \cite{kumeta2015design}:

\begin{equation}
T_{OMC}=\frac{(1-r_1^2-\mathcal{L})(1-r_3^2-\mathcal{L})r_2^2r_4^2}{|1-r_1r_2r_3r_4e^{-im\eta}|^2}
\end{equation}

\begin{equation}
T_{RF}=\frac{(1-r_1^2-\mathcal{L})(1-r_3^2-\mathcal{L})r_2^2r_4^2}{|1-r_1r_2r_3r_4e^{-im\eta+2iL\omega_{RF}/c}|^2}
\end{equation}

where $T_{OMC}$ and $T_{RF}$ are the OMC power transmittance of the higher order modes and RF sideband, respectively; $r_i$ is the reflectivity  of the $i^{th}$ OMC mirror, $\mathcal{L}$ is the optical loss per bounce, $m$ is the mode order, and $\eta$ is the round trip Gouy phase.

For CHRONOS, suppression of higher-order modes is particularly important because the low-frequency signal band is susceptible to excess noise coupling through wavefront distortions, and scattered light. The OMC reduces this coupling pathway and stabilizes the downstream homodyne readout.

\subsection{Optical Design}
The OMC is implemented as a compact optical cavity placed after the interferometer output port, with its specifications for cavity length, mirror reflectivities and radii of curvature chosen to balance mode selectivity and optical losses. The cavity length is inversely related to the free spectral range, or the spacing between adjacent modes, and must be appropriately chosen to transmit only the desired fundamental mode\cite{fritschel2004output}. At the same time, the finesse, which is related to the sharpness of the resonance peak, must be high enough to discriminate against higher-order spatial modes but low enough such that intra-cavity optical losses do not significantly degrade the signal-to-noise ratio.

Mode matching between the interferometer output beam and the OMC eigenmode is a critical design consideration. Imperfect mode matching leads to direct signal loss, as part of the fundamental mode is rejected together with the higher-order content. For this reason, dedicated mode-matching optics are included upstream of the OMC to adapt the beam waist and curvature.

The incidence angle or the geometry of the bow-tie cavity also has a tradeoff, wherein it must be large enough to reduce backscattering but small enough to reduce mode ellipticity\cite{fritschel2004output}.

Mechanical stability of the OMC assembly is another key factor, especially in a cryogenic and low-frequency instrument such as CHRONOS. Length and alignment fluctuations of the cavity can convert into amplitude noise at the transmission port. The cavity is therefore designed to be rigidly mounted, with actuators for fine alignment control and length tuning to maintain resonance with the carrier field.

\begin{figure}[htbp]
 \centering
    \includegraphics[width=0.5\textwidth]{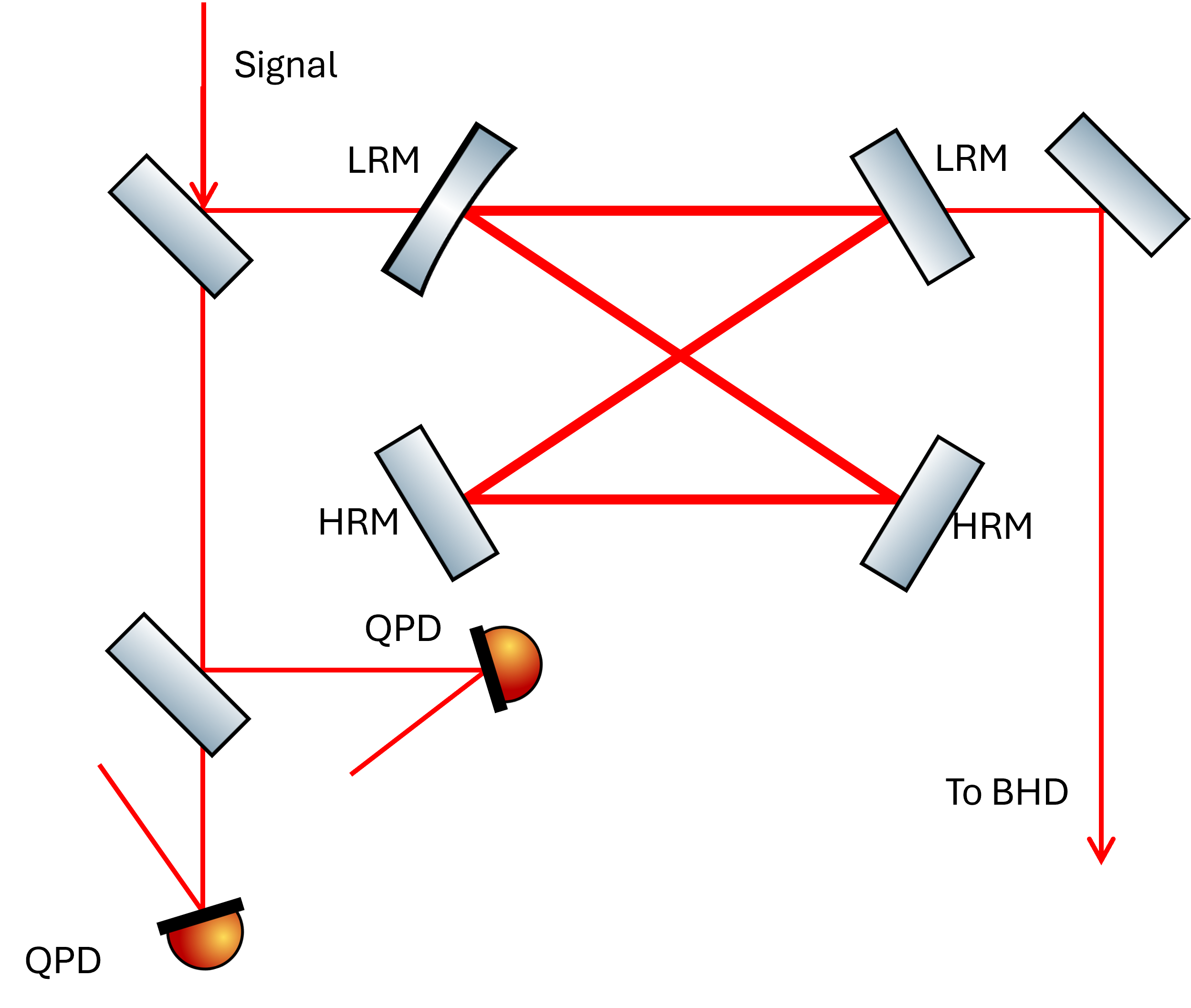}
 \caption{Schematic diagram of output mode cleaner(OMC) composed of mode-matching and alignment optics, followed by the bow-tie cavity with two low-reflectivity mirrors (LRM) and two high-reflectivity mirrors (HRM). A portion of the laser prior to the OMC will be directed to the QPDs for control purposes. The spatially filtered transmitted light, which contains the GW-modulated signal, then proceeds to the BHD readout.}
\label{fig:OMC}
\end{figure}

Fig.~\ref{fig:OMC} shows the layout of the output mode cleaner (OMC), consisting of mode-matching and alignment optics followed by a bow-tie cavity with two low-reflectivity mirrors (LRMs) and two high-reflectivity mirrors (HRMs). A pick-off beam taken before the OMC is used for cavity control. The transmitted, spatially filtered light carrying the GW signal is then sent to the balanced homodyne detection (BHD) readout.

\subsection{Future Upgrade}
In the current configuration, the OMC will be implemented in a fixed aluminum cavity, primarily to validate the optical layout and control scheme. While this approach simplifies integration and commissioning, it also leaves the cavity more exposed to seismic and acoustic disturbances, which can modulate the cavity length and alignment. For a full-scale implementation, a dedicated suspension system is envisioned to provide additional isolation from environmental noise, and the OMC will be constructed on a fused silica breadboard, with components fixed via epoxy.

A suspended OMC would benefit from reduced coupling of ground motion and mechanical vibrations, particularly in the sub-Hz regime relevant for CHRONOS. By designing the suspension to have low resonant frequencies and adequate damping, it is possible to suppress length fluctuations of the cavity and maintain stable resonance conditions. In addition, using a fused silica breadboard minimizes thermal noise.

\section{Balanced Homodyne Detection}
\subsection{Principle}

Achieving further noise suppression in gravitational-wave detectors while maintaining a relatively compact detector size necessitates the use of non-classical optical fields, particularly squeezed states, where the uncertainty in one quadrature is reduced at the expense of increased noise in the conjugate quadrature\cite{Caves1981quantum}. Since future detector schemes may rely on a flexible readout of the quadrature space of the output light field, balanced homodyne detection is expected to replace the currently used DC readout\cite{steinlechner2015local}. Balanced homodyne detection enables the measurement of any chosen quadrature by interfering a weak signal beam with a strong local oscillator. When implemented in interferometric setups, this approach allows high-precision readout that fully exploits the advantages of squeezed light. Because such quantum states play a central role in modern gravitational-wave instrumentation, it has already been demonstrated in a full scale michelson interferometer setup \cite{ligo2011gravitational,aasi2013enhanced} and ongoing research is being done in this field \cite{Stefszky_2012,heinze202210,tsuji2025quantum}. Accordingly, this project focuses on developing the optical hardware and readout electronics required for a balanced homodyne detector suitable for use in cross torsion bar detection experiments involving non-classical light.

Balanced homodyne detection is employed to measure the field quadrature that carries the  signal with high quantum efficiency and low technical noise. In this scheme, the signal beam transmitted by the OMC is interfered with a strong, phase-stable local oscillator (LO) on a 50/50 beamsplitter. The two output ports are detected by a pair of matched photodiodes, and their photocurrents are subtracted to suppress common-mode intensity noise and obtain the differential signal, the BHD readout. This configuration provides a means of selecting the readout quadrature independently of the signal-field carrier power and is particularly useful for interferometers employing non-classical optical states \cite{Caves1981quantum,schnabel2017squeezed}.

Let $\hat a$ and $\hat b$ denote the signal and LO field operators, respectively. For an ideal 50/50 beamsplitter, the output fields may be written as
\begin{equation}
    \hat c=\frac{\hat a+\hat b}{\sqrt{2}},
    \qquad
    \hat d=\frac{\hat b-\hat a}{\sqrt{2}}.
    \label{eq:bhd_beamsplitter}
\end{equation}
The difference in detected photon number is
\begin{equation}
    \hat n_-=
    \hat c^\dagger\hat c-\hat d^\dagger\hat d
    =
    \hat a^\dagger\hat b+\hat b^\dagger\hat a.
    \label{eq:bhd_difference}
\end{equation}

For a strong coherent LO, $\hat b\simeq\beta=|\beta|e^{i\zeta}$, the differential signal becomes
\begin{equation}
    \hat n_-\simeq
    2|\beta|\hat X_\zeta,
    \label{eq:bhd_quadrature_measurement}
\end{equation}
where
\begin{equation}
    \hat X_\zeta=
    \frac{
    \hat a e^{-i\zeta}
    +\hat a^\dagger e^{i\zeta}
    }{2}
    \label{eq:bhd_quadrature}
\end{equation}
is the quadrature selected by the LO phase $\zeta$. 

For classical optical powers, the ideal differential photocurrent is
\begin{equation}
    i_-=2\mathcal{R}_{\mathrm{pd}}
    \sqrt{P_{\mathrm{LO}}P_s}\cos\zeta,
    \label{eq:bhd_classical_current}
\end{equation}
where $\mathcal{R}_{\mathrm{pd}}$ is the photodiode responsivity, and $P_{\mathrm{LO}}$ and $P_s$ are the LO and signal powers. This expression assumes ideal interference visibility and matched photodiode responses.

The one-sided shot-noise current power spectral density is
\begin{equation}
    S_i^{\mathrm{shot}}
    =2e(I_1+I_2)
    \simeq
    2e\mathcal{R}_{\mathrm{pd}}P_{\mathrm{LO}},
    \label{eq:bhd_shot_noise}
\end{equation}
when the LO dominates the detected power. A sufficiently strong LO allows the optical signal and shot-noise level to rise above the electronic-noise floor, but does not itself remove quantum shot noise.

\subsection{Readout Angle}

The principal advantage of BHD is the ability to select the measured quadrature through the relative phase between the LO and signal fields. Using amplitude and phase quadratures $b_1$ and $b_2$, the measured quadrature is
\begin{equation}
    b_\zeta=b_1\cos\zeta+b_2\sin\zeta.
    \label{eq:bhd_readout_angle}
\end{equation}
The optimal readout angle depends on the interferometer response and the frequency-dependent quantum-noise budget. In particular, the selected quadrature determines how the signal and quantum-noise contributions are projected onto the detected output \cite{kimble2001conversion}.

Here $\zeta$ is the local quadrature angle measured from the $b_1$ axis.
It is not the mirror/LO setting $\zeta_{\rm LO}$ in
Table~\ref{tab:sensitivity_operating_points}. Conversion to that setting
must include the carrier propagation phase; a numerical angle cannot be
copied between the two conventions without that conversion.

Fluctuations in the relative LO - signal phase cause mixing between the measured and orthogonal quadratures. For a small phase fluctuation $\delta\zeta$,
\begin{equation}
    \delta b_\zeta
    \simeq
    b_{\zeta+\pi/2}\delta\zeta.
    \label{eq:bhd_phase_noise}
\end{equation}

Careful control of the homodyne angle is therefore required to maintain optimal sensitivity across the measurement band. Stable control of this phase is essential, as fluctuations in the readout angle can lead to mixing of unwanted noise quadratures into the detected signal.

\subsection{Optical Design}

\begin{figure}
 \centering
        \includegraphics[width=0.8\textwidth]{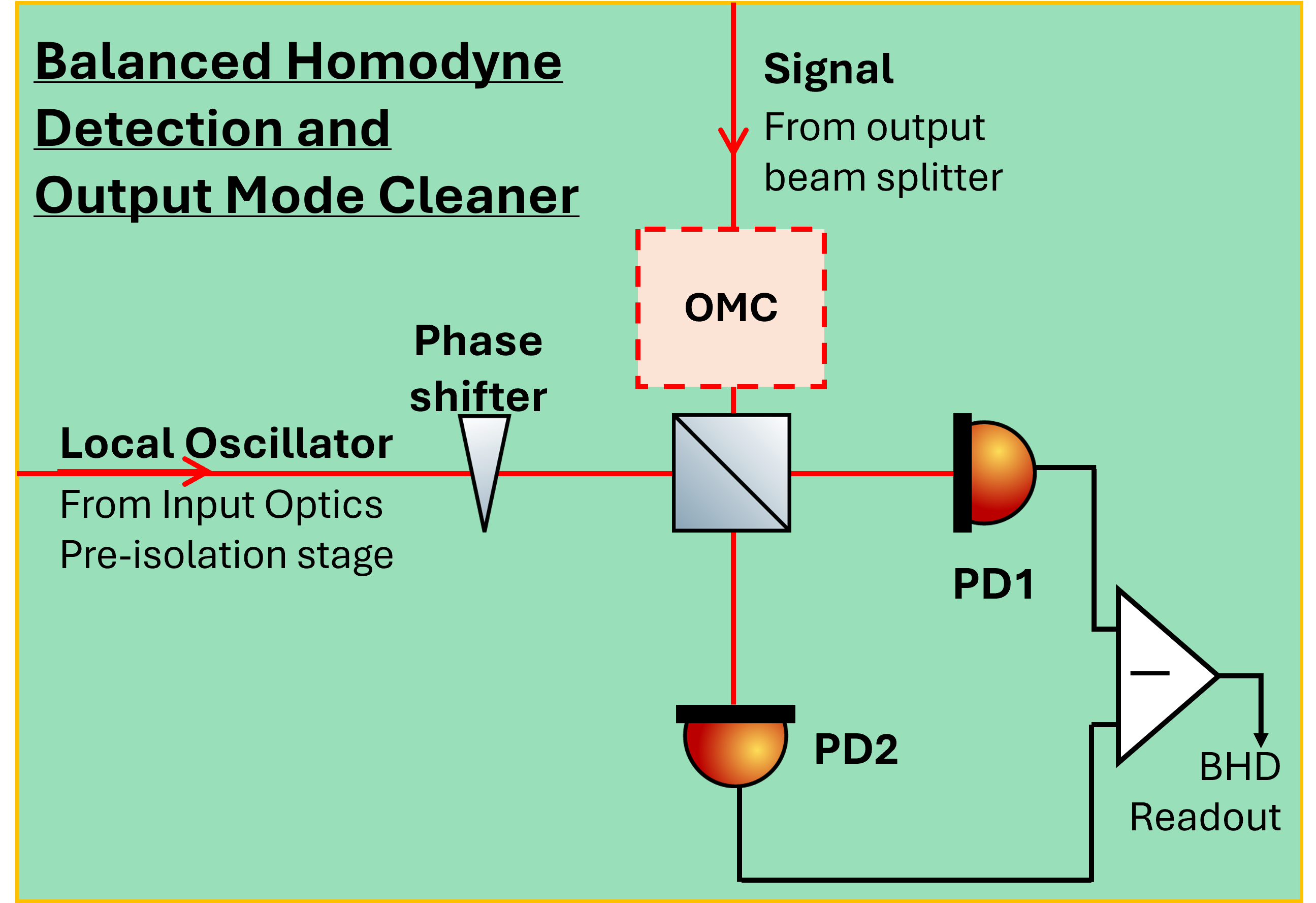}
 \caption{Schematic diagram of balanced homodyne detection(BHD) readout scheme in the scheme of other CHRONOS subsystems to show the overall setup.}
\label{fig:BHD}
\end{figure}

The optical layout of the BHD system is shown in Fig.~\ref{fig:BHD}. The local oscillator is derived from a clean reference beam picked-off before the main interferometer, and routed to the homodyne beamsplitter with carefully controlled path length. The local oscillator and signal beam from the OMC is then mixed at the homodyne beamsplitter, and the two output ports are measured using photodiodes. The photocurrents are then subtracted to obtain the differential signal.

Particular attention is given to balancing optical power and path length on the two photodiodes, as any imbalance reduces the effectiveness of common-mode noise cancellation. 

\subsection{Future Upgrade}

The use of squeezed states of light at the output port offers a method to  reduce quantum noise in the CHRONOS readout. In particular, frequency-dependent squeezing could be employed to simultaneously suppress shot noise at higher frequencies and radiation-pressure noise at lower frequencies \cite{kimble2001conversion}. Implementing such a scheme would require an additional filter cavity to rotate the squeezing angle as a function of frequency before injection into the homodyne detector.

Although not part of the baseline design, the output optics layout may be designed with sufficient flexibility to accommodate future integration of squeezed light sources and associated filter cavities. This forward-compatible approach ensures that upgrades aimed at quantum noise reduction can be incorporated without major redesign of the detection setup.

%% file: Chapters/IFO_control.tex
\chapter{Cavity Control}
\label{sec:cavity_control}

Before discussing the cavity-control architecture of the interferometer, it is useful to clarify the design philosophy that guided the selection of the optical topology, cavity lengths, recycling configuration, and control sidebands adopted in this study. The ultimate goal of CHRONOS is the realization of a large-scale sub-hertz gravitational-wave detector. The interferometer described here is therefore designed to validate the key optical and control concepts required for such a detector while remaining experimentally feasible in a laboratory environment~\cite{Inoue_CHRONOS_Main_2026}.

Stable operation requires simultaneous control of multiple coupled optical cavities, precise mode matching, and robust control of the interferometer operating point. The primary objective of the cavity-control system is consequently to maintain the resonance conditions of the Sagnac cavity and the recycling cavities while providing stable references for balanced-homodyne readout and quantum-noise measurements.

\section{Operation method}

CHRONOS employs a dual-recycled triangular-ring Sagnac topology~\cite{Chen2003}. The system is a laboratory-scale platform for validating optical stability, mode matching, cavity-length control, recycling-cavity phase control, and quantum-noise readout~\cite{Drever1983,Black2001}. Its differential Sagnac observable is velocity-like, or equivalently angular-momentum-like, rather than a direct position-displacement observable. This speed-meter response suppresses radiation-pressure noise at low frequencies and is central to the intended sub-hertz application~\cite{Chen2003}.

The frequency architecture is described first, followed by the sensing ports, degree-of-freedom definitions, and lock-acquisition sequence.

\subsection{Sideband frequencies and frequency architecture}

The RF-sideband frequencies are selected to provide the resonance and anti-resonance conditions needed to sense the power-recycling cavity (PRC), signal-recycling cavity (SRC), and triangular Sagnac cavity independently. The starting point is the round-trip length of the Sagnac ring cavity,
\begin{equation}
	L_{\mathrm{ring}}=5.08~\mathrm{m},
\end{equation}
which gives
\begin{equation}
	\mathrm{FSR}_{\mathrm{ring}}=\frac{c}{L_{\mathrm{ring}}}\simeq59.0~\mathrm{MHz}.
\end{equation}
This frequency is the fundamental reference for the sideband scheme.

\subsubsection*{Signal-recycling cavity}

The effective SRC length is chosen as
\begin{equation}
	L_{\mathrm{src}}\simeq5.08~\mathrm{m},
	\qquad
	\mathrm{FSR}_{\mathrm{src}}\simeq\frac{c}{2L_{\mathrm{src}}}=29.5~\mathrm{MHz}.
\end{equation}
Thus, the SRC free spectral range is one half of the ring-cavity free spectral range.

\subsubsection*{Power-recycling cavity}

The PRC length is set to
\begin{equation}
	L_{\mathrm{prc}}\simeq7.62~\mathrm{m},
	\qquad
	\mathrm{FSR}_{\mathrm{prc}}\simeq\frac{c}{2L_{\mathrm{prc}}}=19.7~\mathrm{MHz}.
\end{equation}
These cavity lengths reproduce the resonance hierarchy used in large-scale gravitational-wave detectors. Although the present interferometer is approximately 2.5~m in scale, the relevant optical phase relationships are equivalent to those of kilometre-scale instruments, permitting a realistic validation of the cavity-control architecture~\cite{meers1988,mizuno1993}.

\subsection{RF sidebands and degree-of-freedom separation}

Two RF sidebands are used for main-interferometer control:
\begin{equation}
	f_1=29.5~\mathrm{MHz},
	\qquad
	f_2=59.0~\mathrm{MHz}.
\end{equation}
The frequency $f_1$ equals the SRC free spectral range, whereas $f_2$ equals the ring-cavity free spectral range. In the coupled-cavity system, $f_2$ is resonant in the PRC and anti-resonant in the SRC. The resulting distinct optical responses provide discrimination among the recycling-cavity and Sagnac degrees of freedom with only two principal modulation frequencies. The sensing assignment is not inferred from the nominal FSR alone; it is fixed by the complete coupled-cavity response and the detection port, as summarized in Table~\ref{tab:control_ports}.

\subsection{Input mode cleaner compatibility}

The input mode cleaner (IMC) is a four-mirror suspended bow-tie cavity that acts as both a spatial-mode filter and a laser-frequency reference~\cite{Kwee2012}. Its round-trip length and free spectral range are
\begin{equation}
	L_{\mathrm{IMC}}=10.2~\mathrm{m},
	\qquad
	\mathrm{FSR}_{\mathrm{IMC}}\simeq14.8~\mathrm{MHz}.
\end{equation}
A dedicated phase-modulation frequency,
\begin{equation}
	f_{\mathrm{IMC}}=44.2~\mathrm{MHz},
\end{equation}
is used to derive the IMC Pound--Drever--Hall (PDH) error signal~\cite{Drever1983,Black2001}. The frequency plan permits the main control sidebands at $f_1$ and $f_2$ to pass through the IMC with minimal attenuation while preserving laser-frequency stabilization.

\subsection{Design philosophy summary}

The cavity-control design is based on the ring FSR, distinct resonance conditions in the PRC and SRC, port-dependent sensing with two phase-coherent RF sidebands, and compatibility with the IMC. Together, these elements provide a laboratory-scale implementation of the optical and control relationships required by the full-scale CHRONOS concept.

\subsection{Control ports and control scheme}
\label{subsec:control_ports}

The longitudinal control degrees of freedom are the input-mode-cleaner length (IMCL), power-recycling-cavity length (PRCL), signal-recycling-cavity length (SRCL), common Sagnac mode (CSAG), and differential Sagnac mode (DSAG). Dedicated sensing ports distributed through the interferometer provide the error signals needed for cavity locking, alignment control, and balanced-homodyne operation. Their physical locations are shown in Fig.~\ref{fig:layout}.

\subsubsection{Initial alignment and control}

Lock acquisition begins by establishing a stable spatial and frequency reference. Mirror positions and orientations for each cavity, together with the input steering mirrors, are adjusted to optimize beam pointing and mode matching. The IMC is the first cavity to be locked. Its reflected field is demodulated at 44.2~MHz to obtain the PDH length signal. The suspended IMC mirrors, denoted $IMC1$--$IMC4$ in Fig.~\ref{fig:layout}, are actuated with coil--magnet actuators, providing low-frequency control authority while retaining mechanical isolation.

The transmitted IMC beam provides the downstream spatial-mode and frequency reference. Its mode quality and intensity are monitored continuously, and its low-frequency component can be used for offset compensation in the laser-frequency stabilization loop~\cite{Kwee2012}. The IMC RF reference is phase-locked to the sources at $f_1=29.5~\mathrm{MHz}$ and $f_2=59.0~\mathrm{MHz}$ so that the REFL and POP demodulation channels share a coherent phase reference. Completion of this stage prepares the system for green pre-lock and subsequent acquisition of the coupled PRC, SRC, and Sagnac cavities.

\subsubsection{Control ports and optical readout scheme}

\paragraph{Reflection port (REFL)}

The $REFL_j$ port, where $j\in\{X,Y\}$, detects light reflected from the input mirror of the PRC. Demodulation at $f_1=29.5~\mathrm{MHz}$ supplies the primary PRCL error signal by the PDH technique. The same port also monitors CSAG, the average optical path length of the clockwise (CW) and counter-clockwise (CCW) fields, and contributes to overall resonance and laser-frequency stabilization.

\paragraph{Pick-off port (POP)}

The $POP_j$ port samples the intracavity PRC field~\cite{Aston2012}. Two photodiodes at distinct Gouy phases separate the relevant spatial and longitudinal responses:
\begin{itemize}
	\item \textbf{$POP_j$-A (near field):} sensitive to CSAG and to PRCL as an auxiliary channel for common-path and laser-frequency stabilization;
	\item \textbf{$POP_j$-B (far field):} displaced by a Gouy phase of approximately $45^{\circ}$--$90^{\circ}$ relative to $POP_j$-A and primarily sensitive to DSAG.
\end{itemize}
Both channels are demodulated at $f_2=59.0~\mathrm{MHz}$. Their Gouy-phase separation helps discriminate CSAG, PRCL, and DSAG.

\paragraph{Antisymmetric port (AS)}

The $AS_j$ port is the nominally dark output port of the Sagnac interferometer. Demodulation at $f_2=59.0~\mathrm{MHz}$ provides the primary SRCL error signal~\cite{Evans2015}. The $AS_X$ and $AS_Y$ fields are combined at the output beam splitter and decomposed into common and differential components. The differential component carries the science observable, is passed through the output mode cleaner, and is read out as DSAG by balanced-homodyne detection at the AS-BHD port~\cite{fricke2012,vahlbruch2010}.

\paragraph{Comparison with a Fabry--Perot Michelson interferometer}

CSAG is analogous to the common-arm-length degree of freedom (CARM) of a Michelson interferometer and represents the common optical path coupled to laser frequency and PRC resonance. DSAG is analogous to the differential-arm-length degree of freedom (DARM), but it does not represent a direct position displacement. Instead, it is an angular-momentum- or velocity-like observable arising from the difference between the CW and CCW propagation responses~\cite{Chen2003}.

\begin{table}[t]
	\centering
	\caption{Major sensing ports, modulation frequencies, and corresponding control degrees of freedom in the CHRONOS dual-recycled Sagnac interferometer. The RF assignments follow the coupled-cavity resonance conditions; AS-BHD is a DC science-readout channel.}
	\label{tab:control_ports}
	\resizebox{\textwidth}{!}{%
		\begin{tabular}{lcccl}
			\hline\hline
			Port & Frequency & Main DOF & Role & Remark \\
			\hline
			IMC-INI & 44.2 MHz & IMCL & IMC length and alignment control & Four-mirror suspended bow-tie cavity \\
			$REFL_j$ & 29.5 MHz & PRCL, CSAG & PRC and common-path control & Also used for frequency stabilization \\
			$POP_j$-A & 59.0 MHz & PRCL (aux.), CSAG & Common Sagnac-mode monitor & Near-field pick-off \\
			$POP_j$-B & 59.0 MHz & DSAG & Differential Sagnac control & Far-field pick-off \\
			$AS_j$ & 59.0 MHz & SRCL & Signal-recycling-cavity control & Output port (dark fringe) \\
			AS-BHD & DC & DSAG velocity (science) & Speed-meter readout & Phase-quadrature homodyne detection \\
			\hline
		\end{tabular}%
	}
\end{table}

\subsubsection{Lock acquisition strategy}

Lock acquisition proceeds in the following stages:
\begin{enumerate}
	\item lock the IMC and complete the initial alignment, establishing the spatial-mode, phase, and laser-frequency reference;
	\item apply the green pre-lock to establish the initial Sagnac-cavity resonance condition;
	\item acquire the coupled cavities, including PRCL and SRCL, using the REFL, POP, and AS error signals while controlling CSAG and DSAG;
	\item bring the interferometer to the dark-fringe operating point and transition the DSAG science channel to DC balanced-homodyne readout.
\end{enumerate}
Loop gains, bandwidths, and demodulation phases are adjusted during the sequence to limit cross-coupling among the strongly coupled degrees of freedom. The final operating state maintains the Sagnac and recycling-cavity resonances while providing the stable local-oscillator phase and dark-port conditions required for quantum-noise and speed-meter measurements.

%% file: Chapters/VIS.tex
\chapter{Vibration Isolation System}
\label{chap:suspension_design}

\section{Design objectives}
\label{sec:suspension_objectives}

The CHRONOS detector measures the differential yaw motion of two orthogonal
torsion-bar end test masses (ETMs) in the sub-Hz band.  Its suspension system must
therefore satisfy two requirements that are unusually difficult to combine.
First, it must provide sufficient low-frequency isolation to prevent ground
translation and rotation from contaminating the differential yaw signal.
Second, it must retain the actuation range required for alignment, damping,
interferometer lock acquisition, and recovery from large disturbances.  The
baseline design separates these functions among a room-temperature seismic
isolation system, a common multistage geometric anti-spring (GAS) chain, and
two low-temperature suspension payloads.

The present chapter defines a preliminary design suitable for a conceptual or
preliminary design review.  Numerical values labeled as requirements are
initial allocations rather than final component specifications.  They will be
updated after the full six-degree-of-freedom mechanical model and the measured
site ground-motion spectrum become available.

The reference science band and minimum-PI strain ASD from
Chap.~\ref{chap:sensitivity} are

\begin{equation}
  0.1~\mathrm{Hz} \leq f \leq 10~\mathrm{Hz},
  \qquad
  S_h^{1/2}(2~\mathrm{Hz}) = 8.31\times10^{-19}/\sqrt{\mathrm{Hz}},
\end{equation}

with the most demanding isolation performance expected between approximately
0.1 and 3 Hz.  Below 0.1 Hz, displacement range, alignment control, and robust
recovery are given priority over aggressive inertial isolation.

\section{Baseline vibration-isolation architecture}
\label{sec:suspension_architecture}

\begin{figure}[t]
  \centering
  \includegraphics[width=0.82\linewidth]{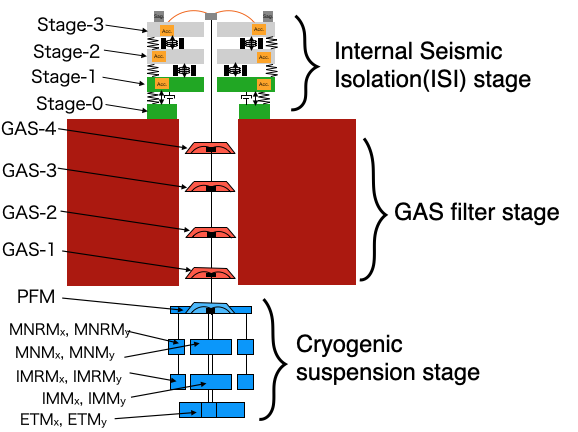}
  \caption{Conceptual architecture and component nomenclature of the CHRONOS
  vibration-isolation and cryogenic suspension system.  The room-temperature
  ISI assembly comprises the coarse-control Stage-0 and the successively
  isolated Stages 1--3.  The yaw Sagnac is mounted on the precision optical
  platform, Stage-3, while Stage-0 supports GAS-4.
  The common GAS-4--GAS-1 chain provides multistage passive isolation and
  supports the cryogenic PFM.  Below the PFM, the suspension divides into
  $x$ and $y$ main chains (MNIMx/MNIMy, IMMx/IMMy, and ETMx/ETMy) and corresponding
  recoil chains (MNRMx/MNRMy and IMRMx/IMRMy).}
  \label{fig:vibration_isolation_concept}
\end{figure}

Figure~\ref{fig:vibration_isolation_concept} defines the physical hierarchy and
nomenclature used throughout this chapter.  Ground motion first enters the
four-stage internal seismic isolation (ISI) assembly.  Stage-0 is the
ground-side, large-range platform and provides coarse position and attitude
control.  It also forms the mechanical suspension head for GAS-4.  Stages 1
and 2 provide successive isolation between Stage-0 and the precision optical
platform, Stage-3.  The yaw-sensitive Sagnac optical system is mounted on
Stage-3 so that its optical reference is stabilized by the most strongly
isolated ISI stage.  Independently of this optical branch, Stage-0 supports
four common geometric anti-spring filters, ordered from top to bottom as
GAS-4, GAS-3, GAS-2, and GAS-1.  GAS-1 supports the common cryogenic platform
mass (PFM).  At the PFM, the load path divides into nominally identical $x$
and $y$ cryogenic payloads.

The baseline architecture is therefore

\begin{equation}
\begin{gathered}
  \text{ground}\rightarrow\text{Stage-0 coarse platform}
  \rightarrow
  \begin{cases}
    \text{Stage-1}\rightarrow\text{Stage-2}\rightarrow
      \text{Stage-3 + Sagnac},\\
    \text{GAS-4}\rightarrow\text{GAS-3}
  \rightarrow\text{GAS-2}\rightarrow\text{GAS-1}\\
  \rightarrow\text{common cryogenic PFM}\\
  \rightarrow
  \left\{
  \begin{aligned}
    \mathrm{MNIM}_{x}\rightarrow\mathrm{IMM}_{x}\rightarrow\mathrm{ETM}_{x},\\
    \mathrm{MNRM}_{x}\rightarrow\mathrm{IMRM}_{x},\\
    \mathrm{MNIM}_{y}\rightarrow\mathrm{IMM}_{y}\rightarrow\mathrm{ETM}_{y},\\
    \mathrm{MNRM}_{y}\rightarrow\mathrm{IMRM}_{y}.
  \end{aligned}
  \right.
  \end{cases}
\end{gathered}
\label{eq:baseline_architecture}
\end{equation}

Here MNIM and IMM denote the marionette intermediate mass and intermediate
main mass; MNRM and IMRM denote the marionette recoil mass and intermediate
recoil mass; and ETM denotes the end test mass, which is the suspended
torsion bar.  Lower-case subscripts $x$ and $y$ identify the two orthogonal
payloads and correspond to the labels MNIMx/MNIMy, IMMx/IMMy, ETMx/ETMy,
MNRMx/MNRMy, and IMRMx/IMRMy in Fig.~\ref{fig:vibration_isolation_concept}.  The
main and recoil chains are mechanically independent below the PFM but share
the four-stage ISI, GAS chain, and PFM above their branch point.  The PFM is
the sole common mechanical support and the unique $x/y$ branching point.  Below
it, the $x$ and $y$ systems are two separate suspensions: there is no mechanical
connection, common suspension element, cross-brace, or lateral link between any
corresponding $x$- and $y$-side component, including MNRMx--MNRMy,
MNIMx--MNIMy, IMRMx--IMRMy, IMMx--IMMy, and ETMx--ETMy.

Two independent GAS towers would provide a direct mechanical separation of
the $x$ and $y$ payloads.  However, such a design would require two heavily loaded
filter chains with closely matched resonances, substantially increase the
available-space and installation requirements, and complicate load tuning.
CHRONOS therefore adopts one common GAS chain and places the X--Y branching
point at the cryogenic PFM.  A common platform motion ideally cancels in the
differential readout,

\begin{equation}
  (\theta_x-\theta_{\rm PFM})-(\theta_y-\theta_{\rm PFM})
  =\theta_x-\theta_y.
  \label{eq:common_pf_cancellation}
\end{equation}

In practice, imperfect symmetry causes residual common-to-differential
coupling.  The $x$ and $y$ suspension parameters and attachment geometry must
therefore be made as symmetric as practical.  The PFM motion and both
PFM-to-ETM transfer functions will be measured, and the remaining common
component will be removed through sensing and actuation diagonalization.

\section{Four-stage internal seismic isolation system}
\label{sec:isi_structure}

The ISI is the room-temperature active-isolation subsystem above GAS-4.  Its
sensor--actuator topology follows principles demonstrated by the Advanced LIGO
BSC isolation system
\cite{Matichard2015AdvancedLIGOSeismic}, but the four-stage
stack and its numbering are CHRONOS-specific.  Each controlled platform is
treated as a rigid body with coordinates $x$, $y$, $z$, pitch, roll, and yaw.
Relative sensors retain position and alignment at low frequency, inertial
sensors suppress motion in and above the blend region, and actuators apply the
forces and torques required for damping and isolation.

The four ISI components are defined as follows:

\begin{description}
  \item[Stage-0] The ground-side, large-range ISI platform and suspension head.
        It provides coarse translation and attitude control, supports GAS-4,
        and carries the combined static load of the GAS chain, PFM, both
        cryogenic payloads, recoil chains, cabling, and thermal hardware.
  \item[Stage-1] The first precision-isolation platform above Stage-0.  It
        reduces residual Stage-0 translation and rotation and provides the
        controlled mechanical reference for Stage-2.
  \item[Stage-2] The second precision-isolation platform.  It further reduces
        motion transmitted from Stage-1 and provides the controlled reference
        for the optical platform.
  \item[Stage-3] The final precision optical platform.  It supports the
        yaw-sensitive Sagnac optical system and is controlled to minimize the
        translation and, especially, yaw motion of that optical reference.
\end{description}

Adjacent platforms are connected by compliant, frictionless suspension
elements and instrumented with collocated relative sensors and actuators where
practical.  Their rigid-body resonances are actively damped.  These ISI elastic
elements are functionally distinct from the softer GAS filters supported by
Stage-0: the ISI controls six-degree-of-freedom platform motion, whereas the GAS chain
provides the main cascaded passive vertical attenuation.

GAS-4 is suspended from the center region of Stage-0.  The suspension head
must distribute this concentrated load into the Stage-0 structure without
excessive static distortion or excitation of internal modes.  The load path
and actuator layout must remain sufficiently symmetric that they do not create
a preferred pitch, roll, or yaw direction.  Mechanical stops at every moving
ISI interface protect the long suspension during installation, earthquakes,
and loss of control.

The CHRONOS-specific ISI provisions are therefore a large-range Stage-0 for
coarse control and suspension support; a Stage-1--Stage-3 precision branch for
the optical system; a yaw Sagnac mounted directly on Stage-3; adequate yaw
inertia and actuator lever arms to limit response to force imbalance; and
loaded structural modes well above the isolation-control band.

\subsection{Stage-3 yaw Sagnac}

The Sagnac interferometer is the dedicated low-frequency yaw sensor for the
precision optical system.  Its optical components and beam geometry are fixed
to Stage-3, so the measured rotation is assigned to the Stage-3 optical
reference rather than to Stage-0, the GAS chain, or the PFM.  The Sagnac signal
is used in the Stage-3 yaw loop, with the lower ISI stages providing the
successive mechanical pre-isolation needed to retain Stage-3 actuator range.
Stage-0 yaw control remains deliberately coarse: it keeps the GAS suspension
head within its alignment and travel envelope but is not the science-quality
optical yaw reference.

\section{Common GAS filter chain}
\label{sec:gas_chain}

The passive vertical-isolation chain shown in
Fig.~\ref{fig:vibration_isolation_concept} consists of four room-temperature
stages named GAS-4, GAS-3, GAS-2, and GAS-1 from top to bottom.  This ordering
is used for all drawings, models, channels, and requirement records.  The exact
distribution along the approximately 15-m suspension length will be determined
by the vacuum-vessel and cryostat layout.  The four filters share a single
central load path and support the full mass below the ISI.  Multiple symmetric
blade packs may be used at each stage to obtain the required load capacity
without constructing two separately tuned towers.

Each GAS stage is tuned to a nominal vertical resonance of
0.15--0.25 Hz.  Local vertical displacement sensors and voice-coil actuators
damp the high-$Q$ filter resonances and provide slow height control.  The local
control gain is concentrated near the filter modes and reduced rapidly above
them, limiting injection of displacement-sensor noise into the science band.
Temperature sensors on the blade assemblies monitor slow changes in stiffness
and equilibrium height.

GAS-1 interfaces the room-temperature isolation chain to the cryogenic
payload.  Heat links, radiation shields, and thermal intercepts near 50 K and
4--10 K are introduced between GAS-1 and the PFM without forming a mechanically
stiff bypass around the suspension.  The PFM and the payloads below it operate
at a target temperature of approximately 10 K.  Thermal-link stiffness and
vibration injection must be included explicitly in the final coupled model.

\section{Cryogenic payload and recoil chains}
\label{sec:cryogenic_payload}

Below the common PFM, the suspension separates into two nominally identical
payloads for the orthogonal $x$ and $y$ torsion bars.  The two payloads are
mechanically independent after their separate attachment points on the PFM.
Each main chain contains a marionette intermediate mass, an intermediate main
mass, and an end test mass.  The two distinct main chains are

\begin{equation}
  \mathrm{PFM}\rightarrow\mathrm{MNIM}_{x}
  \rightarrow\mathrm{IMM}_{x}\rightarrow\mathrm{ETM}_{x},
  \qquad
  \mathrm{PFM}\rightarrow\mathrm{MNIM}_{y}
  \rightarrow\mathrm{IMM}_{y}\rightarrow\mathrm{ETM}_{y}.
\end{equation}

Their corresponding recoil chains are separately suspended from the PFM,

\begin{equation}
  \mathrm{PFM}\rightarrow\mathrm{MNRM}_{x}
  \rightarrow\mathrm{IMRM}_{x},
  \qquad
  \mathrm{PFM}\rightarrow\mathrm{MNRM}_{y}
  \rightarrow\mathrm{IMRM}_{y}.
\end{equation}

This organization follows the demonstrated cryogenic-payload principle of
suspending the main and recoil chains separately from a common platform and
placing relative sensors and actuators on the recoil masses
\cite{Ushiba2021CryogenicSuspension}.  CHRONOS retains that principle but uses
two orthogonal main chains below one PFM and, in the baseline design, omits a
recoil mass at each ETM.  The components in
Fig.~\ref{fig:vibration_isolation_concept} are defined as follows:

\begin{description}
  \item[PFM] The platform mass at the cryogenic boundary and the common
        suspension point for both payloads.  It receives the GAS-1 load,
        distributes it symmetrically to the $x$ and $y$ branches, supports
        cryogenic interfaces, and provides a monitored six-degree-of-freedom
        reference.  It is not treated as an inertial ground.
  \item[MNIMx and MNIMy] The marionette intermediate masses of the $x$ and $y$
        main chains.  Each
        carries the stages below it and provides large-range alignment, modal
        damping, low-frequency control, and DC offloading without direct
        actuation of the associated ETM.
  \item[MNRMx and MNRMy] The marionette recoil masses.  Each is suspended from
        the PFM independently of its MNIM and carries the reaction-side sensors
        and actuators for the corresponding MNIM.  It may also receive compliant
        thermal links, provided that their stiffness and injected vibration are
        included in the coupled model.
  \item[IMMx and IMMy] The intermediate main masses in the two main chains.  Each
        supports its ETM, provides an additional pendulum-isolation stage, and
        supplies fine alignment and damping authority for lower suspension
        modes.
  \item[IMRMx and IMRMy] The intermediate recoil masses.  They are the local
        reaction bodies for IMM sensing and actuation and are suspended through
        the MNRM branch rather than through the corresponding main chain.
  \item[ETMx and ETMy] The orthogonal torsion-bar end test masses and the
        science inertial elements.  Their differential yaw coordinate
        $\Delta\theta=\theta_x-\theta_y$ is the primary readout.  No ETM recoil
        masses are included in the baseline architecture.
\end{description}

Relative sensors and actuators are placed between MNIMx/MNIMy and MNRMx/MNRMy and
between IMMx/IMMy and IMRMx/IMRMy.  The reaction force is thereby closed mainly
inside the suspended payload rather than directly against the PFM or the
ground.  The MNIM stage
provides large-range alignment, low-frequency control, damping, and DC
offloading.  The IMM stage provides finer alignment and damping of the lower
suspension modes.

No recoil mass is placed around either ETM.  This choice preserves the open optical
and gravitational environment of the torsion bar and avoids an additional
nearby suspended structure that could introduce scattering, electrostatic
coupling, or Newtonian gravitational asymmetry.  The absence of an ETM recoil
mass requires separate actuators for lock acquisition and low-noise science
operation, as described in Sec.~\ref{sec:tm_actuation}.

Cryogenic design must treat cooling hardware as part of the mechanical plant.
The KAGRA payload demonstrates why high-conductivity heat links must also be
mechanically compliant, why thermal contraction requires adequate sensor and
alignment range, and why actuator coefficients must be verified at operating
temperature \cite{Ushiba2021CryogenicSuspension}.  For CHRONOS, heat-link
routing is therefore preferentially terminated on recoil structures, and the
stiffness, geometry, and vibration of every thermal link are included in the
PFM-to-ETM transfer functions rather than regarded as ideal thermal-only
connections.

\section{End-test-mass actuation and operating states}
\label{sec:tm_actuation}

The ETM actuation system is divided according to function:

\begin{itemize}
  \item MNIM--MNRM actuation provides alignment, damping, and DC offloading;
  \item IMM--IMRM actuation damps the lower suspension modes and provides fine
        alignment;
  \item a floor-referenced electrostatic actuator (ESA) supplies the large
        torque required during lock acquisition and recovery; and
  \item a photon-pressure actuator (PPA) supplies low-noise torque during
        science operation.
\end{itemize}

For a general electrode geometry the electrostatic force is

\begin{equation}
  F_{\rm ESA}=\frac{1}{2}V^2\frac{\mathrm{d}C}{\mathrm{d}x},
  \label{eq:esa_general}
\end{equation}

and a parallel-plate estimate gives

\begin{equation}
  F_{\rm ESA}\simeq\frac{\epsilon_0 A V^2}{2d^2},
  \qquad
  \tau_{\rm ESA}=r(F_{\rm L}-F_{\rm R}),
  \label{eq:esa_force}
\end{equation}

where $A$ is the effective electrode area, $d$ the gap, $r$ the torque arm,
and $F_{\rm L,R}$ the forces applied on the two sides.  Voltages of several
hundred volts to approximately 1 kV and millimeter-scale gaps can provide a
substantially larger acquisition torque than the PPA.

The optical coating does not prevent electrostatic actuation.  Even when the
ETM substrate and multilayer coating are insulating, the electric field
polarizes the dielectric and changes the capacitance in
Eq.~\eqref{eq:esa_general}.  Electrostatic drives have therefore been used on
dielectric gravitational-wave test masses, and side-mounted concepts have
also been developed for precision interferometers
\cite{Buikema2020AdvancedLIGO,Carbone2015SideESD}.  No conductive pad on the
optical surface is required.  The preferred CHRONOS geometry places electrodes
near a side or rear surface, avoids the high-reflectivity optical area, and
uses a differential electrode pattern to generate yaw torque.  The actuation
coefficient must be measured at the operating temperature because the
permittivity, surface resistance, charge relaxation, and patch-potential
behavior may change during cooldown.

Because the ESA is referenced to the ground, a nonzero bias produces an
electrostatic stiffness of approximate magnitude

\begin{equation}
  k_{\rm ESA}\sim\frac{\epsilon_0 A V^2}{d^3},
  \label{eq:esa_stiffness}
\end{equation}

which can couple electrode or ground motion to the ETM.  Surface charging can
also change the actuation coefficient and introduce excess force noise.
Consequently, the ESA is not used continuously in the nominal science state.
The operational handover is

\begin{enumerate}
  \item acquire the interferometer using the ESA together with MNIM/IMM control;
  \item close the PPA loop after the optical readout is linear and stable;
  \item cross-fade the commanded torque continuously from the ESA to the PPA;
  \item ramp the ESA voltage to zero;
  \item connect the electrodes to a defined low-noise ground; and
  \item remove the ESA command path from the science controller.
\end{enumerate}

The ESA remains available for automatic lock recovery, while the PPA provides
the ETM torque in science mode.  This division preserves acquisition authority
without retaining a floor-referenced electric spring during observation.

\section{Sensors and control topology}
\label{sec:isi_sensors}

The ISI uses relative sensors to control position at very low frequencies and
inertial sensors to suppress absolute motion through the observation band.
Following the active-isolation method reviewed by Matichard et al., calibrated
relative and inertial channels are combined with complementary filters; ground
instruments may additionally supply sensor correction and coherent
feedforward \cite{Matichard2015AdvancedLIGOSeismic}.  CHRONOS applies this
method recursively from Stage-0 through Stage-3 and adds a dedicated
low-frequency Sagnac yaw measurement on Stage-3.

\begin{table}[t]
  \centering
  \caption{Baseline ISI sensor suite.  ``Six-axis'' denotes six appropriately
  oriented single-axis channels used to reconstruct the six rigid-body degrees
  of freedom.}
  \label{tab:isi_sensors}
  \begin{tabular}{p{0.20\linewidth}p{0.47\linewidth}p{0.18\linewidth}}
    \hline
    Location & Sensor & Baseline quantity \\
    \hline
    Facility / Stage-0 & Broadband triaxial seismometer & 3 units \\
    Facility--Stage-0 & Long-range relative position sensors & six-axis \\
    Stage-0 & Inertial sensors for coarse control & six-axis minimum \\
    Stage-0--Stage-1 & Capacitive position sensor (CPS) & six-axis \\
    Stage-1 & Broadband and short-period inertial sensors & six-axis minimum \\
    Stage-1--Stage-2 & Capacitive position sensor (CPS) & six-axis \\
    Stage-2 & Broadband and short-period inertial sensors & six-axis minimum \\
    Stage-2--Stage-3 & Capacitive position sensor (CPS) & six-axis \\
    Stage-3 & Low-noise inertial sensors & six-axis minimum \\
    Stage-3 & Dedicated yaw Sagnac & 1 optical system \\
    All stages & Temperature sensors & at least 6 points per ISI \\
    \hline
  \end{tabular}
\end{table}

At each controlled interface, the relative and inertial signals are blended
according to frequency.  Stage-0 uses facility-relative sensors for coarse
position, attitude, and drift control.  Stages 1--3 progressively suppress the
motion of the precision optical reference.  Ground sensors provide sensor
correction and feedforward, as well as earthquake monitoring.  The Stage-3
Sagnac covers approximately
0.03--0.3 Hz, where yaw reconstructed from translational inertial sensors is
limited by tilt coupling and sensor self-noise.  Above this range, yaw is
reconstructed from the tangential response of the Stage-3 inertial-sensor
array.

Three horizontal and three vertical sensing directions are distributed at
120-degree intervals with a target radius of at least 1 m.  If three
tangential sensors have identical displacement noise $S_{x,n}^{1/2}$, the
equivalent yaw noise is approximately

\begin{equation}
  S_{\theta,n}^{1/2}\simeq
  \frac{S_{x,n}^{1/2}}{R\sqrt{3}}.
  \label{eq:yaw_array_noise}
\end{equation}

Sensors and actuators are collocated and coaligned wherever practical to
reduce non-minimum-phase couplings.  The geometry matrix is measured after the
full payload has been installed.  Frequency-dependent diagonalization is used
if a constant rigid-body matrix is insufficient near structural modes.

Each GAS stage has three vertical displacement sensors arranged at 120-degree
intervals, allowing common vertical motion to be separated from tilt.  GAS-1
also carries horizontal relative sensors, and the PFM is monitored in all six
rigid-body degrees of freedom.  All local GAS loops are shaped so that their
sensor noise contribution decreases above the filter resonances.

\section{Preliminary performance requirements}
\label{sec:suspension_requirements}

\subsection{Top-level angular-noise allocation}

For the differential CHRONOS coordinate, the reference response is

\begin{equation}
  \Delta\theta_{\rm sig}\simeq\eta h,
  \qquad \eta=0.93656.
  \label{eq:strain_to_angle}
\end{equation}

Allocating at most 10\% of the target angular amplitude noise to the combined
ISI and GAS system gives

\begin{equation}
  S_{\Delta\theta,{\rm ISI+GAS}}^{1/2}(f)
  \leq 0.1\eta S_h^{1/2}(f).
  \label{eq:isi_angle_allocation}
\end{equation}

At 2 Hz this corresponds to

\begin{equation}
  S_{\Delta\theta,{\rm ISI+GAS}}^{1/2}(2~\mathrm{Hz})
  \leq 7.8\times10^{-20}~\mathrm{rad}/\sqrt{\mathrm{Hz}}.
  \label{eq:isi_angle_2hz}
\end{equation}

Equation~\eqref{eq:isi_angle_2hz} is a requirement on the residual differential
ETM yaw after the complete suspension, not on Stage-0 yaw itself.  Its
feasibility depends critically on passive attenuation and on the equality of
the two PFM-to-ETM transfer functions.

For a Stage-0 output degree of freedom
$q\in\{x,y,z,R_x,R_y,R_z\}$, the correct displacement or angular requirement is

\begin{equation}
  S_{q,{\rm S0}}^{1/2}(f)
  \leq
  \frac{0.1\eta S_h^{1/2}(f)}
  {\left|T_{q\rightarrow\theta_x}(f)
       -T_{q\rightarrow\theta_y}(f)\right|}.
  \label{eq:stage2_flowdown}
\end{equation}

The denominator is the measured or modeled common-to-differential coupling.
This equation, rather than a generic isolation number, will be used to freeze
the final Stage-0 output specification.

\subsection{Provisional Stage-0 output-motion targets}

Before the coupled transfer functions are available, the values in
Table~\ref{tab:stage2_motion} provide working targets for mechanical design and
sensor selection.  Rotations are expressed as an equivalent displacement
$R_{\rm eff}\theta$ with $R_{\rm eff}=1$ m.

\begin{table}[t]
  \centering
  \caption{Provisional Stage-0 absolute-motion targets.  These values are to be
  replaced by the flowdown in Eq.~\eqref{eq:stage2_flowdown}.}
  \label{tab:stage2_motion}
  \begin{tabular}{ccc}
    \hline
    Frequency & Translation $x,y,z$ & $R_{\rm eff}\theta$ \\
    & $[\mathrm{m}/\sqrt{\mathrm{Hz}}]$ & $[\mathrm{m}/\sqrt{\mathrm{Hz}}]$ \\
    \hline
    0.1 Hz & $1\times10^{-7}$ & $1\times10^{-7}$ \\
    0.2 Hz & $3\times10^{-8}$ & $3\times10^{-8}$ \\
    1 Hz   & $1\times10^{-11}$ & $1\times10^{-11}$ \\
    2 Hz   & $5\times10^{-12}$ & $5\times10^{-12}$ \\
    3 Hz   & $3\times10^{-12}$ & $3\times10^{-12}$ \\
    10 Hz  & $1\times10^{-12}$ & $1\times10^{-12}$ \\
    \hline
  \end{tabular}
\end{table}

The corresponding provisional ground-to-Stage-0 transmissibility limits are

\begin{table}[t]
  \centering
  \caption{Provisional ground-to-Stage-0 transmissibility limits.}
  \label{tab:isi_transmissibility}
  \begin{tabular}{cc}
    \hline
    Frequency & $|T_{{\rm ground}\rightarrow{\rm S0}}|$ \\
    \hline
    0.1 Hz & $\leq0.3$ \\
    0.2 Hz & $\leq0.03$ \\
    1 Hz   & $\leq0.01$ \\
    2 Hz   & $\leq3\times10^{-3}$ \\
    3 Hz   & $\leq1\times10^{-3}$ \\
    10 Hz  & $\leq1\times10^{-4}$ \\
    \hline
  \end{tabular}
\end{table}

Compliance is determined by the more restrictive of the absolute-motion and
transmissibility limits.  Site-specific ground translation, tilt, and rotation
spectra must be used in the final evaluation; the quiet underground
environment cannot be represented adequately by a single translational
spectrum \cite{Akutsu2018KAGRAUnderground}.

\subsection{Mechanical and control requirements}

The preliminary structural and control requirements are summarized in
Table~\ref{tab:isi_mechanical_requirements}.  The 2000-kg nominal load is an
initial mass allocation for the common GAS chain, PFM, both main and recoil
payloads, cryogenic hardware, cables, and installation margin.  It will be
replaced after a controlled mass and inertia budget is completed.

\begin{table}[t]
  \centering
  \caption{Preliminary ISI mechanical and control requirements.}
  \label{tab:isi_mechanical_requirements}
  \begin{tabular}{p{0.55\linewidth}p{0.30\linewidth}}
    \hline
    Parameter & Preliminary requirement \\
    \hline
    Nominal suspended load & 2000 kg \\
    Rated suspended load & 2500 kg \\
    Mechanical connections between $x/y$ branches below PFM & none \\
    Horizontal center-of-mass offset & $\leq\pm50$ mm \\
    Stage-0 coarse translational range & $\geq\pm5$ mm \\
    Stage-0 coarse rotational range & $\geq\pm2$ mrad \\
    Stage-1 and Stage-2 translational range & $\geq\pm1$ mm \\
    Stage-3 translational range & $\geq\pm0.5$ mm \\
    Stage-1 and Stage-2 rotational range & $\geq\pm1$ mrad \\
    Stage-3 rotational range & $\geq\pm0.5$ mrad \\
    Lowest loaded structural mode & $>50$ Hz; goal $>100$ Hz \\
    Effective controlled rigid-body-mode $Q$ & $\leq5$ \\
    Yield safety factor & $\geq2$ \\
    Ultimate safety factor & $\geq3$ \\
    Nominal rigid-body modes & 0.8--2 Hz and 4--8 Hz groups \\
    Normal control band & approximately 0.05--30 Hz \\
    Low-frequency yaw control & available from 0.03 Hz \\
    Unity-gain-frequency design range & 20--40 Hz \\
    Phase margin & $\geq35$ deg; goal 45 deg \\
    Gain margin & $\geq6$ dB \\
    Steady-state DAC use & $\leq30\%$ of full scale \\
    Actuator authority margin & $\geq3$ in nominal operation \\
    Calibrated inter-DOF coupling & $\leq10^{-3}$ in 0.1--3 Hz \\
    \hline
  \end{tabular}
\end{table}

The control-band values are initial design ranges rather than a requirement to
force every loop to the same unity-gain frequency.  Final loop shapes must
respect the measured plant, flexible modes, time delay, sensor blending, and
actuator-noise injection.

\subsection{GAS requirements}

\begin{table}[t]
  \centering
  \caption{Preliminary requirements for the common GAS chain.}
  \label{tab:gas_requirements}
  \begin{tabular}{p{0.55\linewidth}p{0.30\linewidth}}
    \hline
    Parameter & Preliminary requirement \\
    \hline
    Configuration & GAS-4--GAS-1, ordered top to bottom \\
    Installation temperature & room temperature \\
    Nominal vertical resonance & 0.15--0.25 Hz per stage \\
    Normal vertical operating range & $\geq\pm2$ mm \\
    Initial load adjustment & $\geq\pm10\%$ of nominal load \\
    Static $x/y$ branch-load difference & $\leq1\%$ \\
    Effective $x/y$ transfer-function amplitude difference & $\leq1\%$ \\
    Effective $x/y$ transfer-function phase difference & $\leq10$ mrad \\
    Local vertical displacement-sensor noise & $\leq3\times10^{-10}$ m/$\sqrt{\mathrm{Hz}}$ at 0.1--1 Hz \\
    \hline
  \end{tabular}
\end{table}

The transfer-function-matching entries apply to the effective paths from the
common PFM to the $x$ and $y$ ETM yaw coordinates, not merely to the bare GAS
filters.  If the passive equality cannot meet these values, measured
frequency-dependent subtraction must provide the remaining rejection.

\section{Sensor requirements}
\label{sec:sensor_requirements}

The inertial-sensor self-noise should remain below the allowed platform motion,
with a design goal approximately one third of the corresponding motion target.
The preliminary Stage-0 output allocations are listed in
Table~\ref{tab:stage2_sensor_noise}.

\begin{table}[t]
  \centering
  \caption{Preliminary equivalent-displacement self-noise requirement and goal
  for one Stage-0 inertial sensing channel.}
  \label{tab:stage2_sensor_noise}
  \begin{tabular}{ccc}
    \hline
    Frequency & Requirement & Goal \\
    & $[\mathrm{m}/\sqrt{\mathrm{Hz}}]$ & $[\mathrm{m}/\sqrt{\mathrm{Hz}}]$ \\
    \hline
    0.1 Hz & $1\times10^{-8}$ & $3\times10^{-9}$ \\
    0.2 Hz & $3\times10^{-9}$ & $1\times10^{-9}$ \\
    1 Hz   & $1\times10^{-11}$ & $3\times10^{-12}$ \\
    2 Hz   & $5\times10^{-12}$ & $2\times10^{-12}$ \\
    3 Hz   & $3\times10^{-12}$ & $1\times10^{-12}$ \\
    10 Hz  & $1\times10^{-12}$ & $3\times10^{-13}$ \\
    \hline
  \end{tabular}
\end{table}

Using

\begin{equation}
  S_a^{1/2}(f)=(2\pi f)^2S_x^{1/2}(f),
\end{equation}

the 1-Hz requirement corresponds to approximately

\begin{equation}
  S_a^{1/2}(1~\mathrm{Hz})\leq
  4\times10^{-10}~\mathrm{m\,s^{-2}}/\sqrt{\mathrm{Hz}},
\end{equation}

with a goal of $1.2\times10^{-10}~\mathrm{m\,s^{-2}}/\sqrt{\mathrm{Hz}}$.
The final criterion is the sensor-noise contribution at the differential ETM
yaw coordinate after the complete closed-loop transfer function is applied.

The Stage-3 yaw Sagnac has the provisional self-noise targets

\begin{equation}
\begin{array}{c|ccc}
f & 0.03~\mathrm{Hz} & 0.1~\mathrm{Hz} & 0.3~\mathrm{Hz} \\
\hline
S_{\theta,n}^{1/2}
& 1\times10^{-7} & 1\times10^{-9} & 1\times10^{-10}
\end{array}
\quad \mathrm{rad}/\sqrt{\mathrm{Hz}}.
\label{eq:yaw_sensor_targets}
\end{equation}

The CPS channels are used mainly below the inertial-sensor blend and for range
monitoring.  Their preliminary requirements are a self-noise below
$1\times10^{-10}$ m/$\sqrt{\mathrm{Hz}}$, a goal of
$3\times10^{-11}$ m/$\sqrt{\mathrm{Hz}}$, a range of at least $\pm2$ mm at
Stages~2 and 1 and $\pm1$ mm at Stage~0, full-scale nonlinearity below 0.1\%, and
cross-axis sensitivity below $10^{-3}$.  A blend in the approximate range
0.03--0.08 Hz transfers control authority from relative to inertial sensing;
the final blend is optimized against tilt and self-noise.

For the complete sensing and data-acquisition chain, the initial alignment and
calibration requirements are

\begin{itemize}
  \item sensor location error below $\pm0.5$ mm;
  \item sensor-axis error below $\pm0.1$ mrad;
  \item calibrated channel-to-channel gain mismatch below $10^{-3}$;
  \item channel-to-channel phase mismatch below 1 mrad from 0.1 to 3 Hz;
  \item ADC and analog-electronics noise below 30\% of the sensor self-noise;
        and
  \item channel synchronization error below 10 $\mu$s.
\end{itemize}

These values are most important for the reconstructed yaw channels.  They may
be relaxed for auxiliary channels if a full noise projection demonstrates
adequate margin.

\section{Verification and requirement maturation}
\label{sec:suspension_verification}

The numerical allocations above define a starting point, but the Stage-0
motion table cannot be regarded as final until common-to-differential coupling
is quantified.  Requirement maturation will proceed in the following order:

\begin{enumerate}
  \item establish the mass, center-of-mass, and inertia budget for every ISI,
        GAS, PFM, main-chain, recoil-chain, cryogenic, and cabling
        component;
  \item construct a coupled six-degree-of-freedom model of ISI Stage-3 through
        Stage-0, the GAS-4--GAS-1 chain, PFM, and both $x/y$ payloads;
  \item include thermal-link stiffness, cable stiffness, sensor geometry,
        actuator geometry, and realistic damping;
  \item verify by drawing review and as-built inspection that all mechanical,
        thermal, and cable services below the PFM remain branch-specific and do
        not form an unintended $x/y$ mechanical bridge;
  \item calculate
        $T_{q\rightarrow\theta_x}$ and
        $T_{q\rightarrow\theta_y}$ for all six Stage-0 coordinates;
  \item inject measured ground translation, ground rotation, sensor noise,
        actuator noise, and control noise into the closed-loop model;
  \item replace the provisional Stage-0 values using
        Eq.~\eqref{eq:stage2_flowdown}; and
  \item verify that the quadrature sum of ISI and GAS contributions satisfies
        Eq.~\eqref{eq:isi_angle_2hz}, with margin, across the observation band.
\end{enumerate}

Subsystem tests will measure actuator coefficients, sensor self-noise,
cross-axis coupling, loaded structural modes, GAS tuning range, and local-loop
noise injection.  After assembly, the $x$ and $y$ transfer functions will be
measured separately and used to construct the differential diagonalization
filters.  Lock-acquisition tests will then demonstrate continuous ESA-to-PPA
handover and confirm that the grounded ESA does not produce a measurable
science-band coupling.

The design choices that can be retained at the present stage are therefore:
one common four-stage ISI, one common GAS-4--GAS-1 chain, one common
cryogenic PFM, two independent $x/y$ cryogenic payloads
below the PFM, recoil masses at MNIM and IMM, no recoil mass at either ETM, ESA
actuation for acquisition,
and photon-pressure actuation for science operation.  Component-level motion
and sensor-noise values remain controlled preliminary requirements until the
full differential-yaw noise budget is completed.

%% file: Chapters/Substrate.tex
\chapter{Test Mass Mirror}
\label{chap:test_mass_mirror}

\section{Overview of Test Mass}
\label{sec:test_mass_overview}

The test-mass mirror is one of the most important components determining the fundamental performance of CHRONOS.
CHRONOS is a cryogenic torsion-bar gravitational-wave detector designed for precision measurements in the sub-Hz frequency band at a temperature of approximately $4~\mathrm{K}$.
Its test masses therefore play roles that are substantially different
from those of conventional kilometer-scale Fabry--P\'erot gravitational-wave interferometers.

The CHRONOS test mass acts as
\begin{itemize}
\item a high-quality optical mirror,
\item a low-loss cryogenic mechanical oscillator,
\item the inertial body defining the torsional response of the detector,
\item an interface between the optical system and the cryogenic suspension.
\end{itemize}
Consequently, the choice of test-mass material and geometry should be determined by mechanical, optical, thermal, and manufacturing requirements.

The baseline CHRONOS design employs a laser wavelength of
\begin{equation}
\lambda = 1064~\mathrm{\mathrm{nm}},
\end{equation}
together with monocrystalline sapphire test masses operated at temperature of approximately
\begin{equation}
T \simeq 4~\mathrm{K}.
\end{equation}
The 1064~$\mathrm{nm}$ wavelength is selected because mature Nd:YAG laser technology and a wide range of low-noise optical components are available. It also maintains compatibility with optical technologies already developed for precision gravitational-wave interferometers.

At this wavelength, sapphire is the baseline substrate material.
Sapphire provides high mechanical quality factor at cryogenic temperature, high thermal conductivity, good mechanical strength, and optical transparency at 1064~$\mathrm{nm}$.

An additional characteristic of the CHRONOS test mass is its large
mechanical dimension. Because the detector measures the rotational motion of a torsion-bar system, the mass distribution and moment of inertia of the test mass constitute central mechanical design parameters. The required dimensions may exceed those practically achievable by a single high-quality sapphire crystal.

CHRONOS therefore considers a modular mass-extension scheme in which
multiple sapphire components are connected using hydroxide-catalysis
bonding (HCB). This technique allows the total torsion-bar length and moment of inertia to be extended beyond the dimensions of a single available crystal while maintaining a rigid all-sapphire mechanical structure. Consequently, the baseline design philosophy is to make the largest monolithic sapphire blocks with minimum necessary HCB interfaces.

This chapter describes the optical, mechanical, and thermal requirements of the CHRONOS test mass and the engineering strategy for constructing the large-scale cryogenic torsion-bar mirror.

\section{Test-Mass Requirements}
\label{sec:test_mass_requirement}

From the mechanical point of view, the test mass must provide
\begin{itemize}
\item large mass,
\item large moment of inertia,
\item adequate structural strength,
\item compatibility with the cryogenic suspension system.
\end{itemize}

As discussed in the later sections, the final torsion-bar dimensions may exceed those available from a single monolithic sapphire crystal. Therefore, the design must allow the use of HCB for mass extension without introducing excessive mechanical dissipation. The substrate must retain 
\begin{itemize}
\item a high mechanical quality factor $Q$ at cryogenic temperature
\end{itemize}
because mechanical losses in the substrate, bonded interfaces, and attached structures directly contribute to thermal noise. The dimensions of the sapphire components must therefore be chosen not only to realize the required total mass and moment of inertia, but also to preserve sufficiently low mechanical loss in the complete assembled structure.

The optical requirements include
\begin{itemize}
\item low absorption and scattering,
\item high refractive-index homogeneity,
\item small transmitted wavefront distortion,
\item compatibility with high-quality optical polishing.
\end{itemize}
The dielectric coating must likewise provide the required reflectivity
while maintaining low optical absorption and sufficiently low  mechanical loss at cryogenic temperature.

From the aspect of thermal properties, the substrate must have 
\begin{itemize}
\item high thermal conductivity
\end{itemize}
in order to efficiently remove heat deposited by absorbed laser power and to suppress temperature gradients in the test mass. It is also desirable to have 
\begin{itemize}
\item a small thermal expansion coefficient
\end{itemize}
for reducing thermally induced deformation, stress, and long-term drift of the mechanical geometry.

Finally, all materials, bonded interfaces, coatings, and suspension
attachments must remain compatible with operation in
\begin{itemize}
\item ultra-high-vacuum,
\item cryogenic.
\end{itemize}
The test-mass design is therefore determined by a combined  optimization of mass distribution, moment of inertia, mechanical quality factor, optical quality, thermal transport, structural strength, and manufacturability.

\section{Mass and Moment of Inertia}
\label{sec:test_mass_inertia}

CHRONOS measures the rotational response of a suspended torsion-bar system. For the rotational degree of freedom $\theta$, the equation of motion can be written as
\begin{equation}
I\ddot{\theta}(t)+\Gamma_{\theta}\dot{\theta}(t)+\kappa_{\theta}\theta(t)=\tau_{\rm sig}(t)+\tau_{\rm exc}(t),
\label{eq:test_mass_torsion_eom}
\end{equation}
where $I$ is the effective moment of inertia, $\Gamma_{\theta}$ is the rotational damping coefficient, $\kappa_{\theta}$ is the torsional spring constant, and $\tau_{\rm sig}$ and $\tau_{\rm exc}$ denote signal and excess torques.

Above the torsional resonant frequency, the angular displacement generated by an external torque is approximately
\begin{equation}
\theta(\omega)\simeq -\frac{\tau_{\rm exc}(\omega)}{I\omega^{2}}.
\label{eq:test_mass_high_freq}
\end{equation}
A large $I$ therefore suppresses angular motion produced by parasitic
torques such as fluctuations of actuator force and radiation pressure of the main interferometer laser.

The role of a large test mass in CHRONOS is therefore different from that of a large mirror aperture in a kilometer-scale Fabry--P\'erot detector. Although optical clipping loss remains as a constraint, the dimensions of the CHRONOS test mass are primarily determined by its required mechanical mass distribution. The optimum dimensions must consequently be derived from the complete torsion-bar dynamics rather than from the beam radius alone.

\section{Baseline Substrate Material: Sapphire}
\label{sec:sapphire_substrate}

\subsection{Material Properties}
\label{subsec:material_properties}

Monocrystalline sapphire is adopted as the baseline substrate material for the CHRONOS test mass. The primary reason is that sapphire provides a favorable combination of optical, mechanical, and thermal properties for cryogenic operation while remaining compatible with the baseline laser wavelength of 1064~$\mathrm{nm}$.

From the optical point of view, sapphire is transparent at 1064~$\mathrm{nm}$ and can be processed to make high-quality optical surfaces~\cite{kagra_sapphire_lensing}. It is also compatible with dielectric high-reflectivity coating technologies, making it suitable for use as a precision interferometer mirror. This allows CHRONOS to retain the mature 1064~$\mathrm{nm}$ laser and optical technology widely used in gravitational-wave interferometry.

In terms of mechanics, sapphire exhibits a high elastic modulus and high structural strength, which are important for constructing a large torsion-bar test mass~\cite{O3GK}. More importantly, its mechanical quality factor $Q$ can become very high ($\sim 10^8$) at cryogenic temperature~\cite{kagra_sapphire_mechanical}. This property is essential for CHRONOS because mechanical dissipation in the test mass directly contributes to thermal fluctuations expressed by the fluctuation--dissipation theorem.

Sapphire also has excellent cryogenic thermal properties. Its thermal conductivity increases strongly at low temperature, allowing heat generated by optical absorption to be efficiently transported away from the illuminated region~\cite{kagra_sapphire_lensing,kagra_history}. At the same time, its thermal expansion coefficient becomes very small, reducing thermally induced deformation and stress in the test mass~\cite{kagra_history}. These characteristics are particularly advantageous for maintaining the mechanical and optical stability required for sub-Hz measurements.

Table~\ref{tab:sapphire_parameters} summarizes the mechanical and thermal properties of sapphire in 300~K and 20~K~\cite{O3GK,kagra_sapphire_lensing,kagra_history,kagra_sapphire_mechanical}.

\begin{table}[t]
    \centering
    \caption{Parameters of sapphire.}
    \label{tab:sapphire_parameters}
    \begin{tabular}{ccccc}
        \hline
        Parameters ]Unit] & Values (300~K) & Values (20~K) \\
        \hline
        Young’s modulus [Pa] & 4.0$\times$10$^{11}$~Pa & - \\
        Poisson ratio & 0.29 & - \\
        Density [kg~m$^{-3}$] & 4.0$\times$10$^3$ & - \\
        Thermal expansion [K$^{-1}$] & 5.4$\times$10$^{-6}$ & $\sim$10$^{-9}$ \\
        Specific heat [J~K$^{-1}$~kg$^{-1}$] & 7.73$\times$10$^2$ & $\sim$10$^{-1}$ \\
        Thermal conductivity [W~m$^{-1}$~K$^{-1}$] & 46 & 4.3$\times$10$^3$ \\
        Optical absorption [ppm~cm$^{-1}$] & 40-100 & 90\\
        Quality factor & 5$\times$10$^6$ & 2$\times$10$^8$ \\
        \hline
    \end{tabular}
\end{table}

Another important consideration is technological maturity. Sapphire test masses and associated fabrication techniques have already been
developed in the context of cryogenic gravitational-wave detectors, providing experience in crystal production, precision polishing, coating, suspension, and cryogenic operation. This existing technological basis reduces the development risk compared with
a completely new substrate technology.

For these reasons, sapphire provides the most balanced solution for the current CHRONOS design. It enables the interferometer to remain based on mature 1064~$\mathrm{nm}$ optical technology while simultaneously providing the high mechanical quality factor, thermal conductivity, mechanical strength, and cryogenic compatibility required for the large torsion-bar test mass.

\subsection{Crystal Orientation}
\label{subsec:sapphire_orientation}

Sapphire is an anisotropic crystal whose optical, mechanical, and thermal properties depend on the crystallographic orientation. In particular, the crystallographic $c$-axis defines the optical axis of
sapphire and therefore plays an important role in determining the optical birefringence and the propagation of the interferometer beam through the substrate.

In the present CHRONOS design, the sapphire blocks are assumed to be oriented such that the $c$-axis is parallel to the interferometer beam direction. This configuration provides a simple and well-defined reference geometry for the optical design and is also the orientation adopted in the current mass-extension study.

The crystal orientation is also relevant to the mechanical properties of the test mass. Because sapphire has direction-dependent elastic properties, the resonant frequencies and deformation patterns of the internal mechanical modes can change with the orientation of the crystal relative to the torsion and suspension axes. The mode shape and crystal orientation may affect effective mechanical loss. Therefore, the orientation of the sapphire cannot be determined from optical considerations alone.

The thermal properties of sapphire are also anisotropic.
The heat flow from the illuminated region toward the suspension and the resulting temperature distribution can depend on the relative direction of the crystal axes. This becomes important in the cryogenic test mass, where absorbed optical power must be efficiently transported to the suspension system.

For CHRONOS, the baseline configuration therefore adopts the $c$-axis
parallel to the interferometer beam, while the final orientation will be verified together with the complete mechanical and cryogenic design. In particular, the effects of crystal orientation on optical birefringence, internal-mode structure, mechanical dissipation, and heat transport must be evaluated using the final test-mass and suspension geometry.

\section{Fabrication of Crystals}
\label{sec:large_sapphire}

\subsection{Constraint on crystal size}
\label{subsec:crystal_size}

One of the major technological challenges in the development of the CHRONOS test mass is the fabrication of sufficiently large sapphire components while maintaining the optical and mechanical quality required for cryogenic precision measurements. Because the torsion-bar geometry requires a large mass distribution and moment of inertia, the dimensions of the sapphire components are determined
not only by the optical beam size but also by the mechanical design of the test mass.

Large monocrystalline sapphire can be produced using crystal-growth
techniques such as the Heat Exchange Method (HEM). In the HEM process, the thermal environment of the melt and the seed crystal
is carefully controlled during growth and cooling. This makes it possible to produce large sapphire crystals while suppressing
internal stress and reducing crystalline defects. Such large-volume crystal-growth technology is therefore particularly important for CHRONOS, where the dimensions of the test mass can approach or exceed those of conventional optical substrates.

However, the maximum size of the grown sapphire boule does not directly determine the maximum usable size of a CHRONOS test-mass component. The usable volume depends strongly on the crystallographic orientation required for the optical and mechanical design, because the desired optical surface must be cut with a specific orientation relative to the crystal axes. Regions containing inclusions, defects, excessive residual stress, or refractive-index inhomogeneity must also be excluded from the final optical substrate.

Optical requirements further reduce the practically usable dimensions. The selected material must provide sufficiently uniform refractive index and sufficiently low absorption over the region sampled by the interferometer beam. At the same time, additional material must generally be retained during machining to allow the final dimensions, surface orientation, and optical figure to be produced by precision grinding and polishing. The achievable test-mass dimensions are therefore determined by the combination of crystal growth, crystallographic orientation, optical quality,
machining, and polishing rather than by the raw boule diameter alone.

For CHRONOS, the relevant quantity is consequently not the largest sapphire crystal that can be grown, but the largest low-loss and optical-quality sapphire component that can be fabricated with the required crystal orientation. This distinction is particularly important for the mass-extension design, because the practical dimensions of an individual sapphire component determine the numbers of blocks and bonded interfaces required to construct the complete torsion bar.

\subsection{Optical Homogeneity}
\label{subsec:sapphire_homogeneity}

For a substrate through which the optical field propagates,
inhomogeneity of refractive index distort the transmitted wavefront. For a substrate thickness $l$ and refractive-index variation $\delta n$, the optical-path variation is approximately
\begin{equation}
\delta_{\rm OPL}\simeq l\,\delta n.
\end{equation}
On the other hand, we set a requirement of the surface non-uniformity at 1/10 of the laser wavelength $\lambda$. A representative wavefront-quality requirement can therefore be expressed as
\begin{equation}
\delta_{\rm OPL} < \frac{\lambda}{10}.
\label{eq:sapphire_homogeneity}
\end{equation}
For example, where
\begin{equation}
l=0.20~\mathrm{m}, \qquad \lambda=1064~\mathrm{nm},
\end{equation}
this corresponds approximately to
\begin{equation}
\delta n << 5\times10^{-7}.
\end{equation}

The precise requirement depends on the optical role of the mirror. For an input optic through which significant optical power propagates,
bulk homogeneity is particularly important. For a highly reflective end surface, surface figure, scattering, and coating uniformity can instead dominate the optical requirement.

\section{Large-Scale Test-Mass Architecture}
\label{sec:large_test_mass}

Ideally, the CHRONOS torsion-bar test mass would be fabricated as a single monolithic sapphire structure. A monolithic configuration is mechanically attractive because it avoids additional interfaces that can introduce dissipation, alignment errors, and local stress concentrations.

In practice, however, realizing the full CHRONOS torsion-bar geometry from a single sapphire crystal is difficult. The achievable size of a single crystal is limited not only by the size of crucible, but also by the availability of sufficiently large defect-free material with the required crystallographic orientation and optical quality. Even when a sufficiently large raw crystal can be produced,  additional constraints are placed on the dimensions of the final component by machining, handling, precision optical-surface fabrication, and cryogenic integration. The suspension interface must also be incorporated without  compromising the mechanical quality of the torsion bar.

For these reasons, a fully monolithic test mass may not be the most practical solution for the final CHRONOS detector. An alternative is to construct the torsion bar from several sapphire components and assemble them into a single mechanical structure while preserving sufficiently low mechanical loss.

CHRONOS therefore considers hydroxide-catalysis bonding as the baseline technique for such mass extension. This approach allows the total torsion-bar length and mass distribution to exceed the dimensions of an individual sapphire component, while retaining
an all-sapphire structure over most of the mechanical volume. The resulting design strategy is to use the largest practical monolithic sapphire components and introduce bonded interfaces only where they are necessary to achieve the required overall geometry.

\subsection{Hydroxide-Catalysis Bonding}
\label{subsec:hcb_mass_extension}

Hydroxide-catalysis bonding is a precision joining technique that forms a thin and mechanically rigid bond between highly polished surfaces. It has been developed for optical and mechanical systems in which high dimensional stability and low mechanical dissipation are required~\cite{hydrocatalysis}.

For CHRONOS, HCB provides a practical method for extending the dimensions of the torsion-bar test mass beyond those achievable with a single sapphire component. By bonding several sapphire blocks together, the total bar length, mass distribution, and moment of inertia can be adjusted independently of the maximum dimensions of an individual crystal. This is particularly useful when the mechanically preferred torsion-bar geometry exceeds the size of the available optical-quality sapphire substrate.

The mass-extension design is determined by several coupled parameters.
The total torsion-bar length and mass determine the overall geometry, while the moment of inertia about the suspension axis governs the rotational mechanical response. On the other hand, the number of sapphire blocks and cross section are relevant to properties of the bonding layers, thus to the effective mechanical loss of the assembled structure. In particular, the number of sapphire blocks determines the number of bonded interfaces. The bonding area and bonding-layer thickness control how much of the mechanical strain energy is stored in the HCB regions. The elastic energy stored in the HCB and the intrinsic mechanical losses of both substrate and bonding materials determines the effective mechanical loss.

This introduces an important trade-off in the test-mass design.
Increasing the number of bonded blocks makes it possible to construct a longer and mechanically larger torsion bar, but every additional interface introduces a region with substantially larger mechanical loss than the sapphire substrate. The HCB mass-extension scheme must therefore be designed so that the required bar dimensions and moment of inertia are achieved with as the minimal number of bonded interfaces as possible and the least elastic-energy participation in the bonding layers.

In this sense, HCB is not introduced simply as a convenient assembly
technique. It is an integral part of the mechanical design of the CHRONOS test mass, and its geometry must be optimized together with the dimensions of the sapphire blocks, the suspension configuration, and the internal mechanical modes of the complete torsion bar.

\subsection{Reference Sapphire Block}
\label{subsec:reference_sapphire_block}

The current mass-extension study assumes cuboid sapphire blocks whose crystallographic $c$-axis is aligned with the interferometer beam~\cite{tanabe_mass_extension}. This study assume the dimension of a single cuboid as
\begin{equation}
220~\mathrm{mm} \times 220~\mathrm{mm} \times 200~\mathrm{mm}.
\end{equation}
This assumption is based on the maximum available sapphire size of approximately
\begin{equation}
\phi 250~\mathrm{mm} \times 150~\mathrm{mm},
\end{equation}
according to the procurement information used for the mass-extension analysis.

These dimensions should not be interpreted as final procurement
specifications. The available crystal size must be re-evaluated for the final crystallographic orientation and optical-quality requirement.

\subsection{Reference Mass-Extension Geometry}
\label{subsec:mass_extension_geometry}

\begin{figure}[tbp]
    \centering
    \includegraphics[width=0.88\columnwidth]
    {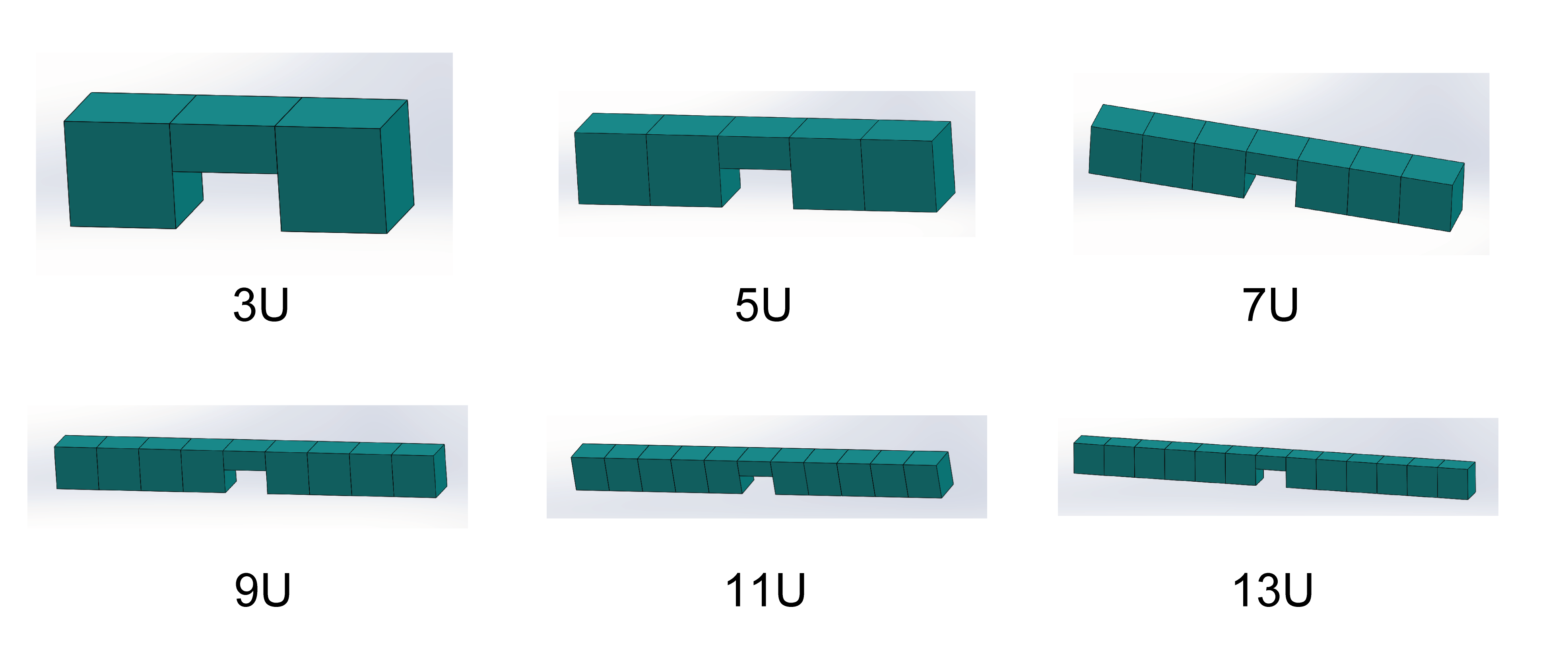}
    \caption{
    Shapes of extended torsion bars of 3, 5, 7, 9, 11, and 13 units.
    }
    \label{fig:longer_bars}
\end{figure}

The current reference CHRONOS torsion bar consists of five sapphire blocks. In the study of extended test masses, structures consisting of
\begin{equation}
N_{\rm block}=3, \, 5, \, 7, \, 9, \, 11, \, 13
\end{equation}
blocks were considered. Appearances of these extended bars are shown in Fig.~\ref{fig:longer_bars}.

Since all blocks have the same nominal longitudinal dimension, the total bar length can also be expressed in units of the elementary block length. The mechanical parameters obtained for the reference geometry are summarized in Table~\ref{tab:test_mass_extension_geometry}.

\begin{table}[tb]
\centering
\caption{
Reference geometrical parameters of HCB-extended sapphire torsion bars.
}
\label{tab:test_mass_extension_geometry}
\begin{tabular}{cccc}
\hline
\begin{tabular}{c}Units\\$N_{\rm block}$\end{tabular} & \begin{tabular}{c}Length\\$L_{\rm bar}$ [cm]\end{tabular} & \begin{tabular}{c} Mass\\$M$ [kg]\end{tabular} & \begin{tabular}{c} Moment of inertia\\$I$ [kg\,m$^{2}$]\end{tabular} \\
\hline
3  & 66  & 94.46  & 4.42   \\
5  & 110 & 171.32 & 19.86  \\
7  & 154 & 248.18 & 53.91  \\
9  & 198 & 325.04 & 113.99 \\
11 & 242 & 401.90 & 207.56 \\
13 & 286 & 478.76 & 342.05 \\
\hline
\end{tabular}
\end{table}

The reference five-block configuration therefore provides
\begin{equation}
L_{\rm bar}=1.10~\mathrm{m},
\end{equation}

\begin{equation}
M=171.32~\mathrm{kg},
\end{equation}
and
\begin{equation}
I=19.86~\mathrm{kg\,m^{2}}.
\end{equation}

Increasing the number of blocks increases the total length, mass, and moment of inertia substantially. However, every additional interface introduces an additional region of mechanical loss. Therefore the optimum structure cannot be determined from $I$ alone.

\section{Mechanical Loss}
\label{sec:test_mass_mechanical_loss}

\subsection{General formalism}
\label{subsec:loss_formalism}

Mechanical dissipation is one of the principal requirements for the
CHRONOS test mass. The mechanical quality factor is related to the loss angle approximately by
\begin{equation}
\phi \simeq \frac{1}{Q}.
\end{equation}
The loss angle represents the structural damping scales of a bulk.

According to the fluctuation--dissipation theorem, the thermal-noise
power spectral density associated with the damping scales
schematically as~\cite{Saulson1990}
\begin{equation}
S_{\rm th}\propto T\phi \simeq \frac{T}{Q}.
\label{eq:T_over_Q}
\end{equation}
The corresponding amplitude spectral density scales as
\begin{equation}
S_{\rm th}^{1/2}\propto \sqrt{\frac{T}{Q}}.
\end{equation}

As a consequence, low temperature alone is therefore not sufficient: the complete suspended mechanical structure must also retain a high mechanical $Q$.

\subsection{Cryogenic Quality Factor of Sapphire}
\label{subsec:sapphire_q}

One of the main advantages of sapphire for CHRONOS is its extremely low mechanical dissipation at cryogenic temperature. Since the mechanical loss angle is inverse of $Q$, a high $Q$ is essential for suppressing thermally driven mechanical fluctuations of the test mass.

Cryogenic measurements of sapphire have demonstrated mechanical quality factors in the range of approximately
\begin{equation}
Q \sim 10^{7} - 10^{8},
\end{equation}
with values as high as about
\begin{equation}
Q \simeq 2.5\times10^{8}
\end{equation}
reported near $4.2~\mathrm{K}$ for internal mechanical modes of sapphire test masses~\cite{kagra_sapphire_mechanical}. These results demonstrate that sapphire can retain exceptionally low mechanical loss in the temperature range relevant to CHRONOS.

However, the quality factor is not a unique material constant.
The measured value depends on the mechanical mode, crystal quality,
crystallographic orientation, surface condition, suspension method, and residual stress. For a large CHRONOS torsion bar, the relevant mechanical dissipation is therefore that of the complete assembled structure rather than that of an isolated sapphire sample.

This distinction becomes particularly important when mass extension is
introduced. Although the sapphire substrate itself can have a very small loss angle, the HCB interfaces and suspension-related structures can store elastic energy and reduce the effective quality factor of the mechanical modes. The final mechanical performance must therefore be evaluated mode by mode, including the sapphire blocks, bonded interfaces, bridge structure, and suspension coupling.

For CHRONOS, the practical requirement is consequently to preserve the
intrinsically low cryogenic loss of sapphire after the complete test mass has been assembled. This requires both finite-element evaluation of the strain-energy distribution and experimental ring-down measurements of the full cryogenic structure.

\subsection{Mechanical Loss of the HCB Extension}
\label{sec:hcb_loss}

The HCB interfaces introduce an additional mechanical-loss contribution. In the current finite-element model, the HCB layer is assumed to have an uniform thickness of
\begin{equation}
t_{\rm bond}=100~\mathrm{nm}
\end{equation}
over the bonded surfaces~\cite{hydrocatalysis_young}.

The representative material properties of sapphire and HCB used in the analysis are shown in Table~\ref{tab:sapphire_hcb_material}~\cite{haughian_phd,hydrocatalysis_young,AZO_silica,hydrocatalysis_loss}.

\begin{table}[tb]
\centering
\caption{
Representative material properties used in the current mass-extension FEA.
}
\label{tab:sapphire_hcb_material}
\begin{tabular}{ccccc}
\hline
Material & Density & Young's modulus & Poisson ratio & Mechanical loss
\\
& [kg/m$^3$] & [GPa] & & \\
\hline
Sapphire & 4000   & 400  & 0.29 & $4\times10^{-9}$ \\
HCB      & 1075.7 & 18.5 & 0.16 & $0.11$ \\
\hline
\end{tabular}
\end{table}

The intrinsic mechanical loss assumed for the HCB material is thus many orders of magnitude larger than that of sapphire. However, the HCB volume is relatively small. Contribution of the bonding to the effective mechanical loss of the whole torsion-bar structure is scaled by its elastic-energy participation.


The elastic energy stored in each material domain is evaluated from the stress tensor $\sigma_{ij}$ and strain tensor $\epsilon_{ij}$ as
\begin{equation}
U=\iiint \frac{1}{2} \sum_{ij} \sigma_{ij}\epsilon_{ij}\,dV.
\label{eq:stored_energy}
\end{equation}

The effective mechanical loss is then expressed as
\begin{equation}
\phi_{\rm eff}=\frac{U_{\rm B}}{U_{\rm S}+U_{\rm B}}\phi_{\rm B}+\frac{U_{\rm S}}{U_{\rm S}+U_{\rm B}}\phi_{\rm S},
\label{eq:effective_hcb_loss}
\end{equation}
where $U_{\rm B}$ and $U_{\rm S}$ are the elastic energies stored in the bonding and sapphire domains, respectively.

Defining the bonding participation ratio
\begin{equation}
p_{\rm B}=\frac{U_{\rm B}}{U_{\rm S}+U_{\rm B}},
\end{equation}
Eq.~(\ref{eq:effective_hcb_loss}) becomes
\begin{equation}
\phi_{\rm eff}=p_{\rm B}\phi_{\rm B}+(1-p_{\rm B})\phi_{\rm S}.
\label{eq:effective_hcb_loss_p}
\end{equation}
The current FEA gives bonding participation ratios approximately of order
\begin{equation}p_{\rm B}
\sim 10^{-5}
\end{equation}
for the relevant internal modes.

Consequently, even though the intrinsic HCB loss is large, the effective loss of the complete structure remains approximately in the
\begin{equation}
\phi_{\rm eff}\sim 10^{-6}
\end{equation}
range in the present model. This demonstrates that the principal design strategy is to minimize the elastic strain energy localized in the HCB regions.

\section{Internal Mechanical Modes}
\label{sec:test_mass_internal_modes}

\subsection{Resonant Frequencies}
\label{subsec:resonant_frequencies}

The mechanical effect of the HCB interfaces depends strongly on the internal mode shape of the torsion bar. Each elastic mode produces a different strain distribution across the sapphire blocks and the bonded interfaces, and therefore stores a different fraction of
its elastic energy in the HCB regions. As a result, the effective mechanical loss of the assembled test mass is inherently mode-dependent.

\begin{figure}[tbp]
    \centering
    \includegraphics[width=0.88\columnwidth]
    {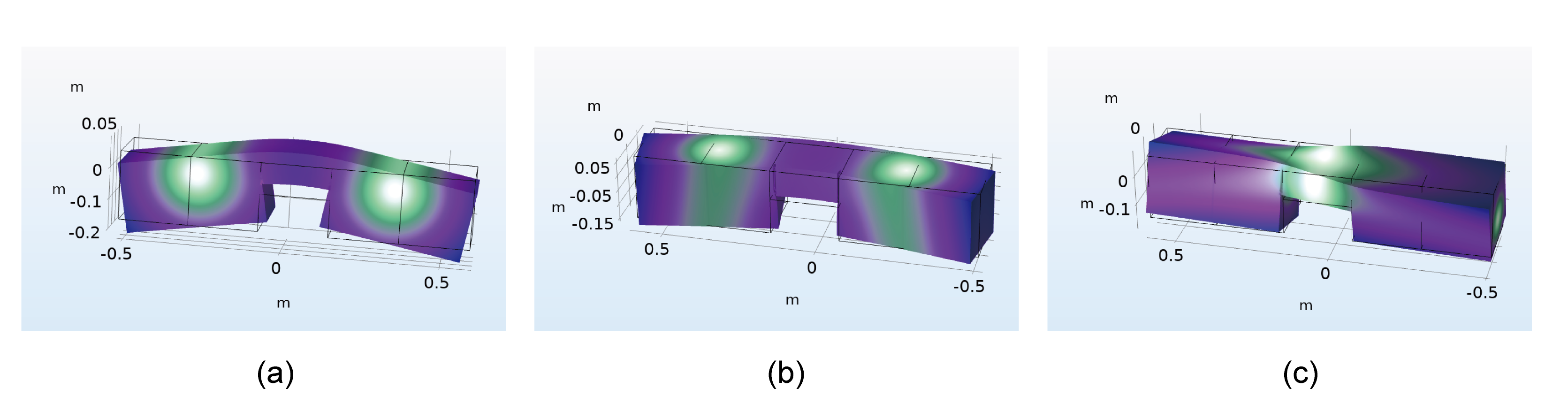}
    \caption{
    Shape of the three lowest internal modes of the CHRONOS torsion bar with five units. (a) vertical bending mode at 666.6~Hz, (b) horizontal bending mode at 1237.7~Hz, (c) twisting mode at 1398.9~Hz. The mode shape was obtained by COMSOL Multiphysics 5.4.
    }
    \label{fig:threemodes}
\end{figure}

The present analysis focuses on the lowest internal resonances
corresponding to vertical bending, horizontal bending, and twisting of the extended torsion bar as shown in Fig.~\ref{fig:threemodes}. For the reference five-units configuration, the calculated resonant frequencies are approximately
\begin{equation}
f_{\rm VB} = 666.6~\mathrm{Hz},
\end{equation}
\begin{equation}
f_{\rm HB} = 1237.7~\mathrm{Hz},
\end{equation}
and
\begin{equation}
f_{\rm TW} = 1398.9~\mathrm{Hz}.
\end{equation}

These resonant frequencies are determined by the geometry of the assembled torsion bar and therefore change as the mass-extension configuration is modified. In particular, increasing the number of sapphire blocks increases the total length of the bar and reduces the frequencies. Thus, extending the test mass changes not only the total mass and moment of inertia, but also the internal elastic dynamics of the structure.


This introduces an important mechanical trade-off in the mass-extension design. A longer torsion bar provides a larger moment of inertia, while at the same time requiring additional bonded interfaces and lowering several internal resonance frequencies. The final geometry must therefore be chosen so that the required mechanical size and inertia are achieved without bringing mechanically relevant internal modes too close to the detector observation or control bands.

The importance of a given internal mode cannot be assessed from its resonance frequency alone. Because the interferometer senses the motion of the optical surface at the beam position, the coupling of each elastic mode to the optical readout depends on its deformation pattern and effective mass. Likewise, the additional dissipation introduced by the HCB depends on the elastic-energy participation of the bonded regions for that particular mode. The internal-mode analysis must therefore evaluate the mode shape, resonant frequency, effective mass, and effective mechanical loss together.

For the mass-extended configurations, these quantities are obtained by
finite-element analysis (FEA) of the complete bonded structure using COMSOL Multiphysics 5.4~\cite{comsol}. The elastic energy stored in the sapphire and HCB regions is evaluated by Eq.~(\ref{eq:effective_hcb_loss}) for each mode, and the corresponding mode-dependent effective loss is calculated from their energy participation. The surface displacement is additionally weighted by the interferometer beam profile in order to determine the effective mass relevant to the optical readout.


\subsection{Effective Mass}
\label{subsec:internal_mode_effective_mass}

The displacement of an internal mode sensed by the interferometer is not equal among the modes. For the $n$-th mechanical mode, an effective mass $m_n$ can be defined by equating the energy of the surface displacement sensed by the optical beam to that of the full elastic mode:
\begin{equation}
\frac{1}{2}
m_n x_n^{2}\omega_n^{2}=\frac{1}{2}\iiint_{\rm bar}\rho\left( \epsilon_{n,x}^{2}+\epsilon_{n,y}^{2}+\epsilon_{n,z}^{2} \right)dV \, \omega_n^{2},
\end{equation}
giving
\begin{equation}
m_n=\frac{\displaystyle \iiint_{\rm bar} \rho \left( \epsilon_{n,x}^{2}+\epsilon_{n,y}^{2}+\epsilon_{n,z}^{2} \right)dV}{x_n^{2}}.
\label{eq:mode_effective_mass}
\end{equation}
Here the $\omega_n$ is the angular resonant frequency of $n$-th mode, $\rho$ is the density of the substrate, and $\varepsilon_{n,i}$ ($i$={$x$,$y$,$z$}) is the strain along $i$-axis at $n$-th mode. The $x_n$ is the optical-weighted surface displacement seen by the interferometer beam which is calculated by
\begin{equation}
x_n=\frac{\displaystyle \iint_{\rm surface} P(\bm{r}) \epsilon_{n,z}\,dS}{\displaystyle \iint_{\rm surface} P(\bm{r}) \, dS},
\label{eq:beam_weighted_displacement}
\end{equation}
where the Gaussian beam profile is represented as
\begin{equation}
P(\bm{r})=\exp \left( -\frac{2\left| \bm{r} \right|^{2}}{\sigma^{2}} \right).
\end{equation}
using the maximum power amplitude $P_0$, the beam radius $\sigma$, and the position vector $\bm{r}$ of the mesh from the center of the beam. 

The resonant peaks that appear in the bar thermal noise spectrum are suppressed by scaling of the inverse effective mass. This formulation makes explicit that the mechanical relevance of an internal mode depends on both its global deformation and the local surface motion sampled by the interferometer beam.

As a summary of the FEA study of the resonant modes of the torsion bar, Table~\ref{tab:resonance_parameters} shows resonant frequencies of the three lowest modes with their energy participation, effective mechanical loss, and effective mass.

\begin{table}[t]
    \centering
    \small
    \caption{Resonant modes, mechanical losses, and effective masses of the torsion bar with different numbers of units bonded by HCB. VB, HB, and TW represent the vertical bending, horizontal bending, and twisting shape, respectively.}
    \label{tab:resonance_parameters}
    \setlength{\tabcolsep}{3pt}
    \begin{tabular}{cccccc}
        \hline
        \begin{tabular}{c} Units \end{tabular} & \begin{tabular}{c} Resonant\\frequency (Hz) \end{tabular} & \begin{tabular}{c} Mode\\shape \end{tabular} & \begin{tabular}{c} Energy ratio\\$p_B$ ($\times$10$^{-5}$) \end{tabular} & \begin{tabular}{c} Mechanical loss\\$\phi_{\rm eff}$ ($\times$10$^{-6}$~rad) \end{tabular} & \begin{tabular}{c} Effective mass\\$m_n$ (kg) \end{tabular}\\
        \hline
        3 & \begin{tabular}{c} 1351.2\\2051.4\\2933.7 \end{tabular} & \begin{tabular}{c} VB\\TW\\HB \end{tabular} & \begin{tabular}{c} 2.40\\1.87\\1.31 \end{tabular} & \begin{tabular}{c} 2.65\\2.06\\1.45 \end{tabular} & \begin{tabular}{c} 2.15$\times$10$^7$\\9.38$\times$10$^2$\\2.16$\times$10$^1$ \end{tabular}\\
        \hline
        5 & \begin{tabular}{c} 666.6\\1237.7\\1398.9 \end{tabular} & \begin{tabular}{c} VB\\HB\\TW \end{tabular} & \begin{tabular}{c} 2.49\\1.31\\1.80 \end{tabular} & \begin{tabular}{c} 2.74\\1.44\\1.98 \end{tabular} & \begin{tabular}{c} 4.69$\times$10$^{11}$\\2.99$\times$10$^1$\\9.79$\times$10$^2$ \end{tabular}\\
        \hline
        7 & \begin{tabular}{c} 400.6\\680.1\\1092.7 \end{tabular} & \begin{tabular}{c} VB\\HB\\TW \end{tabular} & \begin{tabular}{c} 2.47\\1.27\\1.72 \end{tabular} & \begin{tabular}{c} 2.72\\1.40\\1.90 \end{tabular} & \begin{tabular}{c} 7.54$\times$10$^{12}$\\3.92$\times$10$^1$\\9.81$\times$10$^2$ \end{tabular}\\
        \hline
        9 & \begin{tabular}{c} 269.3\\430.3\\902.0 \end{tabular} & \begin{tabular}{c} VB\\HB\\TW \end{tabular} & \begin{tabular}{c} 2.43\\1.23\\1.62 \end{tabular} & \begin{tabular}{c} 2.67\\1.36\\1.79 \end{tabular} & \begin{tabular}{c} 3.96$\times$10$^{12}$\\4.88$\times$10$^1$\\5.87$\times$10$^2$ \end{tabular}\\
        \hline
        11 & \begin{tabular}{c} 194.7\\296.9\\756.3 \end{tabular} & \begin{tabular}{c} VB\\HB\\TW \end{tabular} & \begin{tabular}{c} 2.38\\1.20\\1.43 \end{tabular} & \begin{tabular}{c} 2.62\\1.33\\1.58 \end{tabular} & \begin{tabular}{c} 1.45$\times$10$^{13}$\\5.85$\times$10$^1$\\2.23$\times$10$^2$ \end{tabular}\\
        \hline
        13 & \begin{tabular}{c} 148.0\\217.3\\613.4 \end{tabular} & \begin{tabular}{c} VB\\HB\\TW \end{tabular} & \begin{tabular}{c} 2.32\\1.18\\1.13 \end{tabular} & \begin{tabular}{c} 2.56\\1.30\\1.25 \end{tabular} & \begin{tabular}{c} 5.81$\times$10$^{13}$\\6.82$\times$10$^1$\\9.99$\times$10$^1$ \end{tabular}\\
        \hline
    \end{tabular}
\end{table}

\section{Reduction of Bonding Area}
\label{sec:hcb_area}

The nominal HCB model assumes that the bond covers the entire facing surface between adjacent sapphire blocks. However, full-area bonding is not necessarily optimal from the viewpoint of mechanical dissipation.

For a fixed bonding-layer thickness, the elastic energy stored in the HCB region, and hence the effective mechanical loss, is approximately
proportional to the total bonded area $A_{\rm bond}$. Reducing $A_{\rm bond}$ therefore provides a direct strategy for decreasing $p_{\rm B}$ and hence $\phi_{\rm eff}$. The lower limit of $A_{\rm bond}$ is instead set by mechanical strength.

\subsection{Static Strength Requirement}
\label{subsec:hcb_static_strength}

The present engineering analysis adopts
\begin{equation}
\tau_{\rm max}=0.16~\mathrm{MPa}
\end{equation}
as a reference allowable shear stress~\cite{hydrocatalysis}.

The innermost HCB interface experiences the largest load because it supports all of the outer blocks. We define index of HCB layers by $i=1$ at the innermost interface and ascending to the outer. For an torsion bar made of $N_{\rm block}$ units, the minimum required bonding area of the
$i$-th interface counted from the center is estimated as
\begin{equation}
A_{{\rm bond},i} \geq \frac{ M_{\rm block} \left[ \frac{N_{\rm block}-1}{2}-i+1 \right]g}{\tau_{\rm max}},
\label{eq:hcb_area_requirement}
\end{equation}
where
\begin{equation}
i\leq \frac{n-1}{2}.
\end{equation}

The mass of individual block adopted in the present calculation is
\begin{equation}
M_{\rm block}=38.72~\mathrm{kg}.
\end{equation}
The mass of the bonding material itself was neglected.

\subsection{Required Bonding Area}
\label{subsec:hcb_required_area}

Table~\ref{tab:hcb_area} summarizes the total required bonding area on one side of the torsion bar.

\begin{table}[tb]
\centering
\caption{
Estimated minimum bonding area compared with the nominal full-surface
bonding area.
}
\label{tab:hcb_area}
\begin{tabular}{cccc}
\hline
Units & Required area [mm$^{2}$] & Nominal area [mm$^{2}$] & Required/Nominal
\\
\hline
3  & 2371.6  & 20000  & 0.12 \\
5  & 7114.8  & 64000  & 0.11 \\
7  & 14229.6 & 108000 & 0.13 \\
9  & 23716.0 & 152000 & 0.16 \\
11 & 35574.0 & 196000 & 0.18 \\
13 & 49803.6 & 240000 & 0.21 \\
\hline
\end{tabular}
\end{table}

For the reference five-units test mass,
\begin{equation}
A_{\rm bond}^{\rm req} \simeq 7.1\times10^{3}~\mathrm{mm^{2}},
\end{equation}
whereas the nominal full-area configuration provides
\begin{equation}
A_{\rm bond}^{\rm nominal}=6.4\times10^{4}~\mathrm{mm^{2}}.
\end{equation}

Thus, according to the simple static shear-strength criterion, only
approximately 11\% of the nominal area is required. This suggests that reduced-area or patterned HCB interfaces could substantially reduce the effective mechanical loss.

\subsection{Constraints on Reduced-Area Bonding}
\label{subsec:hcb_reduced_area_constraints}

The minimum bonding area determined from the static shear-strength
requirement does not necessarily correspond to the optimum mechanical
design. Although reducing the bonded area can decrease the elastic-energy participation of the HCB region and thereby lower the effective mechanical loss, excessively small bonding regions may introduce other mechanical effects that are not captured by a simple static-load estimate.

In particular, reducing the bonded area leaves larger portions of the facing sapphire surfaces unbonded.These regions may experience local contact or friction, which could introduce additional dissipation. At the boundaries of the bonded regions, the mechanical load can also become spatially concentrated, producing local stress concentrations and modifying the strain distribution of the assembled structure. As a result, the internal elastic-mode shapes and their effective mechanical losses may differ from those predicted for a uniformly bonded interface.

Cryogenic operation introduces an additional consideration. During cooldown, thermal contraction of the sapphire blocks and the bonded
regions can generate local deformation and stress near the HCB interfaces. Such deformation may affect the relative alignment of the sapphire blocks and may also modify the mechanical boundary conditions of the internal modes. These effects become increasingly important when the bonded area is strongly reduced or when the bonding pattern becomes geometrically asymmetric.

The final bonding geometry must therefore be determined by balancing the reduction of HCB-related mechanical loss against the requirements for mechanical strength, structural stability, and alignment reproducibility. Finite-element analysis should be used to evaluate the stress distribution and internal-mode deformation for candidate bonding patterns, while mechanical-strength tests are required to confirm that the joints retain sufficient load capacity.

The optimized structure must also be verified experimentally through
ring-down measurements of bonded test samples and assembled structures. Measurements at both room temperature and cryogenic temperature are important for determining whether the reduced-area bonding introduces additional dissipation that is not included in the nominal material-loss model. Repeated thermal cycling is also required to confirm that the bond strength, mechanical loss, and relative alignment remain stable after successive cooldown and warm-up cycles.

The reduced-area HCB design should therefore be regarded as a coupled
mechanical optimization problem rather than a simple minimization of the bonding area.

\section{Suspension-Fiber Interface}
\label{sec:test_mass_bridge}

The test mass must provide a mechanically low-loss interface to the
cryogenic suspension. The CHRONOS detector employs crossed torsion bars. Directly attaching the suspension fibers to both bars creates a geometrical problem near the crossing point because the two suspension fibers would occupy overlapping regions.

\begin{figure}[tbp]
    \centering
    \includegraphics[width=0.88\columnwidth]
    {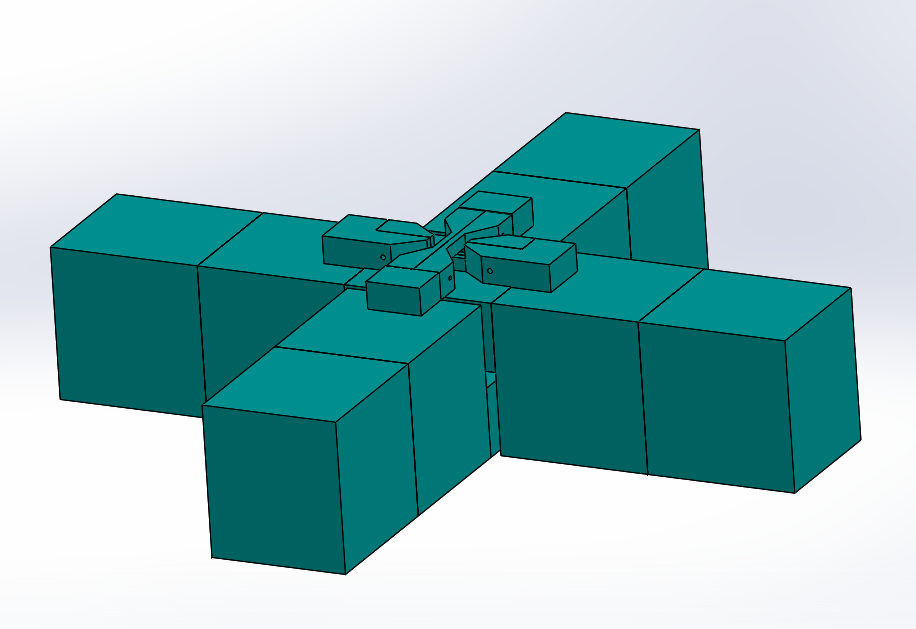}
    \caption{
    Configuration of cross torsion bar with bridge structure to hold suspension fibers.
    }
    \label{fig:bridge_cross}
\end{figure}

A bridge-type sapphire structure is therefore considered as an interface between the torsion bar and its suspension fiber. The bridge is attached to the main sapphire structure using HCB and allows one of the crossed torsion bars to be vertically displaced relative to the other as shown in Fig.~\ref{fig:bridge_cross}.

This configuration makes it possible to realize the crossed geometry while maintaining independent response of each bar.


However, the bridge modifies the elastic boundary conditions of the torsion bar and therefore changes the internal resonant frequencies. The largest change occurs for the vertical-bending mode. For example, at the five-units configuration, the vertical-bending resonant frequency increases from 666.6~Hz to approximately 1007.0~Hz with the bridge.

The horizontal-bending mode is less strongly affected. For five or more blocks, its frequency shift is less than approximately 30~Hz, and the difference decreases for longer bars.

The suspension interface should therefore be included directly in the final FEA model rather than treated as an independent structure added after the torsion-bar geometry has been optimized.

\section{Reduction of Bonded Interfaces}
\label{sec:larger_sapphire_blocks}

The most direct method for reducing HCB mechanical loss is to reduce the number of bonded interfaces. If larger monolithic sapphire components become available, the same dimension of the whole torsion-bar could be achieved by fewer joints.

The present mass-extension analysis considered hypothetical blocks with twice the nominal longitudinal dimension, referred to as ``2U'' blocks. The reduced number of HCB interfaces decreases the effective mechanical loss.

For the horizontal-bending mode, representative results are shown in
Table~\ref{tab:2u_loss}.

\begin{table}[tb]
\centering
\caption{
Comparison of effective mechanical loss for torsion bars constructed from nominal 1U and larger 2U sapphire blocks.
}
\label{tab:2u_loss}
\begin{tabular}{ccc}
\hline
Units & $\phi_{\rm eff}^{\rm 1U}$ & $\phi_{\rm eff}^{\rm 2U}$ \\
& [$\times10^{-6}$] & [$\times10^{-6}$] \\
\hline
5  & 1.44 & 1.32 \\
9  & 1.36 & 1.12 \\
13 & 1.30 & 1.00 \\
\hline
\end{tabular}
\end{table}

The analysis therefore establishes the following mechanical-design
principles:

\begin{enumerate}
\item maximize the dimensions of each monolithic sapphire component;
\item minimize the number of HCB interfaces;
\item minimize the total HCB area consistent with sufficient structural strength;
\item place interfaces in regions with low elastic-energy participation
for mechanically relevant modes;
\item include the bridge and suspension interfaces in the same mechanical optimization.
\end{enumerate}

The HCB concept should therefore not be interpreted as a preference for high segmentation. It is a practical method for achieving the required overall dimensions when a fully monolithic sapphire test mass cannot be manufactured.

\section{Cryogenic Thermal Properties}
\label{sec:test_mass_thermal}

The thermal properties of the substrate affect both intrinsic thermal
fluctuations and the extraction of heat deposited by absorbed laser power.

Sapphire has very high thermal conductivity at cryogenic temperature.
Around several tens of kelvin, the thermal conductivity of high-purity
sapphire can reach approximately~\cite{kagra_sapphire_lensing}
\begin{equation}
\kappa \sim 10^{3}-10^{4}~\mathrm{W\,m^{-1}\,K^{-1}},
\end{equation}
depending on temperature, purity, crystal orientation, and sample size. A representative value around $20~\mathrm{K}$ is
\begin{equation}
\kappa \sim 4\times10^{3}~\mathrm{W\,m^{-1}\,K^{-1}}.
\end{equation}

This high conductivity strongly suppresses temperature gradients across the test mass and facilitates heat extraction through the cryogenic suspension.

\subsection{Thermal Expansion}
\label{subsec:thermal_expansion}

The thermal expansion coefficient of sapphire decreases strongly at low temperature, making it particularly suitable for cryogenic precision measurements. A representative low-temperature expression for the linear thermal expansion coefficient along the crystallographic $c$-axis is
\begin{equation}
\alpha_c=7.5\times10^{-13} T^{3} \quad [\mathrm{K^{-1}}],
\label{eq:sapphire_thermal_expansion}
\end{equation}
within the temperature range of 1-100~K~\cite{kagra_history}. The rapid decrease of $\alpha_c$ toward cryogenic temperature strongly
suppresses dimensional changes of the sapphire test mass produced by
temperature fluctuations.

This small thermal expansion is particularly advantageous when combined with the high thermal conductivity of cryogenic sapphire.
The high thermal conductivity suppresses temperature gradients within the test mass, while the small thermal expansion coefficient reduces the mechanical deformation produced by the remaining temperature anisotropy. As a result, thermally induced changes in the dimensions of the test mass, temperature-gradient-induced stress, and thermoelastic deformation can all be strongly reduced.

These properties are important not only for maintaining the optical surface geometry but also for preserving the mechanical stability of the torsion-bar system. Temperature anisotropy can modify the stress distribution in the test mass and its interfaces and can consequently shift the frequencies of internal mechanical modes. Slow thermal drifts may also change the equilibrium geometry of the suspended torsion bar and its alignment with respect to the interferometer
beam.

For CHRONOS, which targets precision measurements in the sub-Hz frequency band, such slowly varying thermal effects must be carefully controlled. The combination of small thermal expansion and high thermal conductivity therefore provides an important advantage of sapphire for maintaining both the dimensional and mechanical stability of the cryogenic test mass.

\section{Optical Absorption}
\label{sec:test_mass_absorption}

Optical absorption in the test mass converts part of the circulating laser power into heat and can therefore affect both the thermal state and the optical performance of the mirror. For CHRONOS, one should distinguish between absorption in the sapphire substrate and  absorption in the high-reflectivity coating, because the two heat sources are distributed differently within the optic and can produce
different thermal responses.

Bulk absorption deposits heat throughout the optical path inside the
sapphire substrate, whereas coating absorption is concentrated close to the mirror surface. As a result, the corresponding temperature gradients, thermoelastic deformation, and thermo-refractive effects are not identical. The two contributions should therefore be characterized and specified independently when evaluating the cryogenic heat load and the resulting wavefront distortion.

The allowable optical absorption is set by the optical power and cryogenic
thermal budget of CHRONOS. Both substrate and coating absorption contribute
to the heat deposited in the test mass. Their spatial distributions determine
how that heat flows through the sapphire and into the cryogenic suspension.
Optical absorption therefore requires a coupled optical and thermal analysis.

\subsection{Bulk Absorption}
\label{subsec:sapphire_bulk_absorption}

The optical absorption of sapphire depends strongly on crystal purity,
impurity concentration, crystal-growth conditions, and post-growth thermal treatment.

Annealing at approximately 1450-1600$^\circ$C in oxygen or air can reduce optical absorption. This effect is associated with reduction of impurity and defect states which are responsible for absorption at the interferometer wavelength. The CHRONOS design should therefore specify optical absorption together with the material grade and post-growth treatment. 

\subsection{Thermal Lensing}
\label{subsec:thermal_lensing}

Optical absorption in the substrate produces a local heat load and can
generate both thermoelastic deformation of the mirror surface. It leads to thermo-refractive distortion of the transmitted optical wavefront. The magnitude of this thermal-lensing effect depends on the absorbed optical power, the temperature dependence of the refractive index, and the efficiency with which the deposited heat is transported through the substrate.

The thermo-refractive contribution can be represented schematically by
\begin{equation}
\delta s \propto \frac{\alpha_{\rm abs} \left( dn/dT \right)P}{\kappa},
\label{eq:test_mass_thermal_lens}
\end{equation}
where $\alpha_{\rm abs}$ denotes the optical absorption coefficient,
$dn/dT$ represents the temperature dependence of the refractive index,
$P$ is the relevant optical power, and $\kappa$ is the thermal conductivity of the substrate. This scaling shows that thermal lensing is enhanced by optical absorption and thermo-refractive coupling, while efficient thermal transport suppresses the temperature gradients responsible for wavefront distortion.

Sapphire is particularly favorable in this respect at cryogenic temperature. Its thermal conductivity becomes very large at low temperature, while the thermo-refractive coefficient is substantially reduced compared with its room-temperature value as summarized in Table~\ref{tab:sapphire_parameters}. As a result, heat deposited by optical absorption is efficiently distributed through the substrate, and the resulting refractive-index gradient is strongly suppressed. Thermal lensing is therefore expected to be much smaller under cryogenic operation than in a comparable room-temperature sapphire optic.


In addition to the wavefront distortion, a balance between the cooling power and the total heat load is relevant constraint as well. Explicitly writing, the absorbed power must remain below the heat that can be transferred out through the test mass and suspension,
\begin{equation}
P_{\rm abs} < P_{\rm cryo}^{\rm max}.
\end{equation}

The allowable absorption must therefore be evaluated for the actual optical and cryogenic configuration of CHRONOS. The circulating optical power and beam radius determine the spatial heat deposition, while the substrate thickness and bulk absorption determine the amount of heat generated inside the sapphire. Coating absorption provides an additional heat source near the optical surface. The resulting temperature distribution is then determined by the thermal
conductivity of the sapphire, the thermal conductance through the suspension, and the operating temperature.

The final thermal-lensing requirement should therefore be derived from a coupled optical and thermal model of the complete test mass rather than from the substrate absorption coefficient alone.
\section{Alternative Substrate: Crystalline Silicon}
\label{sec:silicon_test_mass}

Crystalline silicon is an important alternative substrate for future
cryogenic upgrades of CHRONOS. At low temperature, silicon can exhibit extremely small mechanical dissipation together with excellent thermal properties, and it is therefore an attractive candidate for a next-generation cryogenic test mass.

The main limitation for the present CHRONOS design is its optical
transparency. Silicon is not sufficiently transparent at the baseline wavelength of 1064~$\mathrm{nm}$, and its use would therefore require operation at a longer wavelength, typically around 1550~$\mathrm{nm}$ or beyond. Such a change would not be limited to the test-mass substrate itself. It would require a corresponding redesign of a substantial fraction of the optical system, including the laser source, frequency-control system, electro-optic modulation, optical isolation, beam splitters,
photodetectors, mode-cleaning optics, dielectric coatings, balanced-homodyne readout, and photon-pressure calibration system.

The substrate material and laser wavelength must therefore be regarded as a coupled design choice rather than as independent parameters. For the present CHRONOS configuration, sapphire operated with a 1064~$\mathrm{nm}$ interferometer remains the baseline because it combines
suitable cryogenic properties with mature optical technology. A silicon test mass operated with a longer-wavelength interferometer remains an important future upgrade path if further reduction of mechanical or coating-related losses becomes necessary.

\section{Baseline Design Strategy}
\label{sec:test_mass_baseline_strategy}

The CHRONOS test-mass design is determined by a coupled optimization of the mechanical geometry, bonded structure, optical requirements, and cryogenic performance. The dimensions of the individual sapphire blocks determine the total number of components required to construct the torsion bar, while the total bar length and mass distribution determine the moment of inertia about the suspension axis. At the same time, increasing the number of blocks introduces additional HCB
interfaces, and the bonding area and layer geometry determine how much
elastic energy is stored in the relatively high-loss bonding regions.

The preferred mechanical strategy is therefore to use the largest practical monolithic sapphire blocks, achieve the required torsion-bar length and moment of inertia with the minimum number of bonded interfaces, and minimize the elastic-energy participation of the HCB regions while preserving sufficient structural strength. The internal mechanical resonances must remain sufficiently separated from the relevant observation and control bands, and the bridge structure and
suspension interfaces must be included directly in the same mechanical
optimization rather than treated as independent attachments.

This mechanical design must simultaneously satisfy the optical and cryogenic requirements of the test mass. The final geometry must provide sufficient optical aperture and wavefront quality, while the sapphire substrate and dielectric coating must maintain sufficiently low absorption and mechanical loss. The absorbed optical power must remain compatible with the available cryogenic heat extraction, and the complete component must remain suitable for precision polishing, coating, handling, and integration with the cryogenic suspension.

The final test mass is therefore not obtained by optimizing a single
parameter such as mass, moment of inertia, or mechanical quality factor. Instead, the block dimensions, total bar length, moment of inertia, number and geometry of HCB interfaces, internal-mode structure, optical quality, and thermal transport must be optimized as a single integrated system.

\section{Development Program}
\label{sec:test_mass_development}

The development of the CHRONOS test mass will proceed from material
selection and component-level characterization toward validation of the complete cryogenic structure. The first stage is to determine the required mass distribution and moment of inertia from the mechanical design of the torsion-bar system and to combine this requirement with the minimum optical aperture defined by the interferometer beam geometry. On this basis, commercially realizable sapphire dimensions and crystallographic orientations will be surveyed, and candidate substrates will be characterized in terms of optical absorption, refractive-index homogeneity, birefringence, and mechanical loss.

The next stage is the experimental validation of the HCB mass-extension concept. Once representative bonded sapphire assemblies are fabricated, their mechanical quality factors will be measured with the ring-down method. These measurements will be compared with finite-element models based on the elastic-energy participation of the sapphire and HCB regions. The bonding area and interface geometry will then be optimized to reduce mechanical dissipation while maintaining sufficient structural strength. Mechanical-strength measurements will be performed in parallel with ring-down measurements in order to establish the relation between bond geometry, durability, and effective mechanical loss.

The suspension interface will subsequently be incorporated into the test-mass prototype. The sapphire bridge structure will be introduced into the mechanical model and experimentally characterized to determine its influence on the internal mode frequencies and losses. The assembled test mass will then be subjected to cryogenic ring-down
measurements and repeated thermal cycling in order to verify that its
mechanical quality factor, structural stability, and alignment are preserved during cooldown and warm-up.

In the final stage, the test mass will be integrated with the cryogenic
suspension and characterized as a complete suspended payload. The mechanical loss and alignment of the assembled structure will be verified under representative operating conditions. After the mechanical design has been validated, the final optical coating
will be applied and the completed mirror will be characterized for optical absorption, surface quality, and cryogenic mechanical performance.

This staged program is intended to ensure that the final CHRONOS test mass satisfies the mechanical, optical, and cryogenic requirements as a complete integrated structure rather than as a collection of independently optimized components.

\section{Summary}
\label{sec:test_mass_summary}

The CHRONOS test mass is not simply an optical mirror.
It is an integrated optical--mechanical--thermal structure that defines the rotational dynamics of the cryogenic torsion-bar detector.

The baseline material is monocrystalline sapphire operated at approximately 4~$\mathrm{K}$ with a 1064~$\mathrm{nm}$ interferometer.

A major engineering challenge is that the required dimensions of the
torsion bar may exceed those practically available from a single
optical-quality sapphire crystal. CHRONOS therefore adopts hydroxide-catalysis bonding as a candidate technology for extending the mass distribution using multiple sapphire components.

The central design trade-off is between increasing the usable dimensions and moment of inertia and, introducing additional mechanical dissipation at the bonded interfaces.

The current reference design consists of five sapphire blocks forming a
1.10~$\mathrm{m}$ long, 171.32~$\mathrm{kg}$ torsion bar with a moment of inertia of 19.86~$\mathrm{kg\,m^{2}}$.

Finite-element studies show that the mechanical loss introduced by HCB is controlled by the elastic-energy participation of the bonding region. This leads directly to the design strategy of maximizing monolithic sapphire block size, minimizing the number and area of bonds, and locating the bonds in mechanically favorable regions.

The bridge structure required for the crossed suspension geometry must also be regarded as an integral part of the test mass and included in the final mechanical optimization.

This integrated approach defines the development strategy for the
large-scale cryogenic test-mass mirror of CHRONOS.

%% file: Chapters/Coating.tex
\chapter{Optical Coatings}
\label{chap:optical-coatings}

\section{Introduction}

The optical coatings of the CHRONOS test masses are a central part of the
instrument design, which determines the optical response of the interferometer with a
contribution to the cryogenic heat load via optical absorption and produces Brownian coating noise through mechanical dissipation. These requirements must be treated together rather than optimized independently.  Coating
thermal noise is an important limitation in precision gravitational-wave
interferometers, while absorption becomes especially restrictive when absorbed
laser power must be extracted through a cryogenic suspension
\cite{ReidMartin2016,LIGOWhitePaper2016,Somiya2012}.

\begin{figure}[!htbp]
  \centering
  \includegraphics[width=0.48\linewidth]{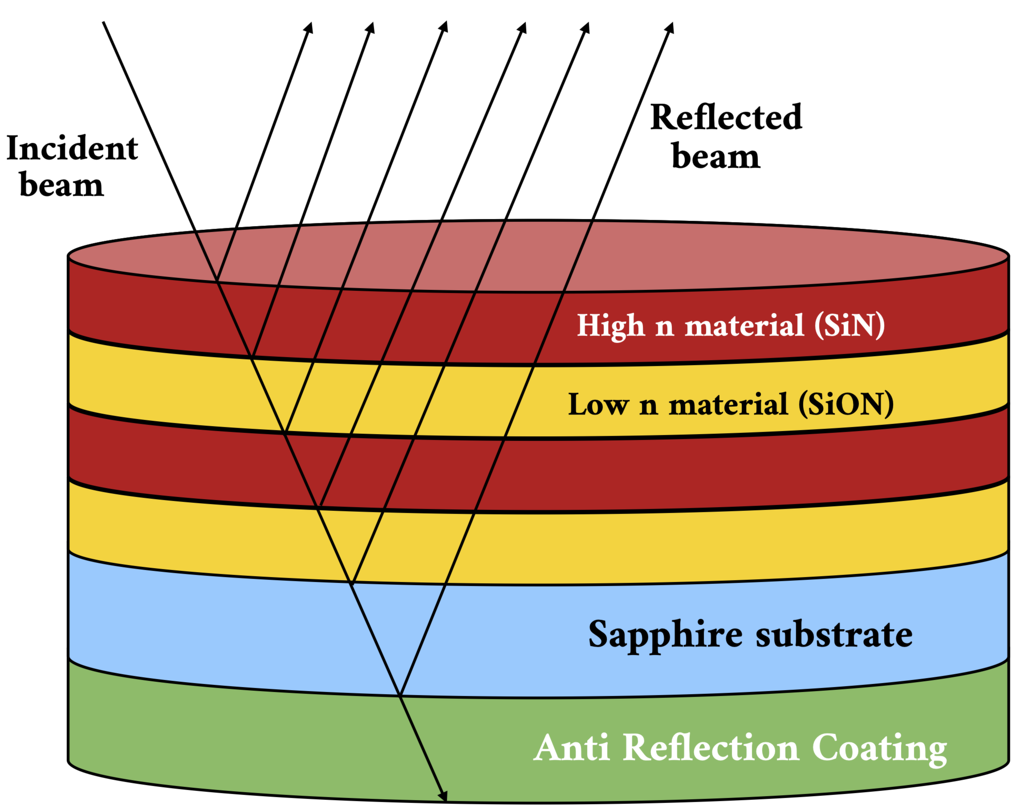}
  \caption{Baseline CHRONOS coating concept. A sapphire test mass carries an HR multilayer on the interferometer-facing surface and an AR coating on the opposite surface.}
  \label{fig:chronos-coating-concept}
\end{figure}

The CHRONOS baseline considered in this chapter is a sapphire substrate with a
high-reflectivity (HR) coating on the interferometer-facing surface and an
anti-reflection (AR) coating on the opposite surface.  The coating material
platform in development consists exclusively of silicon nitride
(SiN$_x$ or SiN) and silicon oxynitride (SiO$_x$N$_y$, abbreviated SiON) as SiN with relatively high-index and SiON in the material with relatively low-index as shown in Fig. \ref{fig:chronos-coating-concept}.
Both HR and AR designs will be based on measured, process-specific optical and
mechanical properties rather than on nominal bulk values.

The target wavelength for the baseline design is \SI{1064}{nm}; an extension to
\SI{1550}{nm} remains of interest.  The final numerical requirements for HR
transmission, AR reflectance, allowable absorption, scattering, and cryogenic
mechanical loss will be fixed by the CHRONOS optical and thermal
budgets at the system-level.  Until those budgets are frozen, values reported in should be read
as material-development results rather than as demonstrated performance of a
detector-ready coating.

\section{Coating Requirements}

An acceptable CHRONOS coating must provide the required HR reflectivity and AR
suppression at the operating wavelength and angle of incidence while
maintaining low absorptance, low wide-angle scatter, and low mechanical loss at
the CHRONOS operating temperature and in the relevant frequency band.  It must
also exhibit acceptable residual stress, adhesion, surface figure, roughness,
and thickness uniformity on sapphire, together with stability under annealing,
handling, vacuum exposure, and repeated room--temperature-cryogenic thermal
cycling.

In a simplified scaling, the coating contribution to the displacement-noise
power spectral density has the form
\begin{equation}
  S_x^{\mathrm{coat}}(f)
  \propto
  \frac{k_{\mathrm B}T}{f w^2}\,\phi_{\mathrm{eff}}(T,f),
  \label{eq:coating-noise-scaling}
\end{equation}
where $T$ is the temperature, $w$ is the beam radius, and
$\phi_{\mathrm{eff}}$ is an effective loss angle that depends on layer
thicknesses, elastic properties, substrate properties, and the distribution of strain-energy.  Equation~\eqref{eq:coating-noise-scaling} is only a scaling
relation; quantitative CHRONOS predictions will use the full multilayer elastic
model.  Cryogenic operation reduces the explicit temperature factor but does
not guaranty a proportional noise improvement because amorphous films can
exhibit temperature-dependent dissipation peaks.  Mechanical loss must
therefore be measured rather than extrapolated from room temperature.

Absorption is a system-level constraint.  It can cause local heating
and thermo-optical distortion and, at cryogenic temperature, adds directly to the
heat that must be removed from the test mass.  The HR and AR coatings must
therefore be evaluated together with sapphire bulk absorption and interface
losses.

\section{SiN--SiON Material Platform}

\subsection{Silicon nitride}

SiN$_x$ is attractive because its refractive index, optical gap, hydrogen
content, defect population, and residual stress can be adjusted through the
deposition parameters and stoichiometry \cite{Pan2018,Tsai2022,Beliaev2022}.
This flexibility is useful, but it also means that ``SiN'' does not imply a
single set of material constants.  Si-rich and N-rich films have markedly different optical losses even when deposited by the same general
method. The conditions during the process must consequently be recorded together with the
composition and measured properties.

Whereas hydrogen can passivate under-coordinated sites, but a large population of
hydrogen-terminated bonds may also accompany network termination, voids, weak
bonds, or structural disorder \cite{LanfordRand1978,Tsu1986,Mei2006}.  Neither hydrogen content nor optical-gap position is therefore expected to be a unique
predictor of near-IR absorption.

\subsection{Silicon oxynitride}

SiON provides a compositionally tunable lower-index partner. The replacment of a
fraction of N by O changes the relative populations of Si--N and Si--O bonding
units, allowing the refractive index to vary between nitride- and
oxide-like values. The dependence is generally non-linear and process
dependent; a linear interpolation of refractive index should not be assumed
without measurement. SiON has been demonstrated as a candidate low-index
coating material for cryogenic gravitational-wave mirrors
\cite{Chao2023,Philipp1973,Albertin2005}.

Using related Si-based networks for both index components may simplify process integration and offer useful control of interfaces and stress. Nevertheless, the actual index contrast, absorption, mechanical loss, and
thermal properties have been established for the specific SiN and SiON recipes
used by CHRONOS.

\section{Deposition Strategy}

\begin{figure}[t]
  \centering
  \begin{minipage}[t]{0.40\linewidth}
    \centering
    \includegraphics[width=\linewidth]{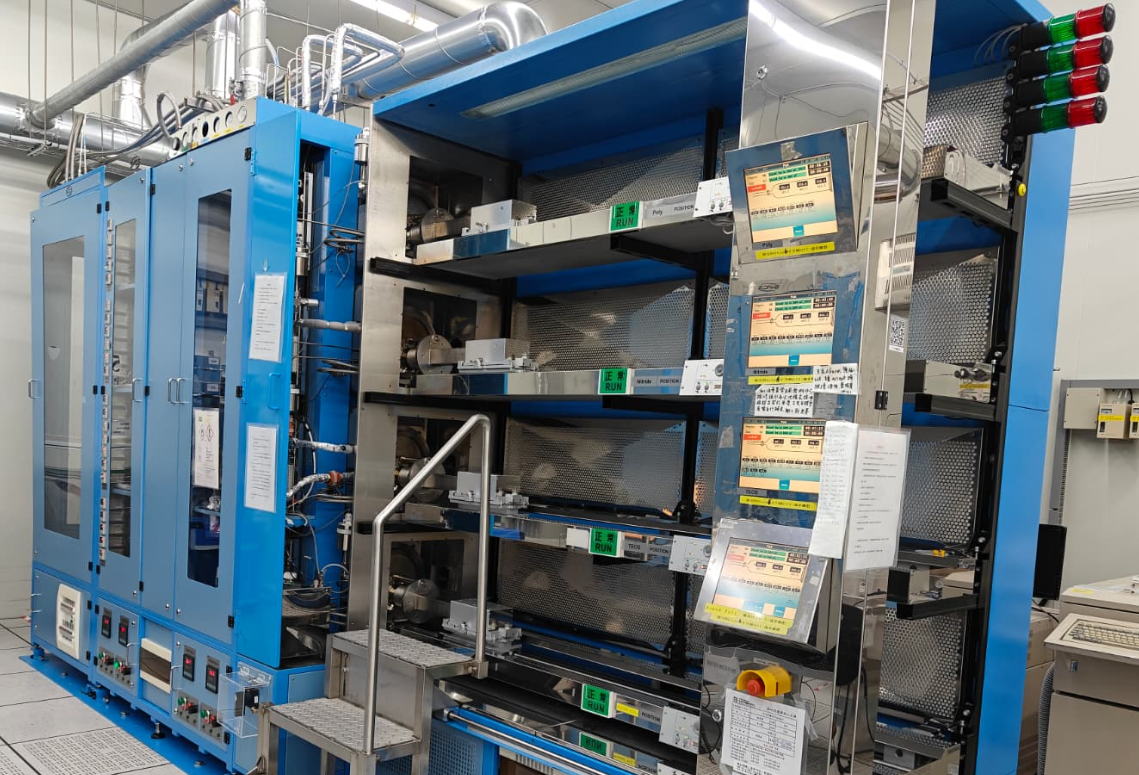}
  \end{minipage}\hfill
  \begin{minipage}[t]{0.56\linewidth}
    \centering
    \includegraphics[width=\linewidth]{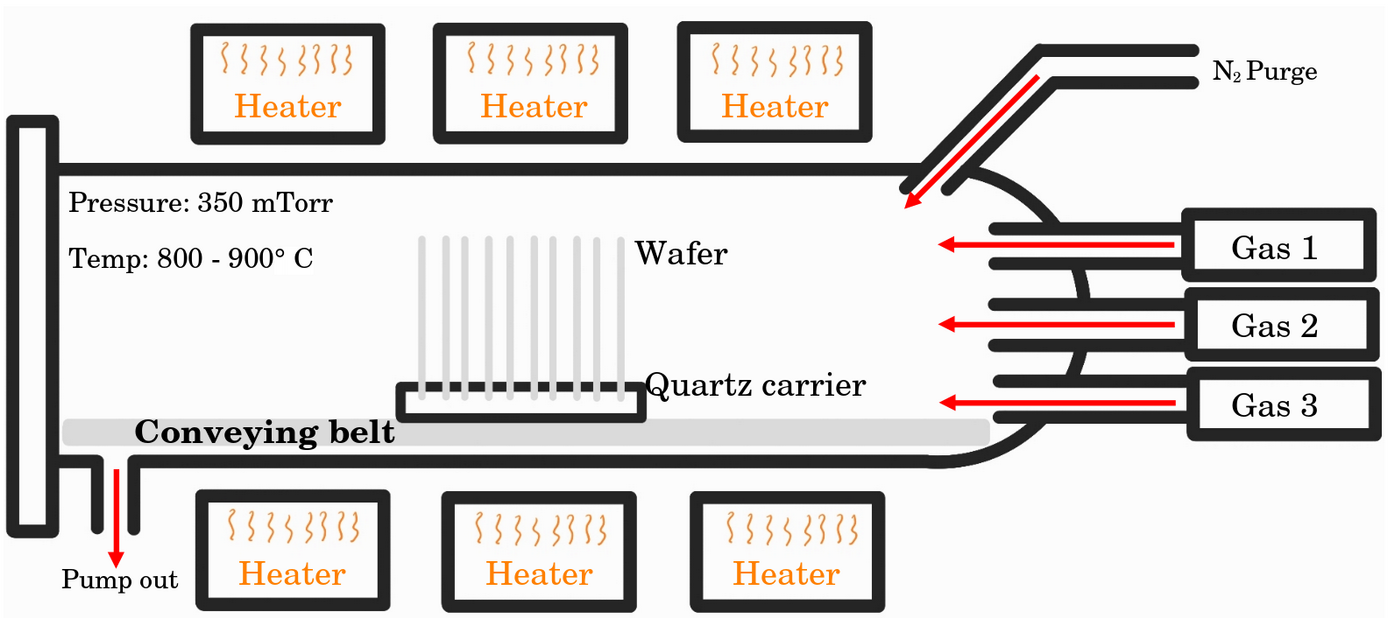}
  \end{minipage}
  \caption{LPCVD setup and reactor design: photograph of the reactor and
  schematic of the horizontal hot-wall system used for film deposition.}
  \label{fig:cvd-process}
\end{figure}

Chemical-vapor-deposition (CVD) routes under investigation includes plasma -
enhanced
CVD (PECVD) and low-pressure CVD (LPCVD) as illustrated in Fig. \ref{fig:cvd-process}.  PECVD permits comparatively low
substrate temperatures and broad compositional control.  LPCVD uses thermally
activated surface reactions at reduced pressure and can produce dense,
uniform films over multiple wafers.  Typical LPCVD SiN growth from
dichlorosilane and ammonia is represented schematically by
\begin{equation}
  3\,\mathrm{SiH_2Cl_2}+4\,\mathrm{NH_3}
  \longrightarrow
  \mathrm{Si_3N_4}+6\,\mathrm{HCl}+6\,\mathrm{H_2}.
\end{equation}
The actual film can depart from the stoichiometric Si$_3$N$_4$ and can contain
hydrogen; therefore, it is expressed as SiN$_x$(:H) unless its composition has been
independently measured.

The primary surface-processing steps are precursor transport, adsorption,
surface reaction, and desorption of volatile by-products
\cite{Peev1990,Kaloyeros2017}.  Temperature, total pressure, precursor ratio,
flow, residence time, and wafer position affect growth rate, composition,
stress, and optical properties.  LPCVD should not be described a priori as the
``best'' process: its higher thermal budget, tensile stress, precursor
chemistry, and compatibility with polished sapphire all require qualification.
The deposition route will be selected from the demonstrated optical, mechanical, and cryogenic performance at the coating-level.

\begin{figure}[t]
  \centering
  \includegraphics[width=0.90\linewidth]{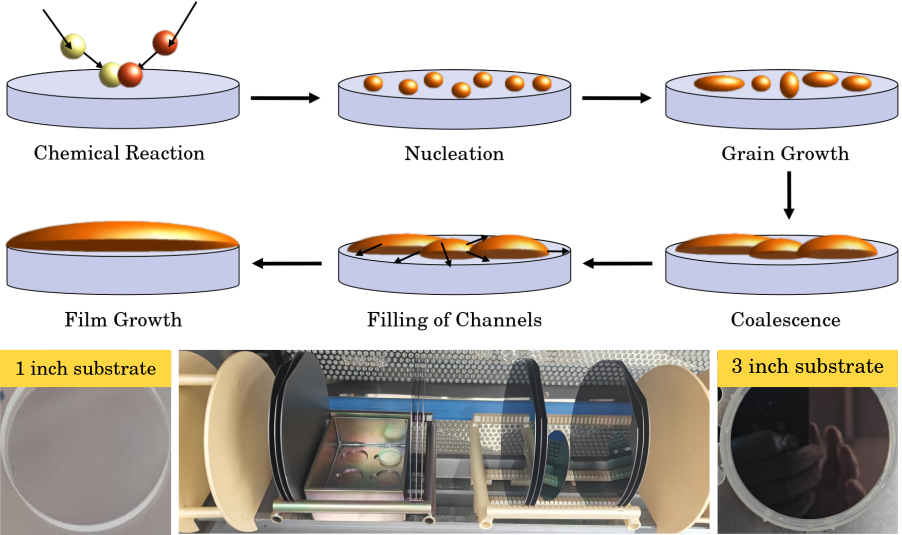}
  \caption{Illustration of the LPCVD growth sequence, including precursor
  transport, adsorption, thermally activated surface reactions, nucleation,
  lateral growth, and formation of a continuous film.}
  \label{fig:lpcvd-growth}
\end{figure}

\section[Optical Design with the Transfer-Matrix Method]{Optical Design with the\protect\linebreak Transfer-Matrix Method}

\subsection{Single-interface reflection}

At normal incidence, the amplitude reflection coefficient between two
non-magnetic media is
\begin{equation}
  r_{12}=\frac{\tilde n_1-\tilde n_2}{\tilde n_1+\tilde n_2},
  \qquad R_{12}=|r_{12}|^2,
  \label{eq:fresnel}
\end{equation}
where $\tilde n=n+i\kappa$ is the complex refractive index. A bare
sapphire--vacuum surface therefore produces a non-negligible Fresnel reflection;
the AR coating is required to suppress that reflection and the associated
parasitic interferometers and scattered-light paths.

\subsection{Quarter-wave starting point}

For a conventional Bragg reflector at vacuum wavelength $\lambda_0$, the
starting design uses layers satisfying
\begin{equation}
  n_i d_i=\frac{\lambda_0}{4},
  \qquad
  \delta_i=\frac{2\pi}{\lambda_0}n_i d_i=\frac{\pi}{2}.
  \label{eq:quarter-wave}
\end{equation}
Alternating high- and low-index layers then add their interface reflections
constructively.  The index contrast controls the number of pairs needed for a
specified transmission.  The familiar closed-form quarter-wave reflectance
depends on the incident index, substrate index, pair ordering, and terminating
layer; it is therefore safer to use the transfer-matrix calculation below than
to apply a substrate-independent approximation.

\subsection{Characteristic matrices}

At normal incidence, the characteristic matrix of the layer $i$ is
\begin{equation}
  \mathbf M_i=
  \begin{pmatrix}
    \cos\delta_i & i\sin\delta_i/\tilde n_i\\
    i\tilde n_i\sin\delta_i & \cos\delta_i
  \end{pmatrix},
  \qquad
  \delta_i=\frac{2\pi}{\lambda}\tilde n_i d_i .
  \label{eq:layer-matrix}
\end{equation}
For $N$ layers in physical order from the incident medium to the substrate,
\begin{equation}
  \mathbf M=\mathbf M_1\mathbf M_2\cdots\mathbf M_N
  =\begin{pmatrix}A&B\\C&D\end{pmatrix}.
\end{equation}
If the incident medium and substrate have indices $n_0$ and $\tilde n_s$, the
input optical admittance and amplitude reflection coefficient are
\begin{equation}
  Y=\frac{C+D\tilde n_s}{A+B\tilde n_s},
  \qquad
  r=\frac{n_0-Y}{n_0+Y},
  \qquad R=|r|^2.
  \label{eq:tmm-reflection}
\end{equation}
The same calculation yields transmission and the field in every layer.  At
non-normal incidence, separate TE and TM admittances and angle-dependent phase
thicknesses will be used.

The local absorption density is proportional to the imaginary permittivity and
the local field intensity.  For a planar stack,
\begin{equation}
  \frac{P_{\mathrm{abs}}}{S}
  =\frac{\omega\epsilon_0}{2}
   \int \operatorname{Im}[\epsilon_r(z)]\,|E(z)|^2\,\mathrm dz .
  \label{eq:field-weighted-absorption}
\end{equation}
Thus, the absorption of a layer embedded in an HR stack is not obtained by
linearly scaling the absorptance of an isolated film by thickness.  Its
contribution depends on $n$, $\kappa$, thickness, wavelength, surrounding
layers, substrate index, and its position in the standing-wave field.  This is
why measured single-layer data must first be converted, with an optical model,
to material parameters and then propagated through the complete HR or AR stack.

The quarter-wave design is only a starting point.  The CHRONOS optimization may
depart from quarter-wave optical thicknesses to reduce field-weighted
absorption, total coating thickness, mechanical loss, or thermo-optic coupling
while retaining the specified HR transmission and AR reflectance.

\subsection{SiN--SiON stack architecture}

The nominal HR architecture is written as
\begin{equation}
  \begin{aligned}
    &\mathrm{vacuum} \\
    &\quad \downarrow \\
    &(HL)^N \\
    &\quad \downarrow \\
    &\mathrm{sapphire}
  \end{aligned}
  \label{eq:nominal-hr-stack}
\end{equation}
where $H$ denotes a relatively high-index SiN layer, $L$ denotes a relatively
low-index SiON layer, and $N$ is the number of layer pairs.  The actual
structure may include a matching layer at the sapphire interface and a cap
layer at the vacuum surface.  Pair ordering, the terminating material, and cap
thickness affect reflectivity, surface electric field, environmental stability,
and absorption.  Equation~\eqref{eq:nominal-hr-stack} therefore represents the
baseline architecture rather than a finalized layer sequence.

For an ideal lossless quarter-wave stack with the ordering $(HL)^N$, the
optical-admittance transformation gives
\begin{equation}
  Y_{\mathrm{in}}=n_s\left(\frac{n_H}{n_L}\right)^{2N},
  \qquad
  R_N=\left|\frac{n_0-Y_{\mathrm{in}}}{n_0+Y_{\mathrm{in}}}\right|^2,
  \label{eq:quarter-wave-stack-reflectance}
\end{equation}
where $n_0$ and $n_s$ are the incident-medium and sapphire indices.  Reversing
the pair ordering reverses the admittance ratio.  The required pair count thus
depends not only on index contrast but also on the substrate, layer ordering,
termination, and specified HR transmission.  The real CHRONOS stack will use
complex indices and the complete transfer-matrix model instead of this ideal
expression.

At \SI{1064}{nm}, the measured SiN indices from the existing campaign
correspond to quarter-wave physical thicknesses of approximately
\SI{114.6}{nm} to \SI{135.8}{nm}.  This range is illustrative.  The SiON
quarter-wave thickness, index contrast, pair count, and total stack thickness
will be calculated after the FY2026 SiON index measurements.  A literature
value will not be substituted as though it were a measured CHRONOS SiON recipe.

The standing-wave intensity is largest near the incident surface and generally
decays into an HR stack.  The same material can consequently make a different
absorption contribution depending on its depth.  The optimized structure may
use non-quarter-wave SiN layers, especially in the high-field region, to reduce
the optical energy stored in the more absorbing material.  The SiON thicknesses
and, if necessary, pair count must then compensate for that change so that the
HR transmission remains within specification.  Numerical optimization will
minimize field-weighted absorption and effective mechanical loss subject to
constraints on reflection, transmission, total thickness, stress, and
manufacturable layer thickness.

The AR coating will be treated as a separate optical-matching stack rather than
as a shortened HR reflector.  A single-layer quarter-wave AR solution generally
requires an index close to $\sqrt{n_0n_s}$.  If the available SiON index does
not provide the required match, a multilayer SiN--SiON design will be used.
Its optimization will include the sapphire index, birefringence, crystal
orientation, polarization, incidence angle, and required wavelength range.

Repeated interfaces introduce effects absent from isolated films.  They can
increase scatter or interface absorption, while accumulated stress can curve
the optic or limit adhesion.  Layer-thickness errors can shift the HR stop band
and raise the residual AR reflectance.  Complete stack demonstrators are
therefore required after single-layer constants are known.  Thickness and index
witnesses distributed across the deposition area will allow measured spectral
reflection and transmission to be compared with an as-built transfer-matrix
model rather than only with the nominal design.

\section{Single-Layer SiN Characterization}

The current LPCVD SiN campaign combines spectroscopic ellipsometry,
photothermal common-path interferometry (PCI), and Fourier-transform infrared
(FTIR) spectroscopy.  These methods probe complementary quantities and should
not be treated as interchangeable.

\subsection{Summary of measurements completed to date}

Table~\ref{tab:existing-measurement-status} separates quantities already
measured in the present LPCVD SiN campaign from measurements planned for
FY2026.  The completed data apply to single SiN films on the substrates used in
the characterization campaign; they are not measurements of a complete
SiN--SiON HR or AR stack on a CHRONOS sapphire test mass.

\begin{table}[t]
  \centering
  \caption{Status of the coating measurements discussed in this chapter.}
  \label{tab:existing-measurement-status}
  \begin{tabular}{p{0.24\linewidth}p{0.30\linewidth}p{0.34\linewidth}}
    \toprule
    Material or structure & Completed measurements & Not yet established in the
    present data set\\
    \midrule
    LPCVD SiN single layers & Thickness and $n(\lambda)$ by ellipsometry;
    coated-sample PCI absorptance at 1064, 1550, and 1950 nm; selected-sample
    N--H stretching data by FTIR; Tauc, Cody, and Urbach parameters &
    Intrinsic $\kappa$ extracted from PCI by TMM; verified elemental and
    hydrogen composition; cryogenic mechanical loss; full thermal-property
    data\\
    SiON single layers & Literature establishes its relevance as a low-index
    material, but no CHRONOS SiON measurement is claimed here & Systematic
    CHRONOS ellipsometry, PCI, FTIR, composition, stress, and mechanical-loss
    measurements scheduled for FY2026\\
    SiN--SiON HR/AR stacks on sapphire & Optical design framework established &
    Fabricated-stack reflection, transmission, absorption, scatter, stress,
    mechanical loss, and cryogenic cycling\\
    \bottomrule
  \end{tabular}
\end{table}

\begin{figure}[!htbp]
  \centering
  \includegraphics[width=0.50\linewidth]{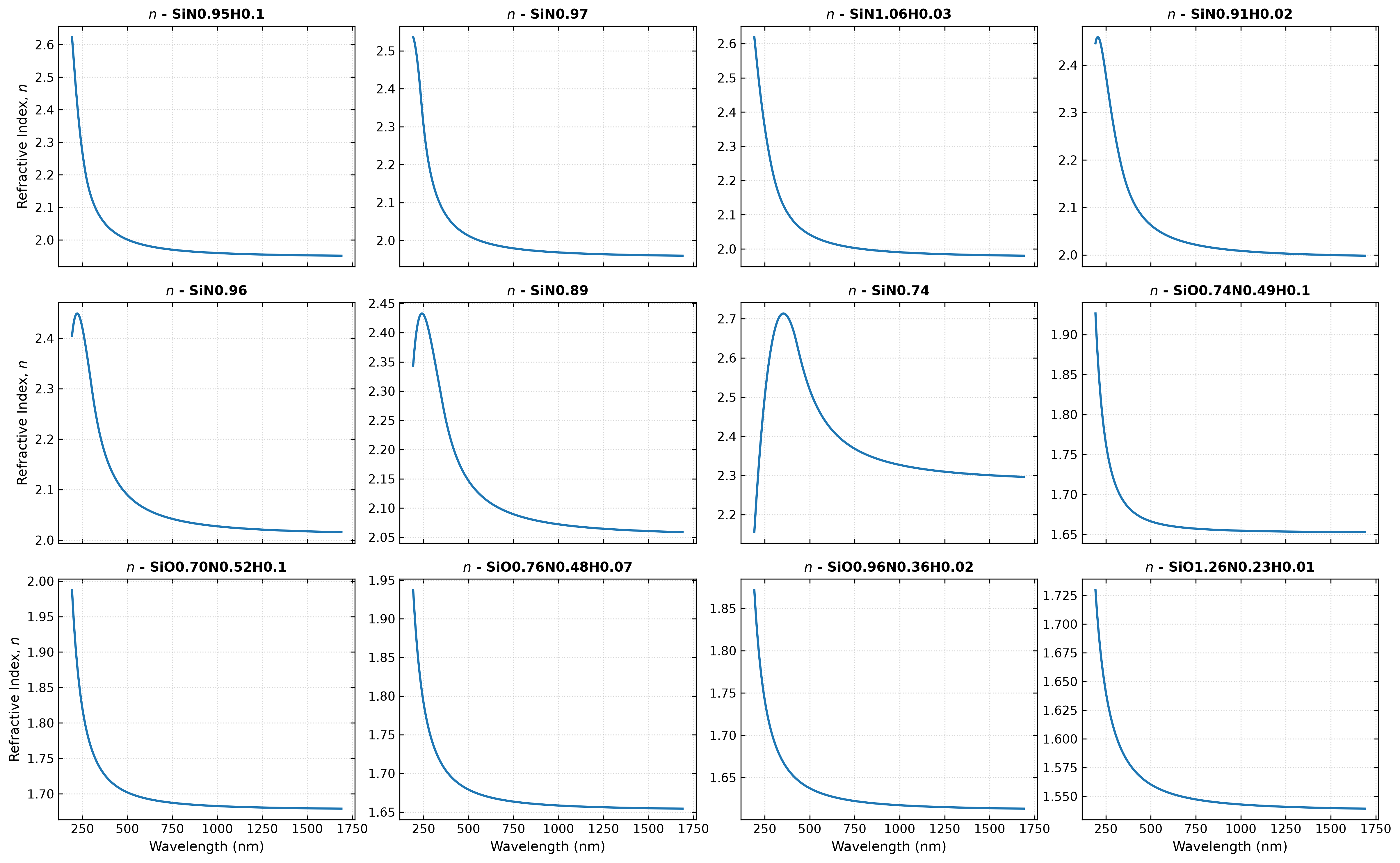}%
  \hfill
  \includegraphics[width=0.50\linewidth]{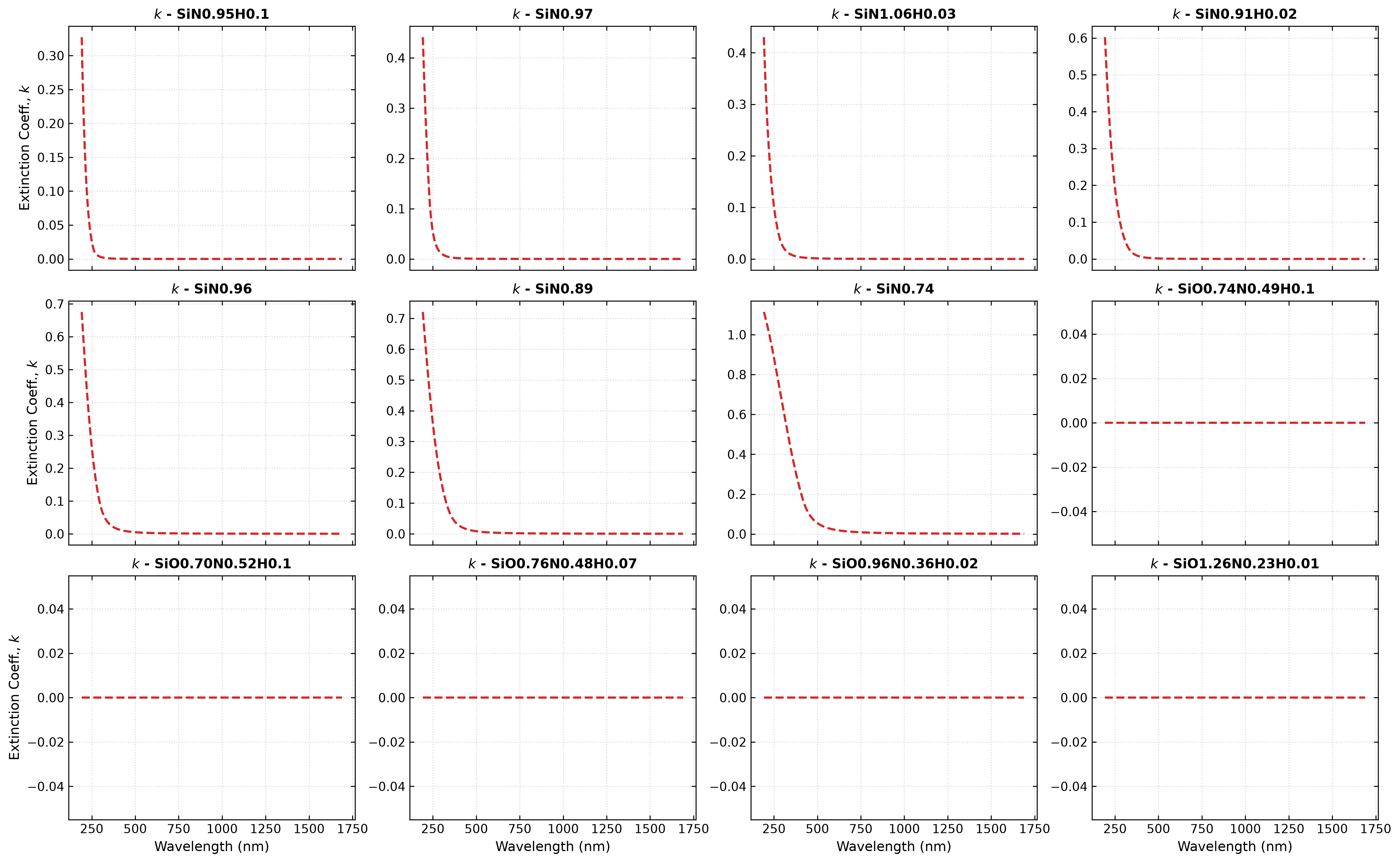}
  \caption{Spectroscopic ellipsometry measurements and extracted optical constants ($n$ and $k$) for the investigated thin-film samples.}
  \label{fig:ellipsometry-results}
\end{figure}

\subsection{Spectroscopic ellipsometry}

Ellipsometry determines film thickness and a model-dependent complex refractive
index,
\begin{equation}
  \tilde n(\lambda)=n(\lambda)+i\kappa(\lambda),
  \qquad
  \alpha(\lambda)=\frac{4\pi\kappa(\lambda)}{\lambda}.
\end{equation}
For the seven reported LPCVD SiN films, $n(1064\,\mathrm{nm})$ ranged from
1.95839 to 2.32114, with thicknesses from \SI{89.3}{nm} to
\SI{425.6}{nm}.  This spread demonstrates useful index tunability, but also the
need for recipe-specific dispersion data.  The very small statistical fit
errors reported by the ellipsometric model do not include systematic errors
from the choice of the model, surface roughness, fitting range or assumed substrate
properties \cite{JellisonModine1996}.

\begin{table}[t]
  \centering
  \caption{Measured thickness and refractive index of the LPCVD SiN and SiON films.
  Thickness uncertainties represent spatial variation among measured
  positions. Refractive-index uncertainties shown in the source are formal
  fit errors of approximately $10^{-5}$ and exclude systematic uncertainty \cite{KumarInPrep}.}
  \label{tab:sin-ellipsometry}
  \begin{tabular}{lrrrr}
    \toprule
    Sample & $d$ (nm) & $n_{1064}$ & $n_{1550}$ & $n_{1950}$\\
    \midrule
    SiN$_{0.95}$H$_{0.10}$ & 251.8 & 1.95839 & 1.95244 & 1.95057\\
    SiN$_{0.97}$H$_{0.07}$ & 283.2 & 1.96703 & 1.96077 & 1.95881\\
    SiN$_{1.06}$H$_{0.03}$ & 425.6 & 1.98844 & 1.98114 & 1.97886\\
    SiN$_{0.91}$H$_{0.02}$ & 278.5 & 2.00662 & 1.99924 & 1.99695\\
    SiN$_{0.96}$           & 257.5 & 2.02525 & 2.01684 & 2.01424\\
    SiN$_{0.89}$           & 192.4 & 2.06966 & 2.05995 & 2.05695\\
    SiN$_{0.74}$           & 89.3  & 2.32114 & 2.29912 & 2.29249\\
    SiO$_{0.74}$N$_{0.49}$H$_{0.1}$  & 261.4 & 1.65429 & 1.65295 & 1.65255\\
    SiO$_{0.7}$N$_{0.52}$H$_{0.1}$   & 309.8 & 1.68207 & 1.67944 & 1.67944\\
    SiO$_{0.76}$N$_{0.48}$H$_{0.07}$ & 290.8 & 1.65769 & 1.65476 & 1.65385\\
    SiO$_{0.96}$N$_{0.36}$H$_{0.02}$ & 141.1 & 1.61655 & 1.61363 & 1.61271\\
    SiO$_{1.26}$N$_{0.23}$H$_{0.01}$ & 182.8 & 1.54222 & 1.53959 & 1.53876\\
    \bottomrule
  \end{tabular}
\end{table}

Tauc, Cody, and Urbach analyzes were applied to the ellipsometric absorption
edge \cite{KumarInPrep, Tauc1966,Cody1981,Urbach1953}.  The reported SiN Tauc gaps span
approximately \SI{2.26} -\SI{4.86}{eV}, while the fitted Urbach energies span
approximately \SI{0.37} -\SI{0.50}{eV} \cite{KumarInPrep}.  The trends are not one-to-one with the
near-infrared PCI results; neither a large fitted gap nor a small Urbach energy
alone guaranties low absorption at the laser wavelength.

\begin{table}[t]
  \centering
  \caption{Optical-gap and Urbach-energy parameters fitted to the
  ellipsometric absorption spectra.  Quoted uncertainties are statistical fit
  errors and exclude model and fitting-range systematics \cite{KumarInPrep}.}
  \label{tab:sin-gap-urbach}
  \begin{tabular}{lrrrr}
    \toprule
    Sample & $E_g^{\mathrm{ideal}}$ (eV) & $E_g^{\mathrm{Tauc}}$ (eV) &
    $E_g^{\mathrm{Cody}}$ (eV) & $E_U$ (meV)\\
    \midrule
    SiN$_{0.95}$H$_{0.10}$ & 4.146 & $4.8585\pm0.0115$ & $4.2652\pm0.0047$ & $412.05\pm0.07$\\
    SiN$_{0.97}$H$_{0.07}$ & 4.179 & $4.5382\pm0.0005$ & $4.3330\pm0.0059$ & $451.96\pm0.08$\\
    SiN$_{1.06}$H$_{0.03}$ & 4.323 & $4.3133\pm0.0112$ & $3.5111\pm0.0032$ & $466.69\pm0.13$\\
    SiN$_{0.91}$H$_{0.02}$ & 4.077 & $3.9767\pm0.0015$ & $3.1250\pm0.0017$ & $373.82\pm0.07$\\
    SiN$_{0.96}$           & 4.163 & $3.6139\pm0.0004$ & $2.7919\pm0.0096$ & $498.32\pm0.12$\\
    SiN$_{0.89}$           & 4.042 & $3.2112\pm0.0003$ & $2.4959\pm0.0036$ & $449.46\pm0.09$\\
    SiN$_{0.74}$           & 3.756 & $2.2628\pm0.0013$ & $2.0748\pm0.0022$ & $372.51\pm0.07$\\
    \bottomrule
  \end{tabular}
\end{table}

\subsection{Photothermal common-path interferometry}

PCI measures the photothermal response produced by absorbed pump power and can
reach ppm sensitivity \cite{Alexandrovski2009,WelschRistau1995}.
The quantity reported is the absorptance of the complete coated-substrate sample
under the measurement conditions.  It is not an intrinsic bulk absorption
coefficient of SiN, and it can include substrate and interface contributions as
well as thin-film interference effects.

\begin{figure}[t]
  \centering
  \includegraphics[width=0.50\linewidth]{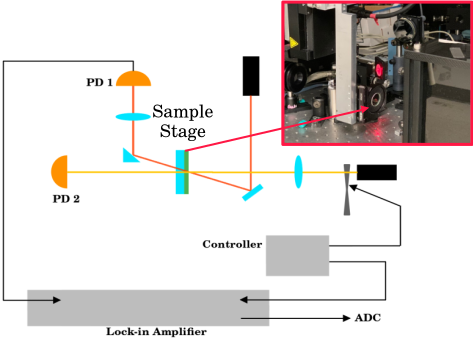}
  \caption{Photothermal common-path interferometry system used to measure the
  absorptance of the deposited film--substrate samples.}
  \label{fig:pci-setup}
\end{figure}

The results of the reported single-layer SiN sample are summarized in
Table~\ref{tab:sin-pci}.  At \SI{1064}{nm}, the absorptance reaches 10.2--682.6 ppm. While $\mathrm{SiN}$ films achieved remarkably low absorption at \SI{1064}{nm} lowering at \SI{10.2}{ppm} for the nominal $\mathrm{SiN}_{0.91}\mathrm{H}_{0.02}$ sample, so by introducing oxygen into the matrix further enhanced optical performance. Shifting to $\mathrm{SiON}$ dropped the \SI{1064}{nm} absorptance into an even lower regime of 1.27--4.54~ppm, reaching a minimum of just \SI{1.27}{ppm} with nominal $\mathrm{SiO}_{0.74}\mathrm{N}_{0.49}\mathrm{H}_{0.1}$. This sudden reduction shows oxygen incorporation mitigating the residual absorption, positioning $\mathrm{SiON}$ as a superior candidate for ultra-low-loss applications. These are directly measured sample absorptances, not predictions for a quarter-wave layer or a complete HR coating.

\begin{table}[t]
  \centering
  \caption{Reported PCI absorptance of the LPCVD SiN and SiON film--substrate samples at 1064 nm. Compositions are nominal labels and require independent verification.}
  \label{tab:sin-pci}
  \begin{tabular}{lrr}
    \toprule
    Sample & $d$ (nm) & $A_{1064}$ (ppm)\\
    \midrule
    SiN$_{0.95}$H$_{0.10}$ & 251.8 & 134.1\\
    SiN$_{0.97}$H$_{0.07}$ & 283.2 & 192.5\\
    SiN$_{1.06}$H$_{0.03}$ & 425.6 & 11.8\\
    SiN$_{0.91}$H$_{0.02}$ & 278.5 & 10.2\\
    SiN$_{0.96}$           & 257.5 & 17.6\\
    SiN$_{0.89}$           & 192.4 & 77.2\\
    SiN$_{0.74}$           & 89.3  & 682.6\\
    SiO$_{0.74}$N$_{0.49}$H$_{0.1}$  & 261.4 & 1.27\\
    SiO$_{0.7}$N$_{0.52}$H$_{0.1}$   & 309.8 & 4.54\\
    SiO$_{0.76}$N$_{0.48}$H$_{0.07}$ & 290.8 & 2.28\\
    SiO$_{0.96}$N$_{0.36}$H$_{0.02}$ & 141.1 & 2.19\\
    SiO$_{1.26}$N$_{0.23}$H$_{0.01}$ & 182.8 & 1.89\\
    \bottomrule
  \end{tabular}
\end{table}

For CHRONOS, these data will be followed by repeated spatial measurements,
uncoated-substrate controls, and transfer-matrix fitting of $\kappa$ at each
wavelength.  That procedure is needed before predicting field-weighted
absorption in realistic SiN--SiON stacks.

\subsection{Correlation between optical absorption and bandgap parameters}

\begin{figure}[htbp]
    \centering
    \includegraphics[width=0.64\linewidth]{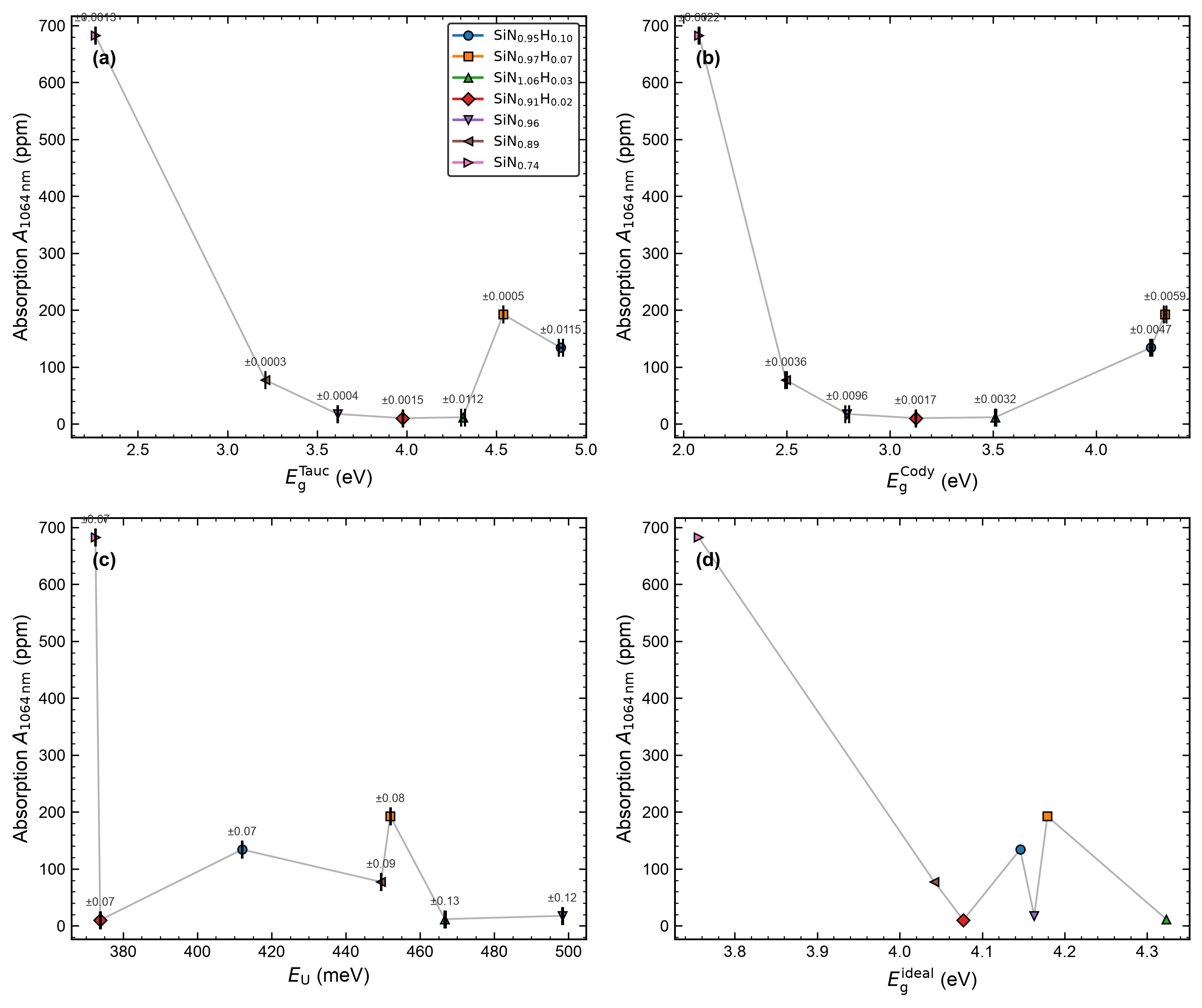}
    \caption{Sub-bandgap optical absorption at $\lambda = 1064\mathrm{~nm}$ ($A_{1064\mathrm{\,nm}}$) as a function of (a) $E_{\mathrm{g}}^{\mathrm{Tauc}}$, (b) $E_{\mathrm{g}}^{\mathrm{Cody}}$, (c) $E_{\mathrm{U}}$, and (d) $E_{\mathrm{g}}^{\mathrm{ideal}}$.}
    \label{fig:absorption_bandgap_correlations}
\end{figure}

For evaluating near IR optical loss mechanisms, the absorption measured at 1064~nm was correlated with the bandgap parameters extracted for the thin films $\mathrm{SiN}_x\mathrm{H}_y$, as shown in Fig.~\ref{fig:absorption_bandgap_correlations}.

The absorption exhibits a non-monotonic trend across the sample set. The minimum loss ($\sim 10\text{--}18\text{~ppm}$) occurs at intermediate band gaps ($E_{\mathrm{g}}^{\mathrm{Tauc}} \approx 3.98\text{--}4.31\text{~eV}$)\cite{KumarInPrep}. In unhydrogenated, Si-rich films like $\mathrm{SiN}_{0.74}$, the sharp drop in bandgap ($E_{\mathrm{g}}^{\mathrm{Tauc}} = 2.26\text{~eV}$) correlates with a drastic increase in absorption ($682.6\text{~ppm}$) due to the mid-gap states introduced by silicon dangling bonds ($\mathrm{Si}^0$) and Si--Si bonds \cite{KumarInPrep, MolinaRuiz2025}. Conversely, elevated absorption in hydrogenated films with wider gaps (e.g., $\mathrm{SiN}_{0.97}\mathrm{H}_{0.07}$ at $192.5\text{~ppm}$ \cite{KumarInPrep}) stems primarily from bond-vibrational overtones or hydrogen-induced defects \cite{Amato2025}. The absence of a linear scaling with Urbach energy ($E_{\mathrm{U}}$) further indicates that the optical loss at $1064\text{~nm}$ is driven by discrete deep-level state absorption rather than simple band-edge tail disorder \cite{Urbach1953}.

\subsection{FTIR spectroscopy and hydrogen-related bonding}

\begin{figure}[t]
  \centering
  \begin{minipage}[t]{0.48\linewidth}
    \centering
    \includegraphics[width=\linewidth]{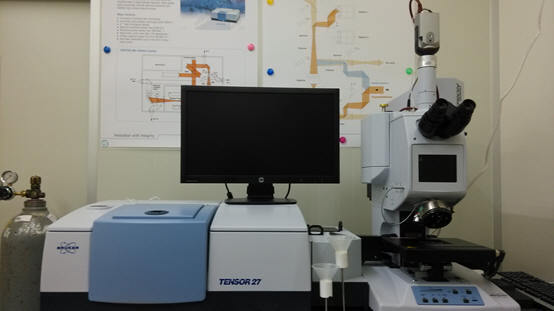}
  \end{minipage}\hfill
  \begin{minipage}[t]{0.48\linewidth}
    \centering
    \includegraphics[width=\linewidth]{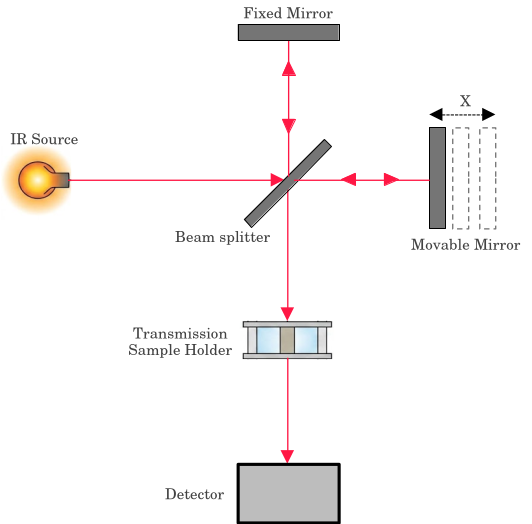}
  \end{minipage}
  \caption{FTIR measurement system: instrument photograph and schematic of the
  transmission-mode Michelson-interferometer configuration.}
  \label{fig:ftir-system}
\end{figure}

Transmission FTIR shown in Fig.~\ref{fig:ftir-system} probes vibrational absorption associated with local bonding. After substrate correction and spectral fitting, the N--H stretching band near \SI{3323} - \SI{3332}{cm^{-1}} provides a comparative measure of N--H bond
population.  Within the measured subset, films with the largest inferred N--H
bond densities also had comparatively large PCI absorptance.  The relationship
was not unique: films with similar gap- or hydrogen-related metrics could still
have different near-infrared loss.

The combined result is therefore multivariable.  Near-infrared loss reflects
the nominal composition, optical-gap position, Urbach-tail width, hydrogen bonding, local disorder, and optical-field distribution.  Independent composition and
hydrogen measurements such as, X-ray photoelectron spectroscopy,
Rutherford backscattering, elastic-recoil detection, or nuclear-reaction
analysis are required before nominal chemical labels are interpreted as
quantitative stoichiometry.

\begin{table}[t]
  \centering
  \caption{Completed FTIR N--H measurements for the five-film subset.  Bond
  densities were inferred using a peak cross section of
  $5.3\times10^{-18}\,\mathrm{cm^2}$ and should be interpreted comparatively \cite{KumarInPrep}.}
  \label{tab:sin-ftir}
  \begin{tabular}{lrr}
    \toprule
    Sample & N--H peak (cm$^{-1}$) &
    $N_{\mathrm{N-H}}$ ($10^{21}\,\mathrm{cm^{-3}}$)\\
    \midrule
    SiN$_{0.91}$H$_{0.02}$ & 3328.66 & 2.704\\
    SiN$_{0.95}$H$_{0.10}$ & 3322.80 & 7.261\\
    SiN$_{0.96}$           & 3326.32 & 2.109\\
    SiN$_{0.97}$H$_{0.07}$ & 3326.28 & 4.577\\
    SiN$_{1.06}$H$_{0.03}$ & 3332.28 & 2.331\\
    \bottomrule
  \end{tabular}
\end{table}

\begin{figure}[!htbp]
  \centering
  \includegraphics[width=0.75\linewidth]{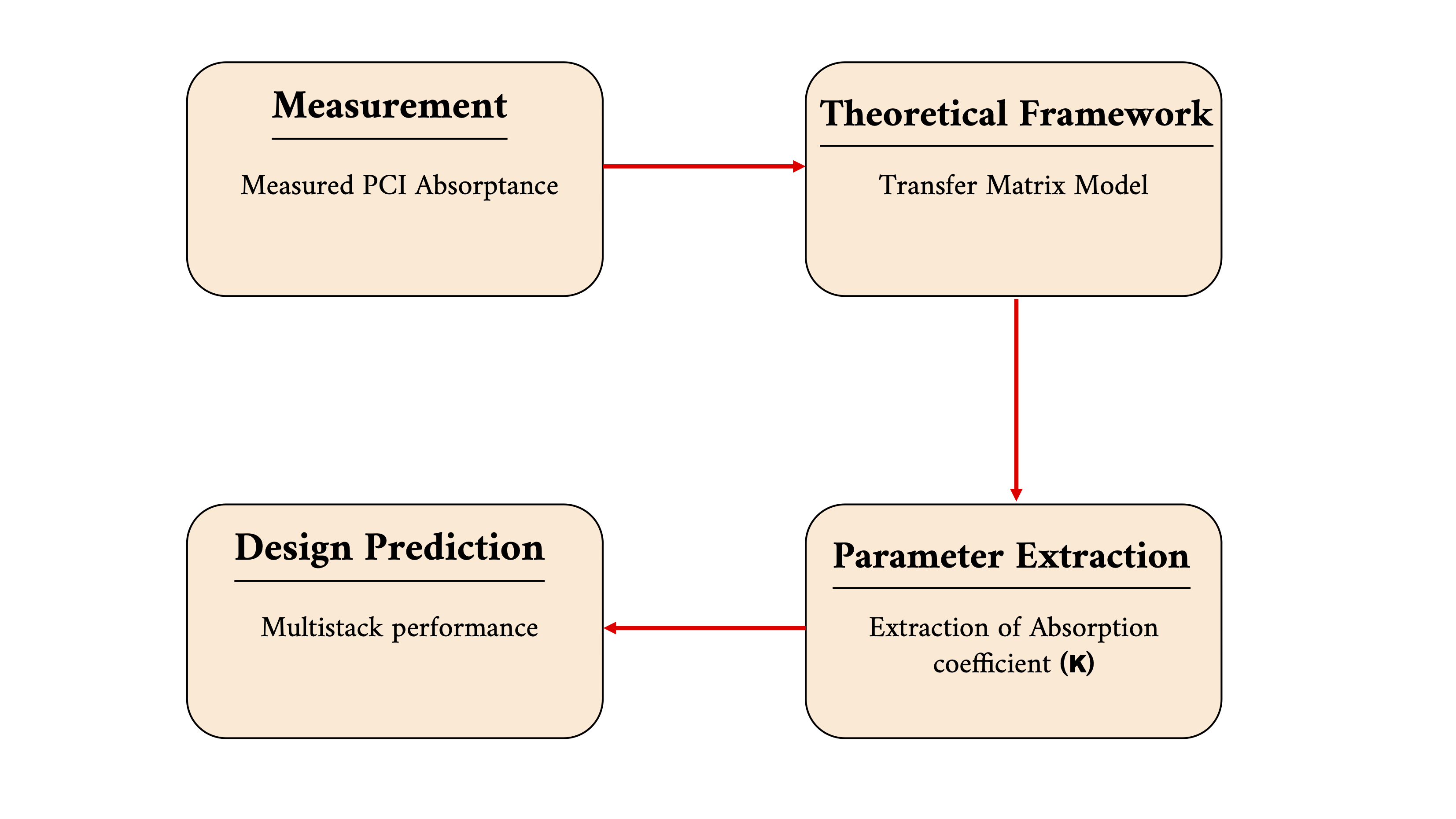}
  \caption{Integrated optical-characterization workflow.}
  \label{fig:characterization-workflow}
\end{figure}

\section{HR and AR Coatings on Sapphire}

\subsection{High-reflectivity coating}

The HR design will use alternating SiN and SiON layers on sapphire. Candidate
stacks will be evaluated with the complex wavelength-dependent indices measured
for the actual deposition recipes.  The optimization variables include pair
count, layer ordering, cap and matching layers, individual optical thicknesses,
and permitted composition changes. Some constraints will include transmission,
field-weighted absorption, total physical thickness, effective mechanical loss,
residual stress, and manufacturability.

For each fabricated HR demonstrator, an as-built model will use measured layer
thicknesses and dispersions to predict spectral reflection, transmission, and
the depth-dependent electric field.  With direct measurements of reflection,
transmission, PCI absorptance, and scatter, the discrepancies will then be test whether discrepancies originate from thickness error, index error, bulk extinction, interfaces, or
roughness.  The result will determine whether the next stack iteration should
change the material recipe, pair count, layer ordering, or individual optical
thicknesses.

The promising single-layer SiN absorptances in Table~\ref{tab:sin-pci} do not
yet demonstrate an acceptable HR coating.  A detector-ready result requires a
complete stack fabricated on sapphire, including the cumulative effects of
interfaces, roughness, stress, scatter, and the standing-wave field.

\subsection{Anti-reflection coating}

The rear surface will use a SiN--SiON AR design with substantially fewer layers
than the HR coating.  Its function is to reduce the sapphire surface reflection
and thereby, limit parasitic cavities and scattered-light coupling.  The design
must use the sapphire index, substrate wedge and incidence geometry, beam
polarization, and required spectral/angular bandwidth.  Although fewer layers
usually reduce their mechanical-noise contribution, AR absorption still adds to
the cryogenic heat load and cannot be neglected.

The HR and AR structures will also be modeled together because light
transmitted by the HR coating propagates through the sapphire and encounters the
rear-surface stack.  The combined model will include both coatings, substrate
thickness and wedge, coherent or incoherent propagation as appropriate to the
measurement geometry, and measured sapphire absorption.

\subsection{Sapphire integration}

The final optical and mechanical properties depend on the coating sapphire
system.  Dedicated witness samples and representative sapphire optics will be
used to measure adhesion, interface absorption, scatter, roughness, figure,
stress, and thickness uniformity.  Uncoated sapphire controls are necessary to
separate the substrate absorption from the PCI response. The optical model will
include the measured sapphire dispersion and any relevant birefringence and
crystal orientation.

Differences in thermal expansion and elastic properties among sapphire, SiN,
and SiON can change stress and curvature during cool-down.  Representative
coatings must therefore undergo repeated thermal cycles to the CHRONOS operating
temperature, followed by optical, mechanical, and surface re-characterization.

\section{FY2026 Measurement Plan and Development Roadmap}

\subsection{Measurements scheduled for FY2026}

Mechanical-loss characterization and the systematic characterization of SiON
is being scheduled for FY2026 (Japanese fiscal year 2026).  They are therefore
listed here as planned measurements, not as completed CHRONOS results.

\paragraph{Mechanical-loss measurements:}
The program will first measure the resonant frequencies and quality factors of
uncoated reference resonators.  Representative single-layer SiN and SiON films
will then be deposited and the ring-down measurements repeated.  Coating
dilution factors will be determined from the measured film thicknesses,
substrate geometry, and elastic modeling in order to infer
$\phi_{\mathrm{coat}}(T,f)$.  The loss will be mapped from room temperature to
the CHRONOS cryogenic operating range with sufficiently dense temperature
sampling to identify possible dissipation peaks.  Several resonant modes will
be measured to test frequency and mode-shape dependence.  Residual stress and
curvature will be recorded before and after coating and after each thermal or
annealing treatment.  The program will subsequently extend from single layers
to representative SiN--SiON multilayers to test interface and stack effects.

\paragraph{SiON measurements:}
The SiON study will establish a controlled series of recipes spanning the
refractive-index range required for the low-index layers.  The O, N, Si, and H
concentrations will be verified independently rather than inferred only from
precursor-flow ratios or nominal sample labels.  Spectroscopic ellipsometry
will determine the thickness and wavelength-dependent $n$ and $\kappa$ at and
around \SI{1064}{nm} and \SI{1550}{nm}.  PCI will measure coated-sample
absorptance at both wavelengths, accompanied by uncoated-substrate controls,
spatial scans, and repeated measurements.  FTIR will characterize
Si--O-, Si--N-, N--H-, and Si--H-related bonding and correlate it with
composition and optical loss.  Residual stress, thickness uniformity,
roughness, and scatter will first be measured on witness substrates and then on
sapphire.  A transfer-matrix model will be used to extract wavelength-specific
extinction coefficients and predict complete SiN--SiON HR and AR stacks.
Selected recipes will subsequently undergo annealing and cryogenic thermal
cycling, followed by repeated optical, stress, and adhesion measurements.

The FY2026 data will provide the first measured mechanical-loss inputs and the
SiON material data required for integrated HR/AR optimization.  A recipe will
not be selected from refractive index or room-temperature absorption alone.

\subsection{Mechanical and thermal parameters required for final design}

Optical optimization will ultimately be combined with measurements of
\begin{equation}
  \phi_i(T,f),\quad Y_i(T),\quad \nu_i(T),\quad
  \alpha_i(T),\quad \kappa_{\mathrm{th},i}(T),
\end{equation}
where $\phi_i$ is mechanical loss, $Y_i$ Young's modulus, $\nu_i$ Poisson's
ratio, $\alpha_i$ thermal expansion, and $\kappa_{\mathrm{th},i}$ thermal
conductivity.  Thermo-refractive coefficients and heat capacity will also be
measured or bounded.  The FY2026 ring-down measurements on coated resonators
will be made
over the relevant temperature and frequency ranges, with dilution-factor
analysis used to infer coating loss.  Single-layer loss measurements will be
supplemented by measurements of realistic HR and AR stacks because interfaces,
stress, and annealing can change multilayer behavior.

Post-deposition annealing is an explicit optimization variable.  It may reduce
some defects or absorption, but it can also change hydrogen bonding, stress,
composition, crystallinity, adhesion, or sapphire figure.  Each annealing
condition will therefore be evaluated with the full optical--mechanical test
set rather than selected from absorption alone.

\subsection{Development sequence}

The development begins in FY2026 by establishing reproducible SiON recipes and
independently verifying their composition and hydrogen content.  Measurements
of $d$, $n$, $\kappa$, coated-sample absorptance, bonding, uniformity, stress,
and room-temperature mechanical loss will follow, together with spatially
repeated absorption measurements and uncoated-sapphire controls.  The resulting
data will be used to infer wavelength specific extinction coefficients with
transfer-matrix models and to predict candidate HR and AR stacks.

The next stage will fabricate SiN--SiON HR and AR demonstrators on sapphire and
measure their transmission, reflection, absorption, scatter, figure, and
stress.  Beginning in FY2026, single-layer and multilayer mechanical loss will
be measured from room temperature to the CHRONOS operating temperature, after
which the additional thermal properties required for the final design will be
determined. Annealing and layer thicknesses will then be optimized jointly.
Representative coated sapphire optics will finally be qualified through
cryogenic cycling and repeated optical and mechanical measurements.

\section{Summary}

The CHRONOS baseline is a sapphire test mass coated with SiN--SiON HR and AR
structures.  Existing LPCVD SiN measurements demonstrate strong
composition-dependent tunability: the reported film--substrate absorptance
ranges from 10.2 to 682.6 ppm at \SI{1064}{nm}, and the lowest reported value at
\SI{1550}{nm} is 7.59 ppm.  These measurements identify promising recipes, but they
do not yet constitute the absorption of a quarter-wave layer or a complete HR
coating.

The principal design lesson is that neither refractive index, optical gap,
Urbach energy, nominal composition, nor hydrogen bonding uniquely predicts
near-infrared loss.  CHRONOS will therefore combine ellipsometry, PCI, FTIR,
independent composition measurements, transfer-matrix modeling, and cryogenic
mechanical-loss measurements.  Final material recipes and layer thicknesses
will be selected from complete HR and AR demonstrations on sapphire under the
optical, mechanical, thermal, and cryogenic conditions relevant to CHRONOS.
The completed coating program will therefore connect the present single-layer
SiN measurements, the FY2026 SiON and mechanical-loss measurements, and the
transfer-matrix stack optimization to a qualified pair of HR and AR coatings
for the sapphire test mass.

%% file: Chapters/Cryogenic.tex
\chapter{Cryogenic System}

The sensitivity of gravitational-wave detectors is fundamentally limited by a combination of intrinsic and environmental noise sources.
Continuous advancements in interferometry, optical design, suspension systems, and control techniques have led to a variety of proposed upgrades aimed at improving detector sensitivity
\cite{Adhikari2014Review,LIGO2015}.
Among these noise sources, thermal noise arising from Brownian motion remains one of the dominant limitations, particularly in the low-frequency regime
\cite{Saulson1990,Levin1998,Harry2002}.

A direct and effective approach to reducing thermal noise is cryogenic operation.
Lowering the temperature suppresses thermally driven fluctuations associated with mechanical dissipation, in accordance with the fluctuation--dissipation theorem
\cite{CallenWelton1951},
thereby improving the fundamental sensitivity of the detector.
For this reason, next-generation and future gravitational-wave detectors,
such as \textit{LIGO Voyager} \cite{Voyager2015},
\textit{KAGRA} \cite{KAGRA2019},
and the \textit{Einstein Telescope} \cite{Punturo2010},
incorporate cryogenic operation as a key design element.
In
However, systems that have been optimized for room-temperature operation must be carefully re-evaluated under cryogenic conditions.
Material properties, mechanical response, optical performance, and control behavior all exhibit temperature dependence
\cite{Yamamoto2006,Steinlechner2018},
requiring comprehensive validation in a low-temperature environment.

In addition to reducing thermal noise in the test mass,
cryogenic operation significantly affects the mechanical loss of optical coatings.
Since coating dissipation is strongly temperature-dependent,
operating at an appropriate cryogenic temperature can lead to substantial improvements in detector sensitivity
\cite{Yamamoto2006,Steinlechner2018}.
Thus, cryogenic technology plays a crucial role not only in thermal noise reduction
but also in enabling improved material performance.

On the other hand, the introduction of a cryogenic system introduces new technical challenges.
Mechanical vibrations originating from cryocoolers,
as well as structural resonances of the cryostat,
can couple to the test mass and degrade detector sensitivity
\cite{KAGRA2019}.
Therefore, minimizing vibration transmission is essential in cryogenic interferometers.
While passive isolation techniques have been widely employed,
they are often insufficient in the low-frequency band.

For future high-sensitivity detectors operating in the sub-Hz regime,
it is necessary to integrate advanced vibration mitigation strategies with thermal design.
In this study, we present the design concept of the cryogenic system for CHRONOS,
focusing on achieving both thermal noise reduction and effective vibration suppression.

\section{Cryocooler and Heat Link}

\begin{figure}[htbp]
	\centering
	\includegraphics[width=\linewidth]{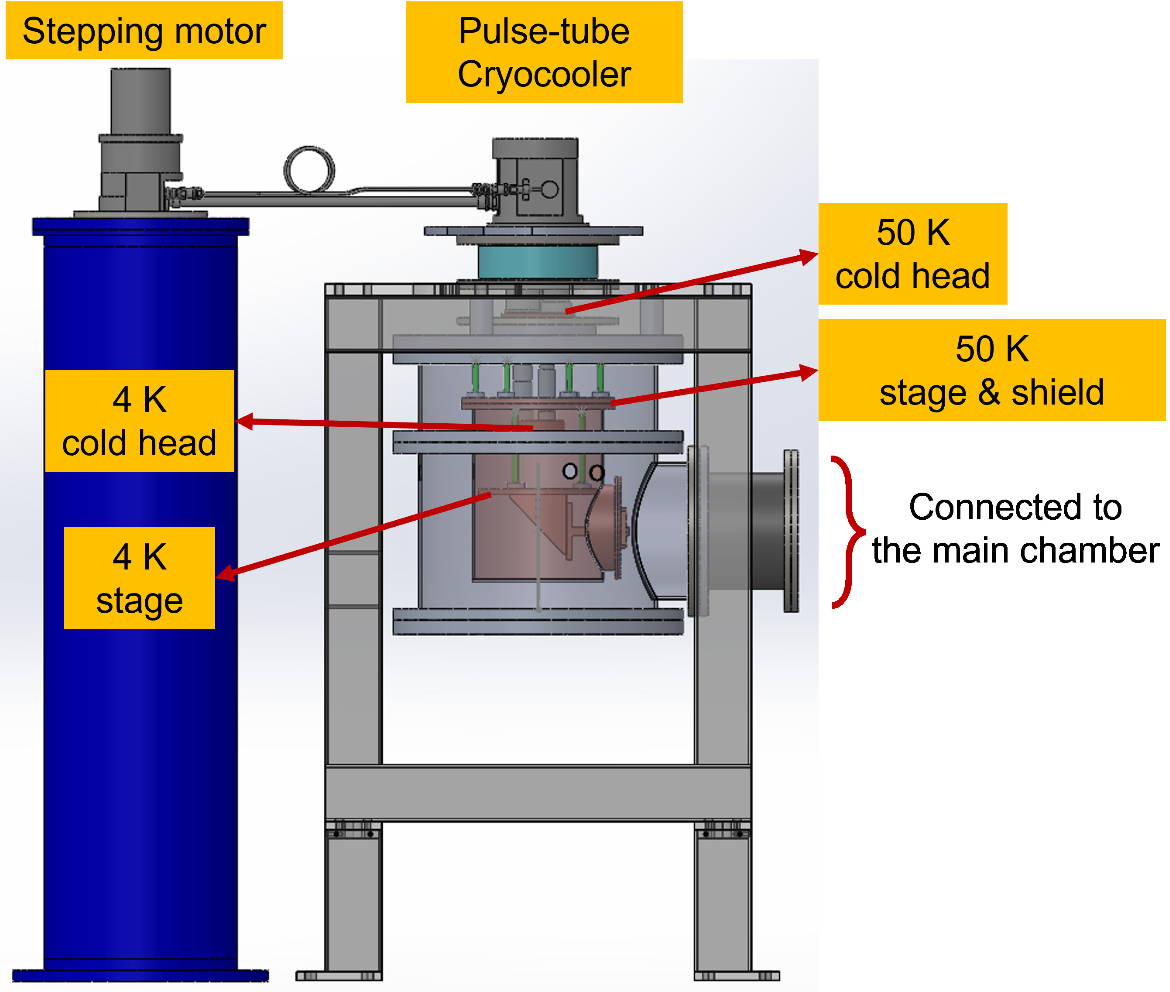}
	\caption{
Overall configuration of the cryogenic chamber.
The system consists of a two-stage pulse-tube cryocooler,
50~K and 4~K stages, and radiation shields at 300~K and 50~K.
The multi-stage shielding structure suppresses radiative heat inflow
from warm components.
	}
	\label{Fig.CRYO_AS_part_mod}
\end{figure}

To realize cryogenic operation for thermal noise reduction,
we developed a dedicated cryogenic system for CHRONOS.
As shown in Fig.~\ref{Fig.CRYO_AS_part_mod},
the system consists of a cryogenic chamber coupled to the main interferometer chamber,
enabling cooling of the suspension system and the test mass.

Inside the chamber, a two-stage pulse-tube cryocooler (PTC)
with nominal 50~K and 4~K stages is installed.
Pulse-tube cryocoolers are widely used in low-vibration cryogenic applications
because they contain no moving parts at the cold head,
significantly reducing vibration transmission compared to conventional
Gifford--McMahon (GM) cryocoolers
\cite{Wang2007PTC,KAGRA2019}.
The radiation shields at 300~K and 50~K suppress radiative heat transfer
from higher-temperature components,
allowing efficient cooling of the inner stages.

Cooling is provided by a commercial two-stage pulse-tube cryocooler
(Sumitomo Heavy Industries, SRP-082B2-F70LP),
which provides cooling capacities at approximately 50~K (first stage)
and 4~K (second stage).
Such cryogenic architectures are also adopted in KAGRA
for low-vibration mirror cooling
\cite{KAGRA2019}.

\begin{figure}[htbp]
	\centering 
	\includegraphics[width=\linewidth]{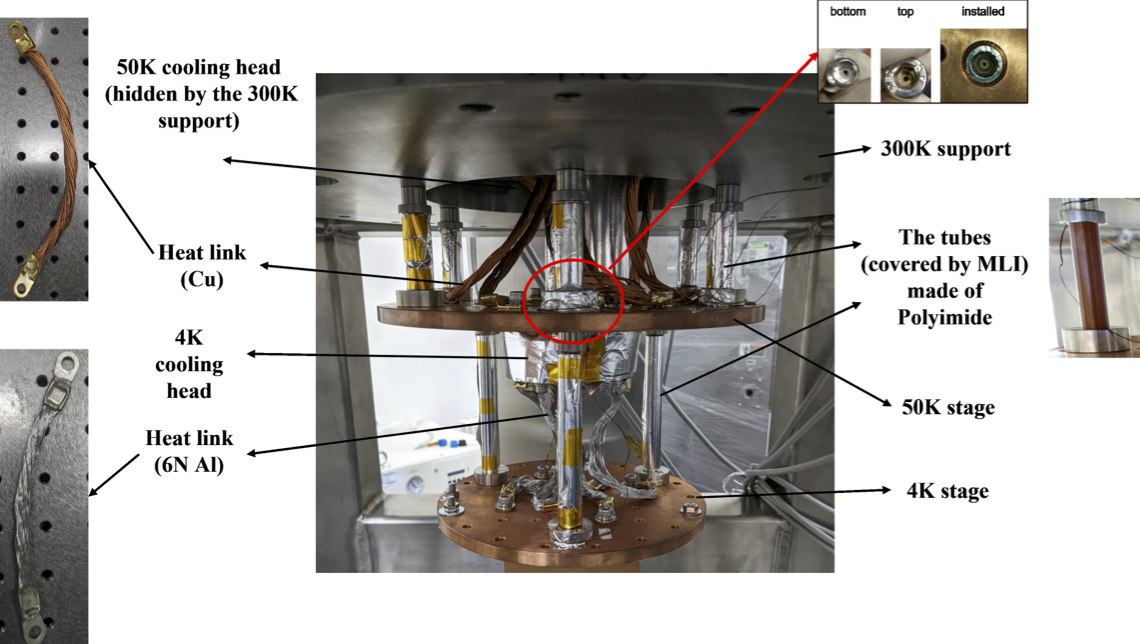} 
	\caption{
Structure of the cold heads and thermal stages.
Low-conductivity Vespel supports reduce parasitic heat flow,
while heat links provide efficient thermal coupling.
This design balances thermal isolation and mechanical stability.
	}
	\label{Fig.cryo_head_stage_mod}
\end{figure}

The cold heads are thermally connected to each temperature stage via heat links,
while mechanical isolation is maintained through low-conductivity supports.
As shown in Fig.~\ref{Fig.cryo_head_stage_mod},
the 50~K stage is suspended from the 300~K structure
using low-thermal-conductivity polyimide (Vespel\textregistered) tubes.
Similarly, the 4~K stage is supported from the 50~K stage using identical structures.
This configuration minimizes parasitic heat conduction
while maintaining mechanical stability.
To suppress residual gas conduction,
vent holes are introduced in the mounting screws
to evacuate trapped gas inside the support tubes.

The heat links play a crucial role in achieving both efficient cooling
and vibration isolation.
Mechanical vibrations generated inside the cryocooler,
such as pressure oscillations of the working gas,
can propagate through rigid thermal connections and couple to the test mass
\cite{Wang2007PTC}.
To suppress this effect, we adopt flexible heat links composed of bundles of thin wires
rather than a single rigid conductor.
For the same cross-sectional area,
such a stranded structure provides significantly lower stiffness,
thereby reducing vibration transmission while maintaining high thermal conductivity.
High-purity copper is used for the 50~K stage,
while high-purity aluminum (99.9999\%, 6N) is used for the 4~K stage,
taking advantage of their excellent cryogenic thermal conductivity
\cite{Pobell2007}.

To further reduce radiative heat transfer,
multi-layer insulation (MLI) is applied to both stages.
MLI consists of multiple reflective layers and is highly effective
in suppressing radiation, which dominates heat transfer in vacuum environments
\cite{Pobell2007}.
In this system, approximately 50 layers are applied to the 50~K stage
and 30 layers to the 4~K stage,
significantly reducing radiative heat load and improving cooling efficiency.

In summary, the cryogenic system is designed to simultaneously achieve
low temperature and low vibration by combining:
(i) multi-stage radiation shielding,
(ii) low thermal-conductivity supports to suppress parasitic heat flow, and
(iii) low-stiffness heat links for vibration isolation.

In addition to passive isolation,
further reduction of residual vibration from the pulse-tube cryocooler
is pursued using active vibration isolation (AVI).
In an active scheme, vibrations are measured using sensors
and suppressed by applying counteracting forces through actuators.
Such feedback-based vibration suppression techniques
are widely employed in precision interferometry
\cite{LIGO2015,KAGRA2019}.
This approach is particularly important for periodic low-frequency vibrations
originating from the cryocooler,
where passive isolation alone is insufficient.

The design and performance of the active vibration isolation system
are described in detail in the following section.

\begin{figure}[htbp]
	\centering 
	\includegraphics[width=0.3\linewidth]{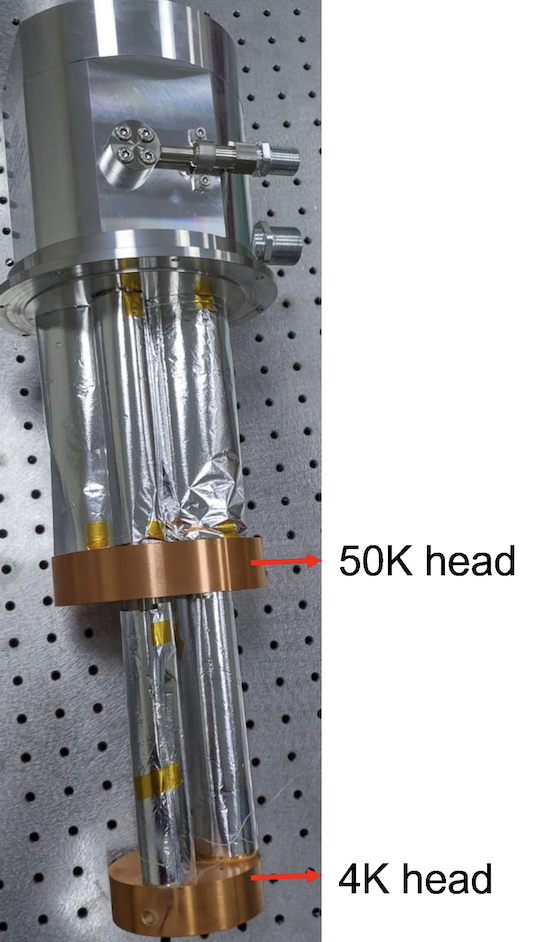} 
	\caption{
Two-stage pulse-tube cryocooler used in this system.
The first stage cools to $\sim$50~K and the second stage to $\sim$4~K.
The absence of moving parts at the cold head provides low-vibration operation suitable for precision interferometry.
	}
	\label{Fig.asgraf_cryocooler}
\end{figure}

\section{Cryogenic Performance}

We evaluated the cryogenic performance of the system,
with particular emphasis on quantifying heat loads arising
from thermal conduction and radiation.
To this end, controlled heat inputs were applied to the 4~K and 50~K stages,
and the corresponding temperature responses were measured.
This evaluation is essential for validating the thermal design
and assessing the cooling margin under cryogenic operation
\cite{Pobell2007}.

During the measurements, the cold heads were thermally decoupled
from each stage to isolate the intrinsic cooling performance
of the cryocooler.
This configuration allows direct evaluation of the response
to externally applied heat loads.
Ceramic heaters were mounted on each stage,
and the injected power was precisely determined
using four-wire measurements of voltage and current.
By varying the applied voltage,
heat loads in the range of approximately
0--25~W for the 50~K stage and
0--2~W for the 4~K stage were introduced.

\begin{figure}[ht!]
\centering
\includegraphics[width=\linewidth]{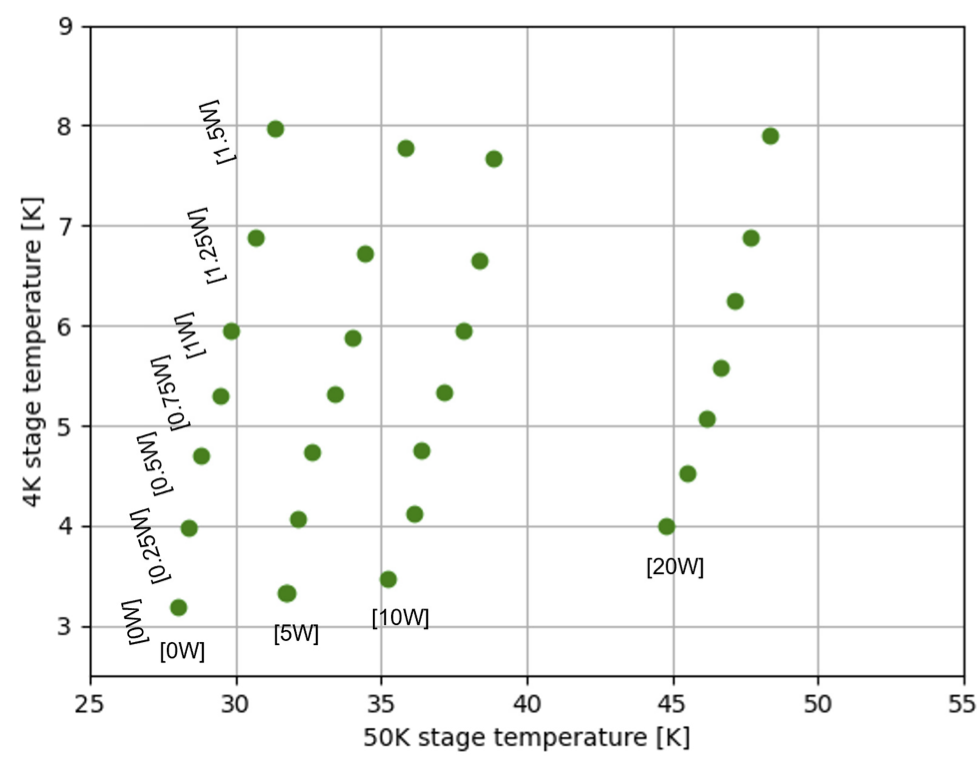}
\caption{
Thermal loading map of the 4~K and 50~K stages.
The horizontal axis shows the applied heater power, and the vertical axis shows the steady-state temperature.
The slope of each curve provides a measure of the effective thermal conductance and radiative heat load.
A stable thermal response is observed over the full range of applied power.
}
\label{Thermal_map}
\end{figure}

Figure~\ref{Thermal_map} shows the relationship
between the applied heater power and the steady-state temperature
of each stage.
From the slope of these curves,
the effective thermal conductance and radiative heat load were evaluated.
The heat balance can be expressed as
\begin{equation}
  Q_{\mathrm{in}}
  =
  Q_{\mathrm{cond}} + Q_{\mathrm{rad}} + Q_{\mathrm{cool}},
\end{equation}
where conductive heat flow follows Fourier's law
and radiative heat load follows the Stefan--Boltzmann relation
\cite{Pobell2007}.

The measured effective parasitic heat loads
at 4.2~K and 30~K are approximately
0.5~W and 5~W, respectively.
These values are consistent with theoretical estimates
based on the number of MLI layers,
the thermal conductivity of the support structures,
and the heat-link design.
They are also comparable to typical performance levels
of cryogenic systems of similar scale,
such as those implemented in KAGRA
\cite{KAGRA2019}.

\begin{figure}[ht!]
\centering
\includegraphics[width=\linewidth]{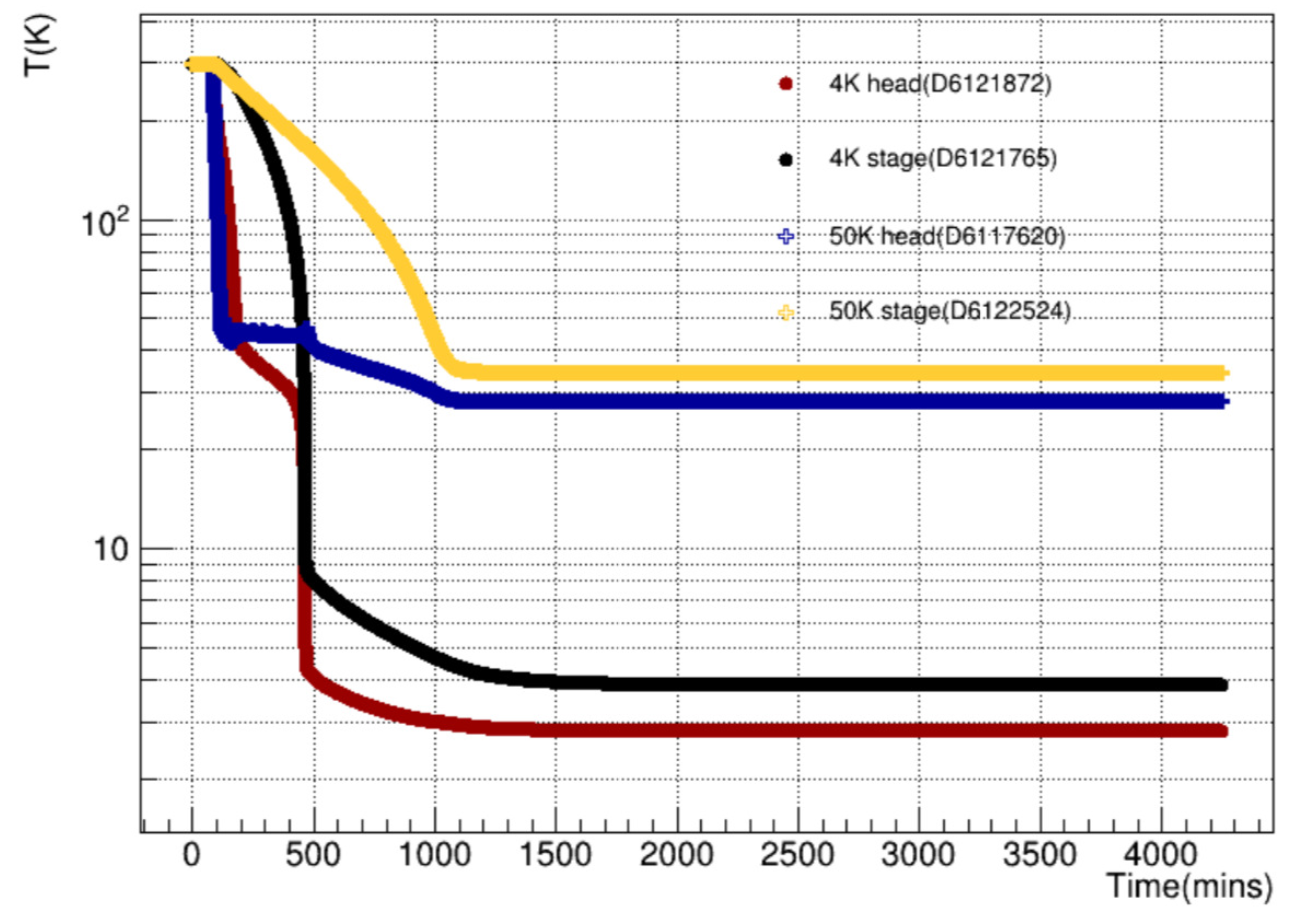}
\caption{
Temperature evolution during the cooldown process.
The 50~K stage cools first, followed by the 4~K stage.
The system reaches operating temperatures in approximately 1000 minutes (17 hours),
indicating efficient thermal design of the cryostat and heat-link system.
}
\label{Readout}
\end{figure}

To characterize the cooling dynamics,
we measured the cooldown process from room temperature
to operating temperature.
Silicon diode thermometers were used to monitor
the temperature evolution of the cold heads and each stage simultaneously.
The temporal temperature response is shown in Fig.~\ref{Readout}.
The system reaches operating temperatures of
4~K and 50~K in approximately 1000 minutes (about 17 hours),
which is consistent with the expected thermal mass
and cooling power of the cryocooler.

Temperature measurements were performed
using calibrated silicon diode sensors
(DT-680B-CU, Lake Shore Cryogenics),
whose temperature dependence is well characterized
in the cryogenic range \cite{LakeShore}.
Four sensors were installed:
one on each of the 4~K and 50~K cold heads,
and one on each of the corresponding stages.
This configuration enables quantitative evaluation
of the temperature gradient between the cold heads and stages,
providing insight into thermal resistance
and heat-transfer efficiency.

These results demonstrate that the cryogenic system
achieves the designed cooling performance
and provides sufficient thermal conditions
for realizing a low thermal-noise environment in CHRONOS.

\subsection{Active Vibration Isolation System}

One of the major challenges in cryogenic systems is periodic mechanical vibration
originating from the cryocooler and compressor.
Although PTC exhibit lower vibration compared to
conventional cryocoolers, residual pressure oscillations and gas flow
inside the system generate periodic disturbances,
particularly in the low-frequency band
\cite{Wang2007PTC,KAGRA2019}.
These vibrations can propagate through heat links and support structures
and couple to the test mass,
potentially limiting the detector sensitivity.

As discussed in the previous section,
passive isolation methods such as bellows and low-stiffness heat links
are effective at higher frequencies,
but their performance degrades in the low-frequency regime
\cite{LIGO2015}.
For CHRONOS, which targets high sensitivity in the sub-Hz band,
active vibration suppression is therefore essential.

\begin{figure}[htbp]
	\centering 
	\includegraphics[width=\linewidth]{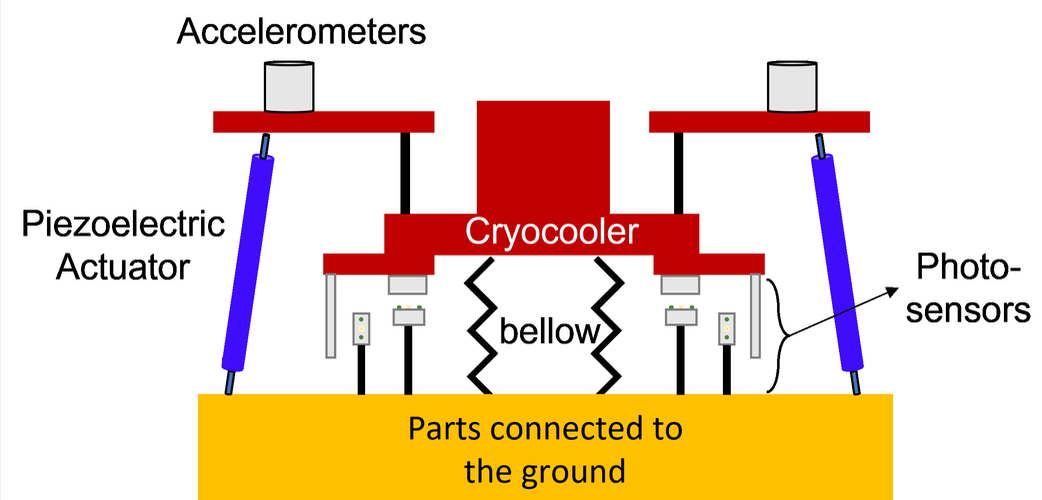}
	\caption{
Schematic of the AVIS.
The yellow structure is fixed to the ground and serves as a reference frame,
while the red structure is connected to the cryocooler.
They are coupled via bellows and piezoelectric actuators,
allowing active control of relative motion.
Multiple sensors (accelerometers and photosensors) enable six-degree-of-freedom control.
	}
	\label{Fig.AVIS_overview}
\end{figure}

To address this issue,
we developed an active vibration isolation system (AVIS).
A schematic overview of the system is shown in Fig.~\ref{Fig.AVIS_overview}.

The AVIS directly controls the relative motion between the cryocooler
and the ground support structure.
The pulse-tube cryocooler is mounted to the 300~K chamber lid via bellows,
which provide passive vibration isolation.
In addition, six piezoelectric actuators (PZTs)
are installed to actively control the relative displacement
between the cryocooler and the support structure.

The system operates based on feedback control theory.
The static equilibrium position of the cryocooler
is defined as the operating point.
Sensor signals measuring displacement and acceleration are continuously monitored,
and the actuators generate counteracting displacements
with opposite phase,
thereby canceling vibration components.
This approach is particularly effective for suppressing
periodic disturbances originating from the cryocooler.

The system is designed to control six degrees of freedom (6-DOF),
consisting of three translational motions ($x$, $y$, $z$)
and three rotational motions (roll, pitch, and yaw).
By appropriately arranging sensors and actuators,
simultaneous control of all degrees of freedom is achieved,
similar to multi-axis active isolation platforms
used in advanced interferometers
\cite{LIGO2015}.

\begin{figure}[htbp]
	\centering
	\includegraphics[width=\linewidth]{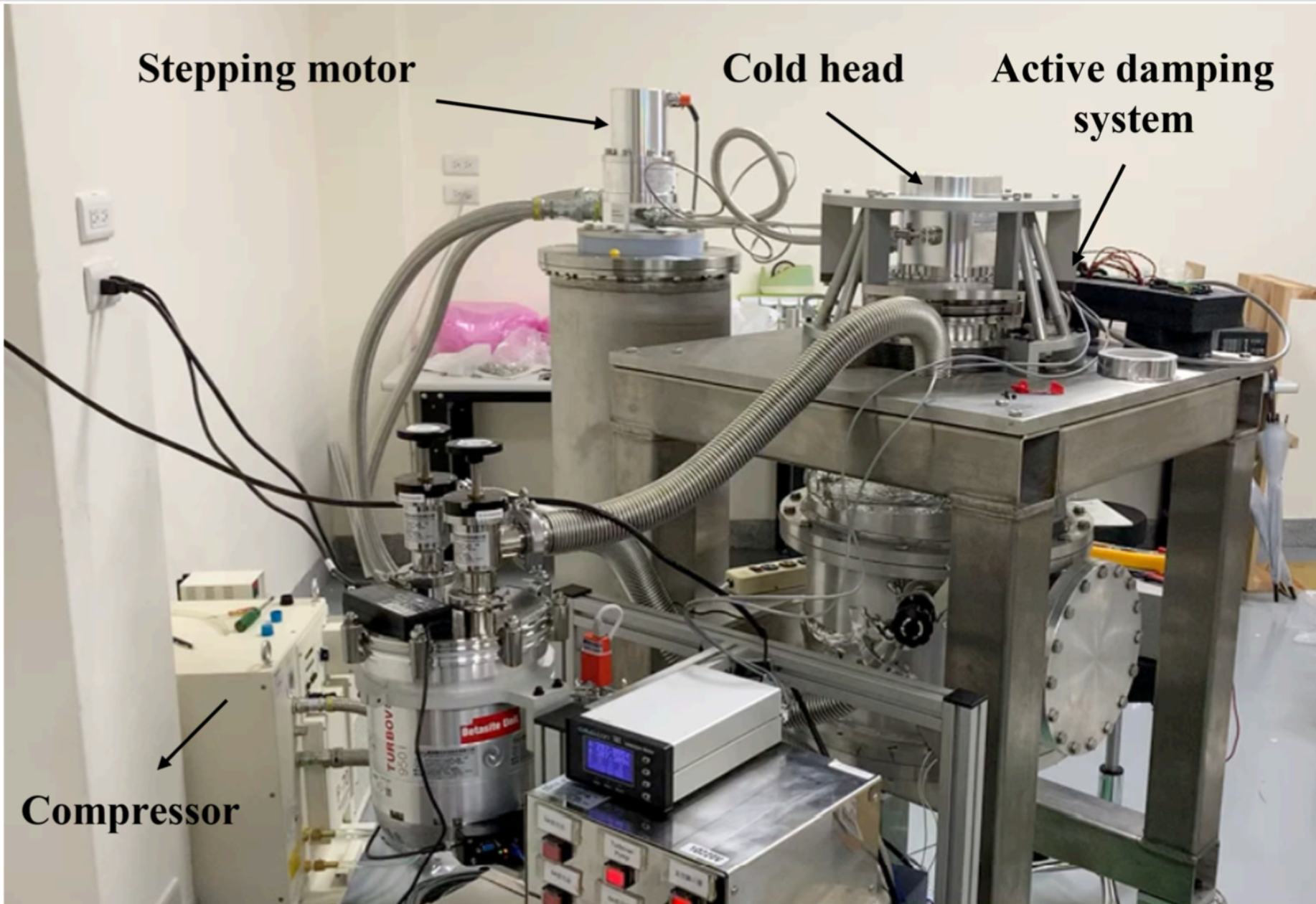}
	\caption{
Photograph of the cryogenic system.
The pulse-tube cryocooler is installed inside the chamber,
and the AVIS is placed between the cryocooler and the ground support structure.
	}
	\label{Fig.AVIS_overview_real}
\end{figure}

A photograph of the installed cryogenic system
is shown in Fig.~\ref{Fig.AVIS_overview_real}.
The cryocooler is located inside the cryogenic chamber,
while the compressor and drive units are installed externally.
The AVIS is positioned between the cryocooler
and the ground support structure,
actively suppressing vibrations originating from the cryogenic system.

\begin{figure}[htbp]
	\centering
	\includegraphics[width=\linewidth]{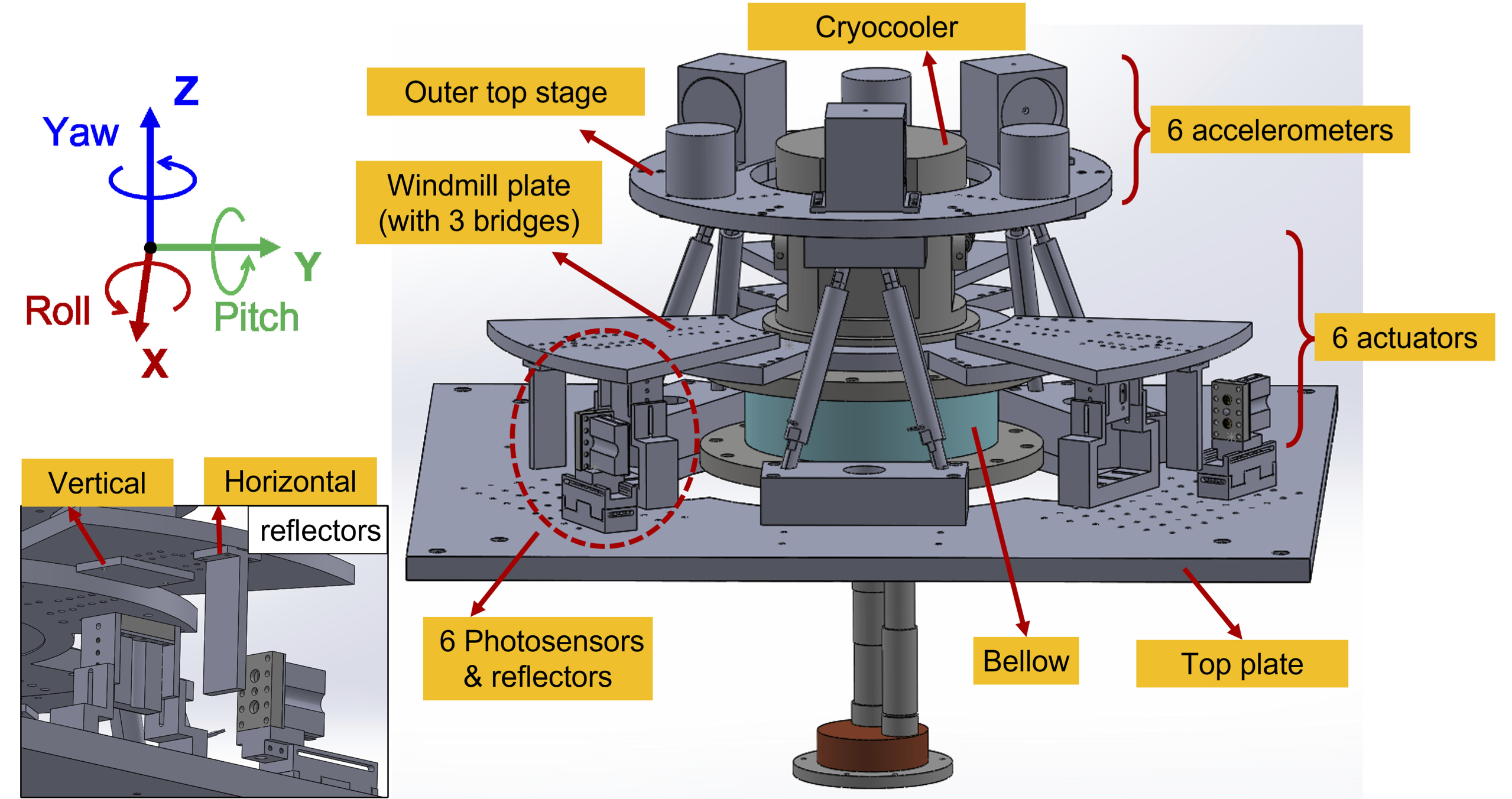}
	\caption{
Integrated setup of the AVIS.
Accelerometers (for high-frequency components) and photosensors (for low-frequency components)
measure vibrations, while six piezoelectric actuators generate compensating motion.
Simultaneous control of all six degrees of freedom is achieved.
	}
	\label{Fig.AVIS_setup}
\end{figure}

Figure~\ref{Fig.AVIS_setup} shows the detailed configuration of the AVIS.
Multiple rigid stages are attached above the cryocooler
to mount the sensors.
Accelerometers are installed on the outer top stage,
while reflective targets for the photosensors
are mounted on a windmill-shaped plate.
This configuration allows simultaneous detection
of horizontal and vertical vibrations.

The six piezoelectric actuators are arranged between
the top plate and the outer stage,
enabling force application in all degrees of freedom.
This configuration realizes real-time feedback control
of the full 6-DOF motion.

In this section, we have described the overall architecture
and operating principle of the AVIS.
The detailed specifications and performance
of the sensors and actuators
are presented in the following section.

\subsection{Photosensors}

To monitor the relative displacement between the cryocooler and the ground support structure,
non-contact photosensors are employed in this system.
Since the measurement is performed optically,
the sensors introduce no mechanical loading on the system,
making them particularly suitable for detecting small displacements
in the low-frequency regime \cite{LIGO2015}.

\begin{figure}[htbp]
	\centering 
	\includegraphics[width=0.6\linewidth]{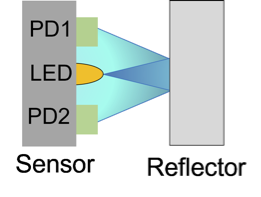} 
	\caption{
Schematic of the photosensor.
Light emitted from the LED is reflected by a target and detected by photodiodes.
Displacement is measured from the change in reflected light intensity.
	}
	\label{Fig.ps_schmtc}
\end{figure}

As shown in Fig.~\ref{Fig.ps_schmtc},
the photosensor is a reflective optical device consisting of one LED and two photodiodes (PDs).
Light emitted from the LED is reflected by a target mirror and detected by the PDs.
The relative displacement between the sensor and the target
is inferred from the change in the reflected light intensity,
following standard optical position-sensing principle.

\begin{figure}[htbp]
	\centering 
	\includegraphics[width=\linewidth]{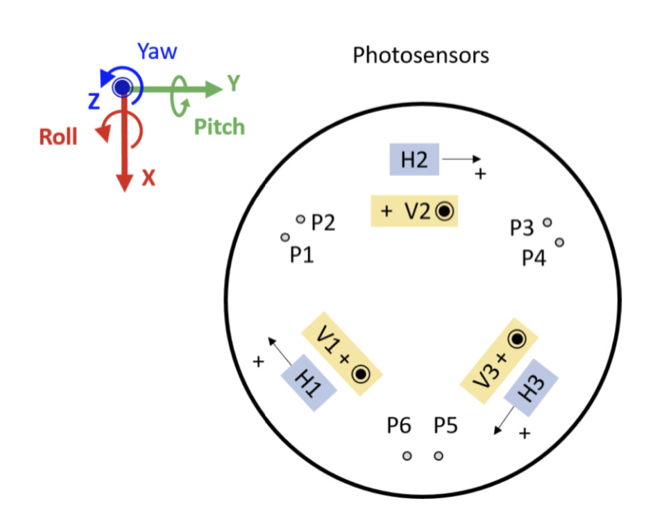} 
	\caption{
Configuration of the photosensors and coordinate system.
Three horizontal and three vertical sensors are used to reconstruct 6-DOF motion.
	}
	\label{Fig.ps_cor}
\end{figure}

The photosensors provide a wide dynamic range
and are effective for monitoring slow drifts and low-frequency vibrations.
In this system, three horizontal and three vertical sensors
are arranged to reconstruct the 6-DOF motion
of the cryocooler,
similar to multi-axis sensing schemes used in active isolation platforms as shown in Fig. ~\ref{Fig.ps_cor}
\cite{LIGO2015}.

Detailed characterization of the sensor performance,
including operation under cryogenic conditions,
is under ongoing study.

\vspace{0.5em}

\subsection{Piezoelectric Actuators}

Active vibration compensation is achieved using PZTs.
Piezoelectric elements convert applied voltage into mechanical displacement,
providing fast response and high resolution suitable for precision control
\cite{UchinoPiezo}.

\begin{figure}[ht!]
\centering
\includegraphics[width=\linewidth]{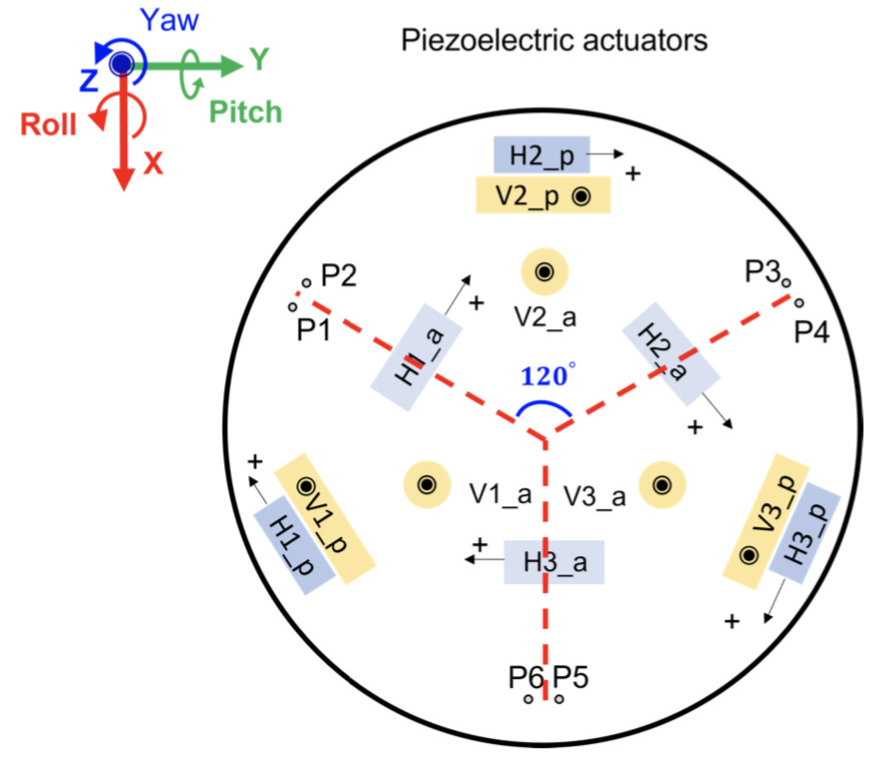}
\caption{
Arrangement of the piezoelectric actuators.
Six PZTs are symmetrically placed,
enabling control of all six degrees of freedom.
}
\label{Piezo_pos}
\end{figure}

The AVIS consists of six supporting legs, each equipped with one PZT.
As shown in Fig.~\ref{Piezo_pos},
the actuators are arranged with an inclination angle of approximately $15^\circ$,
allowing control of three translational and three rotational degrees of freedom.
Such geometric arrangements enable full 6-DOF control
through appropriate transformation between actuator space and Cartesian coordinates.

The actuators are installed between the top plate and the support structure,
and apply forces to compensate for the relative displacement
between the cryocooler and the ground.
Control signals are generated by a digital control system
and independently applied to each actuator,
enabling flexible multi-degree-of-freedom control.

\vspace{0.5em}

\subsection{Control and Transfer Function}

The control performance of the AVIS is characterized
through measurements of the open-loop transfer function.
A swept excitation signal is injected into the control system,
and the system response is analyzed to obtain the frequency response.

The loop dynamics can be described by the standard feedback relation
\begin{equation}
  T(\Omega) = \frac{G(\Omega)P(\Omega)}{1 + G(\Omega)P(\Omega)},
\end{equation}
where $P(\Omega)$ is the plant transfer function
and $G(\Omega)$ is the controller transfer function.

\begin{figure}[ht!]
\centering
\includegraphics[width=\linewidth]{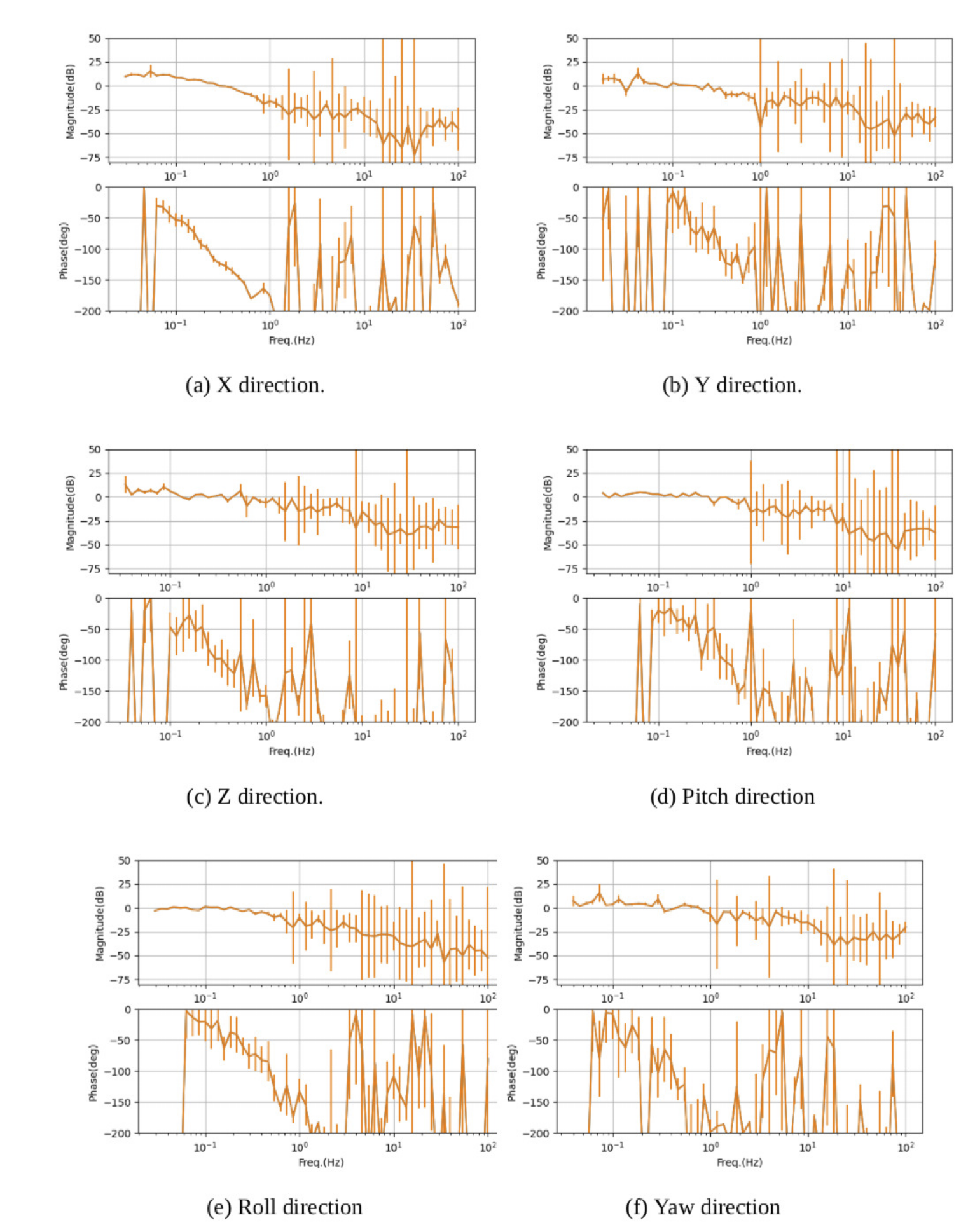}
\caption{
Open-loop transfer function of the control system.
The unity gain frequency is approximately 1~Hz,
and the phase margin indicates stable feedback operation.
}
\label{TF_PD}
\end{figure}

Based on the measured Bode plots,
digital filters are designed to compensate for phase delay
and stabilize the feedback loop.
A representative transfer function is shown in Fig.~\ref{TF_PD}.

The unity gain frequency (UGF) is approximately 0.5~Hz,
indicating that the feedback control is effective
in the sub-Hz to 0.5~Hz range.
A phase margin of about $55^\circ$
ensures stable operation of the control system,
consistent with standard stability criteria.

These results demonstrate that the AVIS provides sufficient control bandwidth
to suppress low-frequency vibrations originating from the cryocooler.

\vspace{0.5em}

\subsection{Noise Performance}

The vibration suppression performance of the AVIS
was evaluated by comparing noise spectra before and after feedback control.

Initially, each degree of freedom was controlled independently,
confirming that noise reduction is achieved
in the frequency range below 1~Hz.
Subsequently, the number of controlled degrees of freedom
was gradually increased,
leading to full 6-DOF simultaneous feedback control.

The system stability during multi-degree-of-freedom operation
was verified through transfer function measurements,
confirming stable control across all degrees of freedom.

\begin{figure}[ht!]
\centering
\includegraphics[width=\linewidth]{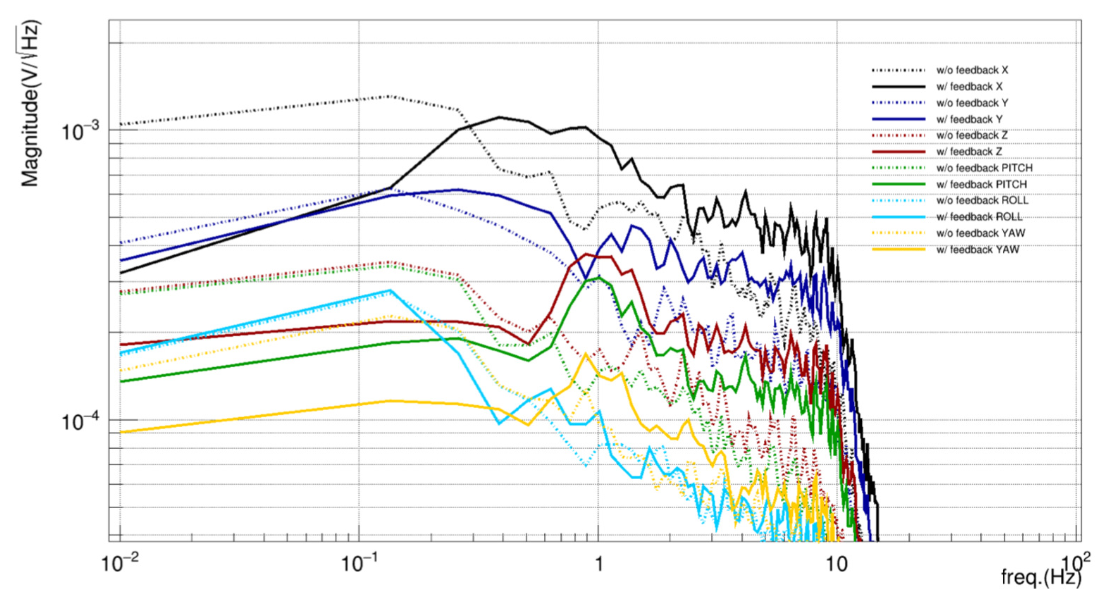}
\caption{
Comparison of noise spectra before and after feedback control.
Significant vibration reduction is achieved below 1~Hz
with full 6-DOF control.
}
\label{Noise}
\end{figure}

Figure~\ref{Noise} shows the measured noise spectra
before and after feedback control.
Significant noise reduction is observed in the low-frequency region below 1~Hz,
with improvements seen in all degrees of freedom.
In particular, a pronounced reduction is observed in the $y$ direction.

These results demonstrate that the AVIS effectively suppresses
cryocooler-induced vibrations over a broad frequency range,
and is particularly effective in the low-frequency band
relevant for CHRONOS.

%% file: Chapters/Photon_Pressure_Actuator.tex
\chapter{Photon-Pressure Actuator}
\label{sec:photon_pressure_actuator}

Low-frequency interferometers require actuators that can maintain
interferometer lock without introducing excess force noise into the test mass. In conventional gravitational-wave detectors, electromagnetic or electrostatic actuators apply forces between the test mass and a nearby recoil mass. Although these schemes provide a large actuation range, motion of the recoil mass can couple seismic motion to the test mass through the actuator. Electromagnetic actuators can additionally couple fluctuations of the environmental magnetic field to the test-mass motion.

CHRONOS therefore adopts a photon-pressure actuator (PPA) as the primary low-noise actuator for the test-mass stage. The PPA transfers photon momentum directly to the torsion-bar test mass using power-controlled laser beams. Since no mechanically-coupled recoil mass is required, the PPA can suppress actuation noise associated with motion of a recoil mass. Photon pressure itself is also intrinsically insensitive to environmental magnetic-field fluctuations. This configuration potentially allows the recoil mass to be omitted from the
test-mass stage, substantially simplifying the final stage of the  CHRONOS suspension.

The same optical system can simultaneously operate as a Photon Calibrator (PCal), because both actuation and calibration rely on the same radiation pressure force. A calibrated sinusoidal modulation can be superimposed on the feedback actuation signal without changing the optical configuration. The PPA and PCal can therefore be realized as a two-in-one system for lock acquisition, lock maintenance, actuator characterization, and absolute calibration~\cite{Karki2016Pcal,virgo_pcal,kagra_pcal, tanabe_photon_pressure_actuator}. A slight discrepancy can be induced by a tail of the internal resonant peaks, although these peaks are outside the science band. Refer Chap.~\ref{chap:calibration} for overview of calibration scheme.

The CHRONOS PPA uses four independently controlled laser beams incident on the back surface of each torsion-bar test mass. Appropriate combinations of the four beam powers generate longitudinal force, yaw torque, and pitch torque independently. A detailed design study predicts a maximum yaw torque amplitude of approximately $1.0\times10^{-8}\,\mathrm{N\,m}$ and an actuation efficiency
of $6.6\times10^{-13}\,\mathrm{rad/V}$ at 1~Hz. The corresponding yaw actuation noise is estimated to be $5.3\times10^{-19}\,\mathrm{rad/\sqrt{Hz}}$ at 1~Hz, whose differential strain conversion is compared with the updated
reference sensitivity in Sec.~\ref{subsec:ppa_noise}. When operated as a PCal, the projected systematic uncertainty is approximately $1.14\%$ \cite{tanabe_photon_pressure_actuator}.

\section{Actuation principle}
\label{subsec:ppa_principle}

The basic principle of the photon-pressure actuator is to control the optical power incident on the test mass and thereby control the radiation-pressure force. To actuate several rigid-body degrees of freedom independently while avoiding interference with the main interferometer beam, four auxiliary laser beams are injected onto the back surface of the torsion-bar test mass.

\begin{figure}[tbp]
    \centering
    \includegraphics[width=0.72\columnwidth]{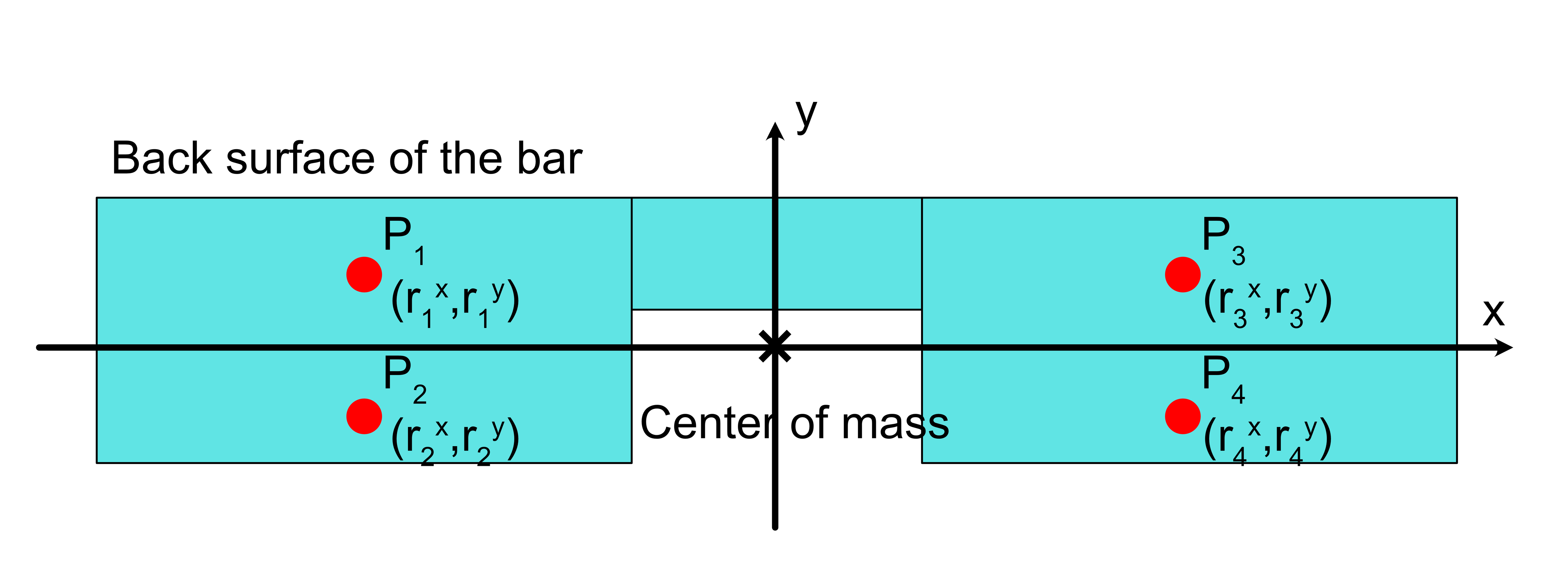}
    \caption{
    Conceptual arrangement and indexing of the four photon-pressure beams incident on the back surface of the CHRONOS torsion-bar test mass. Independent modulation of the four beam powers provides longitudinal, yaw, and pitch actuation.
    }
    \label{fig:4mirrors}
\end{figure}

The conceptual arrangement of the beam positions is shown in Fig.~\ref{fig:4mirrors}. The four injection points are approximately symmetrically distributed around the center of mass.

We label the four beams by the index $i=1,2,3,4$. The back surface of the torsion bar defines the $x$--$y$ plane and the $z$ axis is taken along the surface normal. The corresponding unit vectors are denoted by
$\mathbf{e}_x$, $\mathbf{e}_y$, and $\mathbf{e}_z$. The position of the $i$-th beam relative to the center of mass is 
\begin{equation}
    \mathbf{r}_i =
    \left(r_i^x,r_i^y,0\right).
\end{equation}

We parameterize each beam by its optical power $P_i(t)$, azimuthal incidence angle $\theta_i$, and inclination angle $\phi_i$. Assuming specular reflection, the radiation-pressure force applied by the
$i$-th beam is
\begin{equation}
\label{eq:pcal_force}
\mathbf{F}_i(t)=\frac{2P_i(t)}{c}
\begin{pmatrix}
\cos\phi_i\sin\theta_i \\
\sin\phi_i \\
\cos\phi_i\cos\theta_i
\end{pmatrix},
\end{equation}
where $c$ is the speed of light.

The net force along the translational direction is
\begin{equation}
\label{eq:ppa_translation_force}
F_{\rm T}(t)=\sum_{i=1}^{4}\mathbf{F}_i(t)\cdot\mathbf{e}_z .
\end{equation}

The corresponding yaw and pitch torques are
\begin{equation}
\label{eq:pcal_torque_yaw}
T_{\rm Y}(t)=\sum_{i=1}^{4}\left( \mathbf{r}_i\times\mathbf{F}_i(t) \right)\cdot\mathbf{e}_y ,
\end{equation}
and
\begin{equation}
\label{eq:pcal_torque_pitch}
T_{\rm P}(t)=\sum_{i=1}^{4}\left( \mathbf{r}_i\times\mathbf{F}_i(t)
\right)\cdot\mathbf{e}_x .
\end{equation}

Different linear combinations of the four beam powers therefore correspond to different rigid-body actuation modes. A common-mode modulation primarily generates longitudinal force, whereas an
antisymmetric modulation between the left and right sides of the bar
generates yaw torque. Similarly, a differential modulation between the upper and lower beams produces pitch torque.

\section{Mechanical response}
\label{subsec:ppa_mechanical_response}

The equations of motion for longitudinal translation, yaw, and pitch are
\begin{equation}
\label{eq:trans_t}
M\ddot{z}(t)+\gamma_{\rm T}\dot{z}(t)+k_{\rm T}z(t)=F_{\rm T}(t),
\end{equation}

\begin{equation}
\label{eq:yaw_t}
I_{\rm Y}\ddot{\varphi}_{\rm Y}(t)+\Gamma_{\rm Y}\dot{\varphi}_{\rm Y}(t)+K_{\rm Y}\varphi_{\rm Y}(t)=T_{\rm Y}(t),
\end{equation}
and
\begin{equation}
\label{eq:pitch_t}
I_{\rm P}\ddot{\varphi}_{\rm P}(t)+\Gamma_{\rm P}\dot{\varphi}_{\rm P}(t)+K_{\rm P}\varphi_{\rm P}(t)=T_{\rm P}(t),
\end{equation}
where $M$ is the mass of the torsion bar, $I_{\rm Y}$ and $I_{\rm P}$ are its moments of inertia about the yaw and pitch axes, respectively, $\gamma_{\rm T}$ and $\Gamma_j$ are damping coefficients, and $k_{\rm T}$ and $K_j$ are the corresponding restoring coefficients.

In the frequency domain, with $\omega=2\pi f$, these equations become
\begin{equation}
\label{eq:trans_f}
\left( -M\omega^2+i\omega\gamma_{\rm T}+k_{\rm T} \right) z(\omega)=F_{\rm T}(\omega),
\end{equation}

\begin{equation}
\label{eq:yaw_f}
\left( -I_{\rm Y}\omega^2+i\omega\Gamma_{\rm Y}+K_{\rm Y} \right) \varphi_{\rm Y}(\omega)=T_{\rm Y}(\omega),
\end{equation}
and
\begin{equation}
\label{eq:pitch_f}
\left( -I_{\rm P}\omega^2+i\omega\Gamma_{\rm P}+K_{\rm P} \right) \varphi_{\rm P}(\omega)=T_{\rm P}(\omega).
\end{equation}

The resonant frequencies and quality factors are defined as
\begin{equation}
\Omega_{\rm T}=\sqrt{\frac{k_{\rm T}}{M}}, \qquad \Omega_{\rm Y}=\sqrt{\frac{K_{\rm Y}}{I_{\rm Y}}}, \qquad \Omega_{\rm P}=\sqrt{\frac{K_{\rm P}}{I_{\rm P}}},
\end{equation}
and
\begin{equation}
Q_{\rm T}=\frac{\Omega_{\rm T}M}{\gamma_{\rm T}}, \qquad Q_{\rm Y}=\frac{\Omega_{\rm Y}I_{\rm Y}}{\Gamma_{\rm Y}}, \qquad Q_{\rm P}=\frac{\Omega_{\rm P}I_{\rm P}}{\Gamma_{\rm P}},
\end{equation}

The mechanical responses in displacement and rotation are
\begin{equation}
\label{eq:trans}
z(\omega)=-\frac{F_{\rm T}(\omega)}{M\left( \omega^2-i\Omega_{\rm T}\omega/Q_{\rm T}-\Omega_{\rm T}^2 \right)},
\end{equation}

\begin{equation}
\label{eq:yaw}
\varphi_{\rm Y}(\omega)=-\frac{T_{\rm Y}(\omega)}{I_{\rm Y}\left(
\omega^2-i\Omega_{\rm Y}\omega/Q_{\rm Y}-\Omega_{\rm Y}^2
\right)},
\end{equation}
and
\begin{equation}
\label{eq:pitch}
\varphi_{\rm P}(\omega)=-\frac{T_{\rm P}(\omega)}{I_{\rm P}\left(
\omega^2-i\Omega_{\rm P}\omega/Q_{\rm P}-\Omega_{\rm P}^2
\right)}.
\end{equation}

At frequencies sufficiently above the corresponding rigid-body resonances, the responses approach those of a free mass or free rotor:
\begin{equation}
z(\omega)\simeq -\frac{F_{\rm T}(\omega)}{M\omega^2},
\end{equation}

\begin{equation}
\varphi_{\rm Y}(\omega)\simeq -\frac{T_{\rm Y}(\omega)}{I_{\rm Y}\omega^2},
\end{equation}
and
\begin{equation}
\varphi_{\rm P}(\omega)\simeq -\frac{T_{\rm P}(\omega)}{I_{\rm P}\omega^2}.
\end{equation}
This simple $1/\omega^2$ mechanical response is particularly useful for
calibration because the applied force is determined from the independently measured optical power.

\section{Optical configuration}
\label{subsec:ppa_optics}

\begin{figure}[tbp]
    \centering
    \includegraphics[width=0.95\columnwidth]
    {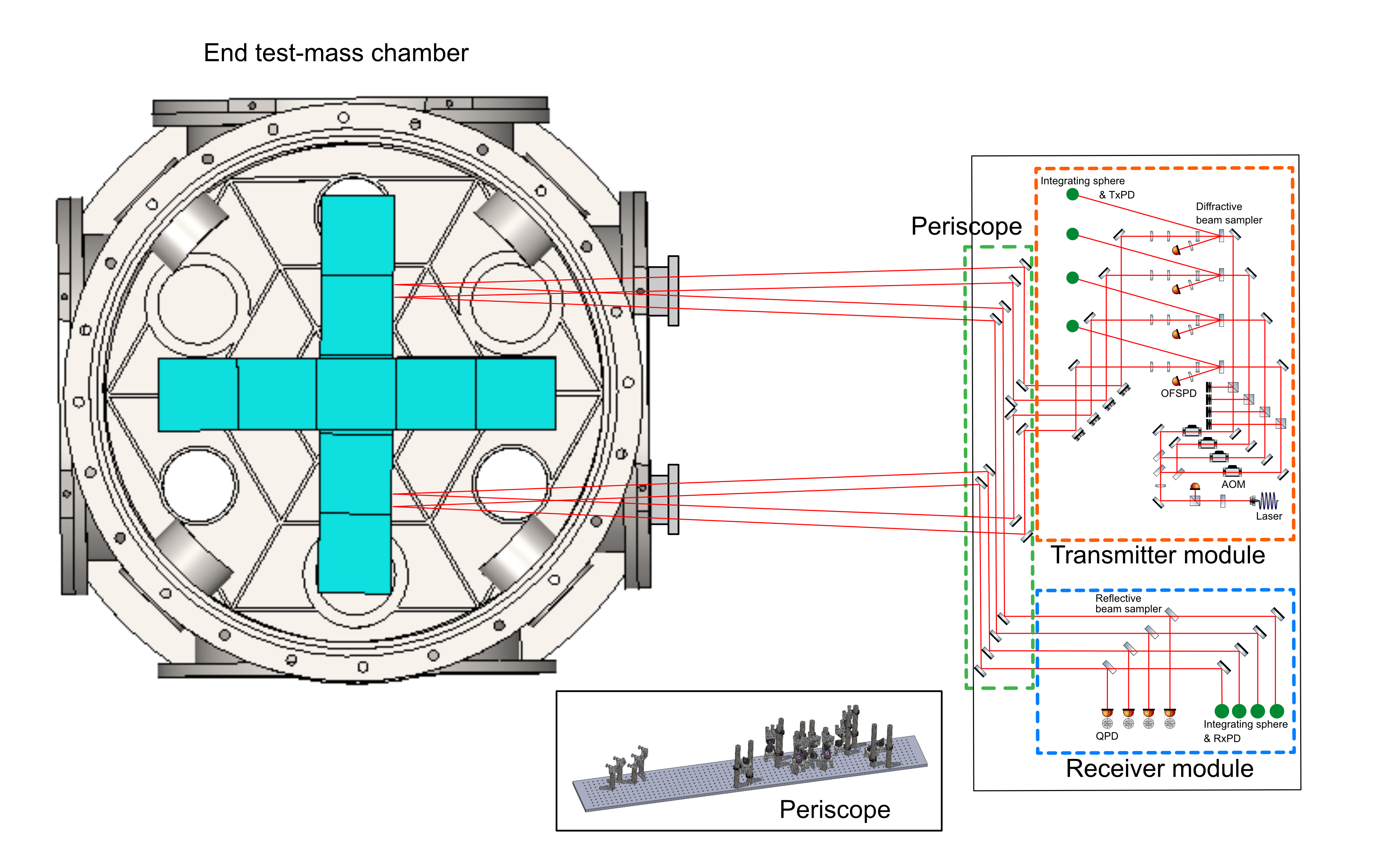}
    \caption{
    Conceptual optical layout of the CHRONOS photon-pressure actuator.
    Four independently controlled beams are injected onto the back surface of one torsion-bar test mass. The system consists of a transmitter module, a beam-steering periscope, and a receiver module.
    }
    \label{fig:overall}
\end{figure}

The optical layout of the four-beam photon-pressure actuator is shown in Fig.~\ref{fig:overall}. The system consists of a transmitter module, a beam-steering periscope, and a receiver module. The transmitter module generates and independently controls four laser beams. The periscope determines their injection positions and angles on the test mass, while the receiver module monitors the reflected powers and beam positions.

A fiber-coupled continuous-wave laser at a wavelength of 1550~nm is used as the optical source. The laser is divided into four independent optical paths. Each path contains an acousto-optic modulator (AOM), whose first-order diffraction power is controlled by an optical follower servo (OFS). An external digital signal is applied to the OFS to provide the actuation command.

\begin{figure}[tbp]
    \centering
    \includegraphics[width=0.68\columnwidth]
    {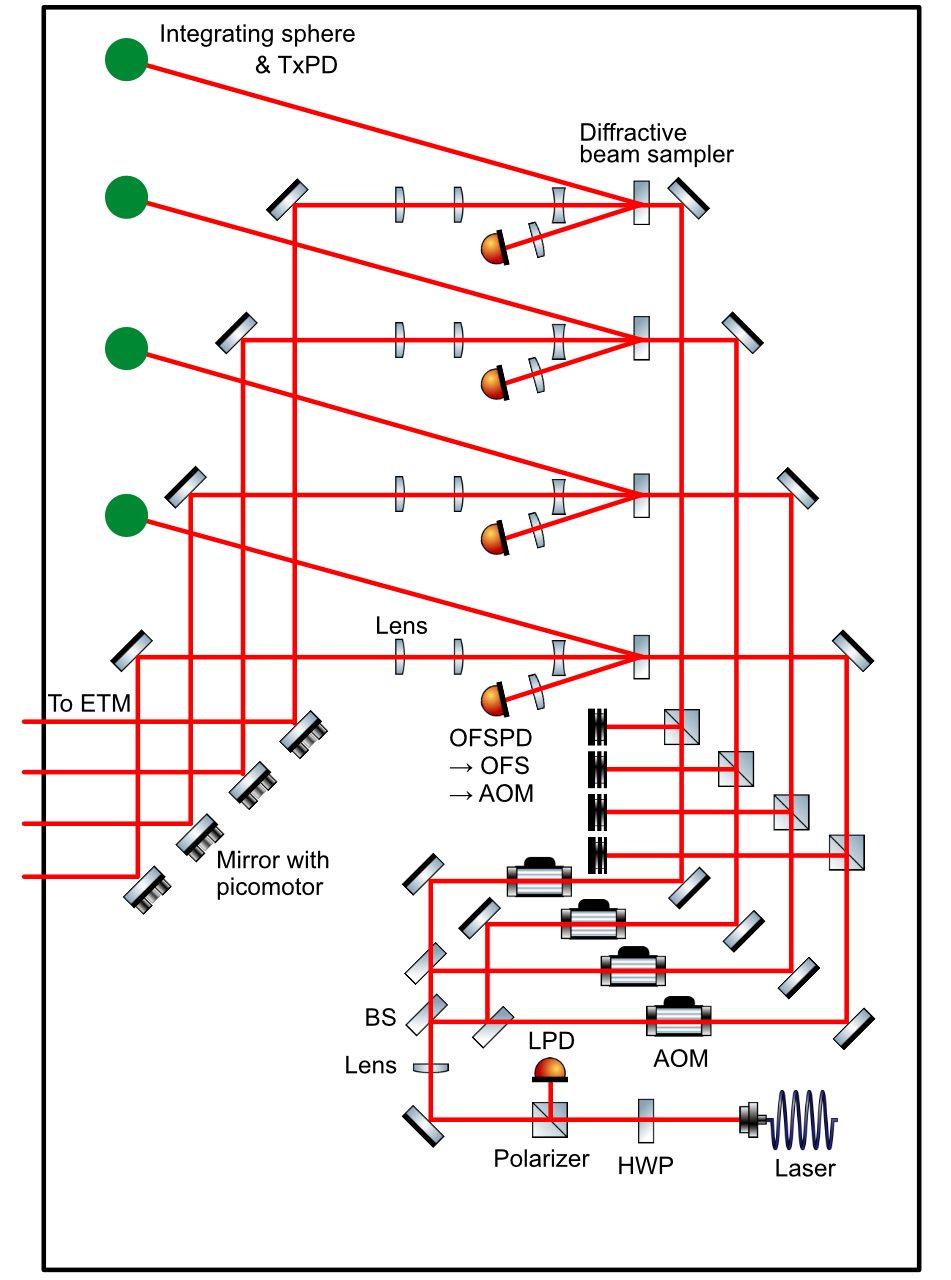}
    \caption{
    Conceptual optical layout of the transmitter module of the CHRONOS
    photon-pressure actuator. The optical powers of the four beams are independently stabilized using AOMs and optical follower servos.
    }
    \label{fig:transmitter}
\end{figure}

Figure~\ref{fig:transmitter} shows the transmitter module. In each beam path, an in-loop photodetector, denoted by OFSPD, measures the optical power and provides the feedback signal for the OFS. An independent out-of-loop photodetector, denoted by TxPD, monitors the actual transmitted optical power. The TxPD is coupled to an integrating sphere so that the total sampled beam power is measured without changing sensitivity by fluctuation of beam shape. A diffractive beam sampler extracts only a small fraction of each beam for these measurements.

The final mirrors of the transmitter module are remotely adjustable, allowing the four beam positions to be aligned to their designated locations on the test mass.

\begin{figure}[tbp]
    \centering
    \includegraphics[width=0.68\columnwidth]
    {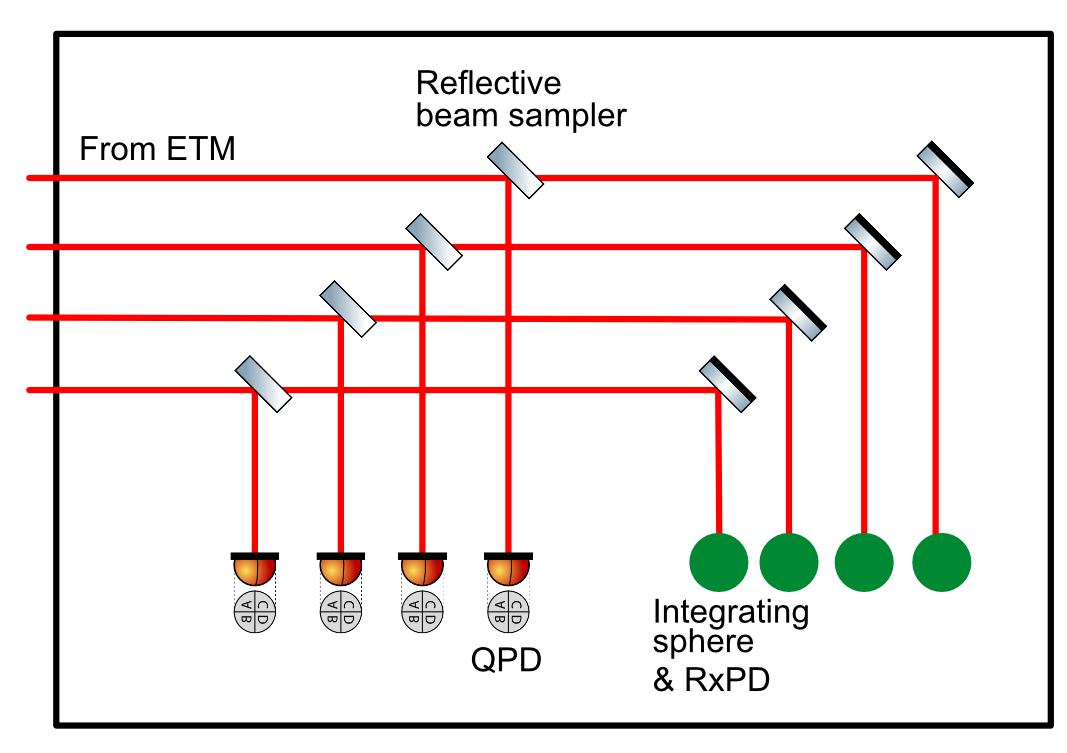}
    \caption{
    Conceptual optical layout of the receiver module. The reflected optical power is monitored using integrating-sphere     photodetectors, while quadrant photodetectors monitor the positions of the four reflected beams.
    }
    \label{fig:receiver}
\end{figure}

The beams reflected by the test mass are collected by the receiver module, shown in Fig.~\ref{fig:receiver}. The returned optical power is monitored by integrating-sphere coupled to photodetectors, denoted by RxPD. A small fraction of each reflected beam is additionally sampled and directed onto quadrant photodetectors (QPDs) to monitor the beam positions.

Beam-position monitoring is especially important for PCal operation because an uncertainty in the injection position directly propagates to an uncertainty in the applied torque~\cite{tanabe_ligo_bulk_deformation}.

\section{Beam-position optimization}
\label{subsec:ppa_beam_positions}

The injection positions cannot be determined only by maximizing the
rigid-body torque. Photon pressure can excite internal elastic modes of the torsion bar and produce surface deformation which potentially couples to the interferometer readout.

\begin{figure}[tbp]
    \centering
    \includegraphics[width=0.88\columnwidth]
    {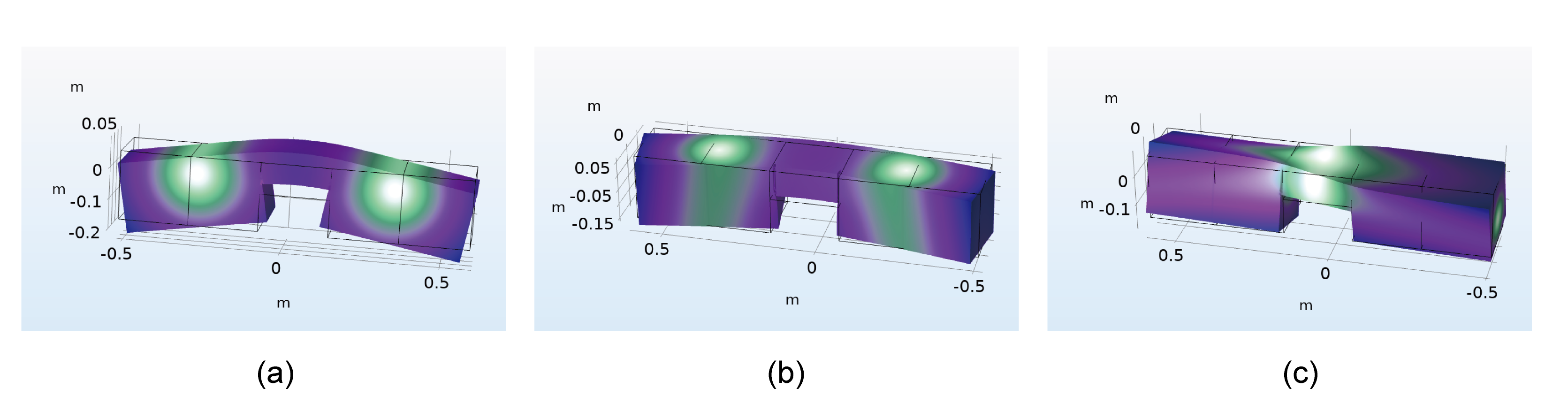}
    \caption{
    Shape of the three lowest internal modes of the CHRONOS torsion bar with five units. (a) vertical bending mode at 666.6~Hz, (b) horizontal bending mode at 1237.7~Hz, (c) twisting mode at 1398.9~Hz. The mode shape was obtained by COMSOL.
    }
    \label{fig:ppa_threemodes}
\end{figure}

The CHRONOS torsion bar has prominent internal resonant modes near
667~Hz and 1238~Hz as shown in Fig.~\ref{fig:ppa_threemodes}. The first is predominantly a vertical bending mode, whereas the latter is a horizontal bending mode. The horizontal mode is particularly relevant to the yaw readout because it causes surface displacement approximately along the axis of the main interferometer beam. This effect is canceled out between the left and right side of the bar when the main beam positions are ideally symmetric, but it can couple to the interferometer in the practical case where the main beams are displaced from the nominal position~\cite{tanabe_chronos_intensity}.

\begin{figure}[tbp]
    \centering
    \includegraphics[width=0.88\columnwidth]
    {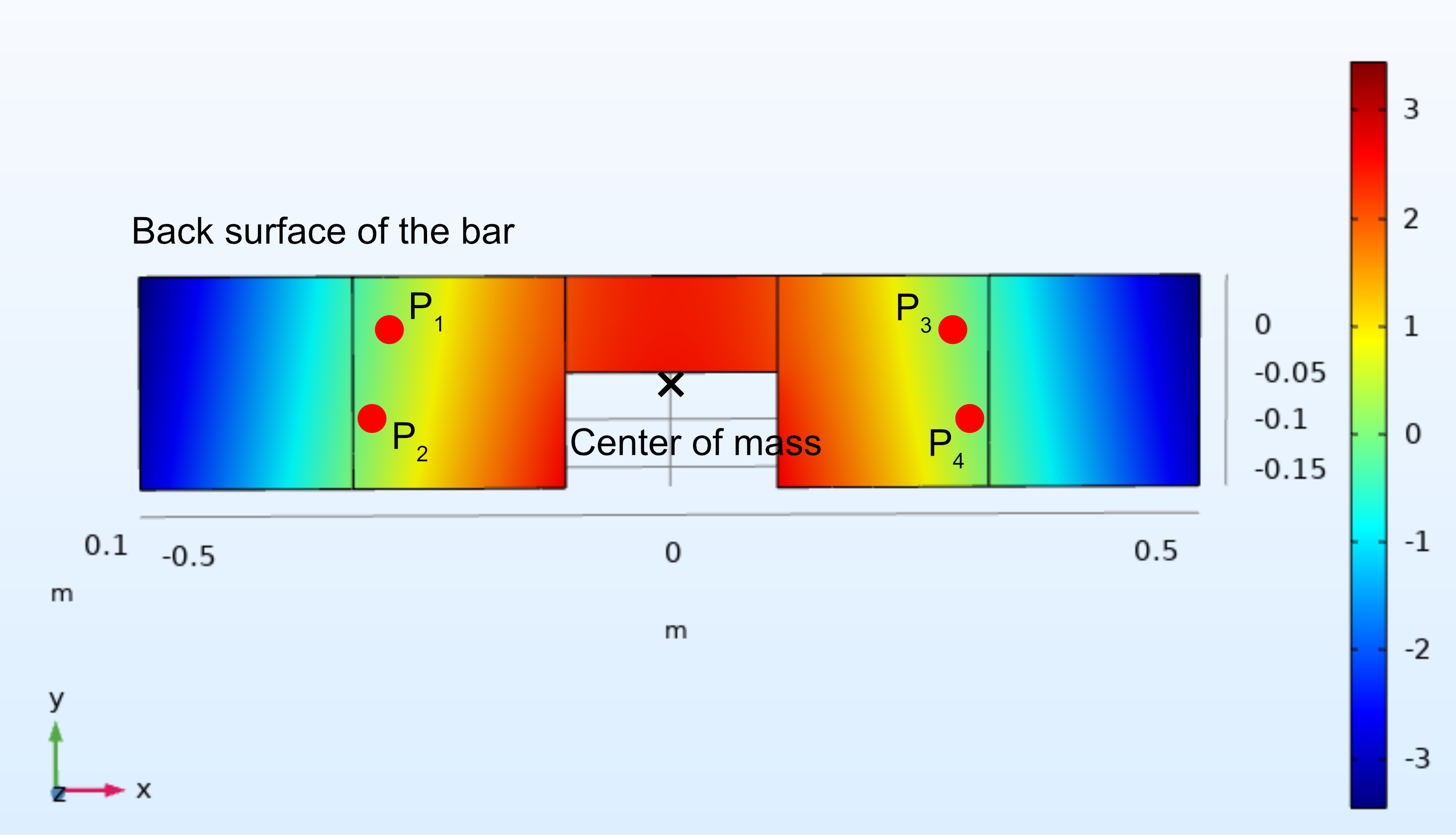}
    \caption{
    Photon-pressure injection positions superimposed on the simulated
    horizontal bending mode of the CHRONOS torsion bar near 1238~Hz.
    The beam positions are selected close to the nodal line in order to
    suppress excitation of the internal elastic mode. The mode shape was obtained by COMSOL.
    }
    \label{fig:injection_node}
\end{figure}

The PPA injection positions are therefore chosen close to the nodal line of the horizontal bending mode. At the same time, the positions are selected to provide the largest yaw lever arm while avoiding bonding layers between the sapphire blocks. The resulting injection points based on the simulated 1238-Hz mode shape are shown in Fig.~\ref{fig:injection_node}.

The representative design parameters are summarized in
Table~\ref{tab:ppa_parameters}.

\begin{table}[tbp]
\centering
\footnotesize
\begin{tabular}{c|c|c}
\hline
Parameter & Symbol & Mean value \\
\hline\hline
Maximum beam-power amplitude & $P_i$ & $1.25~\mathrm{W}$ \\
Azimuthal incidence angle & $\theta_i$ & $\leq 2.9^{\circ}$ \\
Inclination angle & $\phi_i$ & $0^{\circ}$ \\
Upper beam position & $(r_{1,3}^{x},r_{1,3}^{y})$ & $(\mp0.306,\,0.053)~\mathrm{m}$ \\
Lower beam position & $(r_{2,4}^{x},r_{2,4}^{y})$ & $(\mp0.317,\,-0.021)~\mathrm{m}$ \\
Mass of torsion bar & $M$ & $171.32~\mathrm{kg}$ \\
Moment of inertia (yaw) & $I_{\rm Y}$ & $19.86~\mathrm{kg\,m^2}$ \\
Moment of inertia (pitch) & $I_{\rm P}$ & $1.3~\mathrm{kg\,m^2}$ \\
\hline
\end{tabular}
\caption{
Baseline parameters of the CHRONOS photon-pressure actuator.
}
\label{tab:ppa_parameters}
\end{table}

The design values follow the current CHRONOS PPA study, in which a
10-W 1550-nm laser is divided into four paths and an actuation amplitude of approximately $\pm1.25$~W per beam is available after the AOM power-control stage~\cite{tanabe_photon_pressure_actuator}.

\section{Actuation range}
\label{subsec:ppa_range}

Using the parameters in Table~\ref{tab:ppa_parameters}, the maximum
radiation-pressure force generated by one beam is of the order of
\begin{equation}
F_i\simeq \frac{2P_i}{c}\simeq 8.3\times10^{-9}\ {\rm N},
\end{equation}
for $P_i=1.25$~W and near-normal incidence.

Combining the four beams gives the maximum translational force amplitude of
\begin{equation}
F_{\rm T,max}\simeq 3.3\times10^{-8}\ {\rm N}.
\end{equation}

The corresponding maximum torque amplitudes are approximately
\begin{equation}
T_{\rm Y,max}\simeq 1.0\times10^{-8}\ {\rm N\,m},
\end{equation}
for yaw and
\begin{equation}
T_{\rm P,max}\simeq 1.2\times10^{-9}\ {\rm N\,m},
\end{equation}
for pitch.

At 1~Hz, the corresponding actuation efficiencies are approximately
\begin{equation}
A_{\rm T}\simeq 2.5\times10^{-13}\ {\rm m/V},
\end{equation}

\begin{equation}
A_{\rm Y}\simeq 6.6\times10^{-13}\ {\rm rad/V},
\end{equation}
and
\begin{equation}
A_{\rm P}\simeq 1.2\times10^{-12}\ {\rm rad/V}.
\end{equation}
These values provide sufficient actuation authority for maintaining
interferometer lock. The predicted available force is also sufficient for lock acquisition under the present estimate of the CHRONOS seismic motion, although the final requirement must be validated using the measured seismic environment at the experimental site
\cite{tanabe_photon_pressure_actuator}.

This capability is important for the suspension design because it allows the test-mass stage to operate without a dedicated recoil mass.
Local recoil masses can therefore be retained at intermediate suspension stages where required, while the final torsion-bar test mass can be actuated by mechanically contactless photon pressure.

\section{Actuation noise}
\label{subsec:ppa_noise}

The relevant noise of the photon-pressure actuator enters primarily through fluctuations of the optical power delivered to the test mass.
The principal contributions are intrinsic laser intensity noise,
photodetector shot noise, and electronic noise introduced through the
intensity-control system~\cite{tanabe_photon_pressure_actuator}.

The beam power is stabilized by a feedback loop consisting of the OFSPD, OFS, and AOM. The power-control loop suppresses the intrinsic laser intensity noise, but noise from the photodetectors, AOM electronics, and digital-to-analog converters can be introduced through the same control path.

A particularly important contribution is the noise of the DAC used to provide the excitation and offset signals to the OFS. For the baseline design, the DAC-related excitation noise is expected to be the dominant actuation-noise contribution. Analog dewhitening between the DAC and OFS is therefore included in the control concept to reduce this noise.

\begin{figure}[tbp]
    \centering
    \includegraphics[width=0.88\columnwidth]
    {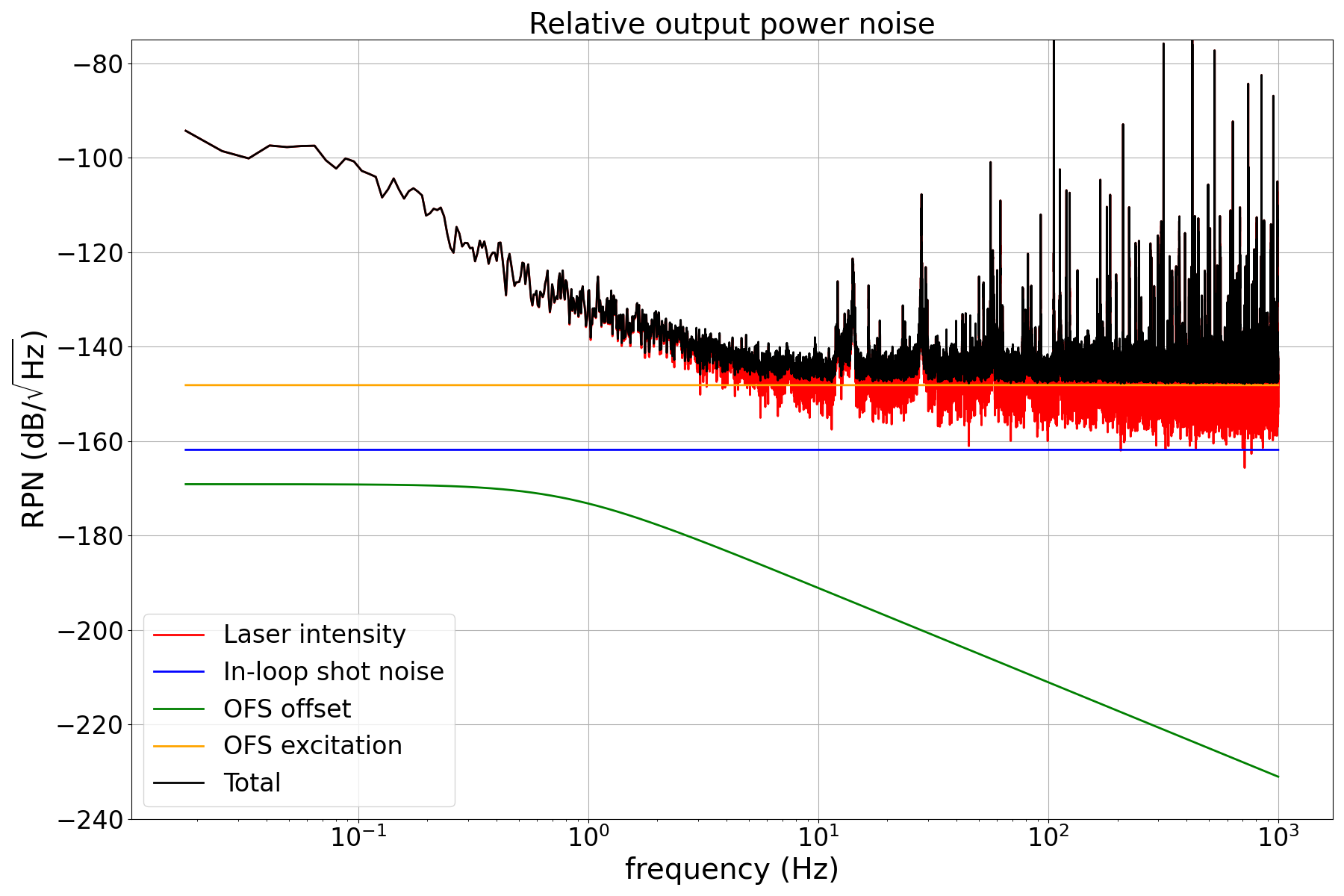}
    \caption{
    Relative power noise budget of the total noise power injected to the ETM. Normalized by the maximum power $P_0=$2.5~W.
    }
    \label{fig:laser_total_noise}
\end{figure}

An important property of the four-beam configuration is that the intrinsic laser intensity noise is largely common to the four optical paths. For yaw actuation, fluctuations of the common power on opposite sides of the torsion bar generate torques of opposite sign and therefore cancel to first order. The remaining coupling is determined by mismatches in beam power and beam position~\cite{tanabe_chronos_intensity}. Independent electronic noise from the four control channels instead adds incoherently. The noise contribution from each component is shown in Fig.~\ref{fig:laser_total_noise}.

At the baseline configuration, the predicted noise amplitudes at 1~Hz are
\begin{equation}
\delta z_{\rm PPA}\simeq 3.9\times10^{-19}\ {\rm m/\sqrt{Hz}},
\end{equation}

\begin{equation}
\delta\varphi_{\rm Y,PPA}\simeq 5.3\times10^{-19}\ {\rm rad/\sqrt{Hz}},
\end{equation}
and
\begin{equation}
\delta\varphi_{\rm P,PPA}\simeq 1.1\times10^{-18}\ {\rm rad/\sqrt{Hz}}.
\end{equation}
The quoted yaw noise is a single-bar angular quantity. If the two bars have
independent equal PPA noise, their differential ASD at 1~Hz is
$\sqrt{2}\,(5.3\times10^{-19})=7.5\times10^{-19}~\mathrm{rad/\sqrt{Hz}}$.
Using $\theta_X-\theta_Y=\eta h$, $\eta=0.93656$, this is
$8.0\times10^{-19}~\mathrm{Hz}^{-1/2}$ in strain, compared with
$3.14\times10^{-18}$ and $1.97\times10^{-18}~\mathrm{Hz}^{-1/2}$ for the
minimum-PI and logarithmic-mean detector budgets at 1~Hz. This establishes
a margin at that frequency under the independence assumption. The full
frequency dependence and inter-bar correlations must be propagated before
claiming a margin across the band. Additional PPA noise is an engineering
allocation, not already included in the six-component reference budget.

The photon-pressure actuator has no direct magnetic coupling. Furthermore, unlike an electromagnetic or electrostatic actuator referenced to a nearby recoil mass, it does not require a mechanically moving local reaction mass. This removes the dominant direct pathways through which seismic motion of recoil-mass and environmental magnetic fluctuations would couple to the actuation force.

Motion of the external PPA optics may nevertheless couple indirectly through beam pointing, beam-position fluctuations, or residual power fluctuations. For this reason, both the transmitted and reflected beam positions are continuously monitored and the optical-power stabilization loop is treated as an essential part of the actuator.

\section{Operation as a Photon Calibrator}
\label{subsec:ppa_pcal}

The same optical system can be used as an absolute Photon Calibrator~\cite{tanabe_photon_pressure_actuator}. In a normal interferometer operation, the dominant component of the beam-power command is used for feedback actuation. A smaller sinusoidal modulation can simultaneously be superimposed on this
signal to generate calibration lines.

Because the radiation-pressure force is directly related to the incident optical power,
\begin{equation}
F=\frac{2P}{c}\cos\theta,
\end{equation}
the applied force can be calculated from independently calibrated optical power and geometry. For rotational calibration, the corresponding torque is determined from the beam positions relative to the center of mass.

The PCal-induced yaw rotation is therefore given by
\begin{equation}
\varphi_{\rm Y}^{\rm PCal}(\omega)=-\frac{T_{\rm Y}^{\rm PCal}(\omega)}{I_{\rm Y}\left( \omega^2-i\Omega_{\rm Y}\omega/Q_{\rm Y}-\Omega_{\rm Y}^2 \right)
}.
\end{equation}
At frequencies sufficiently above the yaw resonance,
\begin{equation}
\varphi_{\rm Y}^{\rm PCal}(\omega)\simeq -\frac{T_{\rm Y}^{\rm PCal}(\omega)}{I_{\rm Y}\omega^2}.
\end{equation}

The conversion from the calibrated torsion-bar rotation to
strain-equivalent detector response is performed using the interferometer response described in Chap.~\ref{chap:sensitivity}. Separating the mechanical PPA model from the interferometer response avoids introducing detector-specific optical factors into the definition of the actuator itself.

The integrated PPA/PCal system provides several operational advantages.
First, the absolute actuation coefficient can be monitored without relying on an electromagnetic actuator model. Second, continuous calibration lines can be applied while the same beams are used for interferometer control. Third, yaw and pitch excitation can be generated independently, allowing measurements of angular response and of the offset of the main interferometer beam from the nominal test-mass center.

\section{Systematic uncertainty}
\label{subsec:ppa_systematic}

When the photon-pressure actuator is used as a PCal, the dominant systematic uncertainties arise from the absolute optical power, injection positions, incident angles, and mechanical parameters of the torsion bar~\cite{tanabe_photon_pressure_actuator}.

For yaw calibration, the relevant quantities are the beam powers $P_i$,
horizontal lever arms $r_i^x$, incident angles, and the yaw moment of inertia $I_{\rm Y}$. The resulting uncertainty in the induced yaw angle can be written schematically as
\begin{equation}
\Delta\varphi_{\rm Y}=\left[ \sum_j \left( \frac{\partial\varphi_{\rm Y}}{\partial x_j}\Delta x_j \right)^2 \right]^{1/2},
\end{equation}
where $x_j$ represents the above parameters.

For the baseline CHRONOS design, the projected total systematic uncertainty when the PPA is operated as a PCal is approximately
\begin{equation}
\frac{\Delta\varphi_{\rm Y}}{\varphi_{\rm Y}}\simeq 1.14\%.
\end{equation}

The uncertainty in the absolute beam power is presently one of the dominant terms. Further improvement is expected from high-accuracy calibration of the integrating-sphere power sensors and from improved measurements of the fabricated torsion-bar geometry and moment of inertia.

Continuous monitoring of the beam position using the QPDs in the receiver module is also essential because an injection-position error changes the lever arm length and can additionally enhance coupling to the internal elastic modes.

\section{Complementarity with the gravitational calibrator}
\label{subsec:ppa_gcal}

The Photon Calibrator and gravitational calibrator (GCal) provide
complementary absolute calibration methods for CHRONOS.

The PCal generates force through photon momentum and can inject calibration signals over a broad frequency range. Its actuation frequency is limited primarily by the optical power-control system and by internal resonances of the test mass. It is therefore particularly suitable for continuous calibration lines and for measurements extending to frequencies well above the sub-hertz band.

In contrast, the GCal produces a purely Newtonian gravitational torque and does not rely on optical-power calibration~\cite{chronos_gcal}. It can therefore provide an independent absolute reference at low frequencies, where its rotating-mass system can operate efficiently. See Chap.~\ref{chap:gcal} for the detail of the GCal specified to CHRONOS.

Cross-calibration between the two systems provides an important consistency test. In particular, the GCal response can be used to constrain the absolute scale of the PCal, while the PCal can extend the calibrated detector response to frequencies inaccessible to practical GCal rotation~\cite{Inoue2018_GCal}. The combination of PPA/PCal and GCal therefore establishes a broadband and redundant calibration strategy for CHRONOS.

\section{Role in the CHRONOS suspension and control system}
\label{subsec:ppa_suspension_role}

The photon-pressure actuator plays a different role from conventional local actuators installed at intermediate stages of the suspension.

During initial suspension damping and alignment, electromagnetic actuators can be retained at the upper suspension stages, where their larger actuation range is advantageous. In particular, recoil masses can be deployed for intermediate-mass and marionette-stage control because these actuators don't directly couple to the test mass.

At the final torsion-bar test-mass stage, however, CHRONOS aims to omit a dedicated recoil mass. Once the interferometer is brought sufficiently close to its operating point, the photon-pressure actuator provides the final longitudinal and angular actuation required for lock acquisition and lock maintenance. The resulting control hierarchy is therefore
\begin{equation}
\begin{array}{c} \mathrm{Actuation of}\\\mathrm{upper \ suspension}\\\mathrm{(EM/ESD)} \end{array} \quad \longrightarrow \quad \begin{array}{c} \mathrm{Lock\ acquisition \ of}\\\mathrm{Test \ mass}\\\mathrm{(PPA)} \end{array} \quad \longrightarrow \quad \begin{array}{c} \mathrm{Low \ noise \ operation\ of}\\\mathrm{Test \ mass}\\\mathrm{(PPA)} \end{array}
\end{equation}

This architecture separates the large-range suspension-control requirement from the ultra-low-noise test-mass actuation requirement.
It is particularly attractive for CHRONOS because the detector targets the sub-hertz band, where seismic and actuator noise become progressively more important.

The photon-pressure actuator is therefore not merely a calibration device, but an integral component of the CHRONOS suspension, interferometer-control, and calibration architecture.

%% file: Chapters/GCAL.tex
\chapter{Torque-coupled Gravitational Calibrator}
\label{chap:gcal}

Precise calibration is particularly challenging for sub-Hz
torsion-bar gravitational-wave detectors, where conventional
force-coupled calibration schemes can provide insufficient
signal-to-noise ratio at low frequencies.
To overcome this limitation, CHRONOS adopts a torque-coupled
gravity field calibrator (GCal)~\cite{Inoue2018_GCal}, in which a rotating quadrupolar
mass distribution generates a deterministic Newtonian gravitational
torque that couples directly to the rotational degree of freedom
of the torsion bars
\cite{InoueYukawaHarmonic2026,Inoue2018_GCal}.

\begin{figure}[t]
    \centering
    \includegraphics[width=0.9\linewidth]
    {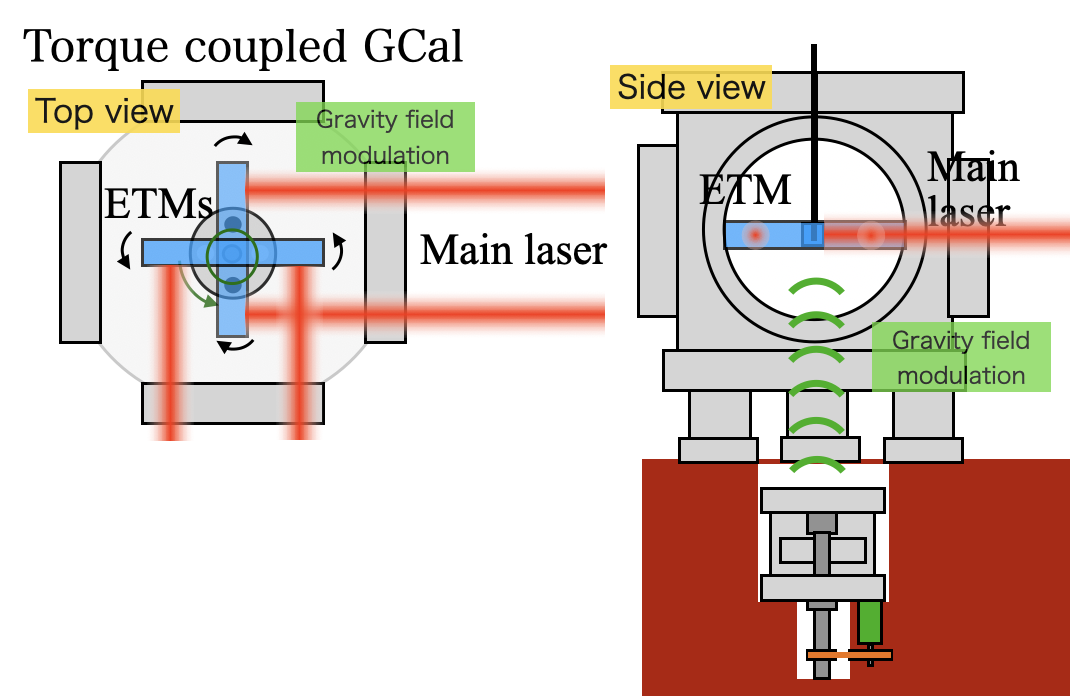}
    \caption{
    Conceptual comparison of a conventional force-coupled GCal
    (left) and the torque-coupled GCal adopted for CHRONOS (right),
    reproduced from the submitted GCal study.
    }
    \label{fig:gcal_concept}
\end{figure}

The central concept is illustrated in
Fig.~\ref{fig:gcal_concept}.
In conventional gravity-based calibration schemes developed for
kilometer-scale interferometers, the time-dependent gravitational field
primarily excites translational motion of the test mass.
In contrast, the torque-coupled configuration places the rotating
quadrupole rotor directly beneath the torsion bar so that the
gravity-gradient interaction directly excites the torsional mode.
This direct coupling substantially enhances the calibration signal
in the sub-Hz band.

Because the rotor has quadrupolar symmetry, the dominant gravitational
excitation appears at twice the rotor rotation frequency,
\begin{equation}
    f_{\rm cal}=2f_{\rm rot}.
\end{equation}
The resulting signal is therefore a narrow spectral calibration line
whose amplitude can be predicted from the gravitational constant,
the calibration masses, and the measured geometry of the system.

Unlike electromagnetic coil actuators, the excitation mechanism
does not require an actuator-force model.
Compared with photon-pressure calibration
\cite{Karki2016Pcal,kagra_pcal}, the applied excitation is determined directly
by Newtonian gravity and the mass geometry.
The observed detector response still contains the mechanical
susceptibility of the torsion bars; however, in the CHRONOS
calibration band the excitation frequency lies well above the
torsional resonance.
The response therefore approaches the inertial limit and becomes
largely insensitive to uncertainties in the torsional spring constant
and damping.

Table~\ref{tab:gcal_parameters} collects the geometric, mass,
frequency, and correction-factor parameters used in the submitted
GCal and Yukawa studies.  The two columns describe distinct source
configurations and should not be combined: the first is the
single-rotor calibration reference, whereas the second is the
long--short differential arrangement used for Newtonian cancellation.

\begin{table}[t]
    \centering
    \caption{Key parameters used in the CHRONOS GCal model
    \cite{chronos_gcal} and the submitted differential-gravity
    analysis \cite{InoueYukawaHarmonic2026}.
    Here $m$ is the mass
    of each rotor endpoint and $M$ is the mass of each effective
    torsion-bar endpoint.}
    \label{tab:gcal_parameters}
    \begin{tabular}{p{0.19\linewidth}p{0.31\linewidth}p{0.36\linewidth}}
        \hline
        Parameter & Calibration GCal & Differential Yukawa GCal \\
        \hline
        $M$ & $85.66~\mathrm{kg}$ & $85.66~\mathrm{kg}$ \\
        $I$ & $19.86~\mathrm{kg\,m^2}$ & $19.86~\mathrm{kg\,m^2}$ \\
        $a$ & $0.30213~\mathrm{m}$ & $0.30213~\mathrm{m}$ \\
        Endpoint mass(es) & $m=1.8842~\mathrm{kg}$ &
            $m_L=m_S=1.8842~\mathrm{kg}$ \\
        Rotor radius/radii & $b=0.160~\mathrm{m}$ &
            $b_L=0.500~\mathrm{m}$, $b_S=0.250~\mathrm{m}$ \\
        Vertical distance(s) & $h=2.00~\mathrm{m}$ &
            $h_L=10.55258~\mathrm{m}$, $h_S=8.000~\mathrm{m}$ \\
        Transverse offset(s) & $\rho_\perp=0.001~\mathrm{m}$ &
            $\rho_L=\rho_S=0.0010~\mathrm{m}$ \\
        Rotor/\allowbreak calibration frequency & $0.1$--$10~\mathrm{Hz}$ calibration band &
            $f_{\rm rot}=0.25~\mathrm{Hz}$,
            $f_{\rm cal}=0.50~\mathrm{Hz}$ \\
        GCal finite-size correction & $\eta_{\rm GCAL}=0.9743$ &
            $\eta_{\rm GCAL}=0.9743$ \\
        Detector response geometry & $\eta_g=I_{\rm eff}/I=0.93656$ &
            $\eta_g=I_{\rm eff}/I=0.93656$ \\
        \hline
    \end{tabular}
\end{table}

\section{Geometrical configuration}
\label{sec:gcal_geometry}

\begin{figure}[t]
    \centering
    \includegraphics[width=0.9\linewidth]
    {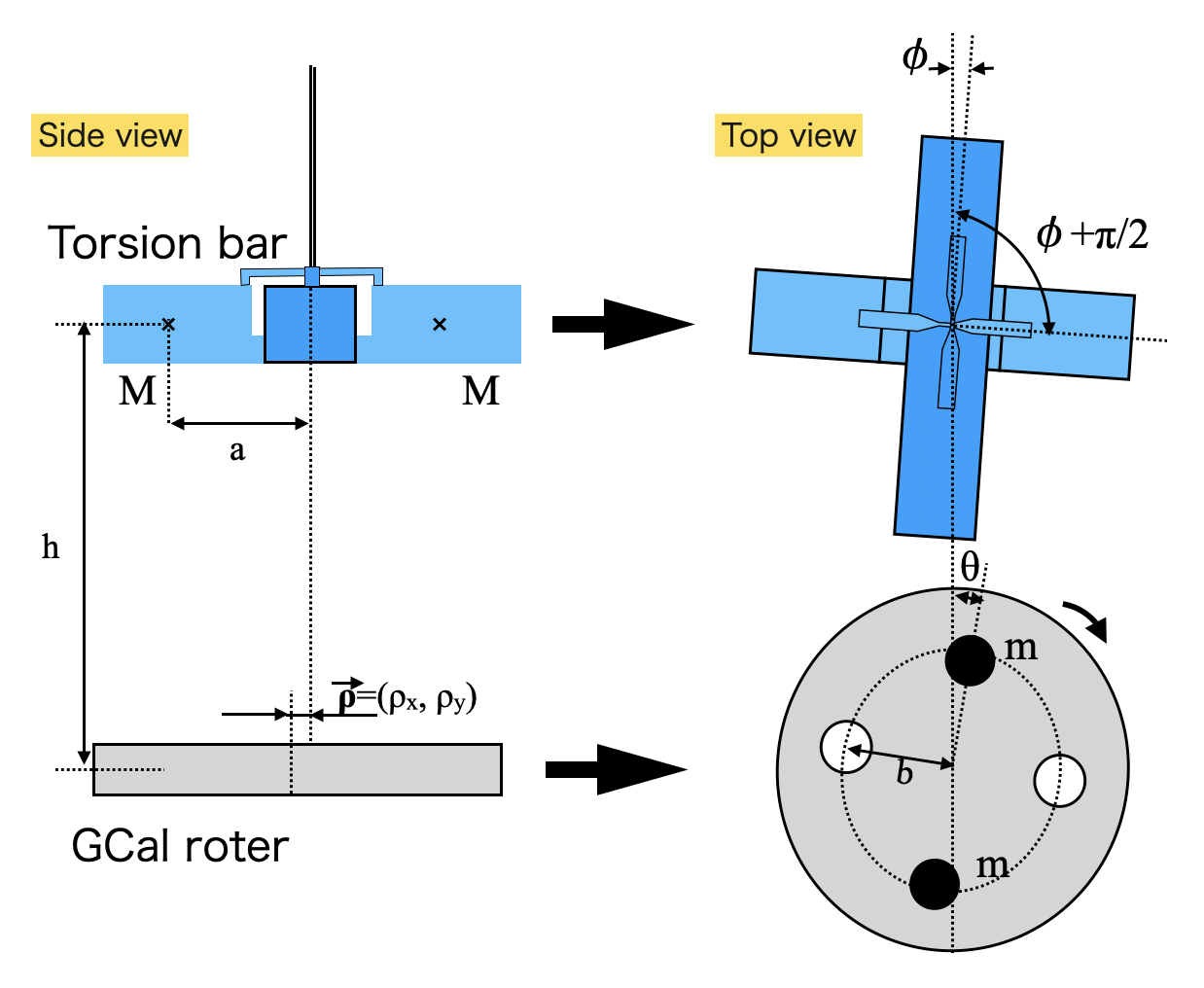}
    \caption{
    Geometrical configuration of the torque-coupled GCal.
    The effective quadrupole arm length of the torsion bar is $a$,
    the rotor radius is $b$, the vertical separation is $h$, and
    $\boldsymbol{\rho}$ denotes the transverse offset between the
    torsion-bar and rotor axes.
    }
    \label{fig:gcal_geometry}
\end{figure}

The geometry of the torque-coupled GCal is shown in
Fig.~\ref{fig:gcal_geometry}.
A Cartesian coordinate system is defined with the center of the
torsion bar at the origin and the torsion axis along the $z$ direction.
The center of the GCal rotor is located below the torsion bar with
vertical separation $h$ and may have a small transverse offset
\begin{equation}
    \boldsymbol{\rho}
    =
    (\rho_x,\rho_y,0),
\end{equation}
with
\begin{equation}
    \rho_\perp^2=\rho_x^2+\rho_y^2.
\end{equation}

To describe the leading quadrupolar interaction, the torsion bar
is represented by two effective masses $M$ located at
$\pm a$ from the torsion axis.
For a distributed torsion bar, $a$ is therefore an
\emph{effective quadrupole arm length}, rather than simply the
physical half-length of the bar.
Similarly, the GCal rotor is represented by two calibration masses
$m$ located at radius $b$ from the rotor axis.

The characteristic geometrical scale is defined as
\begin{equation}
    R_0^2
    =
    a^2+b^2+h^2+\rho_\perp^2,
    \label{eq:gcal_R0}
\end{equation}
and the dimensionless quadrupolar coupling parameter is
\begin{equation}
    \xi
    =
    \frac{2ab}{R_0^2}.
    \label{eq:gcal_xi}
\end{equation}

For the CHRONOS geometry, the hierarchy
\begin{equation}
    \rho_\perp \ll a,b \ll R_0
\end{equation}
allows the gravitational interaction to be expanded perturbatively.
The transverse offset produces higher-order corrections and forms
an important contribution to the systematic uncertainty budget.

\section{Gravitational torque}
\label{sec:gcal_torque}

Let $\phi$ denote the azimuthal orientation of the torsion bar and
\begin{equation}
    \theta(t)=\omega t
\end{equation}
the rotation angle of the GCal rotor.
The relevant relative angle is
\begin{equation}
    \Delta=\phi-\theta .
\end{equation}

Summing the Newtonian interaction between the two effective masses
of the torsion bar and the two rotor masses gives a periodic
gravitational potential.
Because of the left--right symmetry of the quadrupolar mass
distribution, odd harmonics cancel, and the leading oscillatory
component appears at $2f_{\rm rot}$.

The corresponding torque can be written as
\begin{equation}
    \tau_\phi^{(2f)}(t)
    =
    -\tau_2 \sin(2\Delta),
    \label{eq:gcal_torque_time}
\end{equation}
where
\begin{equation}
    \tau_2
    =
    \frac{GmM}{R_0}
    \left[
        \frac{3}{2}\xi^2
        +
        \frac{35}{32}\xi^4
        +
        \frac{3465}{4096}\xi^6
        +\mathcal{O}(\xi^8)
    \right].
    \label{eq:tau2_full}
\end{equation}

The leading term therefore scales as
\begin{equation}
    \tau_2
    \propto
    \frac{GmMa^2b^2}{R_0^5}.
\end{equation}
This relation provides a simple design guideline:
larger calibration masses and rotor radii enhance the gravitational
torque, whereas increasing the rotor--detector separation rapidly
suppresses the calibration signal.

Finite dimensions of the torsion-bar mass distribution introduce
a correction to the idealized point-mass quadrupole model.
For the CHRONOS geometry this effect is represented by the
finite-size correction factor
\begin{equation}
    \eta_{\rm GCAL}=0.9743
    \label{eq:eta_gcal}
\end{equation}
as obtained for the finite-size GCal geometry in the submitted
analyses \cite{InoueYukawaHarmonic2026}.
Thus the physical calibration-torque amplitude is
\begin{equation}
    \tau_{\rm GCAL}^{(2f)}=\eta_{\rm GCAL}\tau_2,
    \label{eq:tau_gcal_corrected}
\end{equation}
and all torque-to-response conversions below use this corrected
amplitude.

CHRONOS employs two orthogonal torsion bars.
For bar orientations separated by $\pi/2$, the GCal produces
equal-amplitude torques with opposite signs,
\begin{align}
    \tau_X(t)
    &=
    \eta_{\rm GCAL}\tau_2\sin(2\omega t),\\
    \tau_Y(t)
    &=
    -\eta_{\rm GCAL}\tau_2\sin(2\omega t).
\end{align}
The calibration signal is therefore enhanced in the differential
torsion channel while common-mode excitation is suppressed.

\section{Mechanical and strain-equivalent response}
\label{sec:gcal_response}

The angular motion of each torsion bar is described by the standard
driven torsional-oscillator equation
\begin{equation}
    I\ddot{\phi}
    +
    \Gamma\dot{\phi}
    +
    \kappa\phi
    =
    \tau_\phi(t),
    \label{eq:gcal_eom}
\end{equation}
where $I$ is the moment of inertia, $\kappa$ is the torsional
spring constant, and $\Gamma$ is the damping coefficient.
Introducing
\begin{equation}
    \omega_0^2=\frac{\kappa}{I},
    \qquad
    \Gamma=\frac{I\omega_0}{Q},
\end{equation}
the exact mechanical susceptibility is
\begin{equation}
    \chi(\Omega)
    =
    \frac{1}
    {I\left(
        \omega_0^2-\Omega^2
        +i\omega_0\Omega/Q
    \right)},
    \label{eq:gcal_chi}
\end{equation}
where
\begin{equation}
    \Omega=2\omega
\end{equation}
is the angular frequency of the calibration signal.

The differential angular response of the two orthogonal torsion bars
is therefore
\begin{equation}
    \Phi_{\rm diff}(\Omega)
    =
    2\eta_{\rm GCAL}\chi(\Omega)\tau_2.
    \label{eq:gcal_phi_diff}
\end{equation}

For gravitational-wave readout, the differential angular response
is related to strain through
\begin{equation}
    \Phi_{\rm diff}
    =
    \eta_g
    \left(
        h_+F_+ + h_\times F_\times
    \right),
\end{equation}
where
\begin{equation}
    \eta_g\equiv\frac{I_{\rm eff}}{I}=0.93656
    \label{eq:eta_g}
\end{equation}
is the geometrical coupling factor of the torsion-bar detector and
$F_+$ and $F_\times$ are the antenna pattern functions.  This
detector-response factor is distinct from the GCal finite-size
correction $\eta_{\rm GCAL}$ defined in
Eq.~\eqref{eq:eta_gcal}.
We therefore define the strain-equivalent GCal amplitude as
\begin{equation}
    h_{\rm GCal}(\Omega)
    =
    \frac{\Phi_{\rm diff}(\Omega)}
    {\eta_g |F_{\rm eff}|},
    \qquad
    |F_{\rm eff}|
    =
    \sqrt{F_+^2+F_\times^2}.
    \label{eq:gcal_strain}
\end{equation}

For CHRONOS, the torsional resonance frequency satisfies
\begin{equation}
    \frac{\omega_0}{2\pi}
    <
    0.01~{\rm Hz},
\end{equation}
whereas GCal operation is considered in the
$0.1$--$10~{\rm Hz}$ band.
Thus,
\begin{equation}
    \Omega\gg\omega_0,
\end{equation}
and the mechanical susceptibility approaches the inertial limit,
\begin{equation}
    \chi(\Omega)
    \simeq
    -\frac{1}{I\Omega^2}.
\end{equation}
The strain-equivalent calibration signal then becomes
\begin{equation}
    h_{\rm GCal}(\Omega)
    \simeq
    -
    \frac{
        2\eta_{\rm GCAL}\tau_2
    }
    {
        \eta_g |F_{\rm eff}| I\Omega^2
    }.
    \label{eq:hgcal_full}
\end{equation}

The two factors have different physical meanings and are not
interchangeable.  The factor $\eta_{\rm GCAL}$ multiplies only the
applied gravitational torque to correct the finite-size GCal model,
whereas $\eta_g$ appears only in the conversion from differential
angle to gravitational-wave strain.  Consequently they enter the
final strain-equivalent amplitude through the ratio
$\eta_{\rm GCAL}/\eta_g$.

The characteristic scaling
\begin{equation}
    |h_{\rm GCal}|
    \propto
    \Omega^{-2}
\end{equation}
is a consequence of the inertial response of the torsion bars.
Importantly, this approximation reduces the sensitivity of the
calibration response to uncertainties in the torsional restoring
force and damping and provides a particularly simple absolute
reference in the sub-Hz observation band.

\section{General Newtonian torque series}
\label{sec:gcal_newtonian_series}

The Newtonian contribution can be written to all multipole orders by
taking the long-range limit, $\beta=R/\lambda\rightarrow0$, of the
kernel expansion derived in the submitted analysis
\cite{InoueYukawaHarmonic2026}.  In this limit the coefficient
$B_n(\beta)$ reduces to the purely Newtonian coefficient $A_n$,
\begin{equation}
    \lim_{\beta\rightarrow0}B_n(\beta)=A_n,
\end{equation}
where
\begin{equation}
    A_n
    =
    \frac{1}{2^{6n-4}}
    \binom{4n}{2n}
    \binom{2n}{n-1}.
    \label{eq:gcal_An}
\end{equation}
The finite-size-corrected Newtonian torque at the fundamental
quadrupolar harmonic is therefore
\begin{equation}
    \tau_{\rm N}(t)
    =
    \eta_{\rm GCAL}\frac{GmM}{R_0}
    \sum_{n=1}^{\infty}
    A_n\xi^{2n}\sin(2\Delta).
    \label{eq:gcal_newtonian_torque_general}
\end{equation}
The first three coefficients are
\begin{equation}
    A_1=\frac{3}{2},
    \qquad
    A_2=\frac{35}{32},
    \qquad
    A_3=\frac{3465}{4096},
\end{equation}
which reproduce Eq.~\eqref{eq:tau2_full}.  Equation
\eqref{eq:gcal_newtonian_torque_general} is thus the general
Newtonian expression underlying the truncated calibration formula.

\section{Expected calibration performance}
\label{sec:gcal_sensitivity}

\begin{figure}[tbp]
\centering
\includegraphics[width=0.95\linewidth]{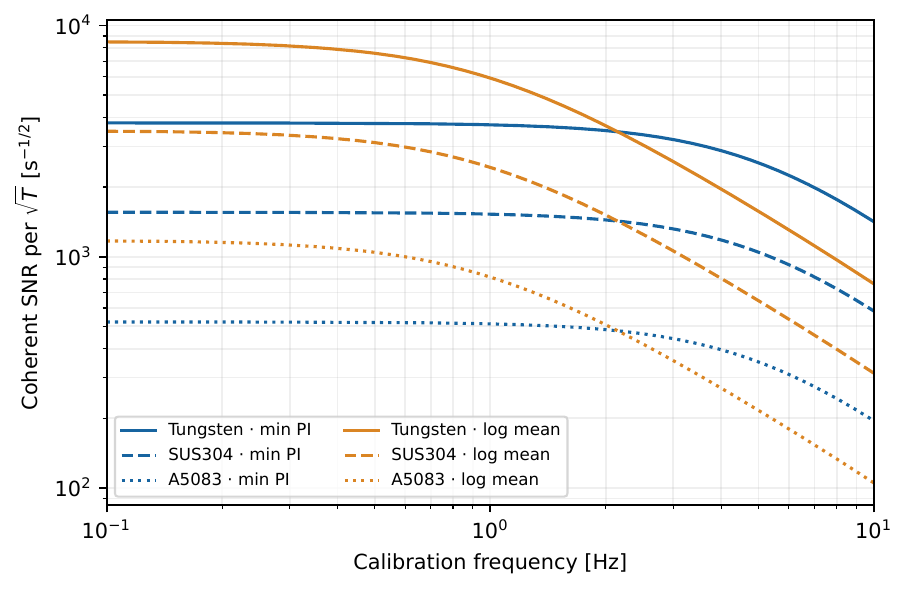}
\caption{Coherent calibration SNR per square root of integration time,
$|h_{\rm GCal}|/\sqrt{S_h}$, using both full-noise operating points from
Chap.~\ref{chap:sensitivity}. A phase-referenced sinusoid of peak amplitude
$|h_{\rm GCal}|$ gives $\rho=|h_{\rm GCal}|\sqrt{T/S_h}$ in stationary
noise. Materials have identical source geometry and density-scaled masses.
The plot does not compare a dimensionless line amplitude directly with an ASD.}
\label{fig:hgcal_vs_noise}
\end{figure}

For the updated bar parameters and the calibration geometry
$a=0.30213$~m, $b=0.160$~m, $h=2.00$~m, $m=1.8842$~kg,
and $\eta_{\rm GCAL}=0.9743$, the tungsten signal at 1~Hz is
\begin{equation}
 |h_{\rm GCal}|=1.17\times10^{-14}.
\end{equation}
The coherent SNR per $\sqrt{T}$ at 1~Hz is approximately
$3.72\times10^3~\mathrm{s}^{-1/2}$ at the minimum-PI point and
$5.92\times10^3~\mathrm{s}^{-1/2}$ at the logarithmic-mean point.
For equal source volumes, SUS304 and A5083 multiply these values by
$7.93/19.3$ and $2.66/19.3$, respectively. This estimate uses the
inertial response, one-sided noise, and known signal phase; practical
integration also requires stable rotor phase and readout calibration.

The systematic uncertainty of the predicted GCal response can be
evaluated by propagating uncertainties in
\begin{equation}
    h_{\rm GCal}
    =
    h_{\rm GCal}
    (M,m,I,a,b,h,\rho_\perp,G).
\end{equation}
A first-order perturbative error analysis gives a total fractional
systematic uncertainty of approximately
\begin{equation}
    \frac{\sigma_{\rm sys}}{|h_{\rm GCal}|}
    \simeq
    2.4\times10^{-3}
    =
    0.24\%.
\end{equation}
The dominant contributions arise from geometrical alignment and
position uncertainties, while uncertainties in the calibration
masses and the gravitational constant are subdominant.
At $1~{\rm Hz}$, the corresponding absolute systematic uncertainty
is of order
\begin{equation}
    \delta h_{\rm GCal}\simeq2.8\times10^{-17}.
\end{equation}

These conditional estimates motivate the torque-coupled GCal as a proposed
absolute calibration scheme for CHRONOS.
Its principal advantages are the direct gravitational coupling to
the torsional observable, a high-SNR monochromatic calibration line,
an analytically predictable excitation amplitude, and a mechanical
response that reduces to a simple inertial form throughout the
observation band.
The technique therefore provides a robust calibration reference for
precision sub-Hz gravitational-wave measurements.

%% file: Chapters/Prototype.tex
\chapter{Prototype Torsion-Bar System at National Central University}
\label{chap:prototype_system}

\section{Overview}

A room-temperature prototype of the CHRONOS torsion-bar detector is currently
under development and commissioning at National Central University (NCU),
Taiwan.
The purpose of this prototype is not to reproduce the final sensitivity of
CHRONOS, but to establish the key experimental technologies required for the
future detector, including torsion-bar suspension, multi-degree-of-freedom
sensing and control, interferometric readout, and stable operation of the
optical system.

The prototype consists of two independently suspended torsion bars arranged
orthogonally in a cross configuration.
This geometry provides a test platform for studying the differential rotational
response that will eventually constitute the gravitational-wave readout of
CHRONOS.
At the present stage, particular emphasis is placed on stabilizing the rigid-body
motion of each torsion bar sufficiently to enable stable Michelson
interferometer operation.

Each suspended torsion bar has six rigid-body degrees of freedom:
three translational motions, Length ($L$), Transverse ($T$), and Vertical ($V$),
and three rotational motions, Pitch ($P$), Yaw ($Y$), and Roll ($R$).
The current NCU prototype focuses on active control of five of these degrees of freedom except for $V$.

The prototype provides an important intermediate step between component-level
tests and the future integrated CHRONOS detector.
In particular, the current commissioning program aims to demonstrate that the
relatively large low-frequency motion of a torsion-bar suspension can be
suppressed sufficiently by local control to permit stable interferometric
readout.

\section{Prototype Instrument}

\begin{figure}[tbph]
    \centering
    \includegraphics[width=0.95\linewidth]{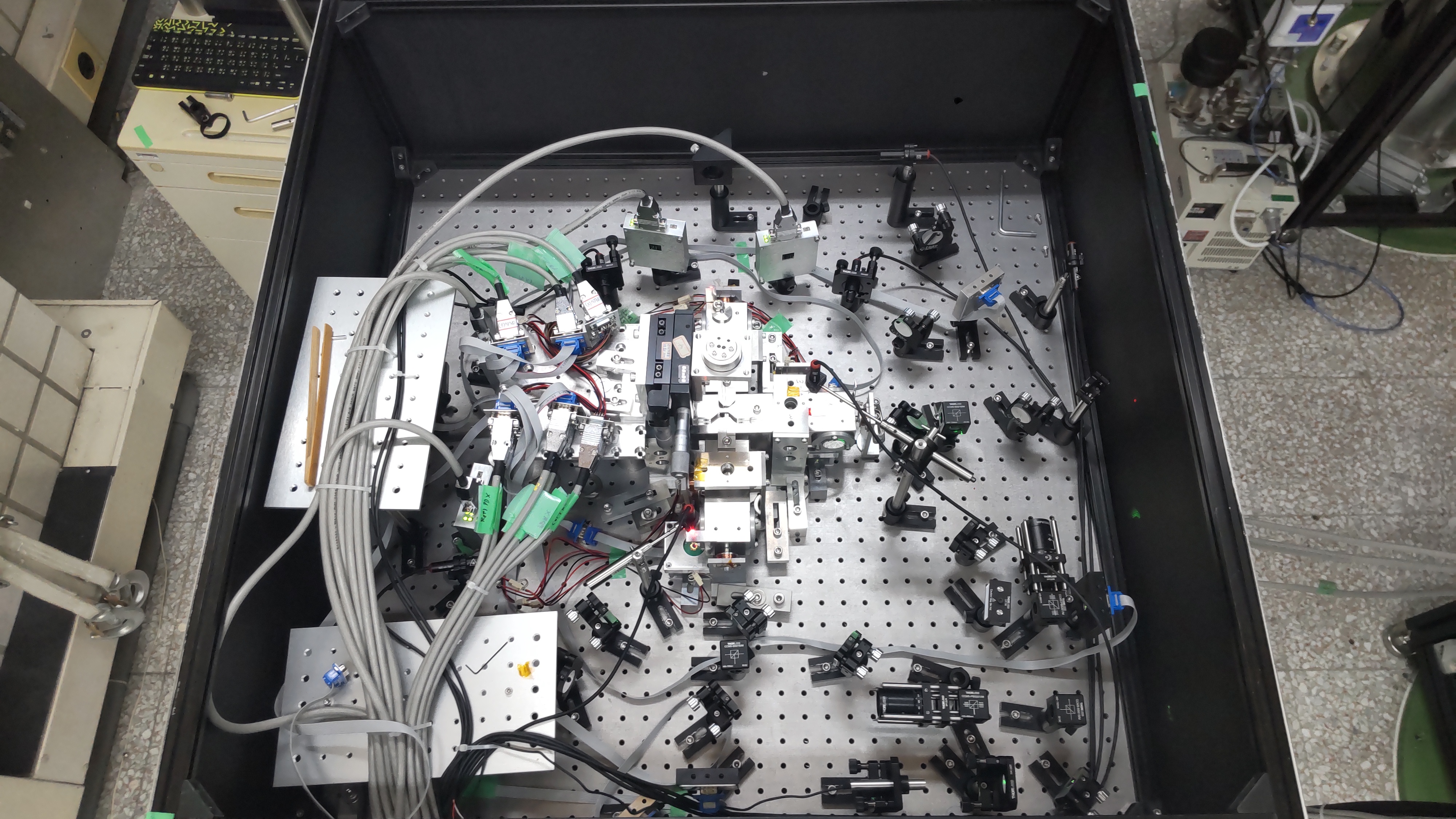}
    \caption{
    Prototype cross-torsion-bar system currently under commissioning at
    National Central University.
    }
    \label{fig:prototype_overview}
\end{figure}

\begin{figure}[tbph]
    \centering
    \includegraphics[width=0.95\linewidth]{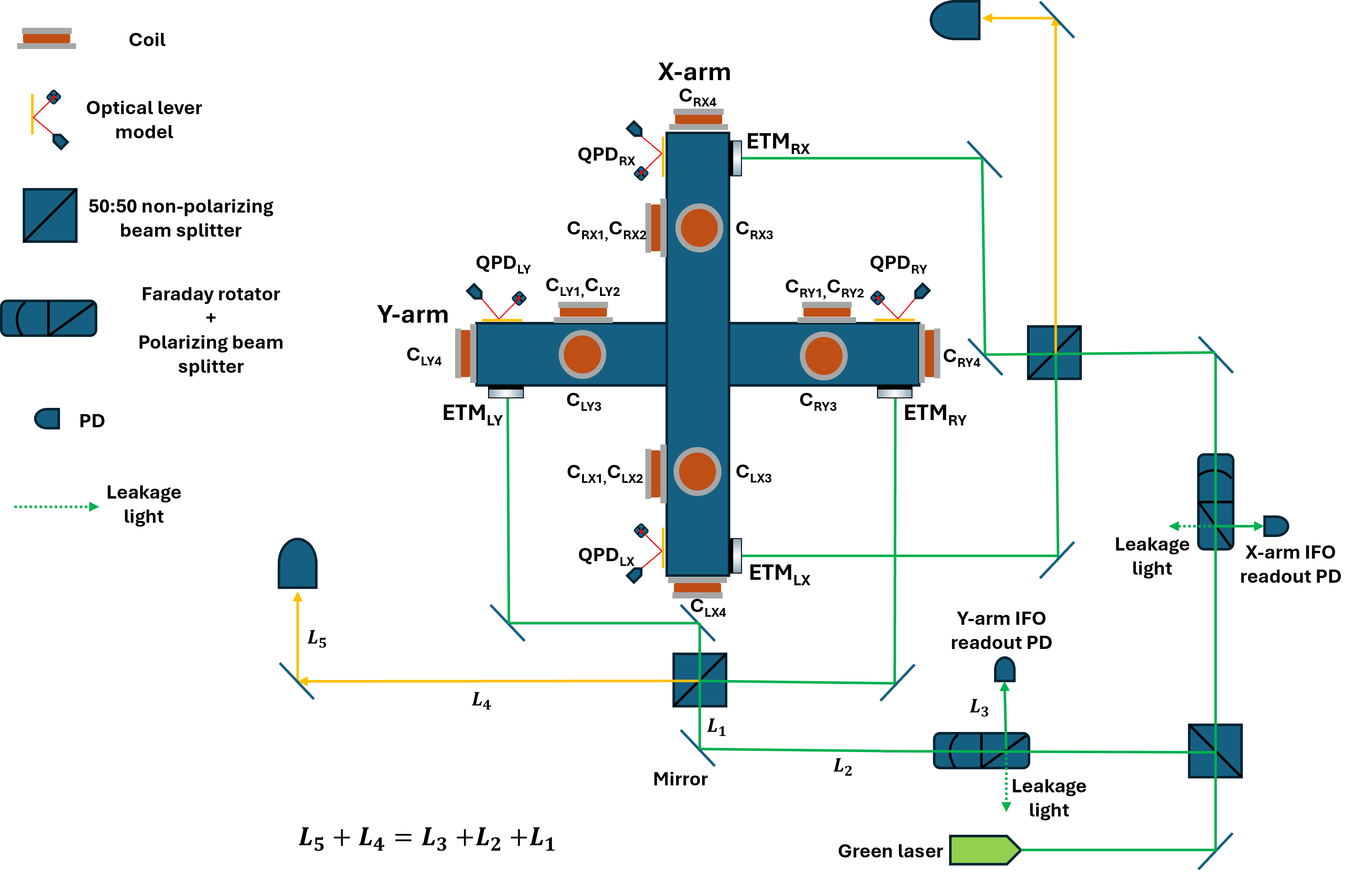}
    \caption{
    Optical and mechanical layout of the NCU prototype, including
    Michelson interferometers, optical levers, shadow sensors,
    and coil--magnet actuators.
    }
    \label{fig:prototype_layout}
\end{figure}

Figure~\ref{fig:prototype_overview} shows the photograph of the prototype setup in NCU, and Fig.~\ref{fig:prototype_layout} shows its schematic drawing. The mechanical system consists of two torsion bars suspended independently and
oriented approximately orthogonally.
Each bar carries optical mirrors for interferometric and local sensing, as well
as permanent magnets that couple to externally mounted coils used for
actuation.

The suspension is designed such that the yaw motion has a substantially lower
resonance frequency than the other rigid-body degrees of freedom.
This is a defining characteristic of torsion-bar suspensions and provides the
mechanical basis for extending gravitational measurements toward the sub-Hz
frequency region. Mechanical and geometrical parameters of the torsion bar are summarized in Table~\ref{tab:prototype_parameters}.

\begin{table}[tb]
\centering
\caption{
Parameters of the torsion bars used in the NCU prototype. Center of mass is measured in vertical direction from the upper surface of each bare where the suspension fiber is attached.
}
\label{tab:prototype_parameters}
\begin{tabular}{cc}
\hline
Quantities (Unit) & Values (X/Y bar) \\
\hline
Material & Stainless Steel SUS304 \\
Density [kg~m$^{-3}$] & 8000 \\
Young's modulus [GPa] & 190 \\
Poisson's ratio & 0.29 \\
Bar length [mm] & 300 \\
Bar width [mm] & 50 \\
Bar height [mm] & 50 \\
Mass [kg] & 5.16 \\
Moment of inertia about yaw [kg~m$^2$] & 0.045 \\
Moment of inertia about pitch [kg~m$^2$] & 0.002 \\
Moment of inertia about roll [kg~m$^2$] & 0.045\\
Arm length [mm]  & 125 \\
Center of mass [mm] & -23.6/26.4 \\
Fiber length [mm] & 129/125 \\
Fiber diameter [mm] & 0.55 \\
Young's modulus of fiber [GPa] & 267.6$\pm$11.3 \\
\hline
\end{tabular}
\end{table}

The optical system contains two classes of readout.
A main laser beam is used to form Michelson interferometers for monitoring the
motion of the torsion bars, while independent optical-lever and shadow-sensor
systems provide local sensing of the individual rigid-body degrees of freedom for lock acquisition.

The main interferometric beam is divided into two optical paths corresponding
to the two torsion bars.
Each path contains polarization optics and a Michelson-type interferometer
formed by a beam splitter and mirrors mounted on the torsion bar.
The returning light is detected by photodetectors and is used to monitor the
relative optical-path variation induced by torsion-bar motion.

\begin{figure}[tbph]
    \centering
    \includegraphics[width=0.75\linewidth]{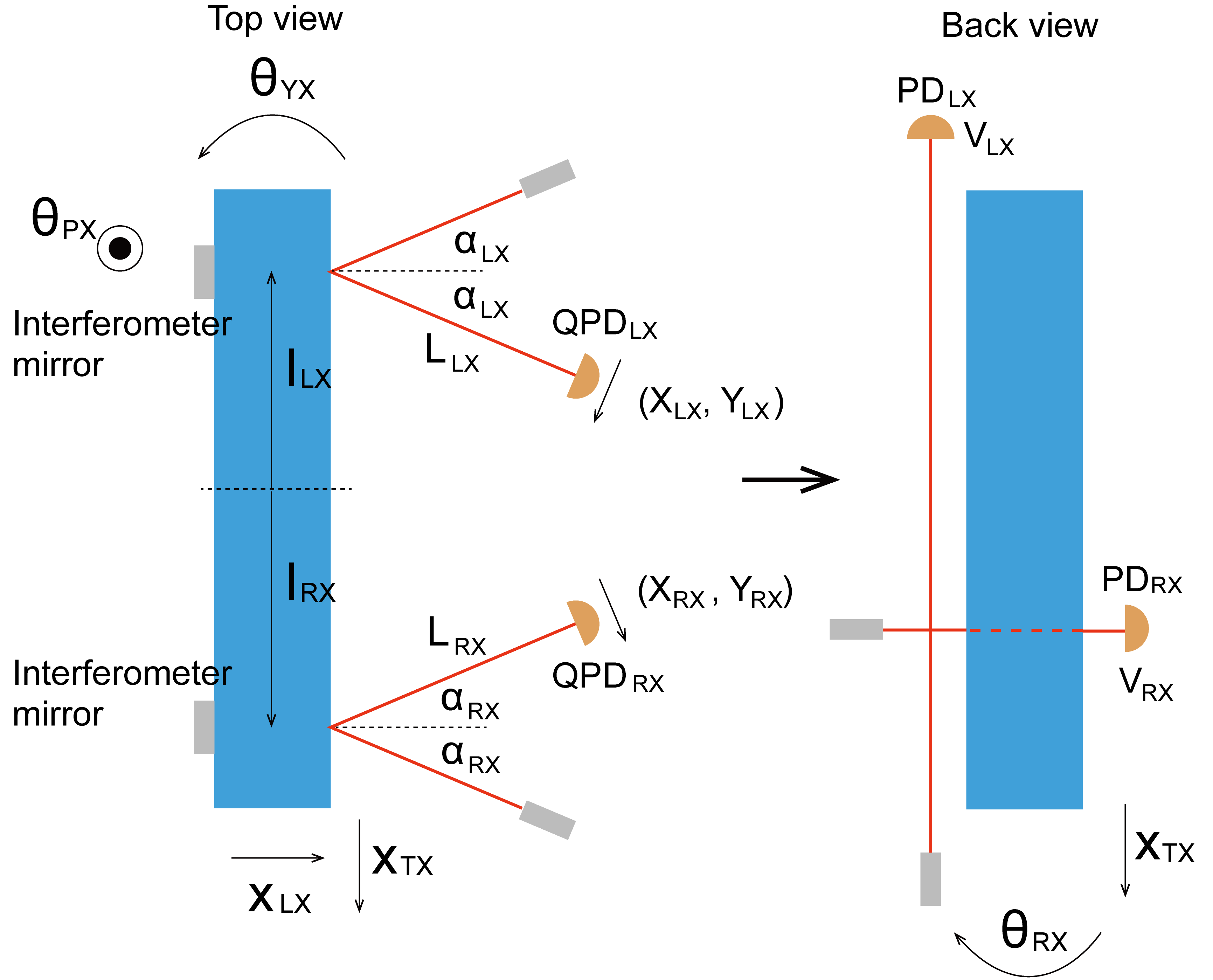}
    \caption{
    Local sensing system of X-arm in the NCU prototype. Left view shows optical lever, and the right view shows shadow sensor.
    }
    \label{fig:sensing_X}
\end{figure}

\begin{figure}[tbph]
    \centering
    \includegraphics[width=0.75\linewidth]{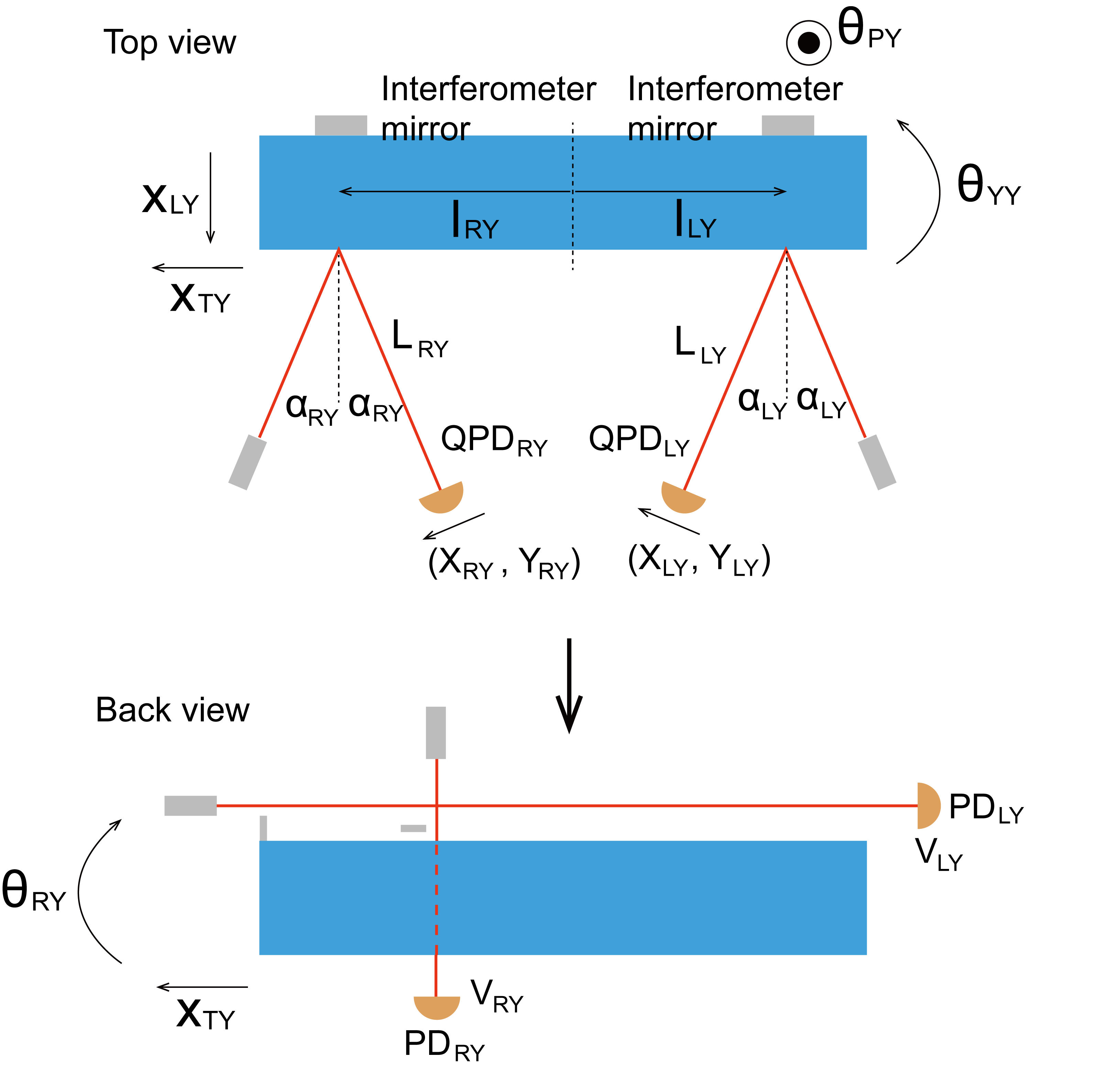}
    \caption{
    Local sensing system of Y-arm in the NCU prototype. Top view shows optical lever, and the bottom view shows shadow sensor.
    }
    \label{fig:sensing_Y}
\end{figure}

Local sensing is performed with optical levers based on quadrant photodiodes
(QPDs) and with shadow sensors.
The optical levers provide angular and translational information relative to
the laboratory frame, while the shadow sensors provide complementary
measurements of selected translational and rotational motions.
Figure~\ref{fig:sensing_X} and Fig.~\ref{fig:sensing_Y} shows geometry of the local sensing systems in X-arm and Y-arm. 

\begin{figure}[tbph]
\centering
\includegraphics[width=0.8\linewidth]{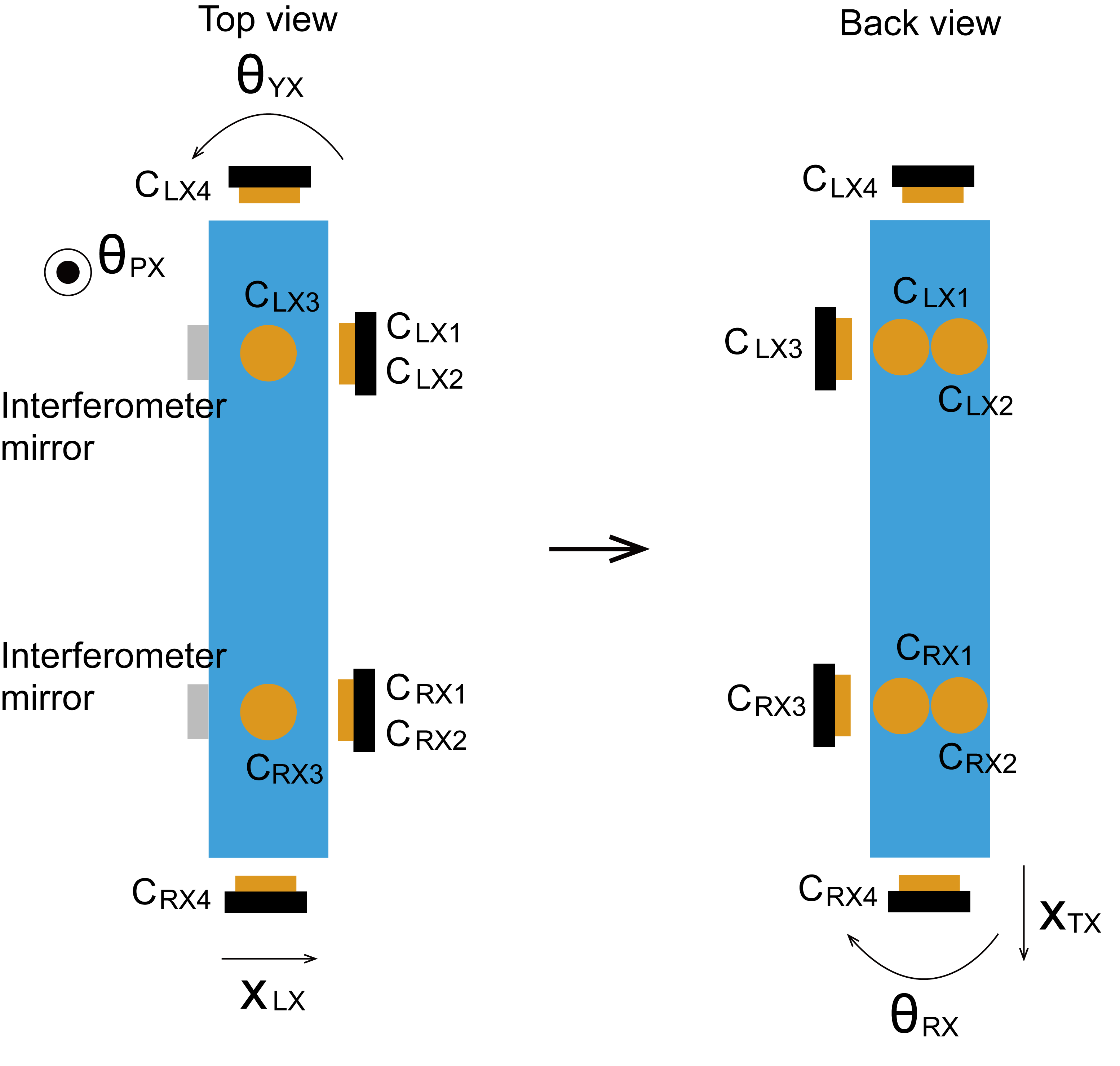}
\caption{Configuration of actuation coils around the X-arm.}
\label{fig:coil_X}
\end{figure}

\begin{figure}[tbph]
\centering
\includegraphics[width=0.8\linewidth]{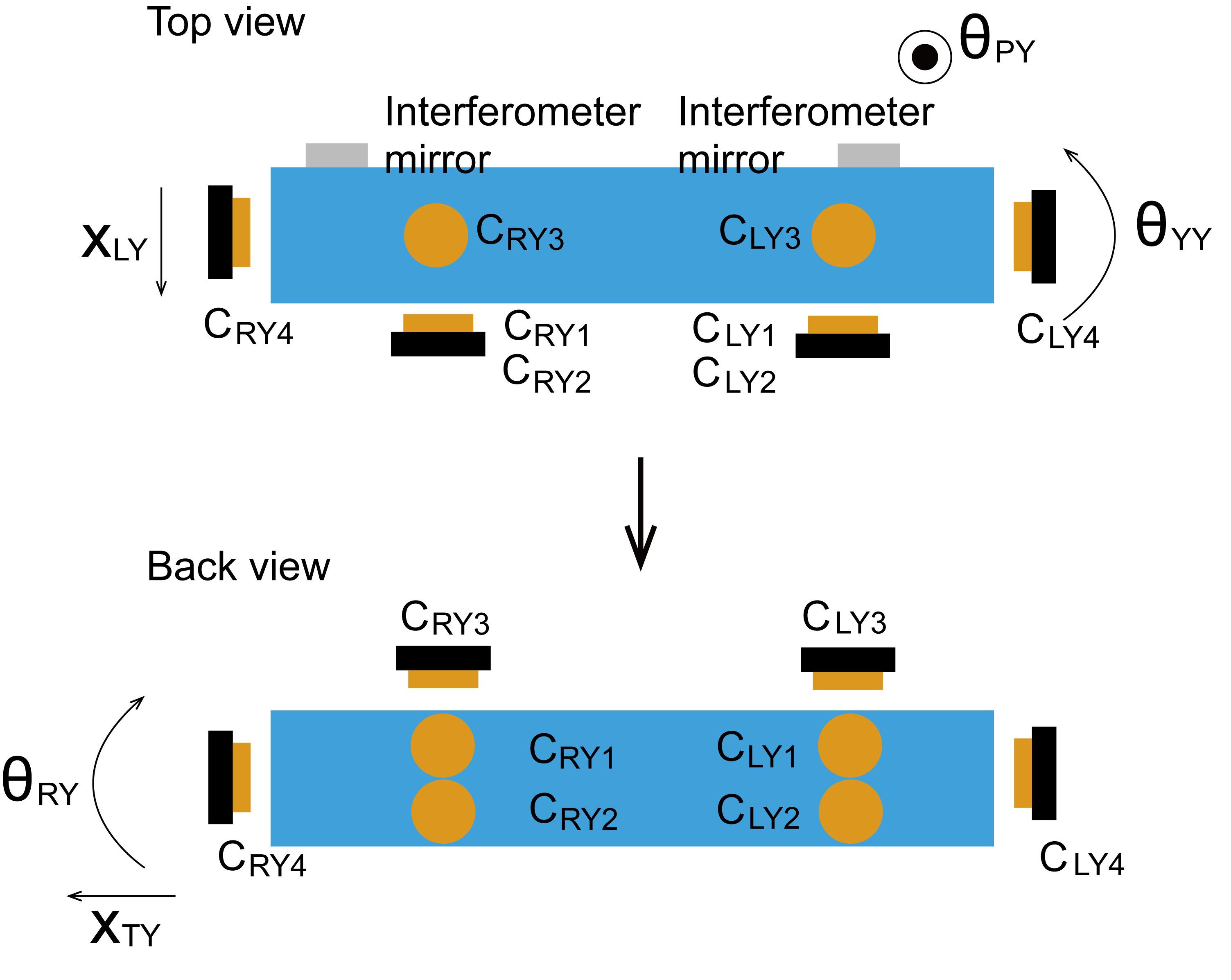}
\caption{Configuration of actuation coils around the Y-arm.}
\label{fig:coil_Y}
\end{figure}

Actuation is provided by coil--magnet systems.
Permanent magnets are mounted on the suspended torsion bars, and forces are
applied by stationary coils attached to the surrounding frames.
The positions of the actuators are selected so that combinations of coil forces
can generate both translational forces and rotational torques.
Figure~\ref{fig:coil_X} and Fig.~\ref{fig:coil_Y} shows positions of the coil around the X-arm and Y-arm bar.

\section{Rigid-Body Modes and Measured Resonance Frequencies}

For the commissioning of the prototype, the rigid-body dynamics of the
suspended torsion bars were first characterized experimentally. The representative measured resonant frequencies of the five degrees of freedom are
\begin{align}
    f_Y &\simeq 0.056~\mathrm{Hz}, \\
    f_R &\simeq 0.93~\mathrm{Hz}, \\
    f_L &\simeq 1.3~\mathrm{Hz}, \\
    f_T &\simeq 1.3~\mathrm{Hz}, \\
    f_P &\simeq 2.3~\mathrm{Hz}.
\end{align}

The particularly low yaw resonance is produced by the weak torsional restoring
force of the suspension fiber.
For a rotational degree of freedom, the resonance frequency can generally be
expressed as
\begin{equation}
    f_i =
    \frac{1}{2\pi}
    \sqrt{\frac{\kappa_i}{I_i}},
\end{equation}
where $\kappa_i$ and $I_i$ are the effective restoring stiffness and moment of
inertia for the corresponding mode.

In the yaw direction, the restoring force is dominated by the torsional
stiffness of the suspension fiber.
In contrast, the pitch and roll resonances are determined primarily by
gravitational restoring torques associated with the relative position of the
suspension point and the center of mass.
The translational resonances are determined by the pendulum-like dynamics of
the suspension.

The separation between the very-low-frequency yaw resonance and the
approximately $1$--$2\,\mathrm{Hz}$ resonances of the other rigid-body modes is
important for the development of the CHRONOS control architecture.
The gravitational-wave-sensitive torsional motion must remain weakly restored,
while the other degrees of freedom must be sufficiently stabilized to prevent
excess coupling into the optical readout.

\paragraph{Material Property of Suspension Fiber}

\begin{figure}[tbp]
\centering
\includegraphics[width=0.7\linewidth]{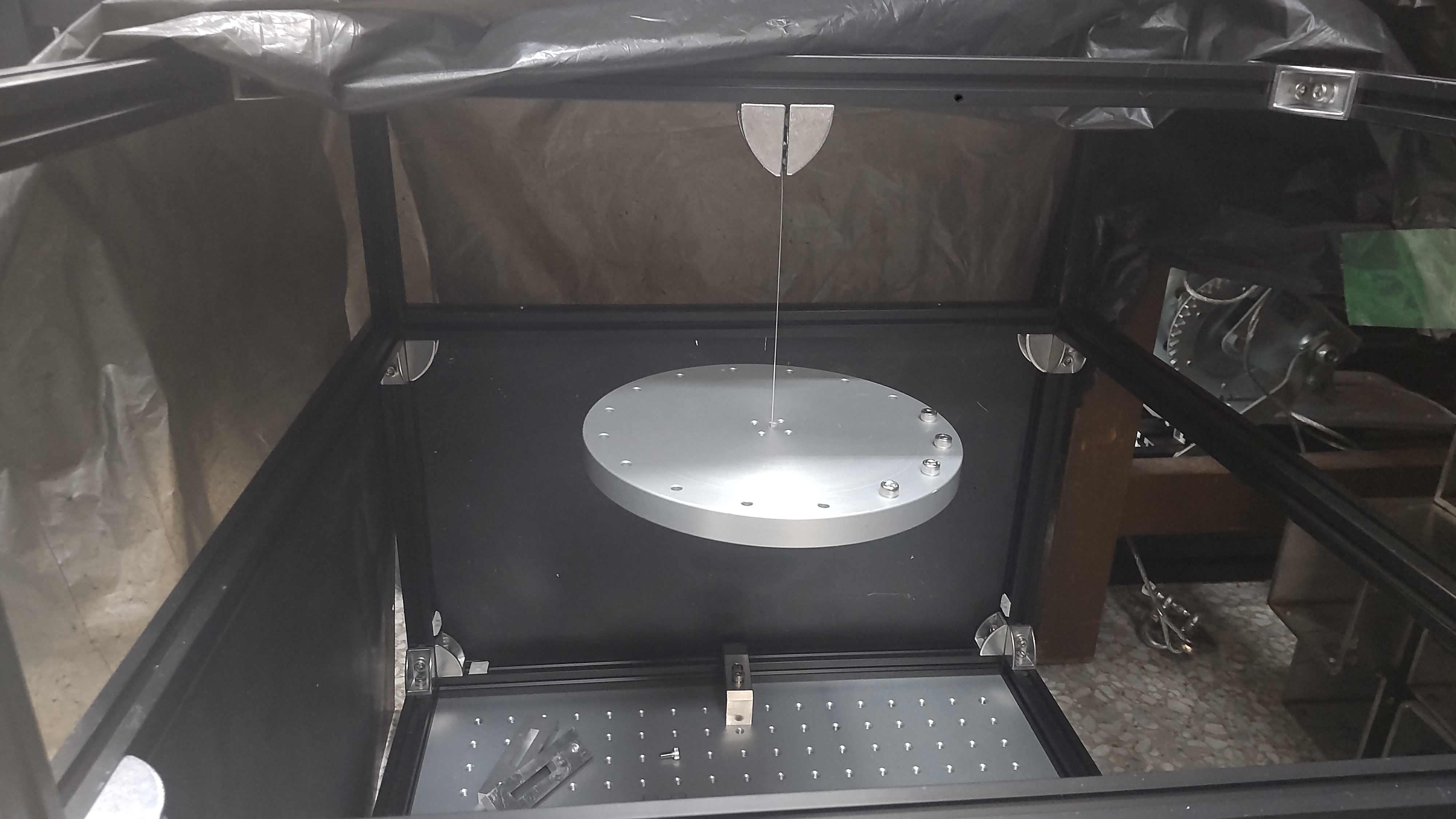}
\caption{Suspended aluminum disk for measurement of Young's modulus of the steel fiber.}
\label{fig:aluminum_disk}
\end{figure}

The torsion bars are suspended by SWG steel wire, and the resonant frequency of the yaw rotation depends on its Young's modulus $E$ and Poisson's ratio $\nu$ by the relationship of
\begin{equation}
\frac{1}{f^2}= \frac{128\pi I}{Gd^4}L, \quad G\equiv \frac{E}{2(1+\nu)},
\end{equation}
where $I$ is torsional moment of inertia, $d$ is diameter of the fiber, and $L$ is the fiber length. The Young's modulus was experimentally determined by measuring the resonant period of a suspended object with various fiber lengths. As a simpler setup without magnetic coupling and conflict with surrounding structures, an aluminum disk with a moment of inertia of 0.0051~kg~m$^2$ was suspended as shown in Fig.~\ref{fig:aluminum_disk}.

\begin{figure}[tbp]
\centering
\includegraphics[width=0.7\linewidth]{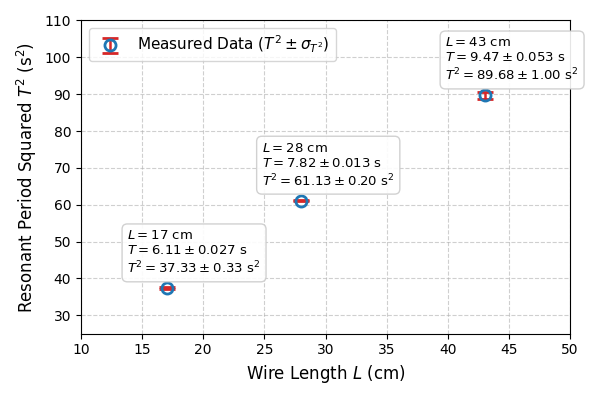}
\caption{Measured relationship between the rotation period of an aluminum disk and the fiber length.}
\label{fig:fiber_young}
\end{figure}

The mean resonant period $T$ in 10 periods was plotted as a function of fiber length $L$ in Fig.\ref{fig:fiber_young}. The best-fit slope was 
\begin{equation}
\frac{T^2}{L}=217.71 \pm 0.65.   
\end{equation}
Using this value and uncertainty, we derived the Young's modulus as $E=$267.6$\pm$11.3~GPa.

\section{Five-Degree-of-Freedom Sensing and Control}

The primary objective of the present commissioning campaign is stabilization of
the five rigid-body degrees of freedom other than the vertical motion.
The local-control system combines measurements, done with shadow sensors and optical levers, and actuation done by forces and torques generated by the coil--magnet actuators.

The control vector is defined as
\begin{equation}
    \mathbf{x}
    =
    \begin{pmatrix}
        L & Y & P & R & T
    \end{pmatrix}^{T}.
\end{equation}

In practice, each physical sensor responds to a combination of several
mechanical degrees of freedom.
The measured sensor vector $\mathbf{s}$ can therefore be written as
\begin{equation}
\label{eq:mat_s_x}
    \mathbf{s}
    =
    \mathbf{M}_{\mathrm{sens}}
    \mathbf{x}.
\end{equation}
Here $\mathbf{M}_{\mathrm{sens}}$ is the sensing matrix, which describes conversion from the real motion to the sensor output. In the prototype setup, it is explicitly derived as
\begin{equation}
\mathbf{M}_{\mathrm{sens}}=
\begin{bmatrix}
-2 \sin \alpha_{L} & -2L_{L} + 2l_{L} \sin \alpha_{L} & 0 & 0 & 0 \\
2 \sin \alpha_{R} & -2L_{R} - 2l_{R} \sin \alpha_{R} & 0 & 0 & 0 \\
0 & 0 & -2 L_{L} \cos \alpha_{L} & 0 & 0 \\
0 & 0 & -2 L_{R} \cos \alpha_{R} & 0 & 0 \\
0 & 0 & 0 & c_{L} & 0 \\
0 & 0 & 0 & 0 & c_{R}
\end{bmatrix}.
\end{equation}

The QPD signals in the optical lever are particularly useful for the pitch, length, and yaw motions.
Shadow sensors provide robust local measurements for transverse and roll
motions.
Because some signals contain significant contributions from more than one
mechanical degree of freedom, the sensing matrix must be experimentally
characterized and diagonalized.

Solving Eq.~(\ref{eq:mat_s_x}), we derive the motion of each degree of freedom from the sensor signal. Since the $\mathbf{M}_\mathrm{sens}$ is generally not regular, the solution is approximately constructed as 
\begin{equation}
\begin{array}{c}
\mathbf{x}_\mathrm{sol}={\mathbf{M}_\mathrm{sens}^\prime}^{-1},\\
{\mathbf{M}_\mathrm{sens}^\prime}^{-1}=
\begin{bmatrix}
\frac{1}{-4 \sin \alpha_{Li}} & \frac{1}{4 \sin \alpha_{Ri}} & 0 & 0 & 0 & 0 \\
\frac{1}{-4L_{Li} + 4l_{Li} \sin \alpha_{Li}} & \frac{1}{-4L_{Ri} - 4l_{Ri} \sin \alpha_{Ri}} & 0 & 0 & 0 & 0 \\
0 & 0 & \frac{1}{-4 L_{Li} \cos \alpha_{Li}} & \frac{1}{-4 L_{Ri} \cos \alpha_{Ri}} & 0 & 0 \\
0 & 0 & 0 & 0 & \frac{1}{c_{Li}} & 0 \\
0 & 0 & 0 & 0 & 0 & \frac{1}{c_{Ri}}
\end{bmatrix}
\end{array}
\end{equation}

In addition, there is a cross coupling, or unwanted excitation, between degrees of freedom due to asymmetry of the hardware system. To compensate such imperfection, we insert a experimentally determined matrix called diagonalization matrix to construct estimators of ``pure'' physical degrees of freedom for control. The estimated physical-degree-of-freedom vector can be constructed as
\begin{equation}
    \mathbf{x}_{\mathrm{est}}
    =
    \mathbf{M}_\mathrm{diag}
    {\mathbf{M}_\mathrm{sens}^\prime}^{-1}
    \mathbf{s},
\end{equation}
where $\mathbf{M}_\mathrm{diag}$ is the diagonalization matrix. Ideally,
\begin{equation}
    \mathbf{M}_\mathrm{diag}
    {\mathbf{M}_\mathrm{sens}^\prime}^{-1}
    \mathbf{M}_{\mathrm{sens}}
    \simeq
    \mathbf{I}.
\end{equation}

\begin{figure}[tbp]
    \centering
    \includegraphics[width=0.95\linewidth]{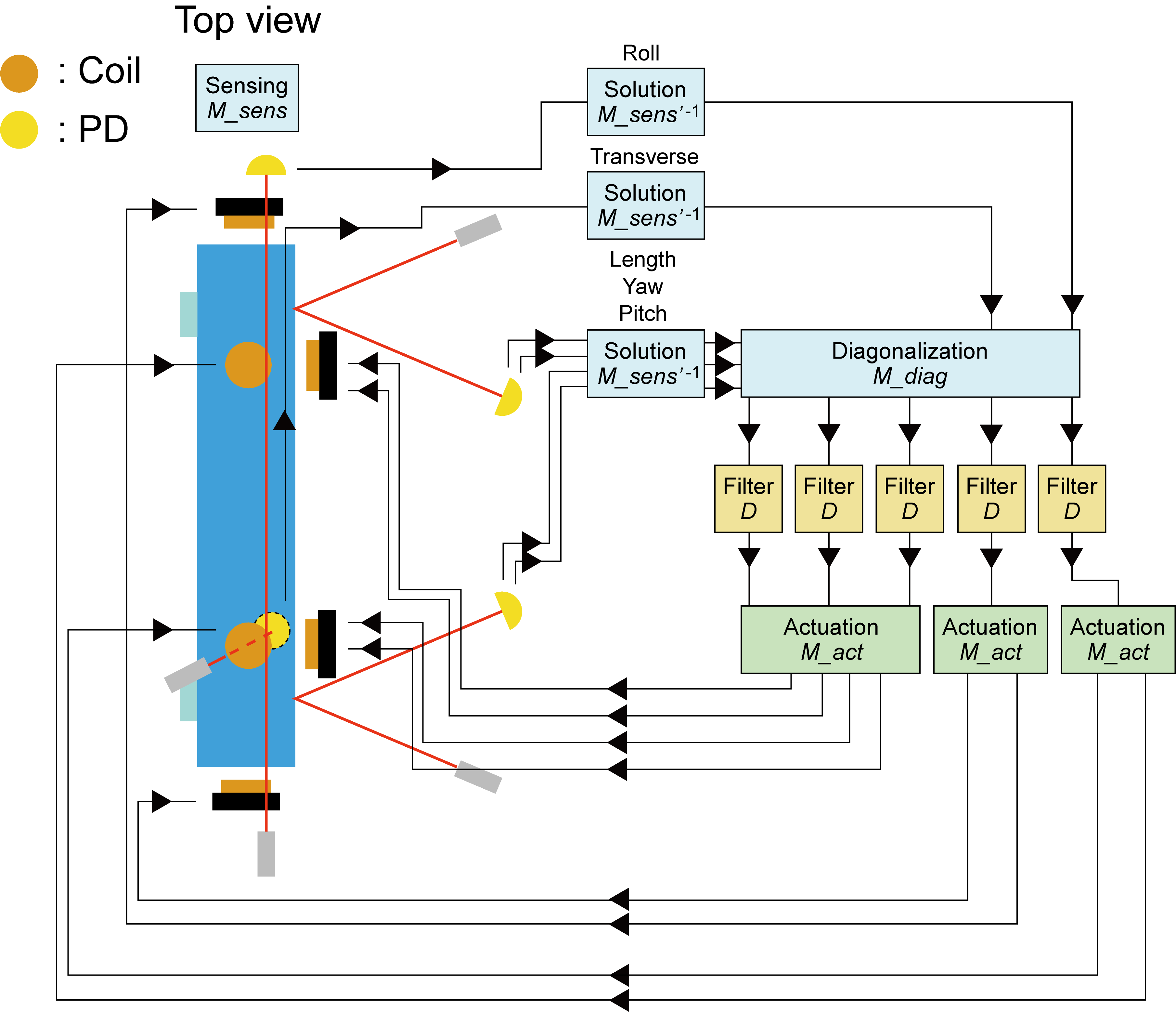}
    \caption{
    Block diagram of the local-control architecture for five degrees of freedom in the NCU prototype.
    }
    \label{fig:prototype_block}
\end{figure}

The actuator system is has a similar multi-input and multi-output structure.
Generally, a given coil generates not only a pure force but also a torque,
depending on its position relative to the center of mass.
The physical forces and torques is represented as
\begin{equation}
    \mathbf{F}
    =
    \mathbf{M}_{\mathrm{act}}
    \mathbf{u},
\end{equation}
where $\mathbf{u}$ is the vector of coil-drive signals and
$\mathbf{M}_{\mathrm{act}}$ is the actuator matrix. The actuator matrix is explicitly written as
\begin{equation}
\mathbf{M}_{\mathrm{act}}=
\begin{bmatrix}
-1 & 1 & -1 & 0 & 0 \\
-1 & 1 & 1 & 0 & 0 \\
0 & 0 & 0 & 1 & 0 \\
0 & 0 & 0 & 0 & 1 \\
-1 & -1 & -1 & 0 & 0 \\
-1 & -1 & 1 & 0 & 0 \\
0 & 0 & 0 & -1 & 0  \\
0 & 0 & 0 & 0 & -1  
\end{bmatrix}.
\end{equation}
The feedback control relates $\mathbf{x}_{\mathrm{est}}$ and $\mathbf{u}$ through the feedback filters $\mathbf{D}$. Figure~\ref{fig:prototype_block} summarizes the feedback loop consist of the sensing, actuation, and filtering operators.

The present control strategy does not require complete simultaneous
matrix control from the beginning.
Instead, the mechanical modes are characterized individually and feedback loops
are introduced progressively.
This approach allows the sign, gain, phase, and cross coupling of each loop to
be determined experimentally before the full five-degree-of-freedom system is
operated.

\begin{figure}[tbp]
    \centering
    \includegraphics[width=0.95\linewidth]{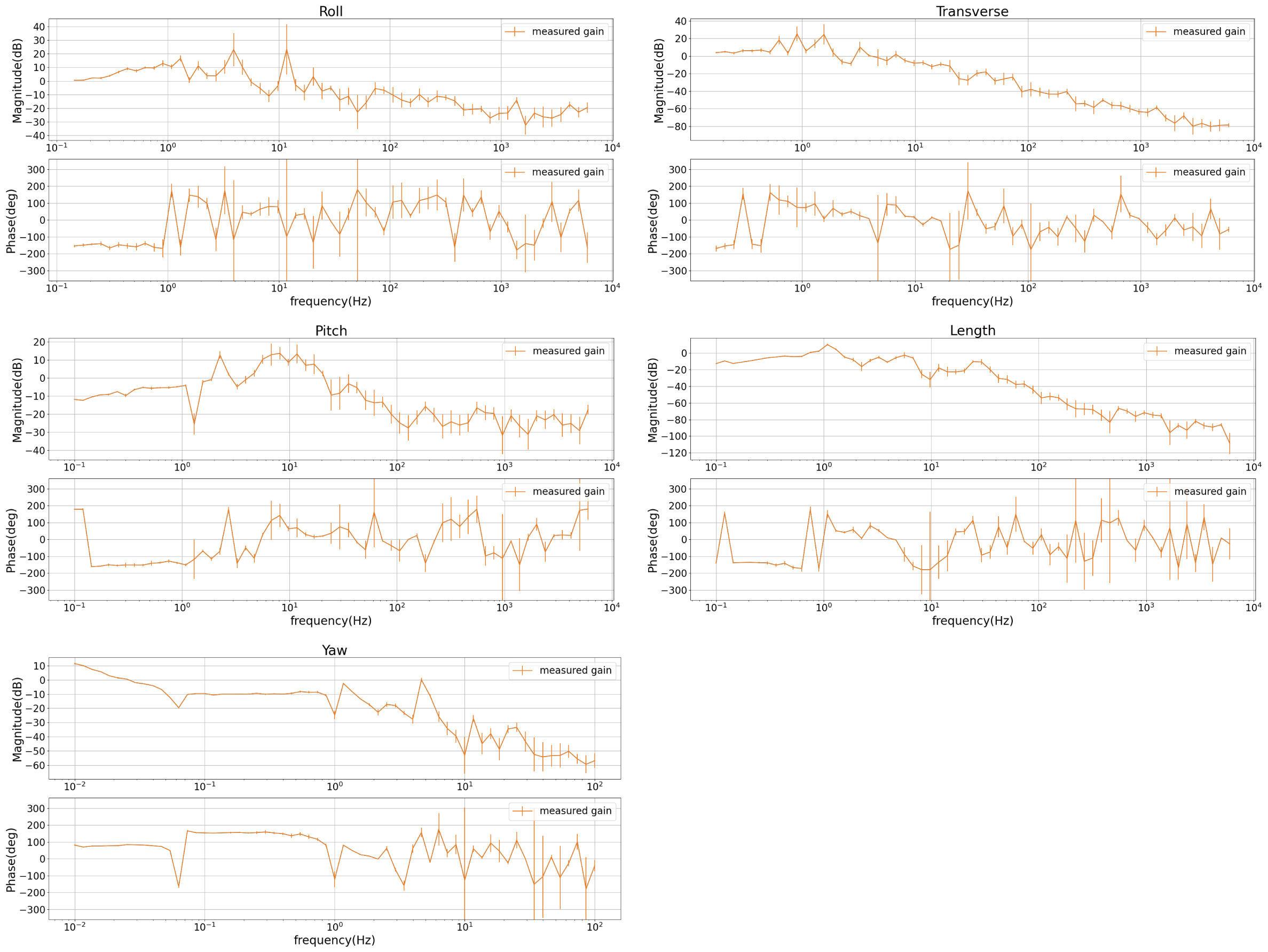}
    \caption{
    Transfer functions of representative rigid-body degrees of freedom measured with the Y-arm of NCU prototype.
    }
    \label{fig:prototype_TF}
\end{figure}

For each degree of freedom, the mechanical transfer function is measured by
injecting a known excitation through the coil actuator and recording the
response of the corresponding local sensor.
These measurements are used to determine the open-loop transfer function,
closed-loop suppression, unity-gain frequency, and phase margin of the
feedback system. Figure~\ref{fig:prototype_TF} shows Bode plots of the open-loop gains $G(f)$ under feedback control in the prototype test of Y-arm. They show a peak in magnitude and a flip in phase at the resonant frequency.

The control performance can be characterized through the standard sensitivity
and complementary-sensitivity functions,
\begin{align}
    S(f) &= \frac{1}{1+G(f)}, \\
    T(f) &= \frac{G(f)}{1+G(f)},
\end{align}
where $G(f)$ is the open-loop gain.

The primary commissioning goal is to reduce the large resonant motions without
introducing excess noise in the frequency region relevant for interferometric
readout.
Particular attention is therefore given to the coupling between different
mechanical degrees of freedom and to the stability of the control loops around
the closely spaced rigid-body resonances.

\section{Suppression of Mechanical Motion}

The uncontrolled torsion-bar system exhibits relatively large motion around
the rigid-body resonances.
This motion causes rapid changes in optical-path length and beam alignment and
therefore makes stable interferometric operation difficult.

\begin{figure}[tbp]
    \centering
    \includegraphics[width=0.95\linewidth]{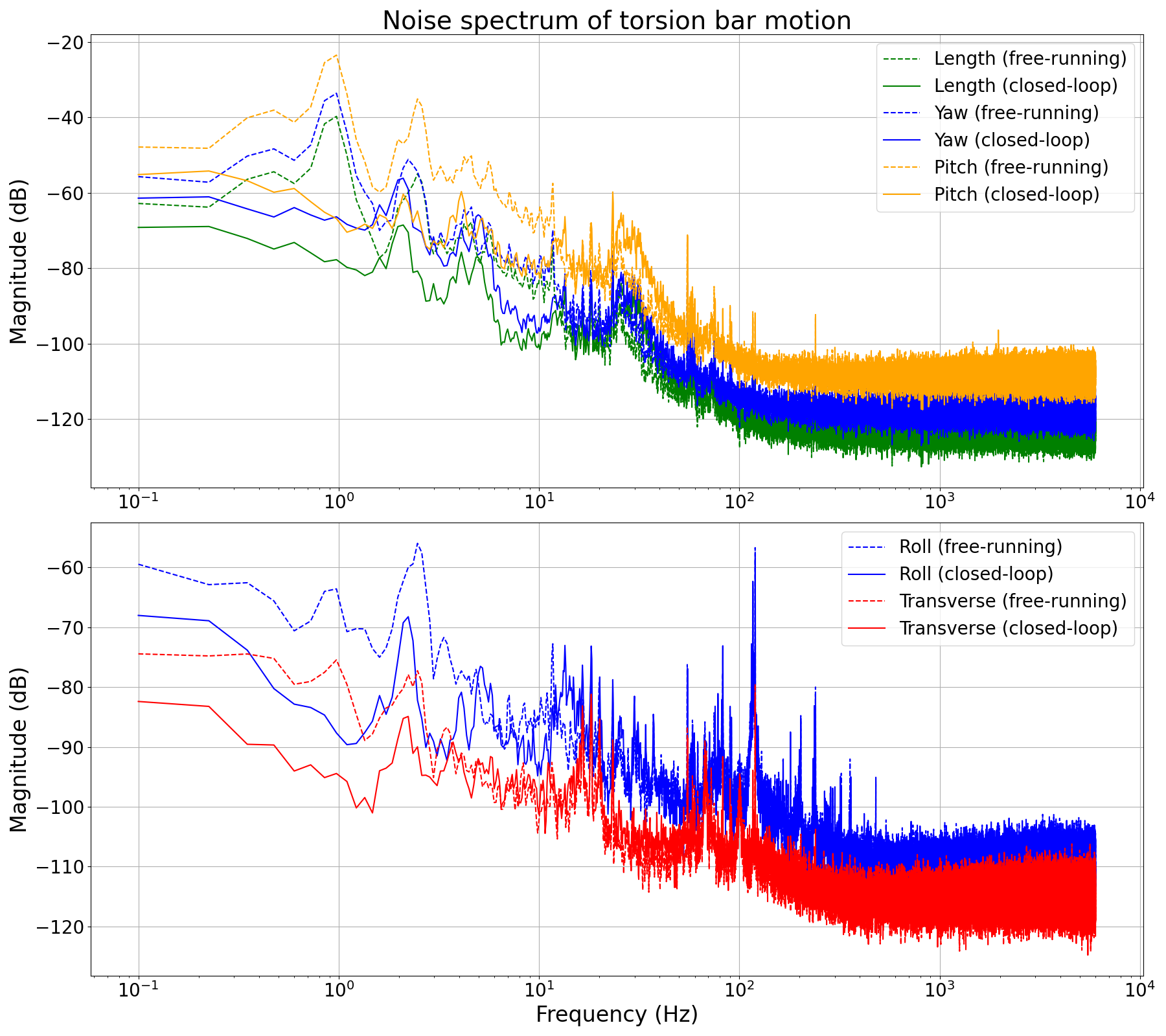}
    \caption{
    Comparison of representative noise spectra before and after
    local feedback control in the five degrees of freedom.
    }
    \label{fig:prototype_spectrum}
\end{figure}

The local-control loops are designed primarily to suppress these resonant
motions.
Noise spectra are measured both with the local controls disabled and enabled as shown in Figure~\ref{fig:prototype_spectrum}.
The comparison provides a direct measure of the suppression achieved for each
mechanical degree of freedom.

For a controlled degree of freedom, the residual displacement or angular
motion is approximately described by
\begin{equation}
    x_{\mathrm{closed}}(f)
    =
    \frac{1}{1+G(f)}
    x_{\mathrm{free}}(f)
\end{equation}
in the frequency region where sensor and actuator noise can be neglected.

The reduction of the rigid-body motion is particularly important for the
Michelson interferometer.
Even mechanical modes that are nominally orthogonal to the interferometric
length degree of freedom can couple into the readout through beam
miscentering, mirror-angle fluctuations, imperfect optical alignment, and
cross coupling in the suspension.

Therefore, stable Michelson operation provides an integrated test of the
performance of all local-control degrees of freedom rather than only the
longitudinal motion.


\section{Michelson Interferometric Readout}

The motion of the torsion bars is monitored globally using Michelson
interferometers.
For a conventional Michelson interferometer, the detected output power can be
written as
\begin{equation}
    P_{\mathrm{PD}}
    =
    \frac{P_{\mathrm{in}}}{2}
    \left[
        1-\cos(2\phi_-)
    \right],
\end{equation}
where $\phi_-$ represents the differential optical phase between the two arms.

Around the middle of an interference fringe, the optical response can be
linearized as
\begin{equation}
    \delta P_{\mathrm{PD}}
    \propto
    \delta\phi_-
    \propto
    \delta L.
\end{equation}

The mid-fringe operating point is therefore used during the present
commissioning phase because it provides a simple and approximately linear
displacement signal.
The interferometer output is converted into an error signal and can be
feedback-controlled to maintain the operating point.

For the torsion-bar system, the Michelson signal is sensitive primarily to the
relative displacement generated by torsional rotation.
For a mirror located at an effective distance $l$ from the
torsion axis, a small yaw rotation produces an effective longitudinal
displacement approximately given by
\begin{equation}
    \delta L_{\mathrm{eff}}
    \simeq
    l\delta\theta_Y.
\end{equation}

The Michelson interferometer therefore provides a global readout that is
directly sensitive to the torsional motion relevant for future gravitational
measurements.

\begin{figure}[tbp]
    \centering
    \includegraphics[width=0.95\linewidth]{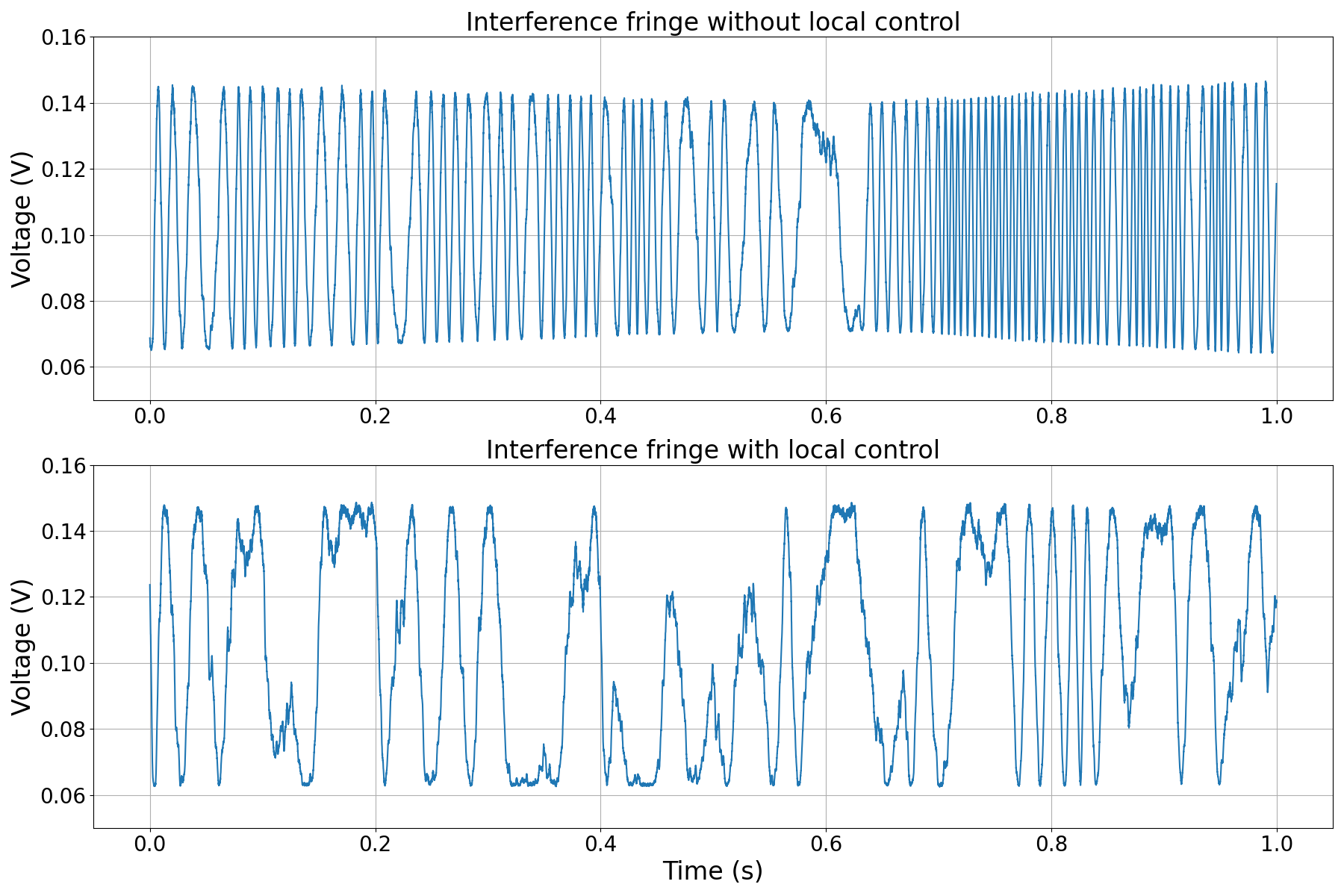}
    \caption{
    Comparison of Michelson fringe signal before and after the local control.
    }
    \label{fig:prototype_fringe}
\end{figure}

At the present stage, the Michelson system is also used as a practical measure
of the overall suspension-control performance.
Without adequate local stabilization, the interferometer rapidly passes
through many fringes because of the free motion of the torsion bars.
As the five-degree-of-freedom control is progressively optimized, the fringe
motion is reduced and the interferometer approaches the regime in which stable
mid-fringe locking can be maintained. Figure~\ref{fig:prototype_fringe} shows the interference fringe pattern before and after the local control.

\section{Michelson Noise}

Once the large rigid-body motion has been sufficiently suppressed, the
remaining interferometer fluctuations can be characterized as an equivalent
displacement noise spectrum,
\begin{equation}
    S_L^{1/2}(f)
    =
    \frac{S_V^{1/2}(f)}
    {\left|dV/dL\right|},
\end{equation}
where $S_V^{1/2}$ is the measured photodetector voltage noise and $dV/dL$ is
the calibrated optical response around the operating point.

The measured Michelson noise represents a combination of residual mechanical
motion and optical sensing noise.
Possible contributions include residual seismic excitation of the suspension,
insufficient suppression of the rigid-body resonances, angular-to-length
coupling, laser intensity noise, laser frequency noise, photodetector noise,
and electronic noise.

Laser intensity noise is especially relevant for a simple mid-fringe readout,
because fluctuations of the incident optical power directly change the
photodetector signal.
A differential mid-fringe readout, in which complementary Michelson output
signals are combined, can suppress common-mode laser-power fluctuations.

Laser frequency noise can also couple to the Michelson signal when the two
optical paths are not perfectly matched.
To first order, the coupling is proportional to the arm-length mismatch
$\Delta L$,
\begin{equation}
    \delta L_{\mathrm{freq}}
    \simeq
    \Delta L
    \frac{\delta\nu}{\nu}.
\end{equation}
Reducing the optical-path mismatch is therefore an important part of the
interferometer commissioning.

At the current prototype stage, the objective of the Michelson noise
measurement is not yet to demonstrate the final CHRONOS displacement
sensitivity.
Instead, it is used to identify the dominant technical noise sources and to
quantify how the local mechanical controls affect the global interferometric
readout.

The comparison between the Michelson noise with local controls disabled and
enabled provides an important diagnostic of residual coupling from the five
controlled mechanical degrees of freedom.

\section{Current Status and Next Milestones}

The NCU prototype has progressed from the characterization of individual
mechanical degrees of freedom toward integrated multi-degree-of-freedom
operation.
The mechanical resonances of the five actively controlled rigid-body modes
have been identified, and local sensing and coil--magnet actuation are being
used to suppress their motion.

The current commissioning effort focuses on quantitative characterization of
the control loops through measurements of the open-loop and closed-loop
transfer functions, optimization of the sensing diagonalization, and reduction
of cross coupling among the rigid-body modes.

A major near-term milestone is stable Michelson interferometer operation with
all five local-control loops simultaneously engaged.
Achieving this condition will demonstrate that the local-control system can
reduce the suspension motion sufficiently to support interferometric readout
of the torsional degree of freedom.

Following stable lock acquisition, the residual Michelson displacement noise
will be characterized systematically.
Noise injections and transfer-function measurements will then be used to
separate contributions from mechanical motion, sensor noise, optical noise,
and control noise.

The NCU prototype therefore serves as an experimental testbed connecting the
fundamental torsion-bar concept to the future CHRONOS detector.
The experience gained from multi-degree-of-freedom control, interferometer
commissioning, and noise characterization will provide direct input to the
design of the larger cryogenic suspension and interferometric system planned
for CHRONOS.

%% file: Chapters/Analysis.tex
\chapter{Data analysis}
\label{sec:data_analysis}

\section{Data-processing framework}

The primary objective of data analysis in laser-interferometric gravitational-wave detectors is to identify gravitational-wave signals embedded in the detector output and to infer their physical properties \cite{Abbott2016Discovery}. The principal targets considered here are compact-binary coalescences (CBCs) and a stochastic gravitational-wave background (SGWB) \cite{allen1999,Christensen1992,RomanoCornish2017}. The CHRONOS analysis follows methods developed for ground-based interferometers \cite{LIGO2015}, with modifications required by its sub-Hz band, torsional mechanical response, long signal durations, and strong susceptibility to environmental disturbances.

At sub-Hz frequencies, stationarity cannot be assumed over arbitrarily long intervals. Newtonian noise, suspension and internal thermal noise, seismic motion, tilt, and instrumental drifts may all be important \cite{Harms2015,Saulson1990}. Consequently, the analysis must retain auxiliary channels and detector-state information, divide the data into intervals over which the response and noise are approximately stationary, and propagate calibration and environmental uncertainties into the final astrophysical results.

The analysis products are separated into the following levels.

\begin{enumerate}
	\item \textbf{Raw and state-vector data.} Photodetector voltages, ADC counts, timing signals, digital control signals, and relevant environmental and instrumental monitors are recorded with a common time reference. The primary torsional degree of freedom is sensed through an angular displacement (or a specified derived quantity such as angular velocity); the choice must be fixed for each channel and recorded in its metadata.
	
	\item \textbf{Calibration and reconstruction.} Error and control signals are combined with measured sensing, actuation, and digital-filter transfer functions to reconstruct the external torsional motion. The torsional motion is subsequently converted to equivalent gravitational-wave strain using the modeled antenna and mechanical response. The response is complex and frequency dependent; treating calibration as a single voltage-to-strain scale factor is generally inadequate near mechanical poles, control-loop features, or optical response features.
	
	\item \textbf{Conditioning and data-quality selection.} Calibrated time series are checked for timing errors, missing data, saturation, control-loop loss, transients, spectral lines, and non-stationarity. Analysis segments are windowed and Fourier transformed with explicitly recorded sampling rate, segment duration, overlap, window, frequency resolution, and PSD-estimation procedure. Vetoes based on auxiliary sensors must be defined without using the gravitational-wave channel in a way that could remove a real signal.
	
	\item \textbf{Analysis-ready products.} The calibrated strain or torsional-motion series, data-quality masks, response functions, PSDs, environmental cross-spectra, and calibration-uncertainty models form the common input to CBC and SGWB pipelines. Versioned configuration files and software environments are retained so that every result can be reproduced from immutable raw data.
\end{enumerate}

\subsection{Control-loop signals and response reconstruction}
\label{sec:calibration_reconstruction}

Here $f$ is frequency in hertz and a tilde denotes a Fourier amplitude,
with time dependence $e^{+2\pi ift}$. This chapter uses $d_{\rm err}$ for
the digitized error signal and $d_{\rm ctrl}$ for the actuator control
signal. Let $\theta_{\rm ext}$ denote the external angular displacement
of the calibrated torsional degree of freedom, and $\theta_{\rm res}$ its
residual physical angle after negative electronic feedback. The
\emph{Sensing Function} $C(f)$ maps $\theta_{\rm res}$ to $d_{\rm err}$,
in error counts/rad, including optics and the readout chain. The digital
filter $D(f)$ maps error counts to control counts; the actuation function
$A(f)$ maps control counts to angle in radians. These functions use the
same mechanical coordinate and radiation-pressure dynamics, as specified in
Sec.~\ref{sec:DARM_CHRONOS}. Define the dimensionless open-loop gain $G(f)$
by the following negative-feedback convention:
\begin{align}
	\tilde d_{\rm ctrl}(f) &= D(f)\tilde d_{\rm err}(f),\\
	G(f) &= A(f)D(f)C(f),
\end{align}
Let $R_\theta$ be the reconstruction response in rad/error count.
Neglecting additive sensing and actuation noise only for the purpose of defining the response, the reconstructed external angular displacement is
\begin{equation}
	\tilde \theta_{\rm ext}(f)
	=R_\theta(f)\tilde d_{\rm err}(f)
	=C^{-1}(f)\tilde d_{\rm err}(f)+A(f)\tilde d_{\rm ctrl}(f),
	\qquad
	R_\theta(f)=\frac{1+G(f)}{C(f)}.
	\label{eq:closed_loop_reconstruction}
\end{equation}
The signs in Eq.~(\ref{eq:closed_loop_reconstruction}) depend on the implemented summing-junction convention and must be verified by injection measurements. If the archived $d_{\rm ctrl}$ is defined at a different point in the digital chain, the corresponding filters and delays must be included in $D(f)$.

Let $\hat\Omega$ be the source-direction unit vector and $\psi$ the
polarization angle. The angular frequency $\Omega=2\pi f$ and sky direction
$\hat\Omega$ are distinct quantities. Let ${\cal H}_{\theta h}(f,\hat\Omega,\psi)$ be the modeled transfer function from incident strain to the torsion-bar angular displacement, including the antenna response, mechanical susceptibility, and any geometrical factors. The equivalent strain is then
\begin{equation}
	\tilde h(f)=\frac{\tilde \theta_{\rm ext}(f)}{{\cal H}_{\theta h}(f,\hat\Omega,\psi)}.
	\label{eq:strain_reconstruction}
\end{equation}
For searches over unknown sky direction $\hat\Omega$ and polarization angle $\psi$, the directional response is applied to the waveform model rather than divided out of the data. A direction-independent strain channel may instead use a documented reference response.

The functions $C$, $A$, and $D$ are measured or verified using swept-sine and broadband injections. Absolute scale and phase are checked with independent photon-calibrator and/or gravitational-calibrator injections where applicable \cite{Karki2016Pcal,Viets2018Calibration,kagra_pcal}. Calibration lines track slow temporal variation between full measurements. Injection frequencies, amplitudes, geometry, timing delay, fit model, validity band, and residuals must be reported. Let $\delta A(f,t)$ be the fractional amplitude error (not the actuator
function $A$), $\delta\phi(f,t)$ the phase error in radians, and $t$ the
calibration epoch. The superscripts true and model identify the physical
and modeled responses. The calibrated response is written as
\begin{equation}
	R_\theta^{\rm true}(f,t)=R_\theta^{\rm model}(f,t)
	\left[1+\delta A(f,t)\right]e^{i\delta\phi(f,t)},
	\label{eq:calibration_uncertainty}
\end{equation}
with correlated amplitude and phase uncertainties propagated by nuisance-parameter marginalization or calibrated-response realizations. Statistical fit errors, reference-standard uncertainties, geometrical uncertainties, timing uncertainty, and model discrepancy should be identified separately.

\section{Compact-binary-coalescence analysis}

\subsection{Detector response and data model}

For a source at direction $\hat\Omega$, the incident strain tensor contains the two tensor polarizations $h_+$ and $h_\times$. Let $F_+$ and $F_\times$ be the dimensionless antenna responses to these
polarizations, and $h_{\rm obs}$ the projected strain. In one channel, the signal model is
\begin{equation}
	\tilde h_{\rm obs}(f)=F_+(f,\hat\Omega,\psi)\tilde h_+(f)
	+F_\times(f,\hat\Omega,\psi)\tilde h_\times(f),
	\label{eq:h_obs_cbc}
\end{equation}
where the response may be frequency dependent. The analysis-ready strain data are
\begin{equation}
	\tilde s(f)=\tilde h_{\rm obs}(f;\boldsymbol\vartheta)+\tilde n(f),
	\label{eq:detector_output_cbc}
\end{equation}
with source parameters $\boldsymbol\vartheta$ and calibrated noise $n$. Equation~(\ref{eq:detector_output_cbc}) applies only after the reconstruction of Sec.~\ref{sec:calibration_reconstruction}; the raw detector output is $d_{\rm err}$, accompanied by $d_{\rm ctrl}$, rather than $s$.

The model assumes linear detector response over the analyzed amplitude range, adequate waveform accuracy in the selected frequency band, and approximate stationarity within each segment. These assumptions are tested with hardware/software injections, residual tests, and analyses repeated over different segment lengths and data selections.

\subsection{Matched filtering and detection statistic}

For real time series $a$ and $b$, using a one-sided noise PSD $S_n(f)$,
let $f_{\min}$ and $f_{\max}$ be the analysis-band edges, $*$ complex
conjugation, and $\mathrm{Re}$ the real part. Define
\begin{equation}
	(a|b)=4\,\mathrm{Re}\int_{f_{\min}}^{f_{\max}}
	\frac{\tilde a^*(f)\tilde b(f)}{S_n(f)}\,df.
	\label{eq:inner_product}
\end{equation}
The complex, time-dependent matched-filter output is used in an actual search; for a fixed, correctly phased template its normalized value is
\begin{equation}
	\rho=\frac{(s|h)}{\sqrt{(h|h)}},
	\qquad
	\rho_{\rm opt}=\sqrt{(h|h)}.
	\label{eq:snr_def}
\end{equation}
Let $\mathcal M_c=(m_1m_2)^{3/5}/(m_1+m_2)^{1/5}$ be the detector-frame
chirp mass for component masses $m_1,m_2$, and $D_L$ the luminosity distance.
For the leading-order stationary-phase inspiral amplitude $|\tilde h(f)|\propto {\cal M}_c^{5/6}f^{-7/6}/D_L$,
\begin{equation}
	\rho_{\rm opt}^2=4\int_{f_{\min}}^{f_{\max}}
	\frac{|\tilde h(f)|^2}{S_n(f)}\,df
	\propto\int_{f_{\min}}^{f_{\max}}\frac{f^{-7/3}}{S_n(f)}\,df.
	\label{eq:snr_scaling}
\end{equation}
The $f^{-7/3}$ numerator does not by itself imply low-frequency dominance: the contribution per frequency interval is set by the full ratio $f^{-7/3}/S_n(f)$, the observation duration, and the valid waveform and calibration bands.

Let $t_c$ be the coalescence time and $t(f)$ the time at which the inspiral
reaches GW frequency $f$. Restoring the gravitational constant $G$ and
speed of light $c$, the leading-order time remaining is
\begin{equation}
	t_c-t(f)=\frac{5}{256}(\pi f)^{-8/3}
	\left(\frac{G{\cal M}_c}{c^3}\right)^{-5/3}.
	\label{eq:time_to_coalescence}
\end{equation}
The lower cutoff is therefore determined by both instrumental sensitivity and the finite observing time. Claims of months-to-years warning must be evaluated for specified component masses, start frequency, duty factor, accumulated SNR, and localization performance; long phase tracking alone does not guarantee a useful alert.

Real data contain glitches, lines, and slowly varying spectra. The PSD used in Eq.~(\ref{eq:inner_product}) is estimated from neighboring data without biasing the on-source statistic. Candidates are ranked using signal-consistency tests and coincidence between genuinely independent instruments or channels where available. Detection significance is obtained empirically from an off-source or time-shift background and reported as a false-alarm rate or false-alarm probability \cite{Usman2016PyCBC}. The template-bank range, waveform approximant, minimal match, sampling rate, frequency cutoffs, gating procedure, veto categories, clustering rule, background livetime, and trials factor must be specified.

For parameter estimation, the likelihood corresponding to Eq.~(\ref{eq:inner_product}) is used only after verifying that residuals are compatible with the adopted noise model. Calibration amplitude and phase, PSD uncertainty, waveform uncertainty, and relevant environmental couplings are included as nuisance parameters. Low-frequency phase evolution can improve chirp-mass measurement and, together with antenna modulation or a detector network, sky localization. Luminosity distance remains correlated with inclination, polarization, sky position, and calibration amplitude, so improved distance measurement should be demonstrated rather than assumed \cite{Veitch2015ParameterEstimation}.

\subsection{CHRONOS validation for CBC searches}

End-to-end simulated-signal and physical-injection campaigns are required to establish detection efficiency and parameter recovery. Injections should cover masses, spins, orientations, distances, sky positions, signal durations, and SNRs relevant to the stated science case. They must pass through the same reconstruction, data-quality, search, and inference stages as observational data. Recovery bias and coverage are reported as functions of source parameters and detector state. Robustness checks include alternative PSD estimates, segment lengths, calibration realizations, waveform families, and veto selections. Search sensitivity is summarized by efficiency curves and sensitive spacetime volume (or an explicitly appropriate long-duration analogue), with uncertainty from finite injection statistics and calibration/systematic effects.

\section{Stochastic gravitational-wave-background analysis}

\subsection{Statistical model and conventions}

The baseline search assumes a tensor-polarized SGWB that is Gaussian, stationary, unpolarized, and statistically isotropic. These are analysis assumptions rather than necessary properties of every background; anisotropic, non-Gaussian, or non-standard-polarization searches require different response models. Let $\rho_{\rm GW}$ and $\rho_c$ denote GW and critical energy densities,
and $H_0$ the present Hubble rate. The dimensionless energy-density spectrum is
\begin{equation}
	\Omega_{\rm GW}(f)=\frac{1}{\rho_c}\frac{d\rho_{\rm GW}}{d\ln f},
	\qquad \rho_c=\frac{3H_0^2c^2}{8\pi G}.
	\label{eq:omega_gw_def}
\end{equation}
With the one-sided strain-PSD convention used here,
\begin{equation}
	S_h(f)=\frac{3H_0^2}{2\pi^2}\frac{\Omega_{\rm GW}(f)}{f^3}.
	\label{eq:ShOmega_relation}
\end{equation}
All reported limits must state the adopted $H_0$ and the precise PSD and overlap-reduction-function (ORF) normalization conventions.

For a finite segment of duration $T$, calibrated channels $A$ and $B$ obey
\begin{equation}
	\left\langle\tilde h_A^*(f)\tilde h_B(f')\right\rangle
	=\frac{1}{2}\delta_T(f-f')\gamma_{AB}(f)S_h(f),
	\label{eq:sgwb_corr}
\end{equation}
where $\delta_T$ is the finite-time delta function and $\gamma_{AB}$ includes the channel geometry and normalized strain response. A response factor must not also be included separately in the data model unless a different, explicitly defined ORF convention is used.

\subsection{Cross-correlation estimator}

Let $s_A=h_A+n_A$ and $s_B=h_B+n_B$ be calibrated strain-equivalent data. A real cross-correlation statistic can be written
\begin{equation}
	Y_{AB}=2\,\mathrm{Re}\int_0^\infty df\,
	\tilde s_A^*(f)\tilde s_B(f)Q_{AB}(f),
	\label{eq:Y_general}
\end{equation}
where $Q_{AB}$ is chosen consistently with the Fourier and PSD conventions. Assuming the signal is uncorrelated with instrumental noise,
\begin{equation}
	\langle Y_{AB}\rangle=T\,\mathrm{Re}\int_0^\infty df\,
	\gamma_{AB}(f)S_h(f)Q_{AB}(f).
	\label{eq:Y_expect}
\end{equation}
In the weak-signal limit, for stationary Gaussian noise with zero cross-power,
\begin{equation}
	\sigma_{Y_{AB}}^2\simeq\frac{T}{2}\int_0^\infty df\,
	P_A(f)P_B(f)|Q_{AB}(f)|^2,
	\label{eq:variance}
\end{equation}
where $P_A$ and $P_B$ are one-sided noise PSDs. Numerical prefactors change with convention, but expectation, variance, and implementation must use one convention throughout.

Let $\Omega_\alpha$ be the target amplitude and $H_\alpha(f)$ a
dimensionless spectral shape, with $\alpha$ labeling the chosen model.
For $\Omega_{\rm GW}(f)=\Omega_\alpha H_\alpha(f)$, the optimal filter is
\begin{equation}
	Q_{AB}(f)={\cal N}\,
	\frac{\gamma_{AB}^*(f)H_\alpha(f)}{f^3P_A(f)P_B(f)},
	\label{eq:optimal_filter}
\end{equation}
where ${\cal N}$ is fixed so that $Y_{AB}$ estimates $\Omega_\alpha$. The corresponding idealized sensitivity scales as
\begin{equation}
	\rho_{AB}^2=2T\left(\frac{3H_0^2}{2\pi^2}\right)^2
	\int_0^\infty df\,
	\frac{|\gamma_{AB}(f)|^2\Omega_{\rm GW}^2(f)}{f^6P_A(f)P_B(f)}.
	\label{eq:SNR_general}
\end{equation}
The explicit $f^{-6}$ factor does not guarantee low-frequency sensitivity because $P_A$, $P_B$, the ORF, usable bandwidth, and correlated-noise contamination may vary steeply with frequency.

Data are analyzed in finite, approximately stationary segments. Cross-spectra and auto-spectra are formed with a documented window and overlap, bins affected by coherent instrumental lines are removed or modeled, and segment estimates are combined with inverse-variance weights. The PSD estimates used for weighting should avoid covariance with the same cross-spectrum that is being measured. Results are checked under changes in segment duration, frequency resolution, data-quality selection, and spectral model. Upper limits or credible intervals must state the likelihood, prior, confidence/credibility convention, and treatment of calibration and correlated-noise uncertainty.

\subsection{Independent detectors and internal channels}

For separated detectors, the ORF is determined by their relative positions, orientations, and antenna patterns \cite{allen1999,RomanoCornish2017}. Their noises are not automatically independent: globally correlated magnetic fields, timing errors, common services, or environmental fields can bias the estimator. Time shifts, environmental witnesses, coherence studies, and null tests are therefore required even for geographically separated instruments.

For two internal CHRONOS channels, write the more general cross-spectrum model
\begin{equation}
	\left\langle\tilde s_a^*(f)\tilde s_b(f')\right\rangle
	=\frac{1}{2}\delta_T(f-f')
	\left[\gamma_{ab}^{\rm(int)}(f)S_h(f)+N_{ab}(f)\right],
	\label{eq:internal_cross_spectrum}
\end{equation}
where $N_{ab}$ is the correlated instrumental/environmental cross-PSD. Internal channels share the site, platform, optics, controls, power, and environment; consequently $N_{ab}=0$ cannot be assumed merely because the readouts are nominally distinct. Moreover, channels derived from the same physical degree of freedom or coupled by feedback may share sensing and control noise. Internal correlation is scientifically useful only if the channel responses are sufficiently linearly independent and $N_{ab}$ is bounded below the target SGWB cross-spectrum or included in a validated joint model.

Environmental witness channels $w_k$ are used to measure transfer functions and coherence with the science channels. Subtraction or marginalization must account for witness noise and avoid removing a true gravitational-wave component. Safe-injection tests, sensor-placement studies, and independent null channels are used to demonstrate this. Newtonian noise is especially important at sub-Hz frequencies because it is a real gravitational coupling and cannot be shielded; seismic and atmospheric arrays are required to constrain its spatial correlations \cite{Harms2015}.

\subsection{Calibration and systematic uncertainty}

If the calibrated response in channel $i$ has fractional complex error $\epsilon_i(f)$, the measured cross-spectrum is multiplied, to first order, by $1+\epsilon_A^*(f)+\epsilon_B(f)$. Thus calibration can alter both the normalization and spectral shape of the inferred $\Omega_{\rm GW}$, and a relative timing/phase error can reduce or rotate the correlated signal. Although independent calibration errors do not normally create an astrophysical ORF-shaped correlation from zero, calibration lines, shared actuators, digital cross-couplings, or common references can inject correlated features. Such frequencies must be modeled or excluded.

The systematic budget should include at least response amplitude and phase, timing, ORF geometry, spectral leakage, PSD estimation, non-stationarity, data-selection bias, residual lines, correlated environmental noise, and uncertainty in any subtraction transfer function. Frequency-dependent nuisance models are propagated to the final estimator; quoting only a single calibration percentage is insufficient unless its validity has been demonstrated across the full analysis band.

\subsection{CHRONOS validation and reproducibility}

The CHRONOS SGWB pipeline is validated using simulated Gaussian and non-Gaussian data, software injections before and after calibration, and physical injections where the actuator geometry permits. Tests must demonstrate unbiased recovery of several spectral shapes, correct statistical coverage, the predicted $T^{1/2}$ scaling only over intervals where stationarity and independence hold, and robustness against realistic correlated disturbances. Null tests include time shifts exceeding the inter-channel signal coherence time where meaningful, sign- or orientation-changing combinations, coherence with witnesses, analysis of null channels, and comparisons among epochs with different detector states.

For every CBC and SGWB result, the following are archived: raw data identifiers and checksums; channel definitions and units; timing standard; $d_{\rm err}$ and $d_{\rm ctrl}$ locations in the loop; response-function version and validity interval; calibration measurements and uncertainty samples; data-quality masks; environmental channels; analysis configuration; software commit and environment; random seeds; intermediate spectra; and injection lists. Blinded hardware or software injections should be used when feasible, with the unblinding criteria fixed in advance. These records provide traceability from the published result to the detector state and raw measurements.

The achievable CBC reach and SGWB sensitivity of CHRONOS must ultimately be computed with measured or design-specific $P_i(f)$, response functions, duty cycle, channel geometry, and validated correlated-noise model. The formal low-frequency weighting in Eqs.~(\ref{eq:snr_scaling}) and (\ref{eq:SNR_general}) is not, by itself, an experimental sensitivity forecast.

%% file: Chapters/Calibration.tex
\chapter{Calibration and Reconstruction}
\label{chap:calibration}

\section{Gravitational Wave Strain}

The dimensionless strain $h(t)$ is reconstructed from the interferometer
output; $t$ denotes time. We use frequency $f$ in hertz and angular
frequency $\Omega=2\pi f$. A tilde denotes a Fourier-domain signal or
transfer function, with the $e^{+i\Omega t}$ time convention.
Slow dependence of a transfer function on $t$ labels its calibration epoch.
Calibration errors propagate into inferred source distance, mass, and
stochastic-background energy density
\cite{abbott2009ligo,vitale2012reconstruction,sun2020characterization}.

CHRONOS measures the rotations of two orthogonal bars. Let
$\theta_{X,\rm ext}$ and $\theta_{Y,\rm ext}$ be their signal angles with
external electronic feedback removed. The differential external angle is
\begin{equation}
 \Delta\theta_{\rm ext}(t)
 =\theta_{X,\rm ext}(t)-\theta_{Y,\rm ext}(t).
 \label{eq:delta_theta}
\end{equation}
With electronic feedback engaged, let $\Delta\theta_{\rm ctrl}$ denote the
angle canceled by the actuator and $\Delta\theta_{\rm res}$ the residual
physical angle seen by the readout. The negative-feedback convention is
$\Delta\theta_{\rm res}=\Delta\theta_{\rm ext}-\Delta\theta_{\rm ctrl}$.
Reconstruction adds the inferred residual and controlled angles.

\section{Sensing Function and control response}
\label{sec:DARM_CHRONOS}

The \emph{Sensing Function} $\tilde C_\theta(\Omega,t)$ converts the
residual differential angle into the digitized error signal
$\tilde d_{\rm err}$, in error-channel counts per radian. It includes the
optical response for that angular coordinate, photodetection, readout
electronics, digitization, and their delays. Its magnitude gives the
conversion gain and its phase gives the frequency-dependent phase shift.
The subscript $\theta$ identifies the differential angular channel.

Let $\tilde d_{\rm ctrl}$ be the digital signal sent to the actuator.
The digital filter $\tilde D_\theta$ converts error-channel counts to
control-channel counts, and the actuation function $\tilde A_\theta$
converts control-channel counts to differential angle in radians. The
actuation function includes the actuator electronics and mechanical plant
for the same angular coordinate used in $\tilde C_\theta$. Neglecting
additive noise only when defining these responses,
\begin{align}
 \tilde d_{\rm err}&=\tilde C_\theta\Delta\tilde\theta_{\rm res},\\
 \tilde d_{\rm ctrl}&=\tilde D_\theta\tilde d_{\rm err},\\
 \Delta\tilde\theta_{\rm ctrl}&=\tilde A_\theta\tilde d_{\rm ctrl}.
\end{align}
Frequency and epoch arguments are suppressed in these three relations.
The dimensionless open-loop gain is
$\tilde G_\theta=\tilde C_\theta\tilde D_\theta\tilde A_\theta$.
The angular reconstruction response $\tilde R_\theta$, in radians per
error-channel count, is therefore
\begin{equation}
 \tilde R_\theta(\Omega,t)
 =\frac{1+\tilde G_\theta(\Omega,t)}{\tilde C_\theta(\Omega,t)}
 =\frac{1}{\tilde C_\theta(\Omega,t)}
  +\tilde A_\theta(\Omega,t)\tilde D_\theta(\Omega).
 \label{eq:Rtheta_full}
\end{equation}
These definitions follow the same convention as
Sec.~\ref{sec:calibration_reconstruction}
\cite{abbott2009ligo,abadie2010calibration,Viets2018Calibration}.

The optical Sensing Function $H_\theta$ in
Sec.~\ref{sec:sensitivity_final_expression} has a power output [W/rad],
whereas $\tilde C_\theta$ has a digital output [counts/rad]. Their angular
inputs must also be matched: $H_\theta$ includes radiation-pressure feedback
on the externally imposed signal angle. In the symmetric reference model,
let $H_{\rm opt}$ be the optical response to a prescribed physical angle
and $\mathcal B_\theta$ the dimensionless radiation-pressure feedback factor;
then $H_\theta=H_{\rm opt}\mathcal B_\theta$
\cite{InoueQuantumTorque2026}. If $K_{\rm read}$ is the electronics and
digitization response in counts/W, the physical-residual-angle convention
used here gives $\tilde C_\theta=K_{\rm read}H_{\rm opt}$.
The corresponding mechanical plant in $\tilde A_\theta$ must include the
same radiation-pressure dynamics. A model using an equivalent free-angle
coordinate may instead assign that factor to its sensing function, but
must transform its actuation and strain responses consistently.

\begin{figure}[htbp]
\centering
\includegraphics[scale=0.5]{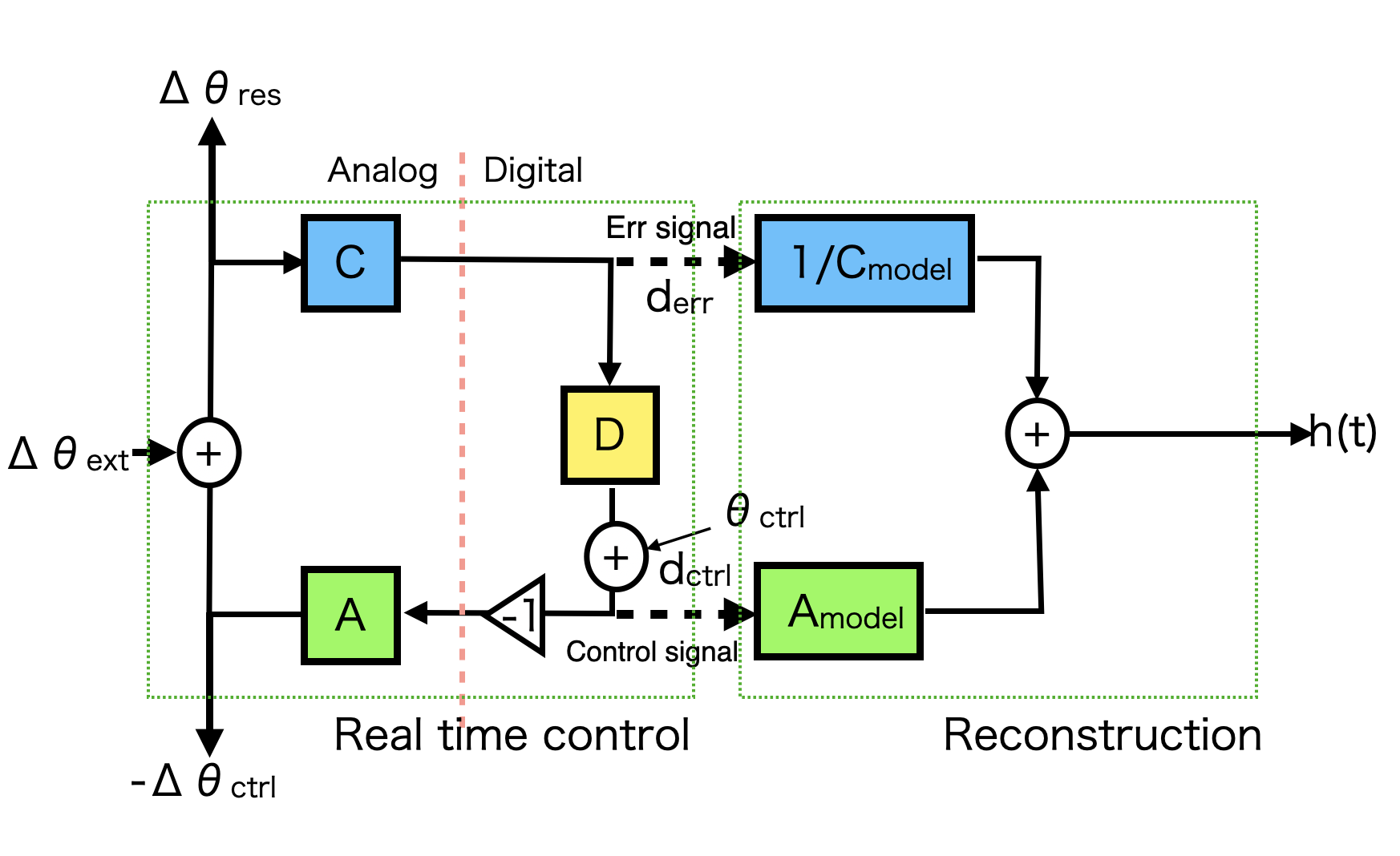}
\caption{Feedback and reconstruction schematic. $C$, $D$, and $A$ denote
 the sensing, digital-filter, and actuation functions defined in the text;
 the ``model'' subscript identifies their calibrated estimates. The right
 summing junction reconstructs external differential angle. Producing the
 strain output labeled $h(t)$ additionally requires the angle-to-strain
 response of Eq.~\eqref{eq:calibration_angle_to_strain}. Any injected
 control signal must be expressed in counts at its injection point.}
\label{fig:CAL}
\end{figure}

The reconstructed angle combines both recorded channels:
\begin{equation}
 \Delta\tilde\theta_{\rm ext}
 =\tilde C_\theta^{-1}\tilde d_{\rm err}
  +\tilde A_\theta\tilde d_{\rm ctrl}
 =\tilde R_\theta\tilde d_{\rm err}.
 \label{eq:freq_delta_theta}
\end{equation}
The last equality assumes $\tilde d_{\rm ctrl}=\tilde D_\theta\tilde d_{\rm err}$;
explicit control injections are retained separately when present.
For a fixed calibration epoch, let $c_{\rm inv}(t)$ and $a_\theta(t)$
be the inverse Fourier transforms of $\tilde C_\theta^{-1}$ and
$\tilde A_\theta$. Their time-domain action is convolution:
\begin{equation}
 \Delta\theta_{\rm ext}(t)
 =(c_{\rm inv}*d_{\rm err})(t)+(a_\theta*d_{\rm ctrl})(t),
 \quad (u*v)(t)=\int_{-\infty}^{\infty}u(t-t')v(t')\,dt'.
 \label{eq:time_delta_theta}
\end{equation}
Here $u$ and $v$ denote generic functions and $t'$ is the integration time.

Let $\mathcal H_{\theta h}$ be the signal-angle response per unit incident
strain, including the chosen antenna and mechanical response, and define
its inverse $\mathcal K_{\theta\rightarrow h}=\mathcal H_{\theta h}^{-1}$.
Then
\begin{equation}
 \tilde h=\mathcal K_{\theta\rightarrow h}\Delta\tilde\theta_{\rm ext}
 =\mathcal K_{\theta\rightarrow h}\tilde R_\theta\tilde d_{\rm err}.
 \label{eq:calibration_angle_to_strain}
\end{equation}
Only in the antenna-projected free-rotation limit with negligible
radiation-pressure modification does the physical-angle response reduce
to $\mathcal H_{\theta h}=\eta$, with $\eta=0.93656$ as defined in
Chap.~\ref{chap:sensitivity}. Elsewhere the full complex response is needed.
The response coordinate, feedback sign, gain, phase, and timing must be
verified using calibrated injections over the observation band.

\section{Calibration Devices}
\label{chap:pcal}

\begin{figure}[tbp]
\centering
\includegraphics[scale=0.5]{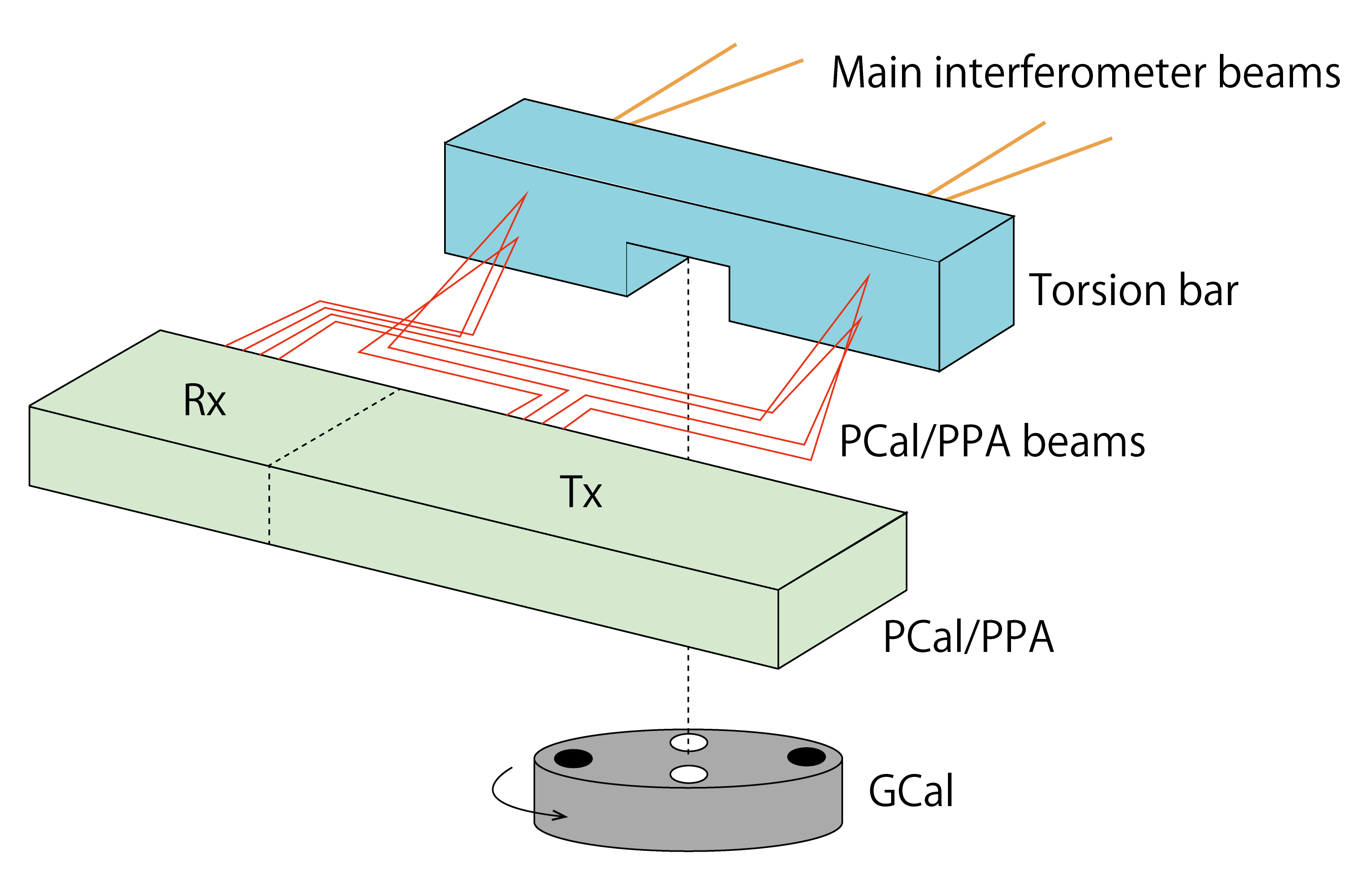}
\caption{Schematic diagram of the combination of PCal and GCal. Only one torsion bar is shown.}
\label{fig:PCal_GCal}
\end{figure}

We need calibration devices which can directly apply force to the test mass to determine the actuation function independently from the sensing function. Conventional GW experiments have developed two types of calibrator; Photon calibrator (PCal) and gravity field calibrator (Gcal). The PCal is a system to actuate a test mass by photon pressure of a calibrated laser, while GCal apply force to the test mass through an artificial gravity gradient generated by mechanical rotors. We deploy these two systems for calibration of different frequency ranges; GCal for 0.1-10~Hz and PCal for above 10~Hz. Figure~\ref{fig:PCal_GCal} shows the configuration of combining PCal and GCal.

\subsection{Four-beams Photon Calibrator}
\label{subsec:pcal}

The PCal is a promising calibration method that has been deployed in the conventional gravitational-wave detectors~\cite{Karki2016Pcal,virgo_pcal,kagra_pcal}. It applies photon force to the test mass by injecting the power-controlled laser beams for generating an artificial harmonic displacement. As a result, it enables us to calibrate the displacement of the test mass against the known force. 

When a beam of power $P$ is injected to the test mass in the incident angle $\theta$, the force applied in the translational direction is $2P\cos\theta/c$ using the speed of light $c$. The latest design of PCal system, which has been deployed by KAGRA, controls power of two beams independently in the transmitter (Tx) module and injects them on the nodal circle of the test mass to suppress excitation of its internal resonant mode~\cite{kagra_pcal,Karki2016Pcal}. Power and position of the reflected beams are monitored in the receiver (Rx) module.

For CHRONOS, we extend the PCal system to four beams. These beams are injected to the back surface of the test mass to avoid interfering the main beam path. This enables us to control degrees of freedom of translation, yaw, and pitch independently. Having these degrees of freedom in control has several advantages; control of inclination of the beams in the triangle cavity, torque-based measurement of the position offset of the main interferometer beam, and application to Photon-pressure actuator for locking the interferometer. 

We utilize the four-beams system for both calibration and interferometer lock. Most of the laser power is assigned to the actuation for locking the cavity, and the rest is to the line injection as a PCal~\cite{tanabe_photon_pressure_actuator}. We call this device as Photon-pressure actuator (PPA) and describe in detail in Chap.~\ref{sec:photon_pressure_actuator}.


\subsection{Torque-coupled Gravitational Calibrator}
\label{subsec:gcal}

The GCAL is a relatively new type of calibrator. Its core component is a rotating disk with distributed masses. Rotation of masses generates periodic gravity gradient which propagates to the test masses. This concept has been independently studied in Virgo and KAGRA~\cite{virgo_gcal,Inoue2018_GCal}. One of the advantages of GCal is that it is free from the intensity fluctuation of laser which constrains low-frequency actuation noise of PCal. It makes GCal a promising tool to calibrate the detector response below 10~Hz.

In contrast to the NCal and GCal placed in front of the test mass in Virgo and KAGRA, the GCal for CHRONOS should be placed below the torsion bar to optimize for applying torque~\cite{chronos_gcal}. See Chap.~\ref{chap:gcal} for detailed description of GCal specified to CHRONOS.

\section{Reconstruction}
\label{sec:reconstruction_pipelines}

This section describes the strain-reconstruction pipelines for CHRONOS.
In contrast to kilometer-scale Michelson interferometers where DARM is a
differential arm length, the primary gravitational-wave readout of CHRONOS is the
\textbf{difference bar angle} (DARM$_\theta$),
\begin{equation}
\Delta \theta(t) \equiv \theta_X(t)-\theta_Y(t),
\end{equation}
where $\theta_X$ and $\theta_Y$ are the two bar angles. The external
part of this differential angle is reconstructed from the digital error signal and auxiliary control signals,
following the standard control-model decomposition into sensing and actuation paths
\cite{abbott2009ligo,abadie2010calibration,vitale2012reconstruction}.
The present implementation is adapted to the torsional (angular) degree of freedom
and to the speed-meter response characteristic of CHRONOS.

We develop two pipelines for reconstruction:
(i) a \textbf{front-end (online)} pipeline for low-latency monitoring during operation, and
(ii) a \textbf{high-latency (offline)} pipeline for the final calibrated strain product.

\subsection{Front-end (online) pipeline}
\label{subsec:frontend_pipeline}

The front-end pipeline runs in real time and provides a continuously updated monitor signal
$h_{\rm mon}(t)$ with minimal latency.
In this stage, the interferometer control-loop signals are partially calibrated online using
a fixed set of infinite-impulse-response (IIR) filters that approximate the nominal sensing and actuation models,
as commonly implemented in low-latency calibration systems
\cite{sun2020characterization}.

\subsection{High-latency (offline) pipeline}
\label{subsec:highlatency_pipeline}

The high-latency pipeline produces the final calibrated strain $h(t)$ using offline raw data
and a more accurate reconstruction model.
This pipeline adopts finite-impulse-response (FIR) filtering applied directly to the recorded error and control signals,
consistent with standard offline calibration frameworks in ground-based detectors
\cite{vitale2012reconstruction,sun2020characterization}.

The time dependence of the calibration parameters is incorporated using
time-dependent correction factors (TDCFs) derived from swept-sine measurements and/or
continuous calibration lines (e.g., photon calibrator, torque-coupled GCal, and reference injections).

In the frequency domain, the angular response function is written as
\begin{equation}
\tilde{R}_{\theta}(\Omega,t)
=
\frac{1}{\tilde{C}_{\theta}(\Omega,t)}
+
\tilde{D}_{\theta}(\Omega)\tilde{A}_{\theta}(\Omega,t),
\end{equation}
and the reconstructed differential angle is
\begin{equation}
\Delta\tilde{\theta}(\Omega)
=
\tilde{R}_\theta(\Omega,t)\,\tilde{d}_{\rm err}(\Omega).
\end{equation}

In practice, the offline pipeline implements as a set of FIR filters for (i) inverse sensing and (ii) actuation reconstruction,
together with slowly varying TDCFs (time-dependent correction factors)
that track drifts in optical gain, effective pole frequency,
actuation efficiency, and timing.

\section{Error estimation}
\label{sec:error_estimation}

Accurate estimation of the response function uncertainty is essential,
because the strain reconstruction process involves both control-loop inversion
and model-dependent angular-to-strain conversion
\cite{sun2020characterization,vitale2012reconstruction}.
In CHRONOS, the primary observable is the differential bar angle
$\Delta\theta$, and the strain is reconstructed through the angular
response function $\tilde{R}_\theta(\Omega,t)$ and the
angle-to-strain conversion factor $\mathcal{K}_{\theta\rightarrow h}$.

\subsection{Residual response function}

The detector response function
$\tilde{R}_\theta(\Omega,t)$
is constructed from the sensing and actuation models
introduced in the previous sections.
In practice, however, the modeled response is not perfectly
accurate because of uncertainties in optical parameters,
control filters, and actuator calibration.
Therefore, it is necessary to compare the model prediction
with direct calibration measurements.

We quantify this discrepancy by defining the
fractional residual response

\begin{equation}
\frac{\delta \tilde{R}_\theta(\Omega)}
{\tilde{R}_{\theta,{\rm model}}(\Omega)}
=
\frac{
\tilde{R}_{\theta,{\rm meas}}(\Omega)
-
\tilde{R}_{\theta,{\rm model}}(\Omega)
}
{\tilde{R}_{\theta,{\rm model}}(\Omega)} .
\label{eq:residual_response}
\end{equation}

Here,
$\tilde{R}_{\theta,{\rm model}}(\Omega)$
is the response predicted from the calibrated sensing
and actuation models,
while
$\tilde{R}_{\theta,{\rm meas}}(\Omega)$
is obtained from calibration measurements such as
swept-sine injections or continuous calibration lines
using photon calibrators or torque-coupled GCal actuators
\cite{sun2020characterization,Karki2016Pcal}.

Equation~\eqref{eq:calibration_angle_to_strain} multiplies the error signal
by the reconstruction response $\tilde R_\theta$. Define
$\delta\tilde R_\theta=\tilde R_{\theta,\rm meas}-\tilde R_{\theta,\rm model}$
and the corresponding strain difference
$\delta\tilde h=\tilde h_{\rm meas}-\tilde h_{\rm model}$ at fixed error signal.
When the angle-to-strain factor is held fixed, their fractional differences
relative to the modeled quantities satisfy

\begin{equation}
\frac{\delta \tilde{R}_\theta}{\tilde{R}_{\theta,\rm model}}
=
\frac{\delta \tilde{h}}{\tilde{h}_{\rm model}} .
\label{eq:strain_error_relation}
\end{equation}

Equation~(\ref{eq:strain_error_relation}) indicates that
both amplitude and phase uncertainties in the response
function translate directly into calibration uncertainty
in the reconstructed strain. Uncertainty in $\mathcal K_{\theta\rightarrow h}$
is an additional contribution and must be propagated with its correlations.
Therefore, accurate measurement and continuous monitoring
of $\tilde{R}_\theta(\Omega,t)$ across the observation band
are essential to control systematic biases in
astrophysical parameter estimation.

%% file: Chapters/Summary.tex
\chapter{Summary}

The CHRONOS initiative --- the Cryogenic sub-Hz cROss torsion-bar detector with quantum NOn-demolition Speed-meter --- aims to establish a new experimental platform for precision measurement in the sub-Hz frequency regime. By combining cryogenic torsion-bar mechanics, precision laser interferometry, quantum measurement techniques, and advanced calibration methods, CHRONOS is designed to explore a frequency band that remains largely inaccessible to existing ground-based gravitational-wave detectors.

This White Paper has presented the scientific and technical program of CHRONOS as an integrated roadmap from the present prototype toward a future cryogenic sub-Hz observatory. The program is driven by several closely connected scientific objectives: the development of new techniques for gravitational-wave detection below the conventional seismic wall, searches for intermediate-mass black-hole binaries and stochastic gravitational-wave backgrounds, precision measurements of time-dependent gravitational fields, tests of gravity at laboratory scales, searches for dark matter, and experimental studies of quantum measurement at macroscopic scales.

A central objective of the present CHRONOS program is to establish the complete measurement chain required for a gravitational-wave detector. This includes the stabilization and control of the torsion-bar system, operation of the interferometer, characterization of the relevant mechanical and optical degrees of freedom, calibration of the detector response, and reconstruction of the physical strain channel \(h(t)\). These developments provide the experimental foundation upon which more advanced cryogenic and quantum measurement techniques can subsequently be implemented.

The quantum measurement program constitutes another defining component of CHRONOS. The triangular Sagnac interferometer provides a natural framework for realizing a speed-meter configuration and for studying measurement schemes in which quantum back-action can be controlled or correlated with measurement noise. The program therefore progresses from the demonstration and characterization of the speed-meter response toward quantum non-demolition measurements and, ultimately, toward experimental investigations of generalized quantum measurement relations. CHRONOS is thus intended not only as a gravitational-wave detector technology demonstrator, but also as a laboratory for macroscopic quantum measurement.

The development strategy is organized as a sequence of experimentally achievable stages. The current prototype focuses on establishing interferometer operation, multidimensional control, detector calibration, and stable reconstruction of the science channels. Subsequent phases will extend the instrument toward measurements of gravitational-field variations and gravity-gradient phenomena, while introducing improved isolation, calibration systems, cryogenic operation, and quantum measurement techniques. In the longer term, these developments provide a pathway toward a detector with genuine astrophysical sensitivity in the sub-Hz gravitational-wave band.

The scientific goals and detector technologies described throughout this White Paper are therefore not independent activities, but components of a single experimental program. Mechanical design, suspension and cryogenic technologies, interferometer topology, optical configuration, sensing and control, calibration, noise characterization, and data analysis must ultimately converge in the calibrated detector output. This system-level approach provides a direct connection between subsystem development and the scientific measurements that CHRONOS seeks to realize.

CHRONOS is also being developed as an international collaborative effort. The collaboration framework provides a structure through which detector development, theoretical studies, data analysis, and scientific interpretation can evolve together while allowing new institutes and research groups to contribute to clearly defined activities. Regular collaboration meetings and international workshops provide opportunities to review experimental progress, refine scientific priorities, and coordinate the long-term development of the project.

The roadmap described in this White Paper should therefore be regarded as an evolving experimental strategy rather than a fixed detector design. Results obtained during prototype operation will determine the priorities and technical choices of subsequent phases, while new scientific opportunities may expand the scope of the program. Maintaining this connection between experimental progress and scientific motivation will be essential as CHRONOS develops.

Through this staged approach, CHRONOS seeks to establish the technologies and measurement principles required for precision interferometry in the sub-Hz regime. The immediate objective is to demonstrate a controlled and calibrated torsion-bar interferometer capable of producing physically meaningful science channels. The longer-term objective is substantially broader: to develop a cryogenic quantum measurement platform capable of exploring gravitational waves, gravitational physics, dark matter, and macroscopic quantum phenomena in a frequency regime that remains largely unexplored.

CHRONOS therefore represents both an instrument-development program and a long-term scientific initiative. By progressively connecting precision mechanics, cryogenic instrumentation, interferometric sensing, quantum measurement, calibration, and astrophysical observation, the collaboration aims to open a new experimental frontier in sub-Hz gravitational physics.